\documentclass[a4paper,fleqn,11pt]{memoir}
\usepackage{dsfont}
\usepackage{comment}

\usepackage{RobeUtiliMem}

\usepackage{definitions}

\usepackage{rotating}

\makeindex

\begin{document}

\copertina{String Theory: exact solutions, marginal deformations and hyperbolic spaces}{Domenico Orlando}

\pagestyle{companion}
\chapterstyle{companion}
\frontmatter

\dedica{A Lei}

\newpage

\thispagestyle{empty}

.\\

\newpage

\tableofcontents*

\pagestyle{empty}

\chapter*{Acknowledgements}
\label{cha:acknowledgements}

\begin{epigraphs}
  \qitem{Never had nay man a wit so presently excellent, as that it
    could raise itselfe; but there must come both matter, occasion,
    commenders, and favourers to it.  }{Il~Volpone\\ \textsc{Ben
      Johnson}}
\end{epigraphs}

While writing these acknowledgements I realized how much I'm indebted
to so many people. I learned a lot from them all -- not just physics
-- and without them this work simply wouldn't have been possible. I'm
pretty sure I'm forgetting someone. I plead guilty. But even if you
don't find your name rest assured that you'll always have my personal
gratitude.

 In any case, I simply can't start without mentioning my
parents for -- although far -- they've always been present in their
peculiar way.

I must thank the members of my thesis committee: \emph{Luis
  Alvarez-Gaum\'e}, \emph{Edouard Brezin}, \emph{Costas Kounnas},
\emph{Marios Petropoulos}, \emph{Alex Sevrin}, \emph{Philippe
  Spindel}, \emph{Arkady Tseytlin} for they extremely precious
feedback. They turned my defence into a valuable learning experience.

I wish to thank \emph{Patrick Mora} for having welcomed me as a PhD
student at the Centre de physique th\'eorique of the \'Ecole
polytechnique. Also I thank \emph{Edouard Brezin} at \'Ecole normale
sup\`erieure, \emph{Dieter L\"ust} at Max Planck Institut (Munich) and
\emph{Alex Sevrin} at Vrije Universiteit Brussel. During the last year
they hosted me for several months at their respective
institutions. Each and all of them have made of my stays invaluable
experiences from both the professional and human points of view.

Then I thank my advisor \emph{Marios Petropoulos} for all what he
taught me and for his continuous support. In these years I learned to
think of him as a friend. \emph{Costas Kounnas} for letting me camp in
his office at \textsc{ens} where he would devote me hours and hours of
his time, sharing his experience and his amazing physical
intuition. \emph{Costas Bachas}, who was among the first professors I
had when I arrived in Paris and (unknowingly) pushed me towards string
theory. Since then he helped me in many an occasion and I'm almost
embarrassed if I think how much I learned from him and in general I
owe him.

Doing one's PhD in Paris is a great opportunity for there are not so
many places in the world with such a physics community. And it's
definitely surprising to a PhD student, like I was, how everybody is
always willing to discuss and makes you feel one of
them. I learned a lot from many people and I hope that those I don't
remember explicitly won't take it too bad. On the other hand I
couldn't not mention \emph{Adel Bilal}, \emph{Eug\`ene Cremmer},
\emph{Pierre Fayet}, \emph{Mariana Gra\~na}, \emph{Jean Iliopoulos},
\emph{Bernard Julia}, \emph{Volodia Kazakov}, \emph{Ruben Minassian},
\emph{Herv\'e Partouche}, \emph{Michela Petrini}, \emph{Boris
  Pioline}, \emph{Ian Troost}, \emph{Pierre Vanhove}.

Sometimes discussions would lead to a publication and I'm honored to
count among my collaborators \emph{Costas Bachas}, \emph{Ioannis
  Bakas}, \emph{Davide Cassani}, \emph{St\'ephane Detournay},
\emph{Dan Isra\"el}, \emph{Marios Petropoulos}, \emph{Susanne
  Reffert}, \emph{Kostas Sfetsos}, \emph{Philippe Spindel}. I really
wish to thank you for what I learned and keep learning from you all.

I must remember my office mates \emph{Luciano Abreu}, \emph{St\'ephane
  Afchain}, \emph{Yacine Dolivet}, \emph{St\'ephane Fidanza},
\emph{Pascal Grange}, \emph{Claudia Isola}, \emph{Liguori Jego},
\emph{Jerome Levie}, \emph{Xavier Lacroze}, \emph{Alexey Lokhov},
\emph{Liuba Mazzanti}, \emph{Herv\'e Moutarde}, \emph{Chlo\'e
  Papineau}, \emph{Susanne Reffert}, \emph{Sylvain Ribault},
\emph{Barbara Tumpach} and especially \emph{Claude de Calan}. With all
of you I shared more than an office and our never-ending discussions
have really been among the most precious moments of my PhD. I hope
none of you will get mad at me if in particular I thank \emph{Sylvain} for
all the advice he gave me, even before starting my PhD. Not to speak
of my brother \emph{Pascal}.

A final thanks (last but not least) goes to the computer people and
secretaries at \textsc{cpht} and \textsc{ens}, without whom life
couldn't be that easy.

\bigskip

Thanks, all of you.


\pagestyle{companion}

\mainmatter


\newdimen\oldindent
\oldindent=\mathindent

\chapter{Introduction}

\setlength{\epigraphwidth}{17em}
\begin{epigraphs}
  \qitem{There is a theory which states that if ever anyone discovers
    exactly what the Universe is for and why it is here, it will
    instantly disappear and be replaced by something even more bizarre
    and
    inexplicable}{The~Restaurant at~the~End of~the~Universe\\ \textsc{Douglas
      Adams}}
\end{epigraphs}

\lettrine{I}{n the last fifty years} theoretical physics has been
dominated by two apparently incompatible models: the microscopic world
being described by quantum field theory and the macroscopic word by
general relativity. \textsc{qft} is by far the most successful theory
ever made, allowing to reach an almost incredible level of accuracy in
its measurable predictions. But gravity is different from all other
interactions; although by far the weakest, it acts on the very
structure of the universe at a more fundamental level. Many attempts
have been made to obtain a consistent quantum theory of gravity and
they all proved unsuccessful so that it is has become clear that
completely new ideas are needed.

To this day, but this has been true for more than twenty years now,
the only promising trail we can follow in the quest for this
unification is string theory. Roughly speaking it postulates that the
fundamental objects are not point particles as in the standard quantum
models but one-dimensional objects -- strings. Although their typical
size is so small that one might even question the very meaning of
distance at this scale, the mere not being pointlike allows to solve
an enormous number of theoretical problems and in particular those
connected with the severe divergencies that gravity presents due to
the local nature of interactions. Field theory particles appear as
vibration modes of the fundamental string, spacetime is a
semiclassical description for a string condensate, supergravity
emerges as a low-energy limit, and the standard model is the result of
a compactification in presence of extended objects (D branes).

Of course this is in many ways a wish-list. In its present state
string theory is largely incomplete. To begin, only a
first-quantization description is known and as such is intrinsically
perturbative, and only the S-matrix elements in a given vacuum are
accessible. So, although in principle the very geometry should emerge
from string dynamics, in practice we are forced to choose a
vacuum, which by itself clips the wings of any hope of attaining a
complete quantum gravity theory. Moreover only the perturbative regime
is in principle available, even though the existence of a web of
dualities can translate strong coupling backgrounds into ones we can
deal with.

One thing nevertheless must be kept in mind. One should stay as close
as possible to the present knowledge of Nature and try to predict the
outcomes of realistic experiments beyond the standard models in
particle physics or cosmology by using phenomenomogical models, but
\emph{string theory is not supergravity}. It is reassuring to find it
as a low energy limit but most, if not all, of the new physics lies in
the regime where the semiclassical approximations break down. From
this point of view an important, almost essential, r\^ole is played by
exact models, \emph{i.e.} systems in which the $\alpha^\prime $
corrections can be kept under control and a conformal field theory
description is possible. Because of technical issues, not many such
backgrounds are known and they are all characterized by a high degree
of symmetry. Hence it is not surprising that they in general do not
have a direct phenomenological impact. But the reason for their
fundamental importance lies elsewhere. They can mostly be used as
laboratories to study the extremal conditions -- black hole dynamics
just to name one -- in which general relativity and field theory show
their limits. The very reason why string theory was introduced.

\paragraph{Plan of the thesis}

This thesis is almost entirely devoted to studying string theory
backgrounds characterized by simple geometrical and integrability
properties. This requires at the same time a good grasp on both the
low-energy (supergravity) description in terms of spacetime and on the
\textsc{cft} side controlling all-order-in-$\alpha^\prime$ effects.

The archetype of this type of system is given by Wess-Zumino-Witten
models, describing string propagation in a group manifold or,
equivalently, a class of conformal field theories with current
algebras. Given their prominent r\^ole we devote the whole
Chapter~\ref{cha:wess-zumino-witten} to recall their properties from
different points of view, trying to outline some of the most important
features.

In Chapter~\ref{cha:deformations} we enter the main subject of these
notes, namely we study the moduli space of \textsc{wzw} models by
using truly marginal deformations obtained as bilinears in the
currents. A vast literature exists on this type of constructions, but
we will concentrate on a particular class, which we dub
\emph{asymmetric deformations}. They actually present a number of
advantages over the more familiar symmetric ones and in particular,
although the \textsc{cft} description is slightly more involved
(Sec.~\ref{sec:partition-functions}), they enjoy a very nice spacetime
interpretation. This can be completely understood in terms of the
always-underlying Lie algebra (Sec.~\ref{sec:geom-squash-groups}) and
can be proven to remain unchanged at all orders in $\alpha^\prime$
(Sec.~\ref{sec:no-renorm-theor}).

The following Chapter~\ref{cha:applications} illustrates some of the
obvious applications for our construction. We then start with the
simplest $SU(2)$ case, leading to a \textsc{cft} on the squashed
three-sphere and on the two-sphere (Sec.~\ref{sec:deformed-su2}). Then
we deal with the considerably richer non compact $SL(2,\setR)$ case
(Sec.~\ref{sec:deformed-sl2}). This leads naturally to the description
of some black hole geometries such as the near-horizon limit for the
Bertotti-Robinson black hole (Sec.~\ref{sec:near-horiz-geom}), and the
three-dimensional electrically charged black string
(Sec.~\ref{sec:blackstring}). Both can be studied in terms of
\textsc{cft}, thus allowing for an evaluation of the spectrum of string
primaries. Further applications regard the possibility of introducing
new compactification manifolds as part of larger, ten-dimensional
backgrounds (Sec.~\ref{sec:new-comp}).

In Chapter~\ref{cha:squashed-groups-type} we consider an alternative
description for the squashed group geometries which are found to be
T-duals of the usual type II black brane solutions.

In Chapter~\ref{cha:renorm-group-flows} we take a slight detour from
what we have seen up to this point: instead of exact \textsc{cft}
backgrounds we deal with off-shell systems. Using a
renormalization-group approach we describe the relaxation towards the
symmetrical equilibrium situation. This same behaviour is studied from
different points of view, \textsc{rg} flow in target space
(Sec.~\ref{sec:target-space-renormalization}), two-dimensional
renormalization (Sec.~\ref{sec:cft-approach-1}) and reading the flow
as a motion in an extra time direction
(Sec.~\ref{sec:rg-flow-friction}), thus obtaining
Freedman-Robertson-Walker-like metrics that in the most simple case
describe an isotropic universe with positive cosmological constant
undergoing a big-bang-like expansion
(Sec.~\ref{sec:friedm-roberts-walk}).

The final Chapter~\ref{cha:hyperbolic-spaces} marks a further
deviation from the construction of exact models: we consider in fact
backgrounds with Ramond-Ramond fields which still elude a satisfactory
\textsc{cft} interpretation. In particular we analyze direct products
of constant-curvature spaces and find solutions with hyperbolic spaces
sustained by \textsc{rr} fields.

The themes we treat here have been the subject of the following
publications:
\nocite{Israel:2004vv,Israel:2004cd,Orlando:2005vt,Detournay:2005fz,Orlando:2005im,Orlando:2005uy}
\begin{itemize}
\item D.~Isra{\"e}l, C.~Kounnas, D.~Orlando and P.~M. Petropoulos,
  \textsl{ Electric / magnetic deformations of S**3 and AdS(3), and
    geometric cosets}, \newblock Fortsch. Phys. \textbf{ 53}, 73--104
  (2005), {hep-th/0405213}.
\item D.~Israel, C.~Kounnas, D.~Orlando and P.~M. Petropoulos,
  \textsl{ Heterotic strings on homogeneous spaces}, \newblock
  Fortsch. Phys. \textbf{ 53}, 1030--1071 (2005), {hep-th/0412220}.
\item D.~Orlando, \textsl{ AdS(2) x S**2 as an exact heterotic string
    background}, \newblock (2005), {hep-th/0502213}, \newblock Talk
  given at NATO Advanced Study Institute and EC Summer School on
  String Theory: From Gauge Interactions to Cosmology, Cargese,
  France, 7-19 Jun 2004.
\item S.~Detournay, D.~Orlando, P.~M. Petropoulos and P.~Spindel,
  \textsl{ Three-dimensional black holes from deformed anti de
    Sitter}, \newblock JHEP \textbf{ 07}, 072 (2005),
  {hep-th/0504231}.
\item D.~Orlando, \textsl{ Coset models and D-branes in group
    manifolds}, \newblock (2005), {hep-th/0511210}. Published in
  Phys.Lett.B
\item D.~Orlando, P.~M. Petropoulos and K.~Sfetsos, \textsl{
    Renormalization-group flows and charge transmutation in string
    theory}, \newblock Fortsch. Phys. \textbf{ 54}, 453--461 (2006),
  {hep-th/0512086}.
\end{itemize}
Some results are on the other hand still unpublished. This concerns in
particular Chapter~\ref{cha:squashed-groups-type} and the second half
of Chapter~\ref{cha:renorm-group-flows} and
Chapter~\ref{cha:hyperbolic-spaces}.

\chapter{Wess-Zumino-Witten Models}
\label{cha:wess-zumino-witten}

\chapterprecis{Wess-Zumino-Witten models constitute a large class of
  the exact string theory solutions which we will use as starting
  points for most of the analysis in the following. In this chapter we
  see how they can be studied from different perspectives and with
  different motivations both from a target space and world-sheet point
  of view.}

\setlength{\epigraphwidth}{17em}
\begin{epigraphs}
  \qitem{``To paraphrase Oedipus, Hamlet, Lear, and all those guys,'' I said, ``I wish I had known this some time ago.''}{Amber Chronicles\\ \textsc{Roger Zelazny}}
\end{epigraphs}

\section{The two-dimensional point of view}
\label{sec:bidim-point-view}

\oldstuff

\subsection{The classical theory}
\label{sec:classical-theory}

\lettrine{W}{ess-Zumino-Witten} models were introduced by Witten in
his seminal paper~\cite{Witten:1983ar} to generalize the usual
bosonization of a free fermion to a system of $N$ fermions.  It has
been known for a long time that the Lagrangian for a free massless
Dirac fermion in two dimensions can be mapped to the Lagrangian for a
free massless boson\marginlabel{Free fermion bosonization} as follows:
\begin{equation}
  L = \psi^\ast \imath \slashed{\partial} \psi \to \frac{1}{2} \partial_\mu
  \phi \partial^\mu \phi ,
\end{equation}
but the generalization to more complex systems is not straightforward.
One of the main motivations for this mapping is given by the fact that
bosonic systems admit a semiclassical limit and then allow for simpler
intuitive interpretations of their physics.

It is of course possible to rewrite the fields in one description as
functions of the fields in the other one but this requires complicated
(non-local) expressions. Other quantities remain simple, in particular
the currents:
\begin{equation}
  J_\mu = \psi^\ast \gamma_\mu  \psi \to \frac{1}{\sqrt{\pi}} \epsilon_{\mu\nu}
  \partial^\nu \phi
\end{equation}
and as such they are the convenient building blocks for a
generalization. A most useful rewriting for their expression is
obtained when going to light-cone coordinates
\begin{align}
  J = \frac{i}{2 \pi} U^{-1} \partial U = - \frac{1}{\sqrt{\pi}} \partial \phi ,&& \bar J = -
  \frac{i}{2 \pi} \bar \partial U U^{-1} = \frac{1}{\sqrt{\pi}} \bar \partial \phi ,
\end{align}
$U = \exp [i \sqrt{4\pi} \phi]$ being the chiral density $\bar \psi ( 1+
\gamma_5) \psi$. One then finds that the currents are chirally conserved:
\begin{equation}
  \partial \bar J = \bar \partial J = 0 ,
\end{equation}
which is equivalent to ask for the bosonic field to be harmonic:
\begin{equation}
  2 \partial \bar \partial \phi = 0 .
\end{equation}

The generalization of this simple system is given by the theory of
$2N$ Dirac fermions.
\begin{equation}
  L = \frac{1}{2} \sum_k \bar \psi^k i \slashed{\partial} \psi^k ,
\end{equation}
This admits a chiral group $U(N) \times U(N)$ with vector and axial
currents written as:
\begin{align}
  V_\mu^a = \bar \psi \gamma_\mu T^a \psi , && A_\mu^a =  \bar \psi \gamma_\mu \gamma_5 T^a \psi .
\end{align}
It is more useful to define the chiral components
\begin{align}
  J^{ij} = -i \psi^i \psi^j , && \bar J^{ij} = - i \bar \psi^i \bar \psi^j ,	
\end{align}
generating the $U(N) \times U(N)$ symmetry and obeying the same
conservation as before:
\begin{equation}
  \partial \bar J = \bar \partial J = 0 ,
\end{equation}
This is obviously not equivalent to a system of $N$ bosons which would
just allow for the diagonal $U(1)^N$ symmetry. What we need is an
object $g \in U(N)$ transforming under a couple $(A,B) \in U(N) \times U(N)$ as
\begin{equation}
  g \to A g B^{-1}, \: \: A,B \in U(N)
\end{equation}
and express the currents as functions of $g$ just as in the Abelian
case we did in terms of the density $U$:
\begin{align}
\label{eq:WZW-currents}
  J = \frac{i}{2\pi} g^{-1} \partial g , && \bar J = - \frac{i}{2\pi} \bar \partial g g^{-1} .
\end{align}
In order for these currents to be conserved as above all we need to
find is a Lagrangian admitting the following equations of motion
\begin{equation}
  \partial ( g^{-1} \bar \partial g ) = 0 .
\end{equation}

The first natural tentative action is given by
\begin{equation}
  S_0 = \frac{1}{4 \lambda^2} \int_\Sigma \di^2 x \: \tr \left[ \partial^\mu g \partial_\mu g^{-1} \right] ,
\end{equation}
since this is the only manifestly chirally invariant choice.
Unfortunately this can't be the right answer for a number of reasons.
In particular it describes an asymptotically free theory with the
wrong equations of motion. \marginlabel{Wess-Zumino term}We are then
forced to add another ingredient, the so-called Wess-Zumino term
\begin{equation}
  \Gamma = \frac{1}{24 \pi} \int_M d^3 y \: \epsilon^{\mu \nu \rho} \tr \left[ \tilde g^{-1} \partial_\mu \tilde g \tilde g^{-1} \partial_\nu \tilde g \tilde g^{-1} \partial_\rho \tilde g \right] ,
\end{equation}
where $M$ is a three-dimensional manifold admitting $\Sigma $ as
border $\partial M = \Sigma $ and $\tilde g $ is the extension of the
mapping $g:\Sigma \to G$ to a mapping $\tilde g : M \to G$. Although
it might appear a bit surprising at first sight, this is precisely what
is needed since the variation of $\Gamma$ gives only a local term on
$\Sigma $ and the equations of motion for the action $S = S_0 + k
\Gamma$ read:
\begin{equation}
  \left( \frac{1}{2 \lambda^2} + \frac{k}{8\pi } \right) \partial \left(g^{-1} \bar \partial g \right) +  \left( \frac{1}{2 \lambda^2} - \frac{k}{8\pi } \right) \bar \partial \left(g^{-1} \partial g \right) = 0 ,
\end{equation}
which in particular for $\lambda^2 = \frac{4 \pi }{k}$ yield precisely the
equations we were expecting. It can be shown that this is an infrared
fixed point for a renormalization group-flow, and we will expand on
this aspect in Ch.~\ref{cha:renorm-group-flows}.

At this point it is on the other hand better to deal more thoughtfully
with the interpretation and the consequences of the Wess-Zumino term
$\Gamma$. First of all it must be remarked that $\Gamma$ can be put in the form
of an ordinary action \emph{ie} an integral over the two-dimensional
space-time of a perfectly respectable although non-manifestly
chiral-invariant action (which changes by a total derivative under a
chiral transformation):
\begin{equation}
  \Gamma = \int_\Sigma d^2 x \: \epsilon^{\mu \nu} B_{ij} (\phi^k) \partial_\mu \phi^i \partial_\nu \phi^j  ,
\end{equation}
where $B$ is a (Kalb-Ramond) two-form. \marginlabel{Level
  quantization}Another important aspect is the fact that an ambiguity
is present in the definition of $\Gamma $ for there are infinite
topologically inequivalent ways to extend $g$ to $\tilde g$,
classified by $\pi_3 (G)$. In the case of a compact group $\pi_3 (G) =
\setZ $ and we are led to the same kind of topological argument
leading to the quantization of the Dirac monopole. In fact if we
consider a different three-manifold $M^\prime$ in the definition of
$\Gamma$, the string amplitude changes by
\begin{equation}
  \exp \left[ \imath \int_M H - \imath \int_{M^\prime} H \right] , 
\end{equation}
where $H = \di B$. This implies that the theory is consistent only if
\begin{equation}
  \frac{1}{2\pi }\int_{S^3} H  \in \setZ ,
\end{equation}
having remarked that $M$ and $M^\prime $ have the same boundary and
then $M-M^\prime \sim S^3 $. Using the normalizations above one can
show that this is equivalent to asking $k \in \setZ $ which can be
read as a quantization condition on the radius of the group
manifold. In particular, the semiclassical limit is obtained when
$k\to \infty $.

For reasons that will appear evident in the following $k$ is called
level of the model and the action is written as:
\begin{equation}
\label{eq:WZW-model}
  S_k [g] = \frac{k}{16 \pi } \int_\Sigma \di^2 z \: \braket{ g^{-1} \partial g, g^{-1} \bar \partial g } + \frac{k}{24 \pi } \int_M \braket{ \tilde g^{-1} \di \tilde g, \comm{\tilde g^{-1} \di \tilde g, \tilde g^{-1} \di \tilde g} } ,
\end{equation}
where $\braket{\cdot,\cdot}$ is the Killing form on $G$.


\subsection{An exact model}
\label{sec:path-integral}

\oldstuff

An important feature of \textsc{wzw} models is that they provide exact
solutions at all orders in $\alpha^\prime$ or, more precisely,
\emph{the effective action is equal to the classical action up to a
  shift in the overall normalization $k \to k + g^\ast
  $}\marginlabel[\textsc{wzw} models!Effective action]{Quantum
  effective action for \textsc{wzw} models}.  The argument goes as
follows~\cite{Knizhnik:1984nr,Leutwyler:1991tv,Tseytlin:1992ri}.
Consider the path integral
\begin{equation}
  \label{eq:eff-act-B}
  \int \mathscr{D} g \: e^{-k S[g] + B \bar J[g]} = e^{-W[B]} ,
\end{equation}
where $S[g]$ is the \textsc{wzw} action at level one and $B \bar J[g]$
is the shorthand for
\begin{equation}
  B  \bar J[g] = \frac{k}{\pi} \int \di z^2 \: \braket{B \bar \partial g g^{-1}} .
\end{equation}
If we write $B$ as $B = u^{-1} \partial u$ we can use the so called
Polyakov-Witten identity
\begin{equation}
  S [a b] = S[a] + S[b] - \frac{1}{\pi} \int \di^2 z \: \braket{a^{-1} \partial a \bar \partial b b^{-1}}  
\end{equation}
and it's then easy to see that $W$ doesn't receive quantum corrections
and is simply given by the classical action evaluated on $u$:
\begin{equation}
  W[B] = - k S[u] .  
\end{equation}
Indeed, $B$ is an external source coupled with the current $\bar J$
so, the effective action for $g$, $\Gamma[g]$, will be given by the
Legendre transformation of $W[B]$, \emph{ie} by the path integral:
\begin{equation}
  \label{eq:Legendre-eff-act}
  \int \mathscr{D}B \: e^{-W[B] + B \bar J [g]} = e^{-\Gamma [g]} .
\end{equation} 
This interpretation of effective action for $\Gamma[g]$ is comforted by
remarking that combining Eq.~(\ref{eq:eff-act-B}) and
Eq.~(\ref{eq:Legendre-eff-act}) one finds that
\begin{equation}
  e^{-\Gamma[g]} = \int \mathscr{D} g^\prime \: e^{- k S[g^\prime ]} \:\delta (\bar J[g^\prime ] - \bar J[g] ) . 
\end{equation}
In order to calculate the Legendre transform in
Eq.~(\ref{eq:Legendre-eff-act}) one can perform a change of variables
from $B = u^{-1} \partial u $ to $u$: the corresponding Jacobian will give
the announced shift in the effective action. More
precisely, as shown in~\cite{Polyakov:1983tt,Polyakov:1988qz} we have:
\begin{equation}
  \frac{\mathscr{D} B}{\mathscr{D}u} = e^{g^\ast S[u]}  
\end{equation}
and putting this back in Eq.~(\ref{eq:Legendre-eff-act})
\begin{equation}
  e^{-\Gamma [g]} =\int \mathscr{D} u \: e^{\left( k + g^\ast \right) S[u] + B[u] \bar J[g]} ,  
\end{equation}
we can use the same technique as above to derive the celebrated result:
\begin{equation}
  \Gamma[g] = \left( k + g^\ast \right) S[g] .
\end{equation}

\subsection{The CFT approach}
\label{sec:cft-approach}

\oldstuff

\subsubsection{\textsc{cft} with current algebras}
\label{sec:textsccft-with-curr}

\marginlabel{\textsc{cft} infinitesimal generators}Let us forget for a
moment the \textsc{wzw} models and consider a more general framework,
\emph{ie} two-dimensional conformal field theories with current
algebras. Given the stress-energy tensor $T_{\mu \nu} (\zeta)$ in two
dimensions one can define
\begin{subequations}
  \begin{align}
    T &= T_{11} - T_{22} + 2 \imath T_{12} \\
    \bar T &= T_{11} - T_{22} - 2 \imath T_{12}
  \end{align}
\end{subequations}
so that the conservation $\partial_{\mu} T^{\mu \nu} = 0$ and the zero
trace condition $T\ud{\mu}{\mu} = 0$ translate into analyticity
conditions
\begin{equation}
  \bar \partial T = \partial \bar T = 0 .
\end{equation}
It is then clear that since the stress-energy tensor is the
infinitesimal generator for translations, $T$ and $\bar T$ have this
role for the conformal transformations.
\begin{subequations}
  \begin{align}
    z &\to z + \epsilon(z) \\
    \bar z & \to \bar z + \bar \epsilon (\bar z)
  \end{align}
\end{subequations}
which is to say that if we take a local field $A(z,\bar z)$, this
transforms under such variations as
\begin{equation}
  \delta_\epsilon A (z,\bar z) = \oint_z T(\zeta) \epsilon(\zeta) A(z, \bar z) \di \zeta    
\end{equation}
where the contour integral is around $z$\footnote{In the following we
  will avoid to write the anti-holomorphic counterpart for each
  relation since it can always be trivially derived.}.

This is just the definition of two-dimensional \textsc{cft} but if the
theory is at the same time invariant under a $G(z) \times G(\bar z)$
action, $G$ being some Lie group, then there are additional generators
$J(z)$ and $\bar J(\bar z)$ allowing to express the variation of $A(z,
\bar z)$ as
\begin{equation}
\label{eq:delta-omega-A}
  \delta_\omega A (z,\bar z) = \oint_z J^a(\zeta) \omega^a(\zeta) A(z, \bar z) \di \zeta    
\end{equation}
where $\omega(z)$ is some element in $\lie G$ (the algebra as usual
parametrizes the infinitesimal group transformations).
  
\marginlabel{Virasoro $\times$ Ka\v c Moody}The variations of those
generators with respect to $\epsilon $ and $\omega$ stem from general
principles and read
\begin{subequations}
  \begin{align}
    \delta_\epsilon T (z) &= \epsilon(z) T^\prime (z) + 2 \epsilon^\prime (z) T(z) + \frac{1}{12} c \epsilon^{\prime\prime} (z) \\
    \delta_\epsilon J (z) &= \epsilon(z) J^\prime (z) + \epsilon^\prime (z) J(z) \\
    \delta_\omega J^a (z) &= \F{a}{bc} \omega^b(z) J^c(z) + \frac{1}{2} k \omega^{a \prime} (z) \label{eq:delta-omega-J}
  \end{align}
\end{subequations}
This is just a way of writing the algebra of the generators. Using the
definition above in terms of contour integrals it can also be put in
terms of operator product expansions
\begin{subequations}
  \label{eq:ope-TJ}
  \begin{align}
    T(z) T(w) &= \frac{c}{2 \left( z - w \right)^4} +
    \frac{2}{\left(z - w \right)^2} T(w) + \frac{1}{z-w} T^\prime
    (w) \\
    T(z) J(w) &= \frac{1}{\left(z - w \right)^2} J(w) +
    \frac{1}{z-w} J^\prime (w) \\
    J^a(z) J^b(w) &= \frac{k \delta^{ab}}{\left(z -w \right)^2} +
    \frac{\F{ab}{c}}{z-w} J^c (w)
  \end{align}
\end{subequations}

Any operator in the theory is characterized by a representation for
the left and right $G$ groups and its anomalous dimensions $\Delta $
and $\bar \Delta$, which depend on the behaviour of the operator under
dilatation $z \to \lambda z$. More precisely an operator of weight
$\left(\Delta, \bar \Delta \right)$ transforms under $z \to \lambda z
$, $\bar z \to \bar \lambda \bar z$ as $A \to \lambda^\Delta \bar
\lambda^{\bar \Delta} A$, and in particular the primary fields are
defined as those who satisfy
\begin{subequations}
  \begin{align}
    T(z) \phi (w, \bar w) &= \frac{\Delta}{\left(z - w \right)^2} \phi (w, \bar w) + \frac{1}{z-w} \frac{\partial }{\partial w} \phi (w, \bar w) \\
    J^a (z) \phi (w, \bar w) &= \frac{t^a}{z-w} \phi (w, \bar w) .
  \end{align}
\end{subequations}
Given these relations it is immediate to write the Ward identities
satisfied by the correlation functions of primary fields:
\begin{subequations}
  \begin{align}
    \label{eq:Ward-Virasoro} 
    \begin{split}
      \braket{ T(z) \phi_1 (z_1, \bar z_1) \dots \phi_N (z_N, \bar
        z_N) } = \sum_{j=1}^N \left(\frac{\Delta_j}{\left(z - z_j
          \right)^2} + \frac{1}{z-z_j} \frac{\partial }{\partial
          z_j}\right) \times \\ \times \braket{\phi_1 (z_1, \bar z_1) \dots \phi_N
        (z_N, \bar z_N)}
    \end{split} \\
    &\braket{ J^a(z) \phi_1 (z_1, \bar z_1) \dots \phi_N (z_N, \bar
      z_N) } = \sum_{j=1}^N \frac{t^a_j}{z-z_j} \braket{\phi_1 (z_1, \bar z_1) \dots \phi_N
      (z_N, \bar z_N)} \label{eq:Ward-KacMoody}
  \end{align} 
\end{subequations}

A further step can be made if one expands the operators $T$ and $J$ in
Laurent series obtaining respectively the modes $L_n$ and $J_n$ which
by definition act on a local operator to give
\begin{subequations}
  \begin{align}
    L_n A (z, \bar z ) &= \oint_z T(\zeta) \left( \zeta - z \right)^{n+1} A (z, \bar z) \di \zeta \\
    J_n^a A (z, \bar z ) &= \oint_z J^a(\zeta) \left( \zeta - z
    \right)^{n+1} t^a A (z, \bar z) \di \zeta
  \end{align}
\end{subequations}
and in particular the primaries will satisfy
\begin{align}
  L_n \phi &= J_n^a \phi = 0 \hspace{3em} \forall n > 0 \\
  L_0 \phi &= \Delta \phi \hspace{3em}  J_0^a \phi = t^a \phi
\end{align}

The commutation relations among the $L_n$'s and the $J_n^a$'s are
natural consequences of the \textsc{ope}s in Eq.~\eqref{eq:ope-TJ} and
read
\begin{subequations}
  \begin{align}
    \comm{L_n, L_m} &= \left( n - m \right) L_{n+m} + \frac{1}{12} c
    \left(n^3 -n \right) \delta_{n+m} \label{eq:Virasoro} \\
    \comm{L_n, J_m^a} &= -m J^a_{n+m} \\
    \comm{J_n^a, J_m^b} &= \F{a}{bc} J^c_{n+m} + \frac{1}{2} k n
    \delta^{ab} \delta_{n+m,0} .\label{eq:Kac-Moody}
  \end{align}
\end{subequations}
This is again a way to write the semi-direct product of the Virasoro
(Eq.~\eqref{eq:Virasoro}) and Ka\v c-Moody algebras
(Eq.~\eqref{eq:Kac-Moody}).

\subsubsection{The \textsc{wzw} model}
\label{sec:textscwzw-model}

\marginlabel{\textsc{wzw cft}}As we emphasized above the currents $J$
and $\bar J$ are the fundamental building blocks for the construction
of \textsc{wzw} models. Their role is even more apparent when we study
the symmetries of the theory, which takes us directly to make contact
with the conformal field theory description. Hence the importance of
these models in giving an explicit realization of the \textsc{cft}
outlined above (among the classical references see
\emph{e.g.}~\cite{Knizhnik:1984nr}).

The key remark is that the action in Eq.~\eqref{eq:WZW-model} is
invariant under the transformation
\begin{equation}
  g (\zeta) \mapsto \Omega (z) g(\zeta) \bar \Omega^{-1}(\bar z)    
\end{equation}
where $\Omega(z)$ and $\bar \Omega (\bar z)$ are $G$-valued matrices
analytically depending on $z $ and $\bar z$. This gives rise to an
infinite set of conserved currents which are precisely those we
defined above in Eq.~\eqref{eq:WZW-currents} and
Eq.~\eqref{eq:delta-omega-A}. Locally this translates into the fact
that for an infinitesimal transformation
\begin{equation}
  \Omega (z) = \uni + \omega(z) = \uni + \omega^a(z) t^a
\end{equation}
the currents change as in Eq.~\eqref{eq:delta-omega-J}
\begin{equation}
  \delta_\omega J = \comm{\omega (z), J(z)} + \frac{1}{2} k
  \omega^\prime (z)
\end{equation}
which is to say that $J$ and $\bar J$ represent an affine Lie algebra
with central charge $k$.

\marginlabel{Sugawara stress-energy tensor}The next step consist in
identifying the stress-energy tensor. In the classical theory this is
a bilinear of the currents, so it is natural to choose the so-called
Sugawara
construction~\cite{Sugawara:1967rw,Sommerfield:1968pk,Knizhnik:1984nr}
\begin{equation}
  T(z) = \frac{1}{2 \left( k + g^\ast  \right)} \sum_a J^a(z) J^a(z) 
\end{equation}
where the constant factor is fixed by asking a current to be a weight
one operator\footnote{It follows easily that a Ka\v c-Moody primary is
  a Virasoro primary but not the other way round. Pictorially
  Eq.~\eqref{eq:Ward-KacMoody} is the ``square root'' of
  Eq.~\eqref{eq:Virasoro}}. Note in particular the fact that the level
$k$ is shifted to $k+g^\ast$ which is precisely the same correction we
found summing the instanton corrections in the previous section. A
simple calculation of the \textsc{ope} of $T$ with itself gives the
value for the central charge of the model
\begin{equation}
  c = \frac{k \dim G}{k+ g^\ast}
\end{equation}

Using the definition of primary field (and in particular the fact that
they are annihilated by $J^a_{-1} t^a_l - L_{-1} $) one can easily
show that each primary is degenerate and has weight
\begin{equation}
  \Delta = \frac{c_l}{g^\ast + k}  
\end{equation}
where $c_l = t^a_l t^a_l$ is the quadratic Casimir and in particular
coincides with the dual Coxeter number $c_l = g^\ast $ if the field
transforms in the adjoint representation.

\subsubsection{Partition function}
\label{sec:partition-function}

\marginlabel{Affine characters}As one would expect, a modular
invariant partition function for a \textsc{wzw} group can be build
entirely out of group theoretical objects. In particular the building
blocks are given by the affine characters, \emph{ie} the generating
functions of the weight multiplicities for a given irrep $\Lambda$
that take into account the conformal dimension of the highest weight
of the representation:
  \begin{equation}
    \chi^{\Lambda} (\tau, \nu, u) = e^{-2\imath \pi k u} 
  \tr_{rep \left(\Lambda\right)} \left[ q^{L_0 -\nicefrac{c}{24}}
    e^{2\imath \pi \kappa(\nu,\mathcal{J})} \right]
  = e^{2\imath \pi \tau \frac{\kappa (\Lambda,\Lambda + 2\rho)}{2\left(k
        +\mathfrak{g}^\ast \right)} 
    - \nicefrac{c}{24}} ch_{\Lambda} \left(\tau,\nu,u\right)
  \end{equation}
  where
  \begin{itemize}
  \item $ch_{\Lambda} \left(\tau,\nu,u\right) $ is the usual character
    for the affine Lie algebra $\hat {\mathfrak{g}}$:
    \begin{equation}
      ch_{\Lambda} \left(\tau,\nu,u\right) = e^{-2\imath \pi k u} 
      \sum_{\hat{\lambda} \in \text{Rep} \left(\Lambda\right)} \text{dim} V_{\hat{\lambda}} 
      \exp \{ 2\imath \pi \tau n + \sum_i \nu_i \kappa \left(e_i , \hat{\lambda}\right) 
      \} ;
    \end{equation}
  \item $V_{\hat \lambda}$ is the multiplicity of the weight $\hat
    \lambda = \left( \lambda, k, n \right)$;
  \item $\braket{e_i}$ is a basis in the root space;
  \item $\rho = \sum_{\alpha>0} \alpha /2$ is the Weyl vector.
  \end{itemize}
  An alternative way of writing the same characters is obtained in terms of theta functions. To each weight one can assign a theta-function defined as:
  \begin{equation}
    \Theta_{\hat{\lambda}} \left(\tau, \nu , u\right) =  e^{-2\imath \pi k u} 
  \sum_{\gamma \in \mathbf{M_{\textsc{l}}} + \frac{\lambda}{k}} 
  e^{\imath \pi \tau k \, \kappa \left(\gamma,\gamma\right)} \ 
  e^{2\imath \pi k \kappa \left(\nu, \gamma\right)}    
  \end{equation}
  where $\mathbf{M_{\textsc{l}}}$ is the long root lattice. Then,
  using the Weyl-Ka\v c formula the characters are written as:
  \begin{equation}
    \chi^{\Lambda} \left( \tau, \nu ,u \right) = \frac{\displaystyle{\sum_{w \in \mathbf{W}}} 
      \epsilon\left(w\right) \Theta_{w\left(\hat{\Lambda} + \hat{\rho}\right)} \left(\tau,\nu,u\right)}
    {\displaystyle{\sum_{w \in \mathbf{W}}} 
      \epsilon\left(w\right) \Theta_{w\left(\hat{\rho}\right)} \left(\tau,\nu,u\right)}\, ,
  \end{equation}
  $\mathbf{W}$ being the Weyl group of the algebra and $\epsilon
  \left(w\right)$ the parity of the element $w$.

  Knowing that the affine Lie algebra is the largest chiral symmetry of the theory it is not surprising that the partition function can be written as
  \begin{equation}
     Z = \sum_{\Lambda , \bar \Lambda} M^{\Lambda \bar \Lambda} 
  \chi^{\Lambda} \left(\tau,0,0 \right)
    \bar{\chi}^{\bar \Lambda} \left(\bar{\tau},0,0\right)
\end{equation}
where the sum runs over left and right representations of $\mathfrak
g$ with highest weight $\Lambda$ and $\bar \Lambda$ and $M^{\Lambda
  \bar \Lambda}$ is the mass matrix which is chosen so to respect the
modular invariance of $Z$.

A generalization that we will use in the following is obtained for
heterotic strings where the $N=\left(1,0 \right)$ local supersymmetry
requires a super-affine Lie algebra for the left sector. The latter
can anyway be decoupled in terms of the bosonic characters above and
free fermion characters as to give:
\begin{equation}
  Z \oao{a}{b} = \sum_{\Lambda , \bar \Lambda} M^{\Lambda \bar \Lambda} 
  \chi^{\Lambda} \left(\tau \right) \ 
  \left( \frac{\vartheta\oao{a}{b} \left(\tau \right)}{\eta \left(\tau \right)} \right)^{
    \mathrm{dim} \left( \mathfrak g \right) /2} 
  \bar{\chi}^{\bar \Lambda} \
\end{equation}
where $\left(a,b\right)$ are the spin structures of the world-sheet fermions.
The characters of the affine algebras can be decomposed according to 
the generalized parafermionic decomposition, by factorizing 
the abelian subalgebra of the Cartan torus. 
For example, we can decompose the left supersymmetric $\mathfrak{g}_k$ characters
in terms of characters of the supersymmetric coset, given 
by the following branching relation (see~\cite{Kazama:1989qp}):
\begin{equation}
  \chi^{\Lambda}  \ 
  \left(
    \frac{\vartheta \oao{a}{b}}{\eta}\right)^{\mathrm{dim} \left(\mathfrak j \right) /2}  = 
  \sum_{\lambda \ \mathrm{mod} \ \left(k + g^* \right) \mathbf{M_\textsc{l}}} 
  \mathcal{P}^{\Lambda}_{\lambda} \oao{a}{b}  
  \frac{\Theta_{\lambda,k+g^*}}{\eta^{\mathrm{dim} \left(\mathfrak k \right)}}
\end{equation}
in terms of the theta-functions associated to $\mathfrak g_k$.

\section{The target space point of view}
\label{sec:target-space-wzw}

Supergravity appears as a low energy description of string theory,
obtained when asking for the Weyl invariance of the $\sigma$-model
Lagrangian. This amounts, at first order in $\alpha^\prime $, to the
following equations of motion for the metric $g$, the Kalb-Ramond
field $B$ and the dilaton $\Phi$~\cite{Callan:1985ia,Tseytlin:1995fh}:
\begin{equation}
\label{eq:sugra-equations}
  \begin{cases}
    \beta_\Phi = - \frac{1}{2} \nabla^\mu \partial_\mu \Phi + \partial_\mu
    \Phi \partial^\mu \Phi - \frac{1}{24}  H_{\mu\rho \sigma}
    H^{\mu \rho \sigma} ,\\
    \beta_{g} = R_{\mu \nu} - \frac{1}{4} H_{\mu\rho \sigma}
    H_\nu^{\phantom{\nu}\rho \sigma} + 2 \nabla_\mu \nabla_\nu \Phi , \\
    \beta _{B} = \nabla_\mu H\ud{\mu}{\nu \rho} + 2 \nabla_\mu \Phi
    H\ud{\mu}{\nu \rho} .
  \end{cases}
\end{equation}
Being a one-loop calculation, the corresponding results should always
be checked against higher order corrections in $\alpha^\prime$. On the
other hand, as we have already stressed many times above, \textsc{wzw}
models (just like the asymmetric deformations we study in this work)
only receive corrections in terms of the level of the algebra (or, in
this language, on the overall volume of the manifold). This implies
that the target space description at one loop in $\alpha^\prime $ is
automatically correct at all orders.  From this point of view,
Wess--Zumino--Witten models describe the motion of a string on a group
manifold geometry. The background fields are completed by a
\textsc{ns-ns} three form $H = \di B$ (Kalb-Ramond field) and a
constant dilaton $\Phi = \Phi_0$.

The target space analysis \danger is greatly simplified by the fact
that the geometric quantities are all naturally expressed in terms of
group theoretical objects. Let us consider for concreteness the case
of a compact group $G$, whose Lie algebra is generated by
$\braket{t^\alpha}$ and has structure constants $\F{\alpha }{\beta
  \gamma }$. The metric for the group manifold can be chosen as the
Killing metric (the choice is unique up to a constant in this case)
and it is then natural to use the Maurer--Cartan
one-forms\index{Geometry!Maurer-Cartan 1-forms} as vielbeins. In our
conventions, then:
\begin{equation}
\label{eq:Killing-metric}
  g_{\mu \nu } = - \frac{1}{2 g^\ast } f\ud{\alpha}{\beta \gamma} f\ud{\gamma}{\delta \alpha} J\ud{\beta}{\mu} J\ud{\delta}{\nu} = \delta_{\beta \gamma}  J\ud{\beta}{\mu} J\ud{\delta}{\nu} ,
\end{equation}
where $g^\ast $ is the dual Coxeter number and
\begin{equation}
\label{eq:MC-1-forms}
  J^\alpha_{\phantom{\alpha}\mu } = \braket{t^\alpha g^{-1} \partial_\mu  g} .
\end{equation}
In this basis the \textsc{ns-ns} 3-form field is written as
\begin{equation}
\label{eq:3-form-wzw}
  H_{[3]} = \frac{1}{3!} f_{\alpha \beta \gamma } J^{\alpha} \land J^{\beta } \land J^{\gamma } .
\end{equation}

\marginlabel[Geometry!Group manifolds]{Geometry of group manifolds}The
connection one-forms $\omega\ud{\alpha}{\beta }$ can be obtained by asking for the
torsion two-form to vanish:
\begin{equation}
  \di J^\alpha + \omega^\alpha_{\phantom{\alpha}\beta } \land J^\beta = T^\alpha = 0 ,
\end{equation}
and out of them one defines the curvature two form $R\ud{\alpha}{\beta}$ as:
\begin{equation}
  R\ud{\alpha}{\beta} = \di \omega\ud{\alpha}{\beta} + \omega\ud{\alpha}{\gamma } \land \omega\ud{\gamma}{\beta} .  
\end{equation}
which in turn is given in terms of the Riemann tensor as:
\begin{equation}
  R\ud{\alpha}{\beta} = \frac{1}{2} R\ud{\alpha}{\beta \gamma \delta } J^{\gamma} \land J^{\delta} .  
\end{equation}
In a Lie algebra with structure constants $\F{\alpha}{\beta \gamma}$ the variation
of the currents is given by the Cartan structure equation:
\begin{equation}
  \di J^\alpha = -\frac{1}{2} \F{\alpha}{\beta\gamma} J^\beta \land J^\gamma    
\end{equation}
whence we can directly read the connection one-forms:
\begin{equation}
\label{eq:MC-structure}
  \omega\ud{\alpha}{\beta} = - \frac{1}{2} \F{\alpha}{\beta \gamma} J^\gamma .
\end{equation}
It is then immediate to write:
\begin{subequations}
  \begin{align}
    \di \omega\ud{\alpha}{\beta} &= \frac{1}{4} \F{\alpha}{\beta \gamma} \F{\gamma}{\delta \epsilon} J^\delta \land J^{\epsilon} \\
    \begin{split}
      \omega \ud{\alpha}{\gamma } \land \omega\ud{\gamma}{\beta} &= \frac{1}{4} \F{\alpha}{\gamma \delta} \F{\gamma }{\beta \epsilon}
      J^\delta \land J^\epsilon = \frac{1}{8} \left( \F{\alpha}{\gamma \delta} \F{\gamma }{\beta \epsilon} - \F{\alpha}{\gamma
          \epsilon} \F{\gamma }{\beta \delta}
      \right)  J^\delta \land J^\epsilon  = \\
      &\hspace{6cm}= - \frac{1}{8} \F{\alpha}{\beta \gamma} \F{\gamma}{\delta \epsilon} J^\delta \land J^{\epsilon}
    \end{split}
  \end{align}
\end{subequations}
where we have antisimmetrized the product of the structure constants
and then used a Jacobi identity. The Riemann tensor, the Ricci tensor
and the scalar curvature are then given respectively by:
\begin{subequations}
  \label{eq:group-curvatures}
  \begin{align}
    R\ud{\alpha}{\beta \gamma \delta } &= \frac{1}{4}  \F{\alpha}{\beta \kappa } \F{\kappa }{\gamma \delta} ,\\
    Ric_{\beta \delta  } &= \frac{1}{4}  \F{\alpha}{\beta \kappa } \F{\kappa }{\alpha  \delta} = \frac{g^\ast}{2} g_{\beta \delta } ,\\
    R &= \frac{g^\ast}{2} \dim G .
  \end{align}
\end{subequations}

\marginlabel[\textsc{wzw} models!Equations of motion]{Equations of
  motion}We are now in a position to show that the metric and $H$
field satisfy the (first order in $\alpha^\prime$) equations of motion
in Eq.~\eqref{eq:sugra-equations}. Of course this result is much less
powerful than what we obtained in Sec.~\ref{sec:bidim-point-view} but
it is nevertheless an interesting example of how these geometrical
calculations are greatly simplified in terms of the underlying
algebraic structure.  For a system without dilaton the equations
reduce to:
\begin{subequations}
  \begin{align}
    \beta_{\alpha \beta }^G &= R_{\alpha \beta } - \frac{1}{4}
    H_{\alpha \delta \gamma} H_\beta^{\phantom{\beta}\delta \gamma} =
    0 , \\
    \beta_{\alpha \beta }^B &= \left( \nabla^\gamma H
    \right)_{\gamma \alpha \beta } = 0 .
  \end{align}
\end{subequations}
The first one is trivially satisfied by using the field in
Eq.~(\ref{eq:3-form-wzw}); for the second one we just need to remark
that in components the Levi-Civita connection is:
\begin{equation}
  \Gamma\ud{\alpha}{\beta\gamma} = \frac{1}{2} f\ud{\alpha}{\beta \gamma}  
\end{equation}
and remember that the covariant derivative of a three-form is
\begin{equation}
  \left( \nabla_\alpha H \right)_{\beta \gamma \delta } = \partial_\alpha H_{\beta \gamma \delta} - \Gamma\ud{\kappa}{\alpha \beta} H_{\kappa \gamma \delta} 
  - \Gamma\ud{\kappa}{\alpha \gamma } H_{\beta \kappa \delta}  - \Gamma\ud{\kappa}{\alpha \delta } H_{\beta \gamma \kappa } .
\end{equation}

In Sec.~\ref{sec:two-loop-equations} we will see from a slightly
different perspective how the normalization for the Kalb-Ramond field
$H$ can be fixed in terms of renormalization-group flow.


\chapter{Deformations}
\label{cha:deformations}

\chapterprecis{In this rather technical chapter we describe marginal
  deformations of Wess-Zumino-Witten models. The main purpose for
  these constructions is to reduce the symmetry of the system while
  keeping the integrability properties intact, trying to preserve as
  many nice geometric properties as possible.}

\setlength{\epigraphwidth}{17em}
\begin{epigraphs}
  \qitem{Mr. Jabez Wilson laughed heavily. ``Well, I never!'' said
    he. ``I thought at first that you had done something clever, but I
    see that there was nothing in it, after all.''
    
    ``I begin to think, Watson,'' said Holmes, ``that I make a mistake
    in explaining. \emph{Omne ignotum pro magnifico}, you know, and my
    poor little reputation, such as it is, will suffer shipwreck if I
    am so candid.''}{The Red Headed League\\ \textsc{Arthur Conan
      Doyle}}
\end{epigraphs}

\lettrine{T}{he power} of \textsc{wzw} models resides in the
symmetries of the theory.  They impose strong constraints which allow
quantum integrability as well as a faithful description in terms of
spacetime fields, whose renormalization properties (at every order in
$\alpha^\prime$) are easily kept under control, as we have seen in the
previous chapter.

It is hence interesting to study their moduli spaces, aiming at
finding less symmetric (and richer) structures, that will hopefully
enjoy analogous integrability and spacetime properties.

This chapter is devoted to introducing the construction of asymmetric
deformations and giving the general results in a formalism adapted to
group manifold geometry. For this reason the stress is put on the more
mathematical aspects. Physical examples and consequences will be
illustrated in greater detail in Ch.~\ref{cha:applications}.

\section{Deformed \textsc{wzw} models: various perspectives}
\label{sec:wzw-deformations}

\subsection{Truly marginal deformations}

\newcommand{\mO}{\mathcal{O}} \marginlabel[\textsc{wzw} models!Truly
marginal deformations]{Truly marginal deformations}In this spirit one
can consider marginal deformations of the \textsc{wzw} models obtained
in terms of $\left(1,1\right)$ operators built as bilinears in the
currents:
\begin{equation}
  \label{eq:Current-current-deform}
  \mO(z, \bar z)= \sum_{ij} c_{ij} J^i \left( z \right) \bar J^j \left( \bar z \right),
\end{equation}
where $J^i \left( z \right)$ and $\bar J^j \left( \bar z\right)$ are
respectively left- and right-moving currents. It is
known~\cite{Chaudhuri:1989qb} that this operator represents a
\emph{truly marginal deformation}, \emph{ie} it remains marginal at
all orders in the deformation parameter, if the parameter matrix
$c_{ij}$ satisfies the following constraints:
\begin{subequations}
\label{eq:marginal-c}
  \begin{align}
    c_{im} c_{jn} f^{ij}_{\phantom{ij}k} &= 0, \\
    c_{mi} c_{nj} \tilde f^{ij}_{\phantom{ij}k} &= 0,
  \end{align}
\end{subequations}
where $f$ and $\tilde f$ are the structure constants of the algebras
generated by $J^i$ and $\bar J^i$. In particular one can remark that
if $J^i$ and $\bar J^j$ live on a torus then the two equations are
automatically satisfied for any value of $c_{mn}$ and hence we get as
moduli space, a $\rank(c)$-dimensional hyperplane of exact
models\footnote{Although for special values of the level $k$ the
  theory contains other operators with the right conformal weights, it
  is believed that only current-current operators give rise to truly
  marginal deformations, \emph{i.e.} operators that remain marginal
  for finite values of the deformation parameter.}.  The proof of this
assertion proceeds as follows: \danger we want to show that
$\mathcal{O}$ keeps its conformal dimensions when a term $\h \mO (z,
\bar z)$ is added to the Lagrangian, $\h$ being a coupling constant.
The two-point function for $\mO(z, \bar z)$ in the interacting theory
with Lagrangian $L + \h \mO$ can be expanded in powers of $\h$ as
follows:
\begin{multline}
  \braket{ \mO (z, \bar z) \mO (w, \bar w) }_{\h} =\\=
  \frac{\displaystyle{\sum_{n=0}^\infty \nicefrac{(-\h)^n}{n!} \int \di^2 z_1 \ldots
      \di^2 z_n \braket{\mO (z, \bar z) \mO (w, \bar w) \mO (z_1, \bar
        z_1) \ldots \mO (z_n, \bar z_n)}}}{\displaystyle{\sum_{n=0}^\infty
      \nicefrac{(-\h)^n}{n!} \int \di^2 z_1 \ldots \di^2 z_n \braket{ \mO (z_1,
        \bar z_1) \ldots \mO (z_n, \bar z_n)}}} ,
\end{multline}
so, in particular, the $\h^2$-order term is:
\begin{multline}
  \braket{\mO (z, \bar z) \mO (w, \bar w)}_g =\\= \frac{\h^2}{2} \int
  \di^2 z_1 \di^2 z_2 \braket{\mO (z, \bar z) \mO (w, \bar w) \mO(z_1,
    \bar z_1) \mO(z_2, \bar z_2 )} + \\- \frac{\h^2}{2} \braket{ \mO
    (z, \bar z) \mO (w, \bar w)} \int \di^2 z_1 \di^2 z_2 \braket{
    \mO(z_1, \bar z_1) \mO(z_2, \bar z_2 )} .
\end{multline}
Only the first term can contain logarithmic divergences that can alter
the scale dependence of $\mO(z, \bar z)$, so let us study it more
closely, by expanding $\mO (z, \bar z)$ in terms of currents:
\begin{multline}
  \frac{\h^2}{2} \int \di^2 z_1 \di^2 z_2 \braket{\mO (z, \bar z) \mO (w,
    \bar w) \mO(z_1, \bar z_1) \mO(z_2, \bar z_2 )} = \\
  =\frac{\h^2}{2} \int \di^2 z_1 \di^2 z_2 \sum_{ghij} \sum_{lmno} c_{gh} c_{hm}
  c_{in} c_{jo} \braket{J_g (z) J_h (w) J_i(z_1) J_j (z_2)} \\
  \braket{ \bar J_l (\bar z) \bar J_m (\bar w) \bar J_n(\bar z_1) \bar
    J_o (\bar z_2)} .
\end{multline}
Rewriting the four-point functions for the currents in terms of their
algebras
\begin{subequations}
\label{eq:deforming-opes}
  \begin{align}
    J_i (z) J_j(w) &= \frac{K_{ij}}{(z-w)^2} + \frac{\imath \F{k}{ij} J_k
      (w)}{z-w} ,\\
    \bar J_i (z) \bar J_j(w) &= \frac{\tilde K_{ij}}{(z-w)^2} +
    \frac{\imath \tilde f^{k}_{\phantom{k}ij} \bar J_k (w)}{z-w} ,
  \end{align}  
\end{subequations}
one can evaluate the integrals passing to momentum space and
introducing some ultraviolet cut-offs $\Lambda_1, \Lambda_2, \Lambda$. In particular,
the terms which are interesting from our point of view are those
diverging as $\abs{ z - w}^{-4}$ and they are:
\begin{subequations}
  \begin{equation}
    \frac{8 \pi^2 \h^2 \log \Lambda_1 \log \Lambda_2}{\abs{z - w}^4} \sum_{ghij} \sum_{lmno} \sum_{kp} 
    c_{gl} c_{hm} c_{in} c_{jo} K_{kk} \tilde K_{pp} \F{g}{hk} \F{i}{jk} \tilde 
    f^{l}_{\phantom{l}mp} \tilde f^{n}_{\phantom{n}op} 
  \end{equation}
  and
  \begin{equation}
    \frac{6 \pi^2 \h^2 \log \Lambda}{\abs{z-w}^4} \sum_{ghij} \sum_{klm} c_{gl} c_{hl} c_{im} c_{jm} K_{kk} 
    \tilde K_{ll} \tilde K_{mm} \F{g}{ik} \F{h}{jk} + c_{lg} c_{lh} c_{mi} c_{mj} 
    \tilde K_{kk} K_{ll} K_{mm} \tilde f^{g}_{\phantom{g}ik} \tilde f^{h}_{\phantom{h}jk} .
  \end{equation}
\end{subequations}
Using the fact that the matrices $K_{ij}$ and $\tilde K_{ij}$ are
positive-definite it is simple to see that they both vanish if and
only if Eq.~(\ref{eq:marginal-c}) are satisfied (the condition is
only sufficient for general semi-simple groups).

Actually there's another piece of information that we learn out of
this construction: \emph{the \textsc{ope} coefficients among the
  currents used for the deformation do not change} with the
deformation. As we will see in the next section, this implies that the
the effect of the deformation is completely captured by a
transformation in the charge lattice of the theory.

\subsection{Algebraic structure of current-current deformations}

The result of the previous section can be recast in more abstract
terms: consider a conformal field theory whose holomorphic and
anti-holomorphic Kac Moody algebras correspond to Lie algebras
$\mathfrak{g} $ and $\bar{\mathfrak{g}}$, which respectively admit the
abelian subalgebras $\mathfrak{h}$ and $\bar{\mathfrak{h}}$. Then each
pair $\mathfrak{u}(1)^d \subseteq \mathfrak{h}, \mathfrak{u}(1)^{\bar d} \subseteq
\bar{\mathfrak{h}}$ gives rise to a new family of conformal field
theory containing those algebras (the ones defined in
Eqs.~\eqref{eq:deforming-opes}).  It is safe to assume (at least in
the compact case) that the \textsc{cft} remains unitary and that its
Hilbert space still decomposes into tensor products of irreducible
highest weight representations of $\mathfrak{h} \times \bar{\mathfrak{h}}$
(from now on $\dim \mathfrak{h} = d $ and $\dim \bar{\mathfrak{h}} =
\bar d$)
\begin{equation}
  \mathcal{H} = \sum_{Q,\bar Q} \mathcal{H}_{Q, \bar Q} \mathcal{V}_Q \otimes \mathcal{V}_{\bar Q} ,
\end{equation}
where we used the fact that those representations are completely
characterized by their charges $(Q , \bar Q ) \in (\mathfrak{h}^\ast,
\bar{\mathfrak{h}}^\ast )$ and the corresponding conformal weights are
given by $h = \nicefrac{1}{2} \kappa ( Q, Q) $ and $\bar h =
\nicefrac{1}{2} \bar \kappa (\bar Q, \bar Q)$, where $\kappa $ and
$\bar \kappa $ are the Killing forms respectively on $\mathfrak{g}$
and $\bar{\mathfrak{g}}$ restricted on $\mathfrak{h}$ and
$\bar{\mathfrak{h}}$.  This set of charges naturally forms a lattice
$\Lambda $ when equipped with the pairing $\braket{,} = \kappa - \bar
\kappa$.  \marginlabel{Boost on the charge lattice}Using in example
deformation theory as in~\cite{Forste:2003km} one can see that the
effect of the deformation is completely captured by an $O (d,\bar d) $
pseudo-orthogonal transformation of this charge lattice $\Lambda
\subset\mathfrak{h}^\ast \times \bar {\mathfrak{h}}^\ast$, \emph{ie}
can be described in terms of the identity component of the group $O
(d, \bar d)$.  Moreover, since the charges only characterise the
$\mathfrak{h} \times \bar{\mathfrak{h}}$ modules up to automorphisms
of the algebras, $O (d) \times O (\bar d) $ transformations don't
change the \textsc{cft}.  Hence the deformation space is given by:
\begin{equation}
  D_{\mathfrak{h}, \bar{\mathfrak{h}}} \sim O (d, \bar d) / \left(O(d)
  \times O(\bar d) \right).
\end{equation}
The moduli space is obtained out of $D_{\mathfrak{h},
\bar{\mathfrak{h}}}$ after the identification of the points giving
equivalent \textsc{cft}s\footnote{Although we will concentrate on
\textsc{wzw} models it is worth to emphasize that this
construction is more general.}.

In the case of \textsc{wzw} models on compact groups, all maximal
abelian subgroups are pairwise conjugated by inner automorphisms. This
implies that the complete deformation space is $D = O (d, d) /
\left(O(d) \times O(d) \right)$ where $d$ is the rank of the group. The
story is different for non-semi-simple algebras, whose moduli space is
larger, since we get different $O (d, \bar d) / \left( O(d) \times O(\bar
  d) \right)$ deformation spaces for each (inequivalent) choice of the
abelian subalgebras $\mathfrak{h} \subset \mathfrak{g}$ and
$\bar{\mathfrak{h}} \subset \bar{\mathfrak{g}}$. We'll see an example of
this in the next chapter where deforming a $SL(2, \setR)$ \textsc{wzw}
model (Sec.~\ref{sec:deformed-sl2} and Sec.~\ref{sec:blackstring})
will give rise to a much richer structure than in the $SU(2)$ case
(Sec.~\ref{sec:deformed-su2}).

\bigskip

\marginlabel[\textsc{wzw} models!Parafermion
decomposition]{Parafermion decomposition}Truly marginal deformations
of \textsc{wzw} model single out abelian subalgebras of the model. It
is then natural that an important tool in describing these
current-current deformations comes from the so-called parafermion
decomposition. The highest-weight representation for a
$\hat{\mathfrak{g}}_k$ graded algebra can be decomposed into
highest-weight modules of a Cartan subalgebra $\hat{\mathfrak{h}} \subset
\hat{\mathfrak{g}}_k$ as follows~\cite{Gepner:1986hr,Gepner:1987sm}:
\begin{equation}
  \mathcal{V}_{\hat \lambda } \simeq \bigoplus_{\mu \in \Gamma_k}
\mathcal{V}_{\hat \lambda, \mu} \otimes \bigoplus_{\delta \in Q_l
(\mathfrak{g})} \mathcal{V}_{\mu + k \delta},
\end{equation}
where $\hat \lambda$ is an integrable weight of
$\hat{\mathfrak{g}}_k$, $\mathcal{V}_{\hat \lambda, \mu}$ is the
highest-weight module for the generalized $\hat{\mathfrak{g}}_k /
\hat{\mathfrak{h}}$ parafermion, $Q_l ( \mathfrak{g} )$ is the
long-root lattice and $\Gamma_k = P ( \mathfrak{g} )/Q_l(
\mathfrak{g} )$ with $P ( \mathfrak{g} ) $ the weight lattice. As
a consequence, the \textsc{wzw} model based on
$\hat{\mathfrak{g}}_k$ can be represented as an orbifold model:
\begin{equation}
  \label{eq:ParafermionRep}
  \hat{\mathfrak{g}}_k \simeq \left( \hat{\mathfrak{g}}_k/ \hat{\mathfrak{h}}
\otimes t_{\Lambda_k} \right) / \Gamma_k,
\end{equation}
where $t_{\Lambda_k}$ is a toroidal \textsc{cft} with charge lattice,
included in the $\hat{\mathfrak{g}}_k$ one, defined as $\Lambda_k =
\set{ \left( \mu, \bar \mu \right) \in P ( \hat{\mathfrak{g}}) \times
  P (\hat{\mathfrak{g}} ) | \mu - \bar \mu = k Q_l
  (\hat{\mathfrak{g}})}$. In our case the advantage given by using this
representation relies on the fact that $\Gamma_k$ acts trivially on
the coset and toroidal model algebras; then, if we identify
$\hat{\mathfrak{h}} $ and $\bar{\hat{\mathfrak{h}}}$ with the graded
algebras of $t_{\Lambda_k}$, the deformation only acts on the toroidal
lattice and the deformed model can again be represented as an
orbifold:
\begin{equation}
  \label{eq:DefParafermionRep}
  \hat{\mathfrak{g}}_k ( \mathcal{O} ) \simeq \left( \hat{\mathfrak{g}}_k/
\hat{\mathfrak{h}} \otimes t_{\mathcal{O}\Lambda_k} \right) /
\Gamma_k,
\end{equation}
where $\mathcal{O}$ is an operator in the moduli space. In other words
this representation is specially useful because it allows to easily
single out the sector of the theory that is affected by the
deformation. As we'll see in the next section this simplifies the task
of writing the corresponding Lagrangian.

\bigskip

\marginlabel{Symmetric and asymmetric deformations}In the following we
will separate this kind of deformations into two categories: those who
give rise to \emph{symmetric} deformations, \emph{i.e.} the ones where
$c_{ij} = \delta_{ij}$ and $J^i \left( z \right)$ and $\tilde J^j
\left( \bar z \right)$ represent the same current in the two chiral
sectors of the theory and the \emph{asymmetric} ones where the
currents are different and in general correspond to different
subalgebras. This distinction is somehow arbitrary, since both
symmetric and asymmetric deformations act as $O \left( d, \bar d
\right)$ rotations on the background fields. It is nonetheless
interesting to single out the asymmetric case. In the special
situation when one of the two currents belongs to an internal $U
\left( 1 \right)$ (coming from the gauge sector in the heterotic or
simply from any $U\left( 1 \right)$ subalgebra in the type II), it is
in fact particularly simple to study the effect of the deformation,
even from the spacetime field point of view; there in fact, the
expressions for the background fields are exact (at all order in
$\alpha^\prime$ and for every value of the level $k$) as we will show
in Sec.~\ref{sec:no-renorm-theor}.

\subsection{Background fields and symmetric deformations}
\label{sec:backgr-fields-symm}

Before moving to the asymmetric deformations we're interested in, let
us consider briefly symmetric deformations (also called
\emph{gravitational}) which are those that have received by far the
most attention in
literature~\cite{Hassan:1992gi,Giveon:1994ph,Forste:1994wp,Forste:2003km,Detournay:2005fz}.
Specialising Eq.~\eqref{eq:Current-current-deform} to the case of one
only current we can write the small deformation Lagrangian as:
\begin{equation}
  S = S_{\textsc{wzw}} + \delta \kappa^2 \int \di^2 z \: J (z ) \bar J ( \bar z )
\end{equation}
This infinitesimal deformation has to be integrated in order to give a
Lagrangian interpretation to the \textsc{cft} described in the previous
section. Different approaches are possible, exploiting the different
possible representations described above.
\begin{itemize}
\item A possible way consists in implementing \emph{an $O(d,d)$
    rotation on the background fields~\cite{Hassan:1992gi}}. More
  precisely, one has to identify a coordinate system in which the
  background fields are independent of $d$ space dimensions and metric
  and $B$ field are written in a block diagonal form. In this way the
  following matrix is defined:
  \begin{equation}
    M = \left( \begin{tabular}{c|c}
        $\hat{g}^{-1}$ &  $-\hat{g}^{-1} \hat{B}$ \\ \hline
        $\hat{B} \hat{g}^{-1}$ & $\hat{g} - \hat{B}\hat{g}^{-1}\hat{B}$
      \end{tabular}
    \right),
  \end{equation}
  where $\hat g $ and $\hat B$ are the pull-backs
  of the metric and Kalb--Ramond field on the $p$ selected
  directions. Then the action of the $O(d,d)$ group on these fields
  and dilaton is given by:
  \begin{subequations}
    \begin{align}
      M & \to M^\prime = \Omega M \Omega^{t}, \label{eq:OddMetric}\\
      \Phi & \to \Phi^\prime = \Phi + \frac{1}{2} \log \left( \frac{\det
          \hat g}{\det \hat g^\prime} \right),\label{eq:OddDilaton}
    \end{align}    
  \end{subequations}
  where $\hat g^\prime$ is the metric after the
  transformation~\eqref{eq:OddMetric} and $\Omega \in O (d,d)$. It
  must be emphasized that this transformation rules are valid at the
  lowest order in $\alpha^\prime$ (but for finite values of the
  deformation parameters).  So, although the model is exact, as we
  learn from the \textsc{cft} side, the field expressions that we find
  only are true at leading order in $\alpha^\prime$.
\item An alternative approach uses the \emph{parafermion
    representation} Eq.~\eqref{eq:DefParafermionRep} (see
  \emph{e.g.}~\cite{Forste:2003km}). In practice this amounts to
  writing an action as the sum of the $G/H$ parafermion and a deformed
  $H$ part and finding the appropriate T-duality transformation
  (realizing the orbifold) such that for zero deformation the
  \textsc{wzw} on $G$ is recovered, in accordance with
  Eq.~\eqref{eq:ParafermionRep}.
\item Finally, another point of view (inspired by the parafermionic
  representation), consists in \emph{identifying the deformed model
    with a $\left( G \times H \right) / H $ coset model}, in which the
  embedding of the dividing group has a component in both
  factors~\cite{Giveon:1994ph}. The gauging of the component in $G$
  gives the parafermionic sector, the gauging of the component in $H$
  gives the deformed toroidal sector and the coupling term
  (originating from the quadratic structure in the fields introduced
  for the gauging) corresponds to the orbifold projection\footnote{An
    instanton-correction-aware technique that should overcome the
    first order in $\alpha^\prime$ limitation for gauged models has
    been proposed in~\cite{Tseytlin:1994my}. In principle this can be
    used to get an all-order exact background when we write the
    deformation as a gauged model. We will not expand further in this
    direction, that could nevertheless be useful to address issues
    such as the stability of the black string (see
    Sec. \ref{sec:blackstring}).}.
\end{itemize}

\section{Background fields for the asymmetric deformation}
\label{sec:backgr-fields-asymm-1}

Let us now consider the less-known case of asymmetric deformations, in
which the two sets of currents $J_i$ and $\bar J_j$ come from distinct
sectors of the theory. The archetype of such construction is what we
get considering an heterotic super-\textsc{wzw} model on a group $G$
at level $k$ and adding an exactly marginal operator built from the
total Cartan currents of $\mathfrak{g}$ (so that it preserves the
local $N=(1,0)$ superconformal symmetry of the theory):
\begin{equation}
  S = S_{\textsc{wzw} } + \frac{\sqrt{kk_g}}{2\pi} \int \di^2 z \  \sum_{a} \h_{a} 
  \left( J^{a} (z) - \frac{i}{k} \F{a}{\textsc{mn}} : 
    \psi^{\textsc{m}} \psi^{\textsc{n}} : \right) \bar J (\bar{z} )
\end{equation}
where the set $\{ \h_{a}\} $ are the parameters of the deformation,
$J^a$ are currents in the maximal torus $T \subset G$ and $\bar J (\bar z
)$ is a right moving current of the Cartan subalgebra of the heterotic
gauge group at level $k_g$. Such a deformation is always truly
marginal since the $J_a $ currents commute.

\marginlabel{Reading the squashed group fields}It is not completely
trivial to read off the deformed background fields that correspond to
the deformed action. A possible way is a method involving a
Kaluza--Klein reduction as in~\cite{Horowitz:1995rf}. For simplicity
we will consider the bosonic string with vanishing dilaton and just
one operator in the Cartan subalgebra $\mathfrak{k}$. After
bosonization the right-moving gauge current $\bar J$ used for the
deformation has now a left-moving partner and can hence be written as
$\bar J = \imath \bar \partial \varphi $, $\varphi \left( z, \bar z
\right) $ being interpreted as an internal degree of freedom.  The
sigma-model action is recast as
\begin{equation}
  \label{eq:KK-action}
  S = \frac{1}{2 \pi} \int \di^2 z \: \left( G_{\textsc{mn}} + B_{\textsc{mn}} \right)
  \partial x^{\textsc{m}} \bar \partial x^{\textsc{n}},
\end{equation}
where the $x^{\textsc{m}}, \textsc{m}=1,\ldots,d+1$ embrace the group coordinates
$x^\mu, \mu = 1,\ldots,d$ and the internal $x^{d+1} \equiv \varphi$:
\begin{equation}
  x^{\textsc{m}} = \left( \begin{tabular}{c}
      $  x^\mu $\\ \hline
      $ \varphi $
    \end{tabular}\right).
\end{equation}
If we split accordingly the background fields, we obtain the following
decomposition:
\begin{align}
  G_{\textsc{mn}} = \left( \begin{tabular}{c|c}
      $G_{\mu\nu } $& $ G_{\varphi \varphi } A_\mu $\\ \hline
      $ G_{\varphi \varphi } A_\mu $& $G_{\varphi \varphi }$
    \end{tabular}\right), &&
  B_{\textsc{m}\textsc{n}} =  \left( \begin{tabular}{c|c}
      $B_{\mu\nu}$ & $B_{\mu \varphi }$ \\ \hline
      $-B_{\mu \varphi }$ & 0
    \end{tabular}\right),
\end{align}
and the action becomes:
\begin{multline}
  S = \frac{1}{2 \pi} \int \di z^2 \left\{ \left( G_{\mu \nu } + B_{\mu \nu} \right)
    \partial x^\mu \bar \partial x^\nu
    + \left( G_{\varphi \varphi } A_\mu + B_{\mu \varphi } \right) \partial x^\mu \bar \partial \varphi \right.\\
  \left. + \left( G_{\varphi \varphi } A_\mu - B_{\mu \varphi } \right) \partial \varphi \bar \partial
    x^\mu + G_{\varphi \varphi } \partial \varphi \bar \partial \varphi\right\}.
\end{multline}
  
We would like to put the previous expression in such a form that spacetime
gauge invariance,
\begin{align}
  A_\mu & \to A_\mu + \partial_\mu \lambda,  \\
  B_{\mu 4} & \to B_{\mu 4 } + \partial_\mu \eta,
\end{align}
is manifest. This is achieved as follows:
\begin{multline}
  S = \frac{1}{2 \pi } \int \di^2 z \: \left\{\left( \hat G_{\mu \nu } + B_{\mu
        \nu }\right) \partial x^\mu \bar \partial x^\nu + B_{\mu \varphi } \left( \partial x^\mu \bar \partial \varphi
      - \partial \varphi \bar \partial x^\mu \right) \right. +\\ \left. + G_{\varphi \varphi } \left( \partial
      \varphi + A_\mu \partial x^\mu \right) \left( \bar \partial \varphi + A_\mu \bar \partial x^\mu
    \right)\right\},
\end{multline}
where $\hat G_{\mu \nu }$ is the Kaluza--Klein metric
\begin{equation}
  \hat G_{\mu \nu } = G_{\mu \nu } - G_{\varphi \varphi } A_{\mu } A_{\nu    }.
\end{equation}
We can then make the following identifications: 
\begin{subequations}
  \label{eq:KK-fields}
  \begin{align}
    \hat G_{\mu \nu } &=  k \left( \mJ_\mu \mJ_\nu - 2 \h^2
      \tilde{\mJ}_\mu \tilde{\mJ}_\nu \right) \label{eq:KK-metric}\\
    B_{\mu \nu } &= k \mJ_\mu \wedge \mJ_\nu     \label{eq:B-field} \\
    B_{\mu \varphi } &= G_{\varphi \varphi } A_\mu  =  \h \sqrt{k k_g} \tilde{\mJ}_\mu, \\
    A_{\mu} &=  2 \h \sqrt{\frac{k}{k_g}} \tilde{\mJ}_\mu, \label{eq:KK-em-field} \\
    G_{\varphi \varphi } & = \frac{k_g}{2}.
  \end{align}
\end{subequations}
where $\tilde{\mJ}$ is the Maurer-Cartan current chosen for the
deformation.  Let us now consider separately the background fields we
obtained so to give a clear geometric interpretation of the
deformation, in particular in correspondence of what we will find to
be the maximal value for the deformation parameters $\h_a$.


\paragraph{The metric.}

\marginlabel{Metric on squashed group and decompactification limit}
\danger According to Eq.~\eqref{eq:KK-metric}, in terms of the target
space metric, the effect of this perturbation amounts to inducing a
back-reaction that in the vielbein (current) basis is written as:
\begin{equation}
\label{eq:def-metric}
  \braket{\di g, \di g}_{\h} = \sum_{M} \mJ_M \otimes \mJ_M - 2 \sum_{a}
  \h_{a}^{2} \mJ_a \otimes \mJ_a = \sum_\mu \mJ_\mu \otimes \mJ_\mu + \sum_a \left( 1 - 2 \h_a^2
  \right) \mJ_a \otimes \mJ_a
\end{equation}
where we have explicitly separated the Cartan generators. From this
form of the deformed metric we see that there is a ``natural'' maximal
value $\h_a = 1/ \sqrt{2}$ where the contribution of the $\mJ_a \otimes
\mJ_a $ term changes its sign and the signature of the metric is thus
changed. One could naively think that the maximal value $\h_a= 1/
\sqrt{2}$ can't be attained since the this would correspond to a
degenerate manifold of lower dimension; what actually happens is that
the deformation selects the the maximal torus that decouples in the
$\h_{a} = \h \to 1/ \sqrt{2}$ limit.

To begin, write the general element $g \in G $ as $g = h t $
where $h \in G/ T, t \in T$. Substituting
this decomposition in the expression above we find:
\mathindent=0em
\begin{multline}
  \braket{\di \left( h t\right), \di \left( h t\right)}_{\h} =
  \braket{ \left( h t\right)^{-1} \di \left( h t\right) \left( h
      t\right)^{-1} \di \left( h t\right) } - \sum_a 2 \h_a^2
  \braket{ T_a \left( h t \right)^{-1} d \left( h t\right)^{-1} } ^{
    2} = \\
  = \braket{ h^{-1} \di h h^{-1} \di h } + 2 \braket{ \di t \ t^{-1}
    h^{-1} \di h } + \braket{ t^{-1} \di t\ t^{-1} \di t} +\\
  - \sum_{a} 2 \h_a^2 \left( \braket{ T_a t^{-1} h^{-1} \di h } +
    \braket{ T_a t^{-1} \di t}\right)^{ 2}
\end{multline}
\mathindent=\oldindent
let us introduce a coordinate system $\left( \gamma_\mu , \psi_a \right)$ such as
the element in $G/ T$ is parametrized as $h = h\left(
  \gamma_\mu \right)$ and $t$ is written explicitly as:
\begin{equation}
  t = \exp \left\{ \sum_{a} \psi_{a} T_{a} \right\}  =  
\prod_{a} e^{ \psi_{a} T_{a}}    
\end{equation}
it is easy to see that since all the $T_a $ commute $t^{-1} \di t = \di t\ 
t^{-1} = \sum_a T_a \di \psi_a $. This allows for more simplifications in the
above expression that becomes:
\begin{multline}
  \braket{\di \left( h t\right), \di \left( h t\right)}_{\h} =
  \braket{ h^{-1} \di h h^{-1} \di h } + 2 \sum_a \braket{ T_a h^{-1} \di
    h} \di \psi_a + \sum_a \di \psi_a \di \psi_a + \\
  - \sum_{a} 2 \h_a^2\left( \braket{ T_a h^{-1} \di h } + \di
    \psi_a\right)^{ 2} = \braket{ h^{-1} \di h h^{-1} \di h } - \sum_{a} 2
  \h_a^2  \braket{ T_a h^{-1} \di h }^{ 2} + \\
  + 2 \sum_a \left( 1 - 2\h_a^2 \right) \braket{ T_a h^{-1} \di h
  } \di \psi_a + \sum_a \left( 1 - 2 \h_a^2\right) \di \psi_a \di \psi_a
\end{multline}
if we reparametrise the $\psi_a $ variables as:
\begin{equation}
  \psi_a = \frac{\hat \psi_a }{\sqrt{1-2 \h_a}}  
\label{rescale}
\end{equation}
we get a new metric $\braket{\cdot, \cdot}_\h^\prime $ where we're free to 
take the $\h_a \to 1/ \sqrt{2}$ limit:
\begin{multline}
  \braket{\di \left( h t\right), \di \left( h t\right)}_{\h}^\prime
  = \braket{ h^{-1} \di h h^{-1} \di h } - \sum_{a} 2
  \h_a^2 \braket{ T_a h^{-1} \di h }^{ 2} + \\ +2
  \sum_a \sqrt{ 1 - 2\h_a^2} \braket{ T_a h^{-1} \di h }
  \di \hat \psi_a + \sum_a \di \hat \psi_a \di \hat \psi_a
\end{multline}
and get:
\begin{equation}
  \label{eq:G/TxT-metric}
  \braket{\di \left( h t\right), \di 
    \left( h t\right)}_{\nicefrac{1}{ \sqrt{2}}}^\prime =
  \left[ \braket{
      h^{-1} \di h h^{-1} \di h } - \sum_{a} \braket{ T_a
        h^{-1} \di h }^{  2} \right] + 
  \sum_a \di \psi_a  \di \psi_a
\end{equation}
where we can see the sum of the restriction of the Cartan-Killing
metric\footnote{This always is a left-invariant metric on $G/H$. A
  symmetric coset doesn't admit any other metric. For a more complete
  discussion see Sec.~\ref{sec:geom-squash-groups}} on $T_h G/ T$
and the metric on $T_tT = T_tU\left( 1\right)^r$. In other words the
coupling terms between the elements $h \in G/T$ and $t \in T$ vanished and
the resulting metric $\braket{\cdot,\cdot}^\prime_{\nicefrac{1}{ \sqrt{2}}} $
describes the tangent space $T_{ht}$ to the manifold $G/T \times T$.

\paragraph{Other Background fields.}
\label{sec:backgound-fields}

The asymmetric deformation generates a non-trivial field strength for
the gauge field, that from Eq.~\eqref{eq:KK-em-field} is found to be:
\begin{equation}
  \label{eq:def-F-field}
  F^a = 2 \sqrt{\frac{k}{k_g}} \h_a \, \di \,\mathcal{J}^a = -
  \sqrt{\frac{k}{ k_g}} \h_a  f\ud{a}{\mu\nu } J^\mu \land J^\nu 
\end{equation}
(no summation implied over $a$).\\
On the other hand, the $B$-field~\eqref{eq:B-field} is not changed, but the
physical object is now the 3-form $H_{[3]}$:
\begin{equation}
  \label{eq:def-H-field}
  H_{[3]} = \di B - G_{\varphi \varphi } A^a \land \di A^a 
  = \frac{1}{3!} f_{\textsc{mnp}} 
  \mathcal{J}^{\textsc{m}} \land \mathcal{J}^{\textsc{n}} \land \mathcal{J}^{\textsc{p}}  
  -  \sum_a  \h_{a}^2 \, f_{a \textsc{np}} \,
  \mathcal{J}^a \land \mathcal{J}^{\textsc{n}} \land \mathcal{J}^{\textsc{p}} ,
\end{equation}
where we have used the Maurer-Cartan structure equations.  At the point
where the fibration trivializes, $\h_a = 1/\sqrt{2}$, we are left with:
\begin{equation}
  H_{[3]} = \frac{1}{3!} f_{\mu \nu \rho}\, 
  \mathcal{J}^{\mu} \land \mathcal{J}^{\nu} \land \mathcal{J}^{\rho}.
\end{equation}
So only the components of $H_{[3]}$ ``living'' in the coset $G/
T$ survive the deformation. They are not affected of course by the
rescaling of the coordinates on $T$.

\paragraph{A trivial fibration.}
\label{sec:trivial-fibration}

The whole construction can be reinterpreted in terms of fibration as
follows. The maximal torus $T$ is a closed Lie subgroup of the Lie
group $G$, hence we can see $G$ as a principal bundle
with fiber space $T$ and base space $G/ T$
\cite{Nakahara:1990th}
\begin{equation}
  G \xrightarrow{T} G/T  
\end{equation}
The effect of the deformation consists then in changing the fiber and the
limit value $\h_a = 1/ \sqrt{2}$ marks the point where the fibration becomes
trivial and it is interpreted in terms of a gauge field whose strength is
given by the canonical connection on $G/ T$
\cite{Kobayashi:1969}.

\section{Geometry of squashed groups}
\label{sec:geom-squash-groups}

\danger In order to describe the squashed group manifolds that we
obtain via asymmetric deformation we need to generalize the discussion
on group manifold geometry presented in
Sec.~\ref{sec:target-space-wzw}.  Let $\set{\hat \theta^\alpha }$ be a set of
one-forms on a manifold $\mathcal{M}$ satisfying the commutation
relations
\begin{equation}
  \comm{ \hat \theta^\beta , \hat \theta^\gamma  } = \F{\alpha }{\beta \gamma } \hat \theta^\alpha  
\end{equation}
as it is the case when $\hat \theta^\alpha $ are the Maurer--Cartan one-forms of
Eq.~(\ref{eq:MC-1-forms}) and $\F{\alpha }{\beta \gamma }$ the structure constants
for the algebra. We wish to study the geometry of the Riemann manifold
$\mathcal{M}$ endowed with the metric
\begin{equation}
  g = g_{\alpha \beta } \hat \theta^\alpha  \otimes \hat \theta^\beta  .
\end{equation}
In general such a metric will have a symmetry $G \times G^\prime$ where $G$ is
the group corresponding to the structure constants $\F{\alpha }{\beta \gamma }$ and
$G^\prime \subset G$. The maximally symmetric case, in which $G^ \prime = G$ is
obtained when $g$ is $G$-invariant, \emph{i.e.} when it satisfies
\begin{equation}
  \F{\alpha }{\beta \gamma } g_{\alpha \delta } + \F{\alpha }{\delta \gamma } g_{\alpha \beta }  = 0 .
\end{equation}
for compact groups this condition is fulfilled by the Killing metric
in Eq.~\eqref{eq:Killing-metric}.

The connection one-forms $\omega\ud{\alpha}{\beta}$ are uniquely determined by the
compatibility condition and the vanishing of the torsion.
Respectively:
\begin{gather}
  \di g_{\alpha \beta } - \omega\ud{\gamma}{\alpha} g_{\gamma \beta } - \omega\ud{\gamma}{\beta} g_{\gamma \alpha } = 0 \\
  \di \hat  \theta^\alpha + \omega^\alpha_{\phantom{\alpha}\beta } \land \hat \theta^\beta = T^\alpha = 0 
\end{gather}
As it is shown in~\cite{Mueller-Hoissen:1988cq}, if $g_{\alpha \beta }$ is
constant, the solution to the system can be put in the form
\begin{equation}
  \omega\ud{\alpha}{\beta}  = -\Di{\alpha }{\beta \gamma } \hat \theta^\gamma 
\end{equation}
where $\Di{\alpha }{\beta \gamma } = 1/2 \F{\alpha }{\beta \gamma } - K\ud{\alpha}{\beta\gamma }$ and
$K\ud{\alpha}{\beta\gamma }$ is a tensor (symmetric in the lower indices) given by:
\begin{equation}
  K\ud{\alpha}{\beta \gamma} = \frac{1}{2} g^{\alpha \kappa } \F{\delta }{\kappa \beta } g_{\gamma \delta } + \frac{1}{2} g^{\alpha \kappa } \F{\delta }{\kappa \gamma } g_{\beta \delta } .  
\end{equation}
\marginlabel{Curvature tensors on squashed groups}Just as in
Sec.~\ref{sec:target-space-wzw} we define the curvature two-form
$R\ud{i}{j}$ and the Riemann tensor which now reads:
\begin{equation}
  R\ud{\alpha}{\beta \gamma \delta} =  \Di{\alpha}{\beta \kappa } \F{\kappa}{\gamma \delta } + \Di{\alpha}{\kappa  \gamma} \Di{\kappa }{\beta \delta } - \Di{\alpha}{\kappa  \delta } \Di{\kappa  }{\beta \gamma }
\end{equation}
and the corresponding Ricci tensor:
\begin{equation}
  Ric_{\beta \delta} = \Di{\alpha}{\beta \kappa } \F{\kappa}{\alpha  \delta } - \Di{\alpha}{\kappa  \delta } \Di{\kappa }{\beta \alpha }
\end{equation}
In particular for $g_{ij} \propto \delta_{ij}$, $K = 0$ so that we recover the
usual Maurer--Cartan structure equation Eq.~(\ref{eq:MC-structure})
and the expressions in Eqs.~(\ref{eq:group-curvatures}).

Let us now specialize these general relations to the case of the
conformal model with metric given in Eq.~(\ref{eq:def-metric}).  The
$\hat \theta^i$'s are the Maurer--Cartan one-forms for the group $G$ and
the metric $g_{\ssc{ab}} $ is
\begin{equation}
  g_{\ssc{ab}} = \begin{cases}
    \delta_{\mu \nu} & \text{if } \mu, \nu \in G/H \\
    \left( 1 - \frac{\h^2}{2} \right) \delta_{ab} & \text{if } a,b \in H
  \end{cases}
\end{equation}
where $H$ is (a subgroup of) the Cartan torus $H \subset G$.  It is quite
straightforward to show that the Ricci tensor is given by\footnote{In
  the $SU(2)$ case this would be
  \begin{align}
      g = \begin{pmatrix}
        1 \\ 
        & 1 \\
        & & 1 - \frac{\h^2}{2}
      \end{pmatrix} && Ric = \begin{pmatrix}
        1 + \frac{\h^2}{2}\\ 
        & 1 + \frac{\h^2}{} \\
        & & \left( 1 - \frac{\h^2}{2} \right)^2
      \end{pmatrix}
    \end{align}
    where we chose $J_3$ as Cartan generator.}:
  \begin{equation}
    Ric_{\ssc{ab}} = \begin{cases}
      \frac{1}{2}\left( g^\ast + \h^2 \right) g_{\mu \nu} & \text{if } \mu,\nu \in G/H \\
      g^\ast /2 \left( 1 - \frac{\h^2}{2} \right) g_{ab} & \text{if } a,b \in H.
    \end{cases}
  \end{equation}
  whence we can read the (constant) Ricci scalar
  \begin{equation}
    R =  \frac{g^\ast}{2} \dim G + \frac{\h^2}{2} \left( \dim G - \rank G \left( 1 + \frac{g^\ast}{2} \right) \right)
  \end{equation}

\bigskip

\marginlabel{Geometry on $G/H$ cosets} Particular attention should be
devoted to the limit case $\h = \sqrt{2} $ in which the Cartan torus
decouples and we are left with the geometry of the $G/T$ coset.  In
this case it is useful to explicitly write down the commutation
relations, separating the generators of $T$ and $G/T$:
\begin{subequations}
  \begin{align}
    \comm{T_m, T_n} &= \F{o}{mn} T_o & \comm{T_m, T_\nu} &=
    \F{\omega}{m \nu} T_\omega    \\
    \comm{T_\mu , T_\nu} &= \F{o}{\mu \nu } T_o + \F{\omega }{\mu \nu
    } T_\omega
  \end{align}
\end{subequations}
Of course there are no $\F{\omega }{mn}$ terms since $T$ is a group.
$G/T$ is said to be symmetric if $\F{\omega }{\mu \nu } \equiv 0$,
\emph{i.e.} if the commutator of any couple of coset elements lives in
the dividing subgroup.  In this case a classical theorem states that
the coset only admits one left-invariant Riemann metric that is
obtained as the restriction of the Cartan-Killing metric defined on
$G$ (see \emph{eg}~\cite{Kobayashi:1969}).  This is not the case when
$T$ is the maximal torus (except for the most simple case $G = SU
\left( 2 \right)$ where maximal torus and maximal subgroup are the
trivial $U(1)$) and the coset manifold accepts different
structures. From our point of view this means that even when
considering deformations and cosets of compact groups where the Cartan
subalgebra is unique (up to inner automorphisms), in general we expect
different possible outcomes depending on how the gauging is performed
(see in particular the $SU(3)$ case studied in detail in
Sec.~\ref{sec:new-comp}).

\marginlabel{K\"ahler structure on $G/H$} These homogeneous manifolds
enjoy many interesting properties. As we pointed out many times
already, the best part of them can be interpreted as consequence of
the presence of an underlying structure that allows to recast all the
geometric problems in Lie algebraic terms.  There's however at least
one intrinsically geometric property that it is worth to emphasize
since it will have many profound implications in the following. All
these spaces can be naturally endowed with complex structures by using
positive and negative roots as holomorphic and anti-holomorphic
generators. This structure doesn't in general correspond to a unique
left-invariant Riemann metric. On the other hand there always exists
such a metric that is also K\"ahler. In fact one can easily show that
the $\left( 1, 1\right)$ form defined as:
\begin{equation}
  \label{eq:two-form}
  \omega = \frac{\imath }{2} \sum_{\alpha >0} c_\alpha \mJ^\alpha \land \mJ^{\bar \alpha }  
\end{equation}
is closed if and only if for each subset of roots $\set{\alpha, \beta,
  \gamma} $ such as $\alpha = \beta + \gamma $, the corresponding real
coefficients $c_\alpha $ satisfy the condition $c_\alpha = c_\beta +
c_\gamma $. Of course this is equivalent to say that the tensor
\begin{equation}
  \label{eq:coset-Kaehler-metric}
  g = \sum_{\alpha >0} c_\alpha \mJ^{\alpha } \otimes \mJ^{\bar \alpha}  
\end{equation}
is a K\"ahler metric on $G/T$ \cite{Borel:1958ch,Perelomov:1987va}.

\section{A no-renormalization theorem}
\label{sec:no-renorm-theor}

As we've said many times, \textsc{wzw} models are exact solutions
keeping their geometrical description at all orders in $\alpha^\prime $, the
only effect of renormalization being a shift in the level
(Sec.~\ref{sec:path-integral}). Here we want to show that this same
property is shared by our asymmetrically deformed models that hence
provide a geometric solution at every order in perturbation.

As we emphasized above, in studying symmetrically deformed
\textsc{wzw} models, \emph{i.e.}  those where the deformation operator
is written as the product of two currents belonging to the same sector
$\mathcal{O} = \lambda J \bar J$, one finds that the Lagrangian formulation
only corresponds to a small-deformation approximation. For this reason
different techniques have been developed so to read the background
fields at every order in $\lambda $~ but, still, the results are in general
only valid at first order in $\alpha^\prime $ and have to be modified so to take
into account the effect of instanton corrections (. This is not the
case for asymmetrically deformed models, for which the background
fields in Eqs.~\eqref{eq:KK-fields} are exact at all orders in $\h_a$
and for which the effect of renormalization only amounts to the usual
(for \textsc{wzw} models) shift in the level of the algebra $k \to k +
g^\ast $.

\marginlabel{No-renormalization for $SU(2)$}Consider in example the
most simple $SU(2)$ case (which we will review in greater detain in
Sec.~\ref{sec:deformed-su2}). In terms of Euler angles the deformed
Lagrangian is written as: \mathindent=0em
\begin{small}
  \begin{multline}
    S = S_{SU(2)} \left( \alpha, \beta, \gamma \right) + \delta S =
    \frac{k}{4 \pi } \int \di^2 z \: \partial \alpha \bar \partial
    \alpha + \partial \beta \bar \partial \beta + \partial \gamma
    \bar \partial \gamma + 2 \cos \beta \partial \alpha \bar
    \partial \gamma  + \\
    + \frac{\sqrt{k k_g} \h}{2 \pi } \int \di^2 z \: \left( \partial
      \gamma + \cos \beta \partial \alpha \right) \bar I .
  \end{multline}
\end{small}
\mathindent=\oldindent
If we bosonize the right-moving current as $\bar I = \bar \partial \phi$ and add
a standard $U(1) $ term to the action, we get:
\begin{multline}
  S = S_{SU(2)} \left( \alpha, \beta, \gamma \right) + \delta S \left( \alpha, \beta, \gamma, \phi \right)
  + \frac{k_g}{4 \pi } \int \di^2 z \: \partial \phi \bar \partial \phi =\\
  = S_{SU(2)} \left( \alpha, \beta, \gamma + 2 \sqrt{\frac{k_g}{k}} \h \phi \right) +
  \frac{k_g \left( 1 -2 \h^2\right)}{4 \pi} \int \di^2 z \partial \phi \bar \partial \phi
\end{multline}
and in particular at the decoupling limit $\h \to 1/\sqrt{2} $,
corresponding to the $S^2$ geometry, the action is just given by $ S=
S_{SU(2)} \left( \alpha, \beta, \gamma + 2 \sqrt{\frac{k_g}{k}} \h \phi \right)$.  This
implies that our (deformed) model inherits all the integrability and
renormalization properties of the standard $SU (2) $ \textsc{wzw}
model. In other words the three-dimensional model with metric and
Kalb--Ramond field with $SU(2) \times U(1)$ symmetry and a $U(1)$ gauge
field is uplifted to an exact model on the $SU(2)$ group manifold (at
least locally): the integrability properties are then a consequence of
this hidden $SU(2)\times SU(2)$ symmetry that is manifest in higher
dimensions.

The generalization of this particular construction to higher groups is
easily obtained if one remarks that the Euler parametrization for the
$g \in SU \left( 2 \right)$ group representative is written as:
\begin{equation}
  g = e^{\imath \gamma t_3} e^{\imath \beta t_1} e^{\imath \alpha t_2} , 
\end{equation}
where $t_i = \sigma_i /2 $ are the generators of $\mathfrak{su}( 2)$ ($\sigma_i$
being the usual Pauli matrices). As stated above, the limit
deformation corresponds to the gauging of the left action of an
abelian subgroup $T \subset SU \left( 2\right)$. In particular here we chose
$T = \set{h | h = e^{\imath \phi t_3}}$, hence it is natural to find (up to
the normalization) that:
\begin{equation}
  h \left( \phi \right) g \left( \alpha, \beta, \gamma \right) = g \left( \alpha, \beta, \gamma + \phi
  \right) .
\end{equation}
The only thing that one needs to do in order to generalize this result
to a general group $G$ consists in finding a parametrization of $g \in
G$ such as the chosen abelian subgroup appears as a left factor. In
example if in $SU(3) $ we want to gauge the $U \left( 1 \right)^2$
abelian subgroup generated by $\braket{\lambda_3, \lambda_8}$ (Gell-Mann
matrices), we can choose the following parametrization for $g \in SU
(3)$~\cite{Byrd:1997uq}:
\begin{equation}
  g = e^{\imath \lambda_8 \phi } e^{\imath \lambda_3 c } e^{\imath \lambda_2 b } e^{\imath \lambda_3 a }
  e^{\imath \lambda_5 \vartheta } e^{\imath \lambda_3 \gamma  } e^{\imath \lambda_2 \beta  } e^{\imath \lambda_3 \alpha  } . 
\end{equation}

The deep reason that lies behind this property (differentiating
symmetric and asymmetric deformations) is the fact that not only the
currents used for the deformation are preserved (as it happens in both
cases), but here their very expression is just modified by a constant
factor. In fact, if we write the deformed metric as in
Eq.~\eqref{eq:KK-metric} and call $\tilde K^\mu $ the Killing vector
corresponding to the chosen isometry (that doesn't change along the
deformation), we see that the corresponding $\tilde \mJ_\mu^{(\h)} $
current is given by:
\begin{equation}
  \tilde \mJ_\nu^{(\h)} = \tilde K^\mu g_{\mu \nu }^{(\h)} = \left( 1 -
    2 \h^2\right) \tilde \mJ_\nu^{(0)}
\end{equation}
The most important consequence (from our point of view) of this
integrability property is that the \textsc{sugra} action in is
actually \emph{exact} and the only effect of renormalization is the $k
\to k + g^\ast $ shift.

\section{Partition functions}
\label{sec:partition-functions}

Studying the algebraic structure of marginal deformations we have
already stressed that they are completely determined by $O (d,\bar d)$
pseudo-rotations on the charge lattice corresponding to the deforming
operator. This means that a modular invariant partition function is
simply obtained once we write the initial \textsc{wzw} one, single out
those charges and apply the boost. This proves to be a relatively
simple exercise for compact groups but presents technical problems
even in the most simple non-compact example $SL(2,\setR)$ which we
will study in greater detail in Sec.~\ref{sec:deformed-sl2}.

\subsection{$SU(2)$}
\label{sec:su2}

Instead of a general construction, for sake of clearness, we can start
with the most simple -- but showing some general features -- example,
taking the $SU(2)$ group (more extensively studied in
Sec.~\ref{sec:deformed-su2}). Our computation will also include the
$S^2$ limiting geometry. To fix the ideas, we will consider the case
$k_G = 2$, \emph{i.e.} a $U(1)$ algebra generated by one right-moving
complex fermion. As we've seen in Sec.~\ref{sec:cft-approach} the
partition function for the supersymmetric $SU(2)$ model can be written
as
\begin{equation}
  Z\oao{a ; h}{b; g} = \sum_{j,\bar{\jmath}=0}^{(k-2)/2} M^{j
    \bar{\jmath}} \ \chi^j \ \frac{\vartheta \oao{a}{b}}{\eta} \ \bar
  \chi^{\bar \jmath} \ \frac{\bar \vartheta \oao{h}{g}}{\bar \eta}.
\end{equation}
where the $\chi^j$'s are the characters of bosonic $SU(2)_{k-2}$,
$(a,b)$ are the $\setZ_2$ boundary conditions for the left-moving
fermions\footnote{We have removed the contribution of the fermion
  associated to $J^3$ since it is neutral in the deformation process.}
and $(h,g)$ those of the right-moving -- gauge-sector -- ones.  We can
choose any matrix $M^{j \bar \jmath}$ compatible with modular
invariance of $SU(2)_{k-2}$. Furthermore, the supersymmetric
$SU(2)_{k}$ characters can be decomposed in terms of those of the
$N=2$ minimal models:
\begin{equation}
  \chi^j (\tau ) \ \vartheta \oao{a}{b} (\tau , \nu ) = \sum_{m \in
    \setZ_{2k}} \mathcal{C}^{j}_{m} \oao{a}{b} \Theta_{m,k} \left( \tau,
    -\frac{2\nu}{k} \right),
\end{equation}
where the $N=2$ minimal-model characters, determined implicitly by
this decomposition, are given
in~\cite{Kiritsis:1988rv,Dobrev:1987hq,Matsuo:1987cj,Ravanini:1987yg}.

\marginlabel{Boost on the charge lattice and partition function}Our
aim is to implement the magnetic deformation in this formalism.  The
deformation acts as a boost on the left-lattice contribution of the
Cartan current of the supersymmetric $SU(2)_k$ and on the right
current from the gauge sector:
\begin{multline}
  \label{defspherespec}
  \Theta_{m,k} \ \bar \vartheta \oao{h}{g} = \sum_{n,\bar{n}}
  \mathrm{e}^{-\imath \pi g\left(\bar{n}+\frac{h}{2}\right)}
  q^{\frac{1}{2} \left(\sqrt{2k} n+\frac{m}{\sqrt{2k}}\right)^2}
  \bar{q}^{\frac{1}{2}\left(\bar{n}+\frac{h}{2}\right)^2}
  \\
  \longrightarrow \sum_{n,\bar{n}} \mathrm{e}^{-\imath\pi
    g\left(\bar{n}+\frac{h}{2}\right)}\ q^{\frac{1}{2} \left[
      \left(\sqrt{2k} n + \frac{m}{\sqrt{2k}} \right)\cosh x
      + \left( \bar n + \frac{h}{2} \right) \sinh x \right]^2} \\
  \times \bar{q}^{\frac{1}{2} \left[ \left( \bar n + \frac{h}{2}
      \right) \cosh x +\left(\sqrt{2k} n + \frac{m}{\sqrt{2k}}
      \right)\sinh x \right]^2}.
\end{multline}
The boost parameter $x$ is related to the vacuum expectation value of
the gauge field as follows:
\begin{equation}
  \label{defsphpar}
  \cosh x = \frac{1}{1-2 \h^2}.
\end{equation}

We observe that, in the limit $\h^2 \to \h^2_{\mathrm{max}}$, the
boost parameter diverges ($x \to \infty$), and the following
constraints arise:
\begin{equation}
  4 \left( k + 2 \right) n+2m + 2 \sqrt{2k} \bar n + \sqrt{2k} h = 0.
  \label{chargcond}
\end{equation}
Therefore, the limit is well-defined only if the level of the
supersymmetric $SU(2)_k$ satisfies a quantization condition:
\begin{equation}
  k = 2p^2 \ , \ \ p \in \setZ.
\end{equation}
This is exactly the charge quantization condition for the flux of the
gauge field, Eq.~(\ref{chargeq}). Under this condition, the
constraints (\ref{chargcond}) lead to
\begin{subequations}
  \begin{align}
    m+ph &\equiv 0 \mod 2p =: 2pN,  \\
    \bar n &= 2  p n + N, \ \ N \in \setZ_{2p}.
  \end{align}
\end{subequations}
\marginlabel{Decoupled partition function}As a consequence, the $U(1)$
corresponding to the combination of charges orthogonal
to~(\ref{chargcond}) decouples (its radius vanishes), and can be
removed. We end up with the following expression for the $S^2$
partition function contribution:
\begin{equation}
  Z_{S^2}\oao{a ; h}{b ;
    g} = \sum_{j,\bar \jmath} M^{j \bar \jmath}
  \sum_{N \in \setZ_{2p}} \mathrm{e}^{\imath\pi g\left(N + \frac{h}{2}\right)}
  \ \mathcal{C}^{j}_{p(2N-h)} \oao{a}{b} \ \bar{\chi}^{\bar
    \jmath}  ,\label{ZS2}
\end{equation}
in agreement with the result found in~\cite{Berglund:1996dv} by using
the coset construction. The remaining charge $N$ labels the magnetic
charge of the state under consideration. As a result, the $R$-charges
of the left $N=2$ superconformal algebra are:
\begin{equation}
  \mathcal{Q}_{R} = n + \frac{a}{2} - \frac{N-h/2}{p} \mod 2.
\end{equation}

We now turn to the issue of modular covariance. Under the
transformation $\tau \to - 1/\tau$, the minimal-model characters
transform as: \mathindent=0em
\begin{equation}
  \mathcal{C}^{j}_{m} \oao{a}{b} \left(-\frac{1}{\tau}\right) =
  \mathrm{e}^{\imath \frac{\pi}{2} ab} \frac{1}{k} \sum_{j'
    =0}^{(k-2)/2} \sin \left( \frac{\pi (2j+1)(2j'+1)}{k} \right)
  \sum_{m' \in \setZ_{2k}} \mathrm{e}^{\imath \pi \frac{mm'}{k}}
  \mathcal{C}^{j'}_{m'} \oao{b}{-a} (\tau).
\end{equation}
On the one hand, the part of the modular transformation related to $j$
is precisely compensated by a similar term coming from the
transformation of $\bar \chi^{\bar \jmath}$, in Eq. (\ref{ZS2}).  On
the other hand, the part of the transformation related to the spin
structure $(a,b)$ is compensated by the transformation of the other
left-moving fermions in the full heterotic string construction. We can
therefore concentrate on the transformation related to the $m$ charge,
coming from the transformation of the theta-functions at level $k$. We
have
\begin{equation}
  \sum_{N \in \setZ_{2p}} \mathrm{e}^{-\imath\pi g\left(N +
      \frac{h}{2}\right)} \ \mathcal{C}^{j}_{p(2N-h)} \oao{a}{b} \to
  \frac{1}{\sqrt{2k}} \sum_{m' \in \setZ_{4p^2}} \sum_{N \in \setZ_{2p}}
  \mathrm{e}^{\frac{\imath\pi}{2} \left(g-\frac{m'}{p}\right)h}
  \mathrm{e}^{2\imath\pi \frac{N(m'+pg)}{2p} } \mathcal{C}^{j}_{m'}
  \oao{b}{-a};
\end{equation}
\mathindent=\oldindent
summing over $N$ in $\setZ_{2p}$ leads to the constraint:
\begin{equation}
  m'+pg \equiv 0 \mod 2p := -2pN' \ , \ N' \in \setZ_{2p}.
\end{equation}
So we end up with the sum
\begin{equation}
  \mathrm{e}^{-\frac{\imath\pi}{2} hg} \sum_{N' \in \setZ_{2p}}
  \mathrm{e}^{-\imath\pi h\left(N'+\frac{g}{2}\right)}
  \mathcal{C}^{j}_{p(2N'-g)} \oao{b}{-a}.
\end{equation}
combining this expression with the modular transformation of the
remaining right-moving fermions of the gauge sector, we obtain a
modular invariant result.

  In a similar way one can check the invariance of the full heterotic
  string under $\tau \to \tau + 1$.

  \subsection{$SU(3)$}
  \label{sec:su3}

  As it is often the case, the $SU(2)$ example is illuminating but not
  exhaustive. In this situation this is due to the fact that $U(1)$ is
  the Cartan torus and at the same time the maximal subgroup. For this
  reason we need to work out in detail the next non-trivial example,
  $SU(3)$. The main difference is that there are two non-equivalent
  construction leading to the same algebraic structure but to the two
  possible different metrics on the $SU(3)/U(1)^2$ coset\footnote{We
    have already pointed out in Sec.~\ref{sec:geom-squash-groups} that
    an asymmetric coset in the mathematical sense in general admits
    more than one left-invariant metric. The two possible choices for
    $SU(3)/U(1)^2$ will be extensively studied in
    Sec.~\ref{sec:new-comp}}.

\subsubsection{The Kazama-Suzuki decomposition of SU(3)}
We would like to decompose our \textsc{wzw} model in terms 
of Kazama-Suzuki (\textsc{ks}) cosets, which are conformal theories 
with extended $N=2$ superconformal 
symmetry~\cite{Kazama:1989qp,Kazama:1989uz}.

The simplest of those models 
are the $N=2$ minimal models that are given by the quotient: 
$\nicefrac{SU(2)_{k-2} \times
  SO(2)_1}{U\left( 1 \right)_{k}}$, 
and their characters come from the branching relation:
\begin{equation}
  \chi^{j}_{k-2} \Xi^{s_2}_{2} = 
  \sum_{m \in \setZ_{2k}} \mathcal{C}^{j\, (s_2)}_{m} \frac{\Theta_{m,k}}{\eta} .
\end{equation}
For convenience, we write the contribution of the world-sheet 
fermions in terms of  $SO(2n)_1$ characters.

Similarly it is possible to construct an $N=2$ coset
\textsc{cft} from $SU(3)$~\cite{Kazama:1989qp,Kazama:1989uz}:\footnote{ 
According to our conventions, the 
weights of a $U\left( 1 \right)$ at level $k$ are
  $\nicefrac{m^2}{4k}$, $m \in \setZ_{2k}$.}
\begin{equation}
  \frac{SU\left(3\right)_{k-3} \times SO(4)_1}{SU(2)_{k-2} \times 
    U\left( 1 \right)_{3k}}.
\end{equation}
The characters of this theory are implicitly defined by the branching
relation:
\begin{equation}
\chi^{\Lambda}_{k-3} \, \Xi^{s_4}_{4}  = 
\sum_{2j=0}^{k-2} \sum_{n \in \setZ_{6k}} \mathcal{C}^{\Lambda \, (s_4 )}_{j\, n} 
\chi^{j}_{k-2} \, \frac{\Theta_{n,3k}}{\eta} .
\end{equation}
Therefore combining the two branching relations, we obtain the decomposition 
of $SU \left( 3 \right)$ in terms of $N=2$ \textsc{ks} models:
\begin{equation}
\chi^{\Lambda}_{k-3} \,  
 \Xi^{s_4}_{4} \Xi^{s_2}_{2}  = 
\sum_{j,m,n}  \mathcal{C}^{\Lambda \, (s_4)}_{j\, n}
\mathcal{C}^{j\, (s_2)}_{m} \ \frac{\Theta_{m,k}}{\eta} \ 
\frac{\Theta_{n,3k}}{\eta}\  
\end{equation}
This decomposition goes along the following pattern:
\begin{multline}
  SU\left(3\right)_{k-3} \times SO(8)_1 \to 
  \frac{SU\left(3\right)_{k-3} \times SO(4)_1}{SU(2)_{k-2} \times 
    U\left( 1 \right)_{3k}} \times \frac{SU(2)_{k-2} \times SO(2)_1}{U\left( 1 \right)_{k}} \times \\
  \times U\left( 1 \right)_{3k} \times U\left( 1 \right)_k \times SO(2)_1
  \label{decompA3}
\end{multline}
and we shall perform the deformation on the 
left lattice of $U\left( 1 \right)_{3k} \times U\left( 1 \right)_k$. However 
the deformation will also act on an appropriate 
sub-lattice of the right-moving gauge sector. The last 
$SO(2)_1$ factor corresponds to the fermions which are 
neutral in the process so they won't be considered 
afterwards.

\paragraph{The gauge sector}
To construct the model we assume that 
the gauge sector of the heterotic strings contain an unbroken 
$SO(6)_1$, whose 
contribution to the partition function is,
written in terms of $SO(6)_1$ free fermionic characters
$\bar{\Xi}^{s_6}_6$.
Since we decompose the characters of the left-moving 
sector according to eq.~(\ref{decompA3}), 
a natural choice for the action of the deformation in the 
right-moving gauge sector is to use a 
similar Kazama-Suzuki decomposition, but for 
$k=3$, in which case the bosonic \textsc{cft} is trivial:
\begin{equation}
SO(8)_1 \to 
\frac{SO(4)_1}{SU(2)_{1} \times 
U\left( 1 \right)_{9}} \times \frac{SU(2)_{1} \times SO(2)_1}{U\left( 1 \right)_{3}}
\times U\left( 1 \right)_{3} \times U\left( 1 \right)_1 \times SO(2)_1
\end{equation}
Since as quoted previously two fermions --~the 
$SO(2)_1$ factor~-- are neutral it is enough 
that the gauge sector contains an $SO(6)_1$ subgroup.
To achieve this decomposition, first we decompose the 
$SO(6)_1$ characters in terms of $SO(4)_1 \times SO(2)_1$:
\begin{equation}
\bar{\Xi}^{\bar{s}_6}_6 = \sum_{\bar{s}_4,\bar{s}_{2} \in \setZ_4} 
C \left[ \bar{s}_6;\bar{s}_4,\bar{s}_{2} \right] 
\bar{\Xi}^{\bar{s}_4}_4 \bar{\Xi}^{\bar{s}_{2}}_2 
\end{equation}
where the coefficients of the 
decomposition $SO(6) \to SO(4) \times SO(2)$ are either zero or one. 
And then we perform a coset decomposition for the 
$SO(4)_1$ characters:
\begin{equation}
\bar{\Xi}^{\bar{s}_4}_4 = \sum_{\ell = 0,1} \sum_{u \in \setZ_{18}} 
\bar{\varpi}^{\bar{s}_4}_{\ell \, u} \bar{\chi}^{\ell}  \ \frac{\bar{\Theta}_{u,9}}{\bar \eta}
\end{equation}
in terms of $SU(2)_1$ characters $\bar{\chi}^{\ell}$ and 
$U\left( 1 \right)$ characters $\bar{\Theta}_{u,9}$. 
It defines implicitly the coset characters 
$\bar{\varpi}^{\bar{s}_4}_{\ell \, u}$.
Then the $SU(2)_1 \times SO(2)_1$ characters are decomposed as:
\begin{equation}
\bar{\chi}^{\ell} \bar{\Xi}^{\bar{s}_{2}}_2 = \sum_{v \in \setZ_6} 
\bar{\varpi}^{\ell,\bar{s}_{2}}_{v} \frac{\bar{\Theta}_{v,3}}{\bar \eta}\, .
\end{equation}
So putting together these branching relations we have the 
following Kazama-Suzuki decomposition for the free fermions 
of the gauge sector:
\begin{equation}
\bar{\Xi}^{\bar{s}_6}_6 = \sum_{\bar{s}_4,\bar{s}_{2} \in 
\setZ_4 } \sum_{\ell = 0,1} \sum_{u \in \setZ_{18}} 
\sum_{v \in \setZ_6} 
C \left[ \bar{s}_6;\bar{s}_4,\bar{s}_{2}\right] \ 
\bar{\varpi}^{\bar{s}_4}_{\ell \, u} \ 
\bar{\varpi}^{\ell,\bar{s}_{2}}_{v} \ 
\frac{\bar{\Theta}_{u,9}}{\bar \eta}\ \frac{\bar{\Theta}_{v,3}}{\bar \eta}.
\end{equation}

\paragraph{The deformation}
Now we are in position to perform the asymmetric deformation 
adding a magnetic field to the model. 
The deformation acts on the following combination of 
left and right theta functions:
\begin{equation}
  \Theta_{n,3k} \, \bar{\Theta}_{u,9} \times 
  \Theta_{m,k} \bar{\Theta}_{v,3}.    
\end{equation}
As for the case of $SU(2)$~\cite{Israel:2004vv}, 
we have to assume that the level obeys the condition: 
\begin{equation}
\sqrt{\frac{k}{3}} = p \in \mathbb{N}\, , \end{equation}
to be able to reach the geometric coset point in the moduli 
space of \textsc{cft}. 
Then we have to perform $O(2,2,\mathbb{R})$ boosts in the lattices 
of the $U\left( 1 \right)$'s, mixing the left Cartan lattice 
of the super-\textsc{wzw} model with the right lattice 
of the gauge sector. These boosts are  
parametrized in function of the magnetic fields as:
\begin{equation}
\cosh \Omega_a = \frac{1}{1-2\textsc{h}_{a}^2}\ , \ \  
a = 1,2 .
\end{equation}
Explicitly we have:
\begin{multline}
  \sum_{N_1,N_2 \in \setZ} q^{3k\left(N_1 + \frac{m}{6k}\right)^2}
  q^{k\left(N_2 + \frac{n}{2k}\right)^2} 
  \ \times 
  \sum_{f_1,f_2 \in \setZ} \bar{q}^{9\left(f_1 + \frac{u}{18}\right)^2}
  \bar{q}^{3\left(f_2 + \frac{v}{6}\right)^2}\\
  \to 
  \sum_{N_1,N_2,f_1,f_2 \in \setZ}
  q^{9 \left[ p \left( N_1 + \frac{m}{18p^2}\right) \cosh \Omega_1 
      + \left( f_1 + \frac{u}{18}\right) \sinh \Omega_1 \right]^2}
  q^{3 \left[ p \left( N_2 + \frac{n}{6p^2}\right) \cosh \Omega_2 
      + \left( f_2 + \frac{v}{6}\right) \sinh \Omega_2 \right]^2}\\
  \times \ 
  \bar{q}^{9 \left[ 
      \left( f_1 + \frac{u}{18}\right) \cosh \Omega_1
      + p \left( N_1 + \frac{m}{18p^2}\right) \sinh \Omega_1\right]^2}
  \bar{q}^{3 \left[ 
      \left( f_2 + \frac{v}{6}\right) \cosh \Omega_2 
      + p \left( N_2 + \frac{n}{6p^2}\right) \sinh \Omega_2 \right]^2} .
\end{multline}
After an infinite deformation, we get the following constraints on the charges:
\begin{subequations}
  \begin{align}
    m &= p \left(18\mu - u \right), \ \mu \in \setZ_p \\
    n &= p \left( 6\nu  - v \right), \ \nu \in \setZ_p 
  \end{align}
\end{subequations}
and the $U\left( 1 \right)^2$ \textsc{cft} that has been deformed marginally decouples from the rest and 
can be safely removed.
In conclusion, the infinite deformation gives:
\begin{multline}
Z^{(s_4,s_2;\bar{s}_6 )}_{F_3} \left(\tau \right) = \sum_{\Lambda} \sum_{j}
  \sum_{\mu,\nu \in \setZ_{p}}\ \sum_{\bar{s}_4,\bar{s}_{2} \in \setZ_4 } C \left[
    \bar{s}_6;\bar{s}_4,\bar{s}_{2}\right]\\ \sum_{\ell = 0,1} \sum_{u \in \setZ_{18}}\ 
  \sum_{v \in \setZ_6}  
  \mathcal{C}^{\Lambda \, (s_4)}_{j\ , \ p (18\mu-u)} \ \mathcal{C}^{j\,
    (s_2)}_{p(6\nu -v) } \times \bar{\chi}^{\Lambda}_{k-3} \ \bar{\varpi}^{\bar{s}_4}_{4;\,
    \ell u} \ \bar{\varpi}^{\ell,\bar{s}_{2}}_{v}  
\end{multline} 
where the sum over $\Lambda, j$ runs over integrable representations.
This is the partition function for the $SU\left(3\right)/U\left( 1
\right)^2$ coset space.  The fermionic charges in the left and right
sectors are summed according to the standard rules of Gepner heterotic
constructions~\cite{Gepner:1988qi}. The modular properties of this
partition function are the same as before the deformation, concerning
the $\setZ_4$ indices of the world-sheet fermions.


\subsubsection{Alternative approach: direct abelian coset}

Here we would like to take a different path, by deforming directly
the Cartan lattice of $\mathfrak{su}_3$ without decomposing the left
\textsc{cft} in terms of \textsc{ks} $N=2$ theories.  It is possible
to perform a generalized (super)parafermionic decomposition of the
characters of the $\hat{\mathfrak{su}}_3$ super-algebra at level $k$
(containing a bosonic algebra at level $k-3$) w.r.t. the Cartan torus:
\begin{equation}
  \chi^{\Lambda}  \ 
  \left(
    \frac{\vartheta \oao{a}{b}}{\eta}\right)^{\mathrm{dim} (\mathfrak j ) /2}  
= 
  \sum_{\lambda \in \mathbf{M}^\ast \mathrm{mod} \ k \mathbf{M}} 
  \mathcal{P}^{\Lambda}_{\lambda} \oao{a}{b}  
  \frac{\Theta_{\lambda,k}}{\eta^{\mathrm{dim} (\mathfrak k )}}
\end{equation}
where the theta function of the $\widehat{\mathfrak{su}}_3$ 
affine algebra reads, 
for a generic weight $\lambda = m_i \lambda^{i}_f$:
\begin{equation}
  \Theta_{\lambda,k} = \sum_{\gamma \in \mathbf{M} 
    + \frac{\lambda}{k} }
  q^{\frac{k}{2} \kappa (\gamma,\gamma)} = 
  \sum_{N^1, N^2 \in \setZ} 
  q^{\frac{k}{2} \| N^1 \alpha_1 + N^2 \alpha_2 
    + \frac{m_1 \lambda_{f}^1 + m_2 \lambda_{f}^2}{k} \|^2} .
\end{equation}
To obtain an anomaly-free model (see the discussion at the beginning
of Sec.~\ref{sec:gauging}) it is natural to associate this model with
an abelian coset decomposition of an $SU(3)_1$ current algebra made
with free fermions of the gauge sector. Thus if the gauge group
contains an $SU(3)_1$ unbroken factor their characters can be
decomposed as:
\begin{equation}
  \bar{\chi}^{\bar{\Lambda}} = \sum_{\bar{\lambda}= \bar{n}_i \lambda^{i}_f  
    \ \in \ \mathbf{M}^\ast 
    \mathrm{mod} \ \mathbf{M}} \bar{\varpi}^{\bar{\Lambda}}_{\bar{\lambda}}
  \bar{\Theta}_{\bar{\lambda}}.
\end{equation}
Again we will perform the asymmetric deformation as a boost between 
the Cartan lattices of the left $\hat{\mathfrak{su}}_3$ algebra 
at level $k$ and the right $\hat{\mathfrak{su}}_3$ lattice algebra at level one 
coming from the gauge sector.
So after the infinite deformation we will get the quantization condition $\sqrt{k}=p$
and the constraint:
\begin{align}
\lambda + p \bar{\lambda} = 0 \mod p \, \textbf{M} = : p \, \mu \ , \ \ 
\mu \in \textbf{M}.
\end{align}
So we get a different result compared to the Kazama-Suzuki construction. It is
so because the constraints that we get at the critical point 
force the weight lattice of the $\hat{\mathfrak{su}}_3$ at level 
$k$ to be projected onto $p$ times the $\hat{\mathfrak{su}}_3$ weight lattice at level 
one of the fermions. This model does not correspond to a
K\"ahlerian manifold and should correspond to the $SU\left(3\right)$-invariant metric on
the flag space. Indeed with the \textsc{ks} method we get instead a projection onto 
$p$ times a lattice of $\hat{\mathfrak{su}}_3$ at level one 
which is dual to the orthogonal sub-lattice defined by $\alpha_1 \setZ 
+ (\alpha_1 + 2 \alpha_2)\setZ
$--~in other words the lattice 
obtained with the Gell-Mann Cartan generators. 
In this case it is possible to decompose the model in \textsc{ks} cosets models 
with $N=2$ superconformal symmetry.\footnote{For the symmetrically 
gauged \textsc{wzw} models, this has been studied in~\cite{Eguchi:2003yy}.}

\section{The deformation as a gauging}
\label{sec:gauging}

In this section we want to give an alternative construction for our
deformed models, this time explicitly based on an asymmetric
\textsc{wzw} gauging.  The existence of such a construction is not
surprising at all since our deformations can be seen as a
generalization of the ones considered in~\cite{Giveon:1994ph}. In
these terms, just like $J \bar J$ (symmetric) deformations lead to
gauged \textsc{wzw} models, our asymmetric construction leads to
asymmetrically gauged \textsc{wzw} models, which were studied
in~\cite{Quella:2002fk}.

A point must be stressed here. The asymmetric deformations admit as
limit solutions the usual geometric cosets that one would have
expected from field theory, as results of a gauging procedure. So, why
do we need to go through this somewhat convoluted procedure? The
reason lays in the fact that string theory is not the usual point
particle field theory. A left and a right sector are present at the
same time and they cannot be considered separately if we don't want to
introduce anomalies. Now, gauging the left action of a subgroup,
\emph{i.e.} the symmetry $G \sim G H$, which would directly give the
geometric coset we are studying, would precisely introduce this kind
of problems. Hence we are automatically forced to condider
the adjoint action $G \sim H^{-1} G H$~\cite{Witten:1991yr}. The key
idea then, as it will appear in this section, is that when $G$ is
semisimple and written as the product of a group and a copy of its
Cartan torus, the left and right action can be chosen such as to act
on the two separate sectors and then be equivalent to two left
actions.

Instead of a general realization, for sake of clearness, here we will
give the explicit construction for the most simple case, the $SU
\left( 2 \right)$ model, then introduce a more covariant formalism
which will be simpler to generalize to higher groups, in particular
for the $SU \left( 3 \right)$ case which we will describe in great
detail in the following.

To simplify the formalism we will discuss gauging of bosonic
\textsc{cft}s, and the currents of the gauge sector of the heterotic
string are replaced by compact $U(1)$ free bosons. All the results are
easily translated into heterotic string constructions.

\subsection{The SU(2)/U(1) asymmetric gauging}
\label{sec:su-left-2}

\marginlabel{Geometric coset as an asymmetric gauging}In this section
we want to show how the $S^2 $ background described in
\cite{Israel:2004vv} can be directly obtained via an asymmetric
gauging of the $SU \left( 2 \right) \times U \left( 1 \right)$
\textsc{wzw} model (a similar construction was first obtained
in~\cite{Johnson:1995kv}).

Consider the \textsc{wzw} model for the group manifold $SU \left( 2 \right)_k \times U
\left( 1 \right)_{k^\prime}$. A parametrisation for the general element of this
group which is nicely suited for our purposes is obtained as follows:
\begin{equation}
  g = 
  \begin{pmatrix}
    z_1 & z_2 & 0 \\
    - \bar z_2 & \bar z_1 & 0 \\
    0 & 0 & z_3
  \end{pmatrix} = \left( \begin{tabular}{c|c}
        $g_2 $& $ 0$\\ \hline
        $ 0 $& $g_1$
      \end{tabular}\right) \in SU \left( 2 \right) \times U \left( 1 \right) 
\end{equation}
where $g_1 $ and $g_2 $ correspond to the $SU \left( 2 \right)$
and $U \left( 1 \right)$ parts respectively and $\left( z_1, z_2, z_3 \right)$ satisfy:
\begin{equation}
  SU \left( 2 \right) \times U \left( 1 \right) = \set{ \left( w_1, w_2,
      w_3 \right) | \abs{w_1}^2 + \abs{w_2}^2 = 1, \abs{w_3}^2 = 1} \subset \setC^3  .
\end{equation}
A possible choice of coordinates for the corresponding group manifold is
given by the Euler angles:
\mathindent=0em
\begin{multline}
\label{eq:su2u1coords}
  SU \left( 2 \right) \times U \left( 1 \right) \\= \Set{\left( z_1, z_2, z_3
    \right) 
= \left( \cos \frac{\beta }{2} e^{ \imath \left( \gamma + \alpha \right)/2 },
      \sin \frac{\beta }{2} e^{\imath \left( \gamma - \alpha \right)/2}, e^{\imath \varphi}\right)|
    0 \leq \beta  \leq \pi , 0 \leq \alpha , \beta , \varphi \leq 2 \pi }
\end{multline}
\mathindent=\oldindent

In order to obtain the coset construction leading to the $S^2 $
background we define two $U \left( 1 \right) \to SU \left( 2 \right) \times
U\left( 1 \right)$ embeddings as follows:
\begin{align}
\label{eq:su2-embeddings}
  \begin{split}
    \epsilon_L :U \left( 1 \right) &\to SU \left( 2 \right) \times U \left( 1 \right) \\
    e^{\imath \tau} &\mapsto \left( e^{\imath \tau}, 0, 1\right)
  \end{split}
  \begin{split}
    \epsilon_R :U \left( 1 \right) &\to SU \left( 2 \right) \times U \left( 1 \right) \\
    e^{\imath \tau} &\mapsto \left( 1, 0, e^{\imath \tau}\right)
  \end{split}
\end{align}
so that in terms of the $z$ variables the action of these embeddings boils
down to:
\begin{align}
  g &\mapsto \epsilon_L \left( e^{\imath \tau } \right) g \epsilon_R \left( e^{\imath \tau }\right)^{-1} \\
  \left( w_1, w_2, w_3 \right) & \mapsto \left( e^{\imath \tau } w_1, 
    e^{\imath \tau } w_2, e^{-\imath \tau } w_3 \right) .
\end{align}
This means that we are free to choose a gauge where $w_2$ is real or, in
Euler coordinates, where $\gamma = \alpha $, the other angular variables
just being redefined.  To find the background fields corresponding to this
gauge choice one should simply write down the Lagrangian where the
symmetries corresponding to the two embeddings in \eqref{eq:su2-embeddings}
are promoted to local symmetries, integrate the gauge fields out and then
apply a Kaluza-Klein reduction, much in the same spirit as in
\cite{Israel:2004vv}.

The starting point is the \textsc{wzw} model, written as:
\begin{equation}
  S_{\textsc{wzw}} \left( g \right) = \frac{k}{4 \pi } \int \di z^2 \: 
  \braket{ g_2^{-1} \partial g_2 g_2^{-1} \bar \partial g_2 } + \frac{k^\prime }{4 \pi } \int \di z^2 \: 
  \braket{ g_1^{-1} \partial g_1 g_1^{-1} \bar \partial g_1 } .
\end{equation}
Its gauge-invariant generalization is given by:
\mathindent=0em
\begin{multline}
  S = S_{\textsc{wzw}}\\ + \frac{1}{2 \pi } \int \di^2 z \left[ k \bar A \braket{
    t_L \partial g g^{-1}} + k^\prime A \braket{ t_R g^{-1} \bar \partial g } +
  \sqrt{k k^\prime }A
  \bar A \left( -2 + \braket{ t_L \, g \, t_R \, g^{-1}} \right) \right]
\end{multline}
\mathindent=\oldindent
where $A $ and $\bar A $ are the components of the gauge field, and 
$t_L$ and $t_R$ are the Lie algebra generators corresponding to 
the embeddings in
\eqref{eq:su2-embeddings}, \emph{i.e.}
\begin{align}
  t_L = \imath \left( \begin{tabular}{c|c}
      $\sigma_3 $& $  0 $\\ \hline
      $ 0 $& $0$
    \end{tabular}\right), &&
  t_R = \imath \left( \begin{tabular}{c|c}
      $0$ & $0$ \\ \hline
      $0$ & $p$
    \end{tabular}\right),
\end{align}
$\sigma_3 $ being the usual Pauli matrix. For such an asymmetric coset to 
be anomaly free, one has the following constraint on the embeddings:
\begin{equation}
  k \braket{t_L}^2 = k' \braket{t_R}^2 \implies 
  k = k' p^2 \ , \ \ \text{with} \ p \in \mathbb{N}.
\label{anomalys2}
\end{equation}
If we pass to Euler coordinates it is
simple to give an explicit expression for the action:
\mathindent=0em
\begin{multline}
  S \left( \alpha, \beta, \gamma, \varphi \right) = \frac{1}{2 \pi } \int \di^2 z \: \frac{k}{4} \left( \partial \alpha \bar \partial \alpha  +
    \partial \beta \bar \partial \beta  + \partial \gamma \bar \partial \gamma + 2 \cos \beta \partial \alpha \bar \partial \gamma
  \right) + \frac{k^\prime }{2} \partial
    \varphi \bar \partial \varphi + \\+ \imath k \left( \partial \alpha + \cos \beta \partial \gamma \right) \bar A +
  \imath k^\prime \sqrt{2} \bar \partial \varphi  A - 2 \sqrt{k k^\prime } A \bar A .
\end{multline}
This Lagrangian is quadratic in $A, \bar A $ and the quadratic part is
constant so we can integrate these gauge fields out and the resulting
Lagrangian is:
\begin{multline}
\label{eq:gauged-SU2}
S \left( \alpha, \beta, \gamma, \varphi \right) = \frac{1}{2 \pi } \int \di^2 z \: \frac{k}{4}
\left( \partial \alpha \bar \partial \alpha + \partial \beta \bar \partial \beta + \partial \gamma \bar \partial \gamma + 2 \cos \beta \partial \alpha \bar
  \partial \gamma \right) + \frac{k^\prime }{2} \partial \varphi \bar \partial \varphi + \\+ \frac{\sqrt{2 k k^\prime
  }}{2} \left( \partial \alpha + \cos \beta \partial \gamma \right) \bar \partial \varphi .
\end{multline}
\mathindent=\oldindent
Now, since we gauged out the symmetry corresponding to the $U \left( 1
\right)$ embeddings, this action is redundant. This can very simply be seen
by writing the corresponding metric and remarking that it has vanishing
determinant: 
\begin{equation}
  \det g_{\mu \nu } =  
  \begin{vmatrix}
    k/4 \\
    & k/4 &k/4 \cos \beta  & \sqrt{2 k k^\prime }/4\\
    & k/4 \cos \beta & k/4 & \sqrt{2 k k^\prime }/4 \cos \beta \\ 
    &  \sqrt{2 k k^\prime }/4 & \sqrt{2 k k^\prime }/4 \cos \beta  & k^\prime/2 
  \end{vmatrix} = 0
\end{equation}
Of course this is equivalent to say that we have a gauge to fix (as we saw
above) and this can be chosen by imposing $\gamma = \alpha $, which leads to the
following action:
\begin{multline}
  S \left( \alpha, \beta, \varphi \right) = \frac{1}{2 \pi } \int \di^2 z \: \frac{k}{4}
  \left( 2 \left( 1 + \cos \beta \right) \partial \alpha \bar \partial \alpha  +
    \partial \beta \bar \partial \beta  \right) + \frac{k^\prime }{2} \partial
  \varphi \bar \partial \varphi + \\
  + \frac{\sqrt{2 k k^\prime }}{2} \left( 1 + \cos \beta
  \right) \partial \alpha \bar \partial \varphi 
\end{multline}
whence we can read a two dimensional metric by interpreting the
$\partial \alpha \bar \partial \varphi $ term as a gauge boson and applying the usual
Kaluza-Klein reduction. We thus recover the two-sphere as expected:
\begin{equation}
  \label{eq:S2-line-element}
  \di s^2 = g_{\mu \nu } - G_{\varphi \varphi } A_\mu A_\nu = \frac{k}{4}\left( \di \beta^2 +
  \sin^2 \beta \di \alpha^2 \right)
\end{equation}
supported by a (chromo)magnetic field
\begin{equation}
\label{eq:S2-magnetic-field}
  A = \sqrt{\frac{k} {k^\prime} } \left( 1 + \cos \beta \right) \di \alpha   
\end{equation}

\subsection{The current formalism}
\label{sec:current-formalism}

\marginlabel{Asymmetric gauging in the current formalism}We now turn
to rewrite the above gauging in a more covariant form, simpler to
generalize. Since we are interested in the underlying geometry, we
will mainly focus on the metric of the spaces we obtain at each step
and write these metrics in terms of the Maurer-Cartan
currents\footnote{One of the advantages of just working on the metrics
  is given by the fact that in each group one can consistently choose
  left or right currents as a basis. In the following we will consider
  the group in the initial \textsc{wzw} model as being generated by
  the left and the dividing group by the right ones.}. As we have
already seen, the metric of the initial group manifold is:
\begin{equation}
  \di s^2 = \frac{k}{2} \sum \mJ_i^2 \otimes \mJ_i^2 + \frac{k^\prime }{2} \mathcal{I} \otimes \mathcal{I}  
\end{equation}
where $\set{\mJ_1, \mJ_2, \mJ_3 }$ are the currents of the $SU \left( 2 \right)$
part and $\mathcal{I}$ the $U \left( 1 \right)$ generator. The effect of the
asymmetric gauging amounts - at this level - to adding what we can see as an
interaction term between the two groups. This changes the metric to:
\begin{equation}
  \di s^2 = \frac{k}{2} \sum \mJ_i^2 \otimes \mJ_i^2 + \frac{k^\prime }{2} \mathcal{I} \otimes \mathcal{I}  +
  \sqrt{k k^\prime } \mJ_3 \otimes \mathcal{I}    .
\end{equation}
Of course if we choose $\braket{\mJ_1, \mJ_2, \mJ_3, \mathcal{I}}$ as a basis we can
rewrite the metric in matrix form:
\begin{equation}
  g = \frac{1}{2} 
  \begin{pmatrix}
    k \\
    & k \\
    & & k & \sqrt{k k^\prime }\\
    & & \sqrt{k k^\prime } & k^\prime 
  \end{pmatrix}
\end{equation}
where we see that the gauging of the axial symmetry corresponds to the
fact that the sub-matrix relative to the $\set{\mJ_3, \mathcal{I} }$
generators is singular:
\begin{equation}
  \begin{vmatrix}
    k & \sqrt{k k^\prime } \\
    \sqrt{k k^\prime } & k^\prime 
  \end{vmatrix} = 0
\end{equation}
explicitly this correspond to:
\begin{equation}
  k \mJ_3 \otimes \mJ_3 + \sqrt{k k^\prime } \mJ_3 \otimes \mathcal{I} + \sqrt{k k^\prime } \mJ_3 \otimes \mathcal{I} + k^\prime \mathcal{I}
  \otimes \mathcal{I} = \left( k + k^\prime \right) \hat \mJ \otimes \hat \mJ
\end{equation}
where
\begin{equation}
  \hat \mJ = \frac{\sqrt{k} \mJ_3 + \sqrt{k^\prime } \mathcal{I}}{\sqrt{k + k^\prime }}  
\end{equation}
is a normalized current. In matrix terms this corresponds to projecting the
interaction sub-matrix on its non-vanishing normalized eigenvector:
\begin{equation}
  \begin{pmatrix}
    \sqrt{\frac{k}{k+k^\prime }} & \sqrt{\frac{k}{k+k^\prime }}
  \end{pmatrix}
  \begin{pmatrix}
     k & \sqrt{k k^\prime } \\
    \sqrt{k k^\prime } & k^\prime
  \end{pmatrix}
  \begin{pmatrix}
    \sqrt{\frac{k}{k+k^\prime }} \\ \sqrt{\frac{k}{k+k^\prime }}
  \end{pmatrix} = k + k^\prime 
\end{equation}
and the resulting metric in the $\braket{\mJ_1, \mJ_2, \hat \mJ }$ basis is:
\begin{equation}
\label{eq:asy-gaug-su2}
  \begin{pmatrix}
    k \\
    & k \\
    & & k+k^\prime 
  \end{pmatrix}
\end{equation}
This manifold $\mathcal{M}$ (whose metric appears in the action
\eqref{eq:KK-action}) corresponds to an $S^1 $ fibration (the fiber being
generated by $\hat \mJ$) over an $S^2 $ base (generated by $\braket{\mJ_1, \mJ_2
}$).
\begin{equation}
  \begin{CD}
    S^1 @>>> \mathcal{M} \\
    @.      @VVV\\
    {} @. S^2
  \end{CD}
\end{equation}

It should now appear natural how to generalize this construction so to
include all the points in the moduli space joining the unperturbed and
gauged model. The decoupling of the $U \left( 1 \right)$ symmetry
(that has been ``gauged away'') is obtained because the back-reaction
of the gauge field (Eq.~\eqref{eq:gauged-SU2}) is such that the
interaction sub-matrix is precisely singular. On the other hand we can
introduce a parameter that interpolates between the unperturbed and
the gauged models so that the interaction matrix now has two non-null
eigenvalues, one of which vanishing at the decoupling point.

In practice this is done by adding to the  
the asymmetrically gauged \textsc{wzw} model an auxiliary 
$U(1)$ free boson $Y$ at radius $R= (k k^\prime )^{\nicefrac{1}{4}}
(\nicefrac{1}{\sqrt{2}\h-1})^{\nicefrac{1}{2}}$. 
This $U(1)$ is coupled symmetrically to the gauge fields such that 
the anomaly cancellation condition is still given by~(\ref{anomalys2}). 
In particular if we choose the gauge $Y=0$, the metric reads:
\begin{equation}
  \begin{pmatrix}
    k & \sqrt{2} \h \sqrt{k k^\prime } \\
    \sqrt{2} \h \sqrt{k k^\prime } & k^\prime 
  \end{pmatrix}
\end{equation}
which is exactly the model studied above. For a generic value of $\h^2
$ the two eigenvalues are given by:
\begin{equation}
  \label{eq:Interact-Eigenv}
  \lambda_{1\atop2} \left( k, k^{\prime }, \h \right) = 
\frac{k + k^\prime \mp  \sqrt{k^2 + {k^\prime
      }^2 + 2 \left( 4 \h^2 - 1 \right)k k^\prime }}{2}
\end{equation}
so we can diagonalize the metric in the $\braket{\mJ_1, \mJ_2, \hat \mJ, \hat
  {\hat \mJ}}$ basis ($\hat \mJ $ and $\hat {\hat \mJ}$ being the two
eigenvectors) and finally obtain:
\begin{equation}
  g = 
  \begin{pmatrix}
    k \\
    & k \\
    & & \lambda_1\left( k, k^{\prime }, \h \right) \\
    & & & \lambda_2\left( k, k^{\prime }, \h \right)
  \end{pmatrix} .
\end{equation}
Of course, in the $\h^2 \to 0$ limit we get the initial \textsc{wzw}
model and in the $\h^2 \to 1/2 $ limit we recover the asymmetrically
gauged model, Eq.~\eqref{eq:asy-gaug-su2}.

It is important to remark that the construction above can be directly
generalized to higher groups with non-abelian subgroups, at least for
the asymmetric coset part. This is what we will further analyse in the
next chapter.


\chapter{Applications}
\label{cha:applications}

\chapterprecis{In this chapter we present some of the applications for
  the construction outlined above. After an analysis of the most
  simple (compact and non-compact) examples, we describe the
  near-horizon geometry for the Bertotti-Robinson black hole, show
  some new compactifications and see how Horne and Horowitz's black
  string can be described in this framework and generalized via the
  introduction of an electric field.}

\lettrine{T}{he technology} we developed in the previous chapter
allows for the construction of a large class of exact string theory
backgrounds which is one of the main motivations of the present
work. This chapter is devoted to the study of some of the most
interesting among them. They can be used to provide new \textsc{cft}
models with clear geometric interpretation
(Sec.~\ref{sec:deformed-su2} and \ref{sec:deformed-sl2}), to describe
near-horizon geometries of four-dimensional black holes
(Sec.~\ref{sec:near-horiz-geom}), as laboratories for the study of
black holes and black strings (Sec.~\ref{sec:blackstring}) or to
provide new physically realistic compactification backgrounds
(Sec.~\ref{sec:new-comp}).

\section{The two-sphere \textsc{cft}}
\label{sec:deformed-su2}

\subsection{Spacetime fields}
\label{sec:spacetime-fields}

The first deformation that we explicitly consider is the marginal
deformation of the $SU \left( 2 \right)$ \textsc{wzw} model. This was
first obtained in \cite{Kiritsis:1995iu} that we will closely follow.
It is anyway worth to stress that in their analysis the authors didn't
study the point of maximal deformation (which was nevertheless
identified as a decompactification boundary) that we will here show to
correspond to the 2-sphere $S^2 \sim SU \left( 2 \right) \slash U \left(
  1 \right)$. Exact \textsc{cft}'s on this background have already
obtained in \cite{Bachas:1993kq} and in \cite{Johnson:1995kv}. In
particular the technique used in the latter, namely the asymmetric
gauging of an $SU \left( 2 \right) \times U \left( 1 \right)$ \textsc{wzw}
model, bears many resemblances to the one we will describe.

Consider a heterotic string background containing the $SU(2)$ group
manifold, times some $(1,0)$ superconformal field theory
$\mathcal{M}$.  The sigma model action is:
\begin{equation}
  S = k S_{SU\left( 2 \right)} (g) + 
  \frac{1}{2\pi} \int d^2 z \ \left\{ 
    \sum_{a=1}^{3} \lambda^a \bar \partial \lambda^a
    + \sum_{n=1}^{g} \tilde{\chi}^n \partial\tilde{\chi}^n 
  \right\} + S( \mathcal{M} ), 
\end{equation}
where $\lambda^i$ are the left-moving free fermions superpartners of the bosonic
$SU(2)$ currents, $\tilde{\chi}^n$ are the right-moving fermions of the
current algebra and $kS_{SU\left( 2 \right)}(g)$ is the \textsc{wzw} action
for the bosonic $SU(2)$ at level $k$. This theory possesses an explicit
$SU(2)_L \times SU(2)_R$ current algebra.

\marginlabel{Gauss decomposition for $SU(2)$}A parametrization of the
$SU(2)$ group that is particularly well suited for our purposes is
obtained via the so-called Gauss decomposition that we will later
generalize to higher groups (see
App.~\ref{cha:geometry-arond-group}). A general element $g \left( z,
  \psi \right) \in SU(2)$ where $z \in \setC $ and $\psi \in \setR $
can be written as:
\begin{equation}
  g \left( z, \psi \right) = 
  \begin{pmatrix}
    1 & 0 \\ z & 1
  \end{pmatrix}
  \begin{pmatrix}
    1/\sqrt{f} & 0 \\ 0 & \sqrt{f}
  \end{pmatrix}
  \begin{pmatrix}
    1 & \bar w \\ 0 & 1
  \end{pmatrix}
  \begin{pmatrix}
    e^{\imath \nicefrac{\psi}{2}} & 0 \\ 0 & e^{- \imath \nicefrac{\psi}{2}}
  \end{pmatrix}
\end{equation}
where $w = - z $ and $f = 1 + \abs{z}^2$. In this parametrisation the
matrix of invariant one-forms $\Omega = g \left( z, \psi \right)^{-1} \di g
\left( z, \psi \right)$ which is projected on the Lie algebra generators
to give the expression for the Maurer-Cartan
one-forms is:
\begin{equation}
  \Omega = \frac{1}{f}\begin{pmatrix}
    \bar z \di z - z \di \bar z   &  - e^{- \imath \psi } \di \bar z \\ 
     e^{\imath \psi } \di z & -\bar z \di z + z \di \bar z 
  \end{pmatrix} + \imath
  \begin{pmatrix}
    \di \psi & 0 \\
    0 & - \di \psi 
  \end{pmatrix}
\end{equation}
(remark that $\Omega$ is traceless and anti-Hermitian since it lives in
$\mathfrak{su} \left( 2 \right)$). From $\Omega $ we can easily derive the
Cartan--Killing metric on $T_g SU(2)_k$ as:
\begin{multline}
  \frac{2}{k} \di s^2 = \braket{ \Omega^\dag{} \Omega } = - \frac{1 }{2
    f^2}\left( \bar z^2 \di z^2 + z^2 \di \bar z^2
    - 2 \left( 2 + \abs{z}^2 \right) \di z  \di \bar z \right) +
  \\+\frac{\imath }{f} \left( z \di \bar z - \bar z \di z \right)  \di \psi
  + \frac{1}{2} \di \psi^2
\end{multline}
The left-moving current contains a contribution from the free fermions
realizing an $SU(2)_2$ algebra, so that the theory possesses (local)
$N=(1,0)$ superconformal symmetry.

The marginal deformation is obtained by switching on a magnetic field in the
$SU(2)$, introducing the following
$(1,0)$\hyph superconformal\hyph symmetry\hyph compatible marginal operator:
\begin{equation}
\label{eq:delta-S-J3}
  \delta S = \frac{\sqrt{k k_g}\h}{2\pi} (J^3 + \lambda^+ \lambda^-) \bar{J}
\end{equation}
where we have picked one particular current $\bar{J}$ from the gauge
sector, generating a $U(1)$ at level $k_g$. For instance, we can
choose the level-two current: $\bar J = i \tilde{\chi}^1 \tilde{\chi}^2$.
As a result the solutions to the deformed $\sigma
$-model~\eqref{eq:def-metric}, \eqref{eq:def-F-field} and
\eqref{eq:def-H-field} read\footnote{This type of structure is common
  to $U(1)$ fibrations over K\"ahler spaces. In example, the line
  element for $S^5$ which can be seen as a $U(1)$ fiber over $\setC P^2$
  is written as
  \begin{equation}
    \di s^2 = \di s^2 (\setC P^2) + \left( \di \psi + A \right)^2    
  \end{equation}
  where $\di A$ is the K\"ahler form on $\setC P^2$. We will encounter the
  same structure again in Sec.~\ref{sec:new-comp} for $SU(3)$ written
  as the (principal) fibration $U(1)^2 \to SU(3)/U(1)^2$.}:
\begin{equation}
  \begin{cases}
    \frac{1}{k} \di s^2 = \frac{\di z \di \bar z}{\left( 1 +
        \abs{z}^2\right)^2 } + \left( 1 - 2 \h^2 \right) \left( \frac{\imath
        z \di \bar z - \imath \bar z \di z}{f} + \di \psi \right)^2
    \label{eq:deformed-su2-metric}  \\
    \di B = \frac{\imath k}{2} \frac{1}{\left( 1+ \abs{z}^2\right)^2} \di z
    \land
    \di \bar z \land \di \psi \\
    A = \sqrt{\frac{k}{2 k_g}}\h \left( - \frac{\imath }{f}\left( \bar z \di
        z - z \di \bar z \right)+ \di \psi\right)  .
  \end{cases}
\end{equation}
It can be useful to write explicitly the volume form on the manifold and the
Ricci scalar:
\begin{align}
  \sqrt{\det g} \ \di z \land \di \bar z \land \di \psi &= \frac{k}{2} \frac{\sqrt{k
      \left( 1 - 2 \h^2 \right)}}{\left( 1+ \abs{z}^2 \right)^2} \ \di z \land
  \di \bar z \land \di  \psi \\
  R &= \frac{6 + 4 \h^2}{k}
\end{align}

It is quite clear that $\h = \h_{\text{max}}= 1/\sqrt{2}$ is a special point. In general the three-sphere $SU \left( 2\right)$ can be seen as
a non-trivial fibration of $U \left( 1\right) \sim S^1 $ as fiber and $SU
\left( 2\right)/ U\left( 1\right) \sim S^2$ as base space: the
parametrization in (\ref{eq:deformed-su2-metric}) makes it clear that
the effect of the deformation consists in changing the radius of the
fiber that naively seems to vanish at $\h_{\max}$. But as we already
know the story is a bit different: reparametrizing as in
Eq.~\eqref{rescale}:
\begin{equation}
  \psi \to  \frac{\hat \psi }{\sqrt{1-2\h^2}}
\end{equation}
\marginlabel{K\"ahler structure on $\mathbb{CP}^1$ model}one is free
to take the $\h \to 1/ \sqrt{2}$ limit where the background fields
assume the following expressions:
\begin{equation}
  \begin{cases}
    \frac{1}{k} \di s^2 \xrightarrow[\h \to \nicefrac{1}{\sqrt{2}}]{}
    \frac{ \di z \di \bar z}{\left( 1 + \abs{z}^2\right)^2 } + \di \hat
    \psi^2  \\
    F \xrightarrow[\h \to \nicefrac{1}{\sqrt{2}}]{} \sqrt{\frac{k}{4
        k_g}}\frac{ \imath \di z \land \di \bar z }{\left( 1 +
        \abs{z}^2\right)^2} \\
    H \xrightarrow[\h \to \nicefrac{1}{\sqrt{2}}]{} 0
  \end{cases}
\end{equation}

Now we can justify our choice of coordinates: the $\left( z, \bar
  z \right)$ part of the metric that decouples from the $\psi $ part is
nothing else than the K\"ahler metric for the manifold $\mathbb{CP}^1$ (which
is isomorphic to $SU \left( 2\right)/U\left( 1\right)$). In these terms the
field strength $F$ is proportional to the K\"ahler two-form:
\begin{equation}
  F = \imath \, \sqrt{\frac{k}{k_g}} g_{z \bar z  } \ 
  \di z \land \di \bar z  .
\end{equation}
This begs for a remark. It is simple to show that cosets of the form
$G/H$ where $H$ is the maximal torus of $G$ can always be endowed with
a K\"ahler structure. The natural hope is then for this structure to pop
up out of our deformations, thus automatically assuring the $N=2$
world-sheet supersymmetry of the model. Actually this is not the case.
The K\"ahler structure is just one of the possible left-invariant
metrics that can be defined on a non-symmetric coset (see
Sec.~\ref{sec:geom-squash-groups}) and the natural generalization of
the deformation considered above leads to $\setC$-structures that are not
K\"ahler. From this point of view this first example is an exception
because $SU(2) / U \left( 1 \right)$ is a symmetric coset since $U
\left( 1 \right)$ is not only the maximal torus in $SU(2)$ but also
the maximal subgroup. It is nonetheless possible to define an exact
\textsc{cft} on flag spaces but this requires a slightly different
construction, already outlined in Sec.~\ref{sec:gauging}.

We conclude this section observing that the flux of the gauge field on the
two-sphere is given by:
\begin{equation}
  \mathcal{Q} = \int_{S^2} F = \sqrt{\frac{k}{k_g}} \int d\Omega_2 = 
  \sqrt{\frac{k}{k_g}} 4 \pi
\label{flux}
\end{equation}
However one can argue on general grounds that this flux has to be
quantized, \emph{e.g.} because the two-sphere appears as a factor of
the magnetic monopole solution in string theory~\cite{Kutasov:1998zh}.
This quantization of the magnetic charge is only compatible with
levels of the affine $SU(2)$ algebra satisfying the condition:
\begin{equation}
  \frac{k}{k_g} =p^2 \ , \ \ p \in \setZ.
  \label{chargeq}
\end{equation}

\section{$SL(2, \setR)$}
\label{sec:deformed-sl2}

Anti-de~Sitter space in three dimensions is the (universal
covering of the) $SL(2,\mathbb{R})$ group manifold. It provides
therefore an exact string vacuum with \textsc{ns} background,
described in terms of the $SL(2,\mathbb{R})_k$ \textsc{wzw} model,
where time is embedded in the non-trivial geometry. We will
consider it as part of some heterotic string solution such as
$\mathrm{AdS}_3 \times S^3 \times T^4$ with \textsc{ns} three-form
field in $\mathrm{AdS}_3 \times S^3$ (near-horizon \textsc{ns}5/F1
background). The specific choice of a background is however of
limited importance for our purpose.

The issue of $\mathrm{AdS}_3$ deformations has been raised in several
circumstances. It is richer\footnote{As we have already stressed
  before the Cartan subgroups are not conjugated by inner
  automorphisms if the group is not simple.} than the corresponding
$S^3$ owing to the presence of elliptic, hyperbolic or parabolic
elements in $SL(2,\mathbb{R})$. The corresponding generators are
time-like, space-like or light-like.  Similarly, the residual symmetry
of a deformed $\mathrm{AdS}_3$ has $U(1)$ factors, which act in time,
space or light direction.

\marginlabel{$SL(2,\setR)$ marginal symmetric deformations}Marginal
\emph{symmetric} deformations of the $SL(2,\mathbb{R})_k$ \textsc{wzw}
are driven by bilinears $J\bar J$ where both currents are in
$SL(2,\mathbb{R})$ and are of the same
kind~\cite{Forste:1994wp,Israel:2003ry}.  These break the
$SL(2,\mathbb{R})_{\mathrm{L}}\times SL(2,\mathbb{R})_{\mathrm R}$
affine symmetry to $U(1)_{\mathrm L}\times U(1)_{\mathrm R}$ and allow
to reach, at extreme values of the deformation, gauged
$SL(2,\mathbb{R})_k/U(1)$ \textsc{wzw} models with an extra free
decoupled boson. We can summarize the results as follows:
\renewcommand{\labelenumi}{(\alph{enumi})}
\begin{enumerate}
\item $J^3\bar J^3$ These are time-like currents (for conventions see
  App.~\ref{cha:geometry-arond-group}) and the corresponding
  deformations connect $SL(2,\mathbb{R})_k$ with $\left.U(1) \times
    SL(2,\mathbb{R})_k / U(1)\right|_{\text{axial or vector}}$.  The
  $U(1)$ factor stands for a decoupled, non-compact time-like free
  boson\footnote{The extra bosons are always non-compact.}.  The
  gauged \textsc{wzw} model $\left. SL(2,\mathbb{R})_k/U(1)
  \right|_{\text{axial}}$ is the \emph{cigar} (two-dimensional
  Euclidean black hole) obtained by gauging the $g\to hgh$ symmetry
  with the $h=\exp{i\frac{\lambda}{2}\sigma^2}$ subgroup, whereas
  $\left. SL(2,\mathbb{R})_k/U(1)\right|_{\text{vector}}$ corresponds
  to the $g\to hgh^{-1}$ gauging. This is the \emph{trumpet} and is
  T-dual to the cigar\footnote{Actually this statement holds only for
    the vector coset of the \emph{single cover} of
    $SL(2,\mathbb{R})$. Otherwise, from the n-th cover of the group
    manifold one obtains the n-th cover of the
    trumpet~\cite{Israel:2003ry}.}.  The generators of the affine
  residual symmetry $U(1)_{\mathrm L}\times U(1)_{\mathrm R}$ are both
  time-like (the corresponding Killing vectors are not orthogonal
  though). For extreme deformation, the time coordinate decouples and
  the antisymmetric tensor is trade for a dilaton. The isometries are
  time-translation invariance and rotation invariance in the
  cigar/trum\-pet.
\item $J^2\bar J^2$ The deformation is now induced by space-like currents.
  So is the residual affine symmetry $U(1)_{\mathrm L}\times U(1)_{\mathrm R}$ of
  the deformed model.  Extreme deformation points are T-dual: $U(1) \times
  SL(2,\mathbb{R})_k / U(1)$ where the $U(1)$ factor is space-like, and the
  $U(1)$ gauging of $SL(2,\mathbb{R})_k$ corresponds to $g\to hgh^{(-1)}$
  with $h=\exp{-\frac{\lambda}{2}\sigma^3}$~\cite{Dijkgraaf:1992ba}.
  The corresponding manifold is (some sector of) the
  Lorentzian two-dimensional black hole with a non-trivial dilaton.
  \item $(J^1 + J^3)(\bar J^1 + \bar J^3)$ This is the last alternative, with
  both null currents. The deformation connects $\mathrm{AdS}_3$ with
  $\mathbb{R} \times
  \mathbb{R}^{1,1}$ plus a dilaton linear in the first factor. The $U(1)_{\mathrm
    L}\times U(1)_{\mathrm R}$ left-over current algebra is
    light-like\footnote{The isometry is actually richer by one (two
    translations plus a boost), but the extra generator (the boost) is not
    promoted to an affine symmetry of the sigma-model.}. Tensorized with 
    an $SU(2)_k$ CFT, this background describes the decoupling limit of the 
    NS5/F1 setup~\cite{Israel:2003ry}, where the fundamental strings 
    regularize the strong coupling regime.
\end{enumerate}

Possible choices for the coordinate systems and the resulting fields
are reported in App.~\ref{cha:symm-deform-su2}.

Our purpose here is to analyze \emph{asymmetric} deformations of
$\mathrm{AdS}_3$. Following the similar analysis of the previous
section for $SU(2)$, we expect those deformations to preserve a
$U(1)_{\mathrm L}\times SL(2,\mathbb{R})_{\mathrm R}$ symmetry appearing as
affine algebra from the sigma-model point of view, and as isometry
group for the background. The residual $U(1)_{\mathrm L}$ factor can
be time-like, space-like or null depending on the current that has
been used to perturb the \textsc{wzw} model.

It is worth to stress that some deformations of $\mathrm{AdS}_3$ have
been studied in the past irrespectively of any conformal sigma-model
or string theory analysis. In particular it was observed
in~\cite{Rooman:1998xf}, following~\cite{Reboucas:1983hn} that the
three-dimensional\footnote{In fact, the original G\"odel solution is
  four-dimensional, but the forth space dimension is a flat spectator.
  In the following, we will systematically refer to the
  three-dimensional non-trivial factor.}  G\"odel solution of Einstein
equations could be obtained as a member of a one-parameter family of
$\mathrm{AdS}_3$ deformations that precisely enters the class we
discuss here.  G\"odel space is a constant-curvature Lorentzian
manifold. Its isometry group is $U(1) \times SL(2,\mathbb{R})$, and the
$U(1)$ factor is generated by a time-like Killing vector.  These
properties hold for generic values of the deformation parameter. In
fact the deformed $\mathrm{AdS}_3$ under consideration can be embedded
in a seven-dimensional flat space with appropriate signature, as the
intersection of four quadratic surfaces.  Closed time-like curves as
well as high symmetry are inherited from the multi-time maximally
symmetric host space.  Another interesting property resulting from
this embedding is the possibility for changing the sign of the
curvature along the continuous line of deformation, without
encountering any singular behaviour (see Eq. (\ref{curnecogo})).

It seems natural to generalize the above results to \emph{new}
$\mathrm{AdS}_3$ deformations and promote them to \emph{exact} string
backgrounds. Our guideline will be the requirement of a $U(1) \times
SL(2,\mathbb{R})$ isometry group, with space-like or light-like
$U(1)$'s.

We will first review the time-like (elliptic) deformation of
$\mathrm{AdS}_3$ of~\cite{Rooman:1998xf} and recently studied from a
string perspective in~\cite{Israel:2003cx}. Hyperbolic (space-like)
and parabolic (light-like) deformations will be analyzed in the
following.  All these deformations are of the type presented in the
previous chapter; further generalizations will be obtained in
Sec.~\ref{sec:blackstring}. We show in the following how to implement
these deformations as exact marginal perturbations in the framework of
the $SL(2,\mathbb{R})_k$ \textsc{wzw} model embedded in heterotic
string.

\subsection{Elliptic deformation: magnetic background}

Consider $\mathrm{AdS}_3$ in $\left(t, \rho, \phi \right)$
coordinates, with metric given in (\ref{eq:ads-rhotphi-metric}). In
these coordinates, two manifest Killing vectors are $L_3
\sim \partial_t$ and $R_2 \sim \partial_\phi$, time-like and
space-like respectively (see App.~\ref{cha:geometry-arond-group},
Tab.~\ref{tab:currents-timelike}).

The deformation studied in~\cite{Rooman:1998xf} and quoted
as ``squashed anti de Sitter'' reads, in the above coordinates:
\begin{equation}
  \di s^2= \frac{L^2}{4} \left[ \di \rho^2 + \cosh^2 \rho  \di \phi^2 -
    \left( 1 + 2 \h^2\right) \left( \di t + \sinh \rho \di
      \phi \right)^2 \right].
  \label{dsnecogo}
\end{equation}
It preserves a $U(1)\times SL(2,\mathbb{R})$ isometry group. The $U(1)$ is
generated by the \emph{time-like} vector $L_3$ of one original
$SL(2,\mathbb{R})$, while the right-moving $SL(2,\mathbb{R})$ is unbroken
(the expressions for the $\set{L_3,R_1,R_2,R_3}$ Killing vectors in
Tab.~\ref{tab:currents-timelike} remain valid at any value of the
deformation parameter). The Ricci scalar is constant
\begin{equation}
  R=-{2\over L^2}(3 - 2\h^2), \label{curnecogo}
\end{equation}
while the volume form reads:
\begin{equation}
  \omega_{[3]} = \frac{L^3}{8} \sqrt{\left| 1+2\h^2\right| }\, \cosh \rho\,  \di \rho
  \land \di \phi \land \di t.
\end{equation}
For $\h^2 = 1/2$, this deformation coincides with the G\"odel
metric. It should be stressed, however, that nothing special
occurs at this value of the deformation parameter. The properties
of G\"odel space are generically reproduced at any $\h^2>0$.

From a physical point of view, as it stands, this solution is pathological
because it has topologically trivial closed time-like curves through each
point of the manifold, like G\"odel space-time which belongs to this family.
Its interest mostly relies on the fact that it can be promoted to an exact
string solution, with appropriate \textsc{ns} and magnetic backgrounds. The
high symmetry of (\ref{dsnecogo}), is a severe constraint and, as was shown
in~\cite{Israel:2003cx}, the geometry at hand does indeed coincide with the
unique marginal deformation of the $SL(2,\mathbb{R})_k$ \textsc{wzw} that
preserves a $U(1)_{\mathrm L}\times SL(2,\mathbb{R})_{\mathrm R}$ affine algebra with
time-like $U(1)_{\mathrm L}$.

\marginlabel{Closed time-like curves in the $SL(2,\setR)$ elliptic
  deformation}It is interesting to observe that, at this stage, the
deformation parameter $\h^2$ \emph{needs not be
  positive.}:~(\ref{dsnecogo}) solves the Einstein-Maxwell-scalar
equations~\cite{Reboucas:1983hn} for any $\h^2$.  Furthermore, for
$\h^2 <0$, there are no longer closed time-like curves\footnote{As
  mentioned previously, the geometry at hand can be embedded in a
  seven-dimensional flat space, with signature $\varepsilon ---+++$,
  $\varepsilon = {\mathrm sign}(-\h^2)$~\cite{Rooman:1998xf}.  This
  clarifies the origin of the symmetry as well as the presence or
  absence of closed time-like curves for positive or negative
  $\h^2$.}. This statement is based on a simple argument\footnote{This
  argument is local and must in fact be completed by global
  considerations on the manifold (see~\cite{Rooman:1998xf}).}.
Consider a time-like curve $x^\mu = x^\mu \left( \lambda \right)$. By
definition the tangent vector $\partial_\lambda$ is negative-norm,
which, by using Eq.  (\ref{dsnecogo}), translates into
\begin{equation}
  \left( {\di\rho \over \di\lambda}\right)^2 + \cosh^2 \rho
  \left( {\di\phi \over \di\lambda}\right)^2 -
  \left( 1 + 2 \h^2\right) \left( {\di t \over \di\lambda} + \sinh \rho
    {\di\phi \over \di\lambda} \right)^2 <0.
\end{equation}
If the curve is closed, $\di t /\di \lambda$ must vanish somewhere. At
the turning point, the resulting inequality,
\begin{equation}
  \left(2\h^2 \sinh^2 \rho  -1\right) \left({\di\phi \over \di\rho}\right)^2
  >1
\end{equation}
is never satisfied for $\h^2<0$, whereas it is for large enough\footnote{This
  means $\rho > \rho_{\mathrm c}$ where $\rho_{\mathrm c}$ is the radius where the norm of
  $\partial_\phi$ vanishes and switches to negative ($\| \partial_\phi \|^2 = L^2\left(
    1-2\h^2\sinh^2 \rho \right)/4 $).  This never occurs for $\h^2<0$.} $\rho$
otherwise.

This apparent \emph{regularization of the causal pathology},
unfortunately breaks down at the string level. In fact, as we will
shortly see, in order to be considered as a string solution, the
above background requires a (chromo)magnetic field. The latter
turns out to be proportional to $\h$, and becomes
\emph{imaginary} in the range where the closed time-like curves
disappear. Hence, at the string level, unitarity is trade for
causality. It seems that no regime exists in the magnetic
deformation of $\mathrm{AdS}_3$, where these fundamental
requirements are simultaneously fulfilled.

We now turn to the string realization of the above squashed sphere.
In the heterotic backgrounds considered here, of the type
$\mathrm{AdS}_3 \times S^3 \times T^4$, the two-dimensional $N=(1,0)$
world-sheet action corresponding to the $\mathrm{AdS}_3$ factor is:
\mathindent=0em
 \begin{equation}
    S_{SL(2,\mathbb{R})_k} =\frac{1}{2\pi} \int {\mathrm d}^2 z
    \left\{{k\over 4}
      \left( \partial \rho \bar \partial \rho - \partial t \bar \partial t + \partial \phi \bar \partial \phi - 2 \sinh \rho \,
        \partial \phi \bar \partial t \right) + \eta_{ab}\, \psi^a \bar \partial \psi^b\right\},
    \label{SL2RWZW}
  \end{equation}
\mathindent=\oldindent
where $\eta_{ab}={\mathrm diag \ }(++-)$, $a=1,2,3$ and $\psi^a$ are
the left-moving superpartners of the $SL(2,\mathbb{R})_k$ currents
(see Tab.~\ref{tab:currents-timelike}). The corresponding
background fields are the metric (Eq.
(\ref{eq:ads-rhotphi-metric})) with radius $L=\sqrt{k}$ and the NS
B-field:
\begin{equation}
 \label{eq:ads-rhotphi-B}
 B = - \frac{k}{4} \sinh \rho \di \phi \land \di t.
\end{equation}
The three-form field strength is $H_{[3]} = \di
B=-\frac{2}{\sqrt{k}}\,
 \omega_{[3]}$ with $\omega_{[3]}$ displayed in Eq.~\eqref{eq:ads-rhotphi-vf}.

 The asymmetric perturbation that preserves a $U(1)_{\mathrm L}\times
 SL(2,\mathbb{R})_{\mathrm R}$ affine algebra with time-like
 $U(1)_{\mathrm L}$ is $\delta S$ given in
 Eq.~\eqref{eq:delta-S-J3}, where $J^3$ now stands for the left-moving
 time-like $SL(2,\mathbb{R})_k$ current given in App.~\ref{antids},
 Tab.~\ref{tab:currents-timelike}. This perturbation corresponds to
 switching on a (chromo)magnetic field, like in the $SU(2)_k$ studied
 in Sec.~\ref{sec:deformed-su2}. It is marginal and can be integrated for finite
 values of $\h$, and is compatible with the $N=(1,0)$ world-sheet
 supersymmetry. The resulting background fields, extracted in the
 usual manner from the deformed action are the metric (\ref{dsnecogo})
 with radius $L=\sqrt{k}$ and the following gauge field:
\begin{equation}
  A = \h \sqrt{\frac{2k}{k_g}} \left(\di t + \sinh \rho \di \phi
  \right).
  \label{adsmag}
\end{equation}
The NS $B$-field is not altered by the deformation, (Eq.
(\ref{eq:ads-rhotphi-B})), whereas the three-form field strength
depends explicitly on the deformation parameter $\h$, because of
the gauge-field contribution:
\begin{equation}
  H_{[3]} = \di B - \frac{k_G}{4} A \land \di A =
 - \frac{k}{4}\left( 1+ 2\h^2\right) \cosh \rho \di \rho \land \di \phi \land \di
 t.
  \label{adsmagH}
\end{equation}

One can easily check that the background fields (\ref{dsnecogo}),
(\ref{adsmag}) and (\ref{adsmagH}) solve the lowest-order equations of
motion. Of course the solution we have obtained is exact, since it has
been obtained as the marginal deformation of an exact conformal
sigma-model. The interpretation of the deformed model in terms of
background fields $\{ G_{ab}, B_{ab}, F_{ab}^G \}$ receives however
the usual higher-order correction summarized by the shift $k \to k + 2
$ as we have already explained in Sec.~\ref{sec:no-renorm-theor}.

Let us finally mention that it is possible to extract the spectrum
and write down the partition function of the above
theory~\cite{Israel:2003cx}, since the latter is an exact
deformation of the $SL(2,\mathbb{R})_k$ \textsc{wzw} model. This
is achieved by deforming the associated elliptic Cartan
subalgebra. The following picture emerges then from the analysis
of the spectrum. The short-string spectrum, corresponding to
world-sheets trapped in the ``center'' of the space--time (for some
particular choice of coordinates) is well-behaved, because these
world-sheets do not feel the closed time-like curves which are
``topologically large''. On the contrary, the long strings can
wrap the closed time-like curves, and their spectrum contains many
tachyons. Hence, the caveats of G\"odel space survive the string
framework, at any value of $\h^2>0$. One can circumvent them by
slightly deviating from the G\"odel line with an extra purely
gravitational deformation, driven by $J^3 \bar{J}^3$. This
deformation isolates the causally unsafe region, $\rho > \rho_{\mathrm
c}$ (see~\cite{Israel:2003cx} for details). It is similar in 
spirit with the supertube domain-walls of~\cite{Drukker:2003sc} 
curing the G\"odel-like space-times with RR backgrounds.

\subsection{Hyperbolic deformation: electric background}

\subsubsection{The background and its CFT realization}

We will now focus on a different deformation. We use coordinates
(\ref{eq:ads-rxt-coo}) with metric (\ref{eq:ads-rxt-met}), where the
manifest Killing vectors are $L_2 \sim \partial_x$ (space-like) and $R_3 \sim \partial_\tau$
(time-like) (see App.~\ref{antids}, Tab.~\ref{tab:currents-spacelike}). This
time we perform a deformation that preserves a $U(1) \times SL(2,\mathbb{R})$
isometry. The $U(1)$ corresponds to the space-like Killing vector $L_2$,
whereas the $SL(2,\mathbb{R})$ is generated by $R_1, R_2, R_3$, which are
again not altered by the deformation. The resulting metric reads:
\begin{equation}
  \di s^2= \frac{L^2}{4}\left[ \di r^2 - \cosh^2 r \di \tau^2 +
    \left( 1-2\h^2\right) \left( \di x + \sinh r \di \tau \right)^2\right].
  \label{dsnecoma}
\end{equation}
The scalar curvature of this manifold is constant
\begin{equation}
  R=-\frac{2}{L^2}\left(3+2\h^2\right)\label{Rel}
\end{equation}
and the volume form
\begin{equation}
  \omega_{[3]} = \frac{L^3}{8}\sqrt{\left|1-2\h^2 \right|}\, \cosh^2 r\,  \di r
  \land \di \tau \land \di x.\label{vfelec}
\end{equation}

Following the argument of the previous section, one can check whether
closed time-like curves appear. Indeed, assuming their existence, the
following inequality must hold at the turning point \emph{i.e}. where
$\di t/\di \lambda$ vanishes ($\lambda$ being the parameter that
describes the curve):
\begin{equation}
  \left( 2\h^2 -1 \right)\left( {\di x \over \di r} \right)^2>1 .
\end{equation}
The latter cannot be satisfied in the regime $\h^2<1/2$. Notice
that the manifold at hand is well behaved, even for negative
$\h^2$.

Let us now leave aside these questions about the classical
geometry, and address the issue of string realization of the above
background. As already advertised, this is achieved by considering
a world-sheet-supersymmetric marginal deformation of the
$SL(2,\mathbb{R})_k$ \textsc{wzw} model that implements
(chromo)electric field. Such a deformation is possible in the
heterotic string at hand:
\begin{equation}
  \delta S = \frac{\sqrt{k k_G}\h}{2\pi} \int {\mathrm
    d}^2 z \left(J^2 + i \psi^1 \psi^3\right) \bar J_G,
\label{actelecdef}
\end{equation}
($\bar J_G$ is any Cartan current of the group $G$ and $J^2$ is
given in App.~\ref{antids}, Tab.~\ref{tab:currents-spacelike}),
and corresponds, as in previous cases, to an integrable marginal
deformation. The deformed conformal sigma-model can be analyzed in
terms of background fields. The metric turns out to be
(\ref{dsnecoma}), whereas the gauge field and three-form tensor
are
\begin{align}
  A &= \h \sqrt{\frac{2k}{k_g}} \left( \di x + \sinh r \di \tau
  \right),  \\
  H_{[3]} &= \frac{k}{4} \left(1-2\h^2\right) \cosh r \di r \land \di \tau \land \di
  x.
\end{align}
As expected, these fields solve the equations of motion.

The background under consideration is a new string solution generated
as a hyperbolic deformation of the $SL(2,\mathbb{R})_k$ \textsc{wzw}
model. In contrast to what happens for the elliptic deformation above,
the present solution is perfectly sensible, both at the classical and
at the string level.

\subsubsection{The spectrum of string primaries}

\marginlabel{String primaries for $SL(2,\setR)$ hyperbolic
  deformation}The electric deformation of $\mathrm{AdS}_3$ is an exact
string background.  The corresponding conformal field theory is
however more difficult to deal with than the one for the elliptic
deformation. In order to write down its partition function, we must
decompose the $SL(2,\mathbb{R})_k$ partition function in a hyperbolic
basis of characters, where the implementation of the deformation is
well-defined and straightforward; this is a notoriously difficult
exercise. On the other hand the spectrum of primaries is
known\footnote{In the following we do not consider the issue of the
  spectral-flow representations. The spectral-flow symmetry is
  apparently broken by the deformation considered here.}  from the
study of the representations of the Lie algebra in this basis (see
\emph{e.g.}~\cite{Vilenkin}, and~\cite{Dijkgraaf:1992ba} for the
spectrum of the hyperbolic gauged \textsc{wzw} model, \emph{i.e.} at
the extreme value of the deformation parameter).  The part of the
heterotic spectrum of interest contains the expression for the
primaries of $N=(1,0)$ affine $SL(2,\mathbb{R})$ at purely bosonic
level\footnote{More precisely we consider primaries of the purely
  bosonic affine algebra with an arbitrary state in the fermionic
  sector.} $k+2$, together with some $U(1)$ from the lattice of the
heterotic gauge group:
\begin{align}
  L_0 &= -\frac{j(j-1)}{k} - \frac{1}{2} \left(n+\frac{a}{2}\right)^2,
  \label{leftspecads} \\
  \bar L_0 &= -\frac{j(j-1)}{k} +   \frac{1}{2}
  \left(\bar{n}+\frac{h}{2}\right)^2  ,
  \label{rightspecads}
\end{align}
where the second Casimir of the representation of the $SL(2,\mathbb{R})$
algebra, $-j(j-1)$, explicitly appears. The spectrum contains
\emph{continuous representations}, with $j = \frac{1}{2} + \imath s$, $s \in
\mathbb{R}_+$. It also contains \emph{discrete representations}, with $j \in
\mathbb{R}_{+}$, lying within the unitarity range $1/2 < j < (k+1)/2$
(see~\cite{Maldacena:2000hw, Petropoulos:1990fc}). In both cases the
spectrum of the hyperbolic generator $J^2$ is $\mu \in \mathbb{R}$. The
expression for the left conformal dimensions, Eq.~(\ref{leftspecads}), also
contains the contribution from the world-sheet fermions associated to the
$\imath \psi^1 \psi^3$ current. The sector (\textsc{r} or \textsc{ns}) is labelled
by $a \in \setZ_2$. Note that the unusual sign in front of the lattice is the
natural one for the fermions of the light-cone directions. In the
expression~(\ref{rightspecads}) we have similarly the contribution of the
fermions of the gauge group, where $h$ labels the corresponding sector.

We are now in position to follow the procedure, familiar from the previous
examples: we have to (\emph{i}) isolate from the left spectrum the lattice
of the supersymmetric hyperbolic current $J^2 + \imath \psi^1 \psi^3$ and
(\emph{ii}) perform a boost between this lattice and the fermionic lattice
of the gauge field. We hence obtain the following expressions:
\begin{subequations}
\label{eq:J2spectrum}
  \begin{align}
    \begin{split}
      L_0 &= -\frac{j(j-1)}{k} - \frac{\mu^2}{k+2} - \frac{k+2}{2k}
      \left( n+ \frac{a}{2} + \frac{2\mu}{k+2} \right)^2 +\\
      & \hspace{6em}+ \frac{1}{2} \left[ \sqrt{\frac{2}{k}} \left(\mu
          + n + \frac{a}{2} \right) \cosh x + \left(\bar{n} +
          \frac{h}{2} \right) \sinh x \right]^2,
    \end{split}\label{leftspecadsdef}\\
    \bar L_0 &= -\frac{j(j-1)}{k} + \frac{1}{2} \left[ \left(\bar{n} +
        \frac{h}{2} \right) \cosh x +\sqrt{\frac{2}{k}} \left(\mu + n
        + \frac{a}{2} \right) \sinh x \right]^2.
    \label{rightspecadsdef}
  \end{align}
\end{subequations}
The relation between the boost parameter $x$ and the deformation
parameter $\h^2$ is given in Eq. (\ref{defsphpar}), as for the case of
the $SU(2)_k$ deformation. In particular it is worth to remark that
the first three terms of~(\ref{leftspecadsdef}) correspond to the left
weights of the supersymmetric two-dimensional Lorentzian black hole,
\emph{i.e.} the $SL(2,\mathbb{R})/ O(1,1)$ gauged super-\textsc{wzw}
model.

This result is less striking that the whole partition function we
obtained for the compact $SU(2)$. It is worthwhile to remark that the
difference is only due to technical reasons: in principle the very
same construction could be applied for the case at hand but it would
require the decomposition of the $SL(2,\setR)$ partition function in
terms of hyperbolic characters that at present is not yet known.

\subsection{Parabolic deformation: the AdS-wave background}

In the deformations encountered in the previous sections one
$SL(2,\mathbb{R})$ isometry breaks down to a $U(1)$ generated either
by a time-like or by a space-like Killing vector.  Deformations which
preserve a light-like isometry do also exist and are easily
implemented in Poincar\'{e} coordinates.

We require that the isometry group is $U(1) \times SL(2,\mathbb{R})$
with a null Killing vector for the $U(1)$ factor. Following the by now
familiar for the particular case of light-like residual isometry, we
are lead to
\begin{equation}
  \di s^2 =L^2\left[\frac{ \di u^2 }{u^2}+ \frac{\di x^+ \di x^-
    }{u^2}-2\h^2 \left(\frac{\di x^+}{u^2}\right)^2 \right].
  \label{dsemdef}
\end{equation}
The light-like $U(1)$ Killing vector is $L_1 + L_3 \sim \partial_-$ (see
App.~\ref{antids}, Tab.~\ref{tab:currents-poincare}). The remaining
$SL(2,\mathbb{R})$ generators are $\set{R_1+R_3,R_1-R_3,R_2}$ and
remain unaltered after the deformation.

The above deformed anti-de-Sitter geometry looks like a superposition
of $\mathrm{AdS}_3$ and of a plane wave (whence the AdS-wave name).
As usual, the sign of $\h^2$ is free at this stage, and $\h^2<0$ are
equally good geometries. In the near-horizon region ($\abs{u} \gg \abs{
  \h^2}$) the geometry is not sensitive to the presence of the wave.
On the contrary, this plane wave dominates in the opposite limit, near
the conformal boundary.

The volume form is not affected by the deformation, and it is still
given in (\ref{eq:ads-poinc-volume}); neither is the Ricci scalar
modified:
\begin{equation}
  R=-\frac{6}{L^2} . \label{Remdef}
\end{equation}
\marginlabel{Parabolic discrete deformation}Notice also that the
actual value of $\abs{ \h }$ is not of physical significance: it can
always be absorbed into a reparametrization $x^+ \to x^+ /\abs{ \h}$
and $x^-\to x^- \abs{ \h}$. The only relevant values for $\h^2$ can
therefore be chosen to be $0, \pm 1$.

We now come to the implementation of the geometry (\ref{dsemdef}) in a
string background. The only option is to perform an asymmetric exactly
marginal deformation of the heterotic $SL(2,\mathbb{R})_k$
\textsc{wzw} model that preserves a $U(1)_{\mathrm L}\times
SL(2,\mathbb{R})_{\mathrm R}$ affine symmetry.  This is achieved by
introducing
\begin{equation}
  \delta S_{\mathrm electric-magnetic} = -4{\sqrt{k k_G}\h} \int
  \di^2 z \left(J^1 + J^3 + i \left( \psi^1 + \psi^3 \right)
    \psi^2 \right) \bar J_G,
\label{actemdef}
\end{equation}
($J^1 + J^3$ is defined in App.~\ref{antids},
Tab.~\ref{tab:currents-poincare}). The latter perturbation is
integrable and accounts for the creation of a (chromo)electromagnetic
field
\begin{equation}
  A = 2 \sqrt{2k\over k_G} \h {\di x^+\over u^2}.\label{adsem}
\end{equation}
It generates precisely the deformation (\ref{dsemdef}) and leaves
unperturbed the \textsc{ns} field, $H_{[3]}= \di B = -
\frac{2}{\sqrt{k}} \, \omega_{[3]}$.

As a conclusion, the $\mathrm{AdS}_3$ plus plane-wave gravitational
background is described in terms of an exact conformal sigma model,
that carries two extra background fields: an \textsc{ns} three-form and
an electromagnetic two-form. Similarly to the symmetric parabolic
deformation~\cite{Israel:2003ry}, the present asymmetric one can be
used to construct a space--time supersymmetric background. The
$SL(2,\mathbb{R})_k$-\textsc{cft} treatment of the latter deformation
would need the knowledge of the parabolic characters of the affine
algebra, not available at present.

\subsection{Quantum point particles in the AdS-wave background}
\label{sec:point-part-adsw}

Further insights of the physics of the AdS-wave background can be
gathered if we look at the motion of point particles. Let us start
with the sigma model Lagrangian where we keep the $\h$ parameter
explicitly for sake of consistency:
\begin{equation}
  S = \int \di z^2 \: \frac{1}{u^2} \partial u \bar \partial u + \frac{1}{u^2} \partial x^+ \bar \partial x^- - 2 \frac{\h}{ u^2 } \partial x^- \bar \partial \varphi + \frac{1}{2} \partial \varphi \bar \partial \varphi , 
\end{equation}
where all the fields are function of $\sigma $ and $\tau$. The point particle
limit can be obtained if we let the $\sigma $-dependence drop. This leads to:
\begin{equation}
  S_{\text{point}} = \frac{1}{2 } \int \di \tau  \left\{ \frac{1}{u^2} \dot u^2 
    + \frac{1}{ u^2}  \dot x^- \dot x^+ + \frac{1}{2} \dot \varphi^2  -
     2 \frac{\h }{u^2}  \dot x^- \dot \varphi   \right\} ,
\end{equation}
\marginlabel{Kaluza-Klein reduction as a partial integration}where the
dot stands for the time derivative. The fourth dimension $\varphi$ was
introduced as a fake direction along which perform a Kaluza--Klein
reduction. In this framework the same result is obtained if we
consider $\varphi $ as an auxiliary variable and then substitute its
equation of motion:
\begin{equation}
  \dot \varphi = 2 \frac{\h \dot x^-}{2 u^2}.
\end{equation}
The resulting effective action is then written as:
\begin{equation}
  S_{\text{point}} = \frac{1}{2 } \int \di \tau  \left\{ \frac{1}{u^2} \dot u^2 
    + \frac{1}{ u^2} \dot x^- \dot x^+ - \frac{2\h^2}{ u^4} \left( x^-\right)^2    \right\}
\end{equation}
that is exactly the action for a free particle in the $3$\textsc{d}
AdS-wave metric in Eq.~\eqref{dsemdef}. Now, out of this we can derive the Hamiltonian:
\begin{equation}
  H_{\text{point}} = \frac{1}{16} \left( c^2 p_+^2 + 2 u^2 \left( p_- p_+ +
      p_u^2 \right)\right) ,
\end{equation}
and with the usual rules of quantization this naturally translates to
the Laplacian of the AdS-wave geometry:
\begin{equation}
  \triangle = \nabla_\mu \nabla^\mu = 8 \h^2 \partial_+^2 + 4 u^2 \partial_+ \partial_- -  u
    \partial_u +  u^2 \partial_{u}^2 .
\end{equation}
A quantum point particle described by the wave function $\Psi \left(
  u, x^-, x^+ \right)$ must then obey the Klein-Gordon equation:
\begin{equation}
  -\triangle \Psi \left( u, x^-, x^+ \right) = m^2 \Psi \left( u, x^-, x^+ \right) . 
\end{equation}
The fact that only the $u$ variable appears explicitly suggests that
we can write the solution as:
\begin{equation}
  \Psi \left( u, x^-, x^+ \right) = \int \di p_- \di p_+ \: e^{\imath \left( p_- x^-
      + p_+ x^+ \right)} \tilde \Psi \left( u, p_-, p_+ \right)
\end{equation}
so that the wave equation becomes:
\mathindent=0em
\begin{equation}
  u^2 \partial^2_u  \tilde \Psi \left( u, p_-, p_+
  \right) - u \partial_u  \tilde \Psi
  \left( u, p_-, p_+ \right)  = \left( -m^2 + 8 \h^2 p_+^2 + 4 u^2
    p_- p_+  \right) \tilde \Psi \left( u, p_-, p_+ \right) .
\end{equation}
\mathindent=\oldindent
This is a modified Bessel equation whose canonical form is:
\begin{equation}
  z^2 y^{\prime \prime } \left( z \right) + z y^\prime \left( z \right) - \left( z^2 +
    \nu^2 \right) y \left( z \right) = 0
\end{equation}
and after some algebra we can write the general solution as:
\begin{equation}
   \tilde \Psi \left( u, p_-, p_+ \right) = u I_\nu \left( 2 \sqrt{p_- p_+ }
     u \right) C_1 \left( p_-, p_+ \right) + u K_\nu \left( 2 \sqrt{p_- p_+ }
     u \right) C_2 \left( p_-, p_+ \right)
\end{equation}
where $C_1 \left( p_-, p_+ \right)$ and $C_2 \left( p_-, p_+ \right)$ are
arbitrary functions, $I_\nu \left( z\right)$ and $K_\nu \left( z \right)$ are
modified Bessel functions of the first and second kind and $\nu = \sqrt{1- m^2
  + 8 \left(\h  p_+ \right)^2}$.

\subsection{Limiting geometries: $\mathrm{AdS}_2$ and $H_2$}
\label{gencos}

We have analyzed in Sec.~\ref{sec:deformed-su2} the behaviour of the
magnetic deformation of $SU(2)_k$, at some critical (or boundary)
value of the modulus $H^2$, where the background factorizes as
$\mathbb{R}\times S^2$ with vanishing \textsc{ns} three-form and
finite magnetic field. We would like to address this question for the
asymmetric deformations of the $SL(2,\mathbb{R})_k$ model and show the
existence of limiting situations where the geometry indeed factorizes,
in agreement with the expectations following the general analysis of
Sec.~\ref{sec:wzw-deformations}

What can we expect in the framework of the $SL(2,\mathbb{R})_k$
asymmetric deformations? Any limiting geometry must have the generic
$U(1) \times SL(2,\mathbb{R})_k$ isometry that translates the affine
symmetry of the conformal model. If a line decouples, it accounts for
the $U(1)$, and the remaining two-dimensional surface must be
$SL(2,\mathbb{R})$-invariant.  Three different situations may arise:
$\mathrm{AdS}_2$, $H_2$ or $\mathrm{dS}_2$.  Anti de Sitter in two
dimensions is Lorentzian with negative curvature; the hyperbolic plane
$H_2$ (also called Euclidean anti de Sitter) is Euclidean with
negative curvature; de Sitter space is Lorentzian with positive
curvature.

Three deformations are available for $\mathrm{AdS}_3$ and these have
been analyzed in Sec.~\ref{sec:deformed-sl2}. For unitary string
theory, all background fields must be real and consequently $\h^2>0$
is the only physical regime. In this regime, only the hyperbolic
(electric) deformation exhibits a critical behaviour at
$\h^2_{\text{max}} = 1/2$. For $\h^2 < 1/2$, the deformation at hand
is a Lorentzian manifold with no closed time-like curves.  When $\h^2
> 1/2$, $\det \mathbf{g}>0$ and \emph{two} time-like directions
appear. At $\h^2_{\vphantom m} = \h^2_{\text{max}}$, $\det \mathbf{g}$
vanishes, and this is the signature that some direction indeed
decompactifies.

\marginlabel{Decompactifying to $\mathrm{AdS}_2$}We proceed therefore
as in Sec.~\ref{sec:wzw-deformations}, and define a rescaled
coordinate in order to keep the decompactifying direction into the
geometry and follow its decoupling:
\begin{equation}
  y=\sqrt{\frac{k}{2}\left(\frac{1}{2}  -\h^2\right)} \, x\ .
\end{equation}
The metric and volume form now read:
\begin{equation}
  \di s^2 = \di y^2 + \frac{k}{4} \left[ \di r^2 - \left( 1 + 2 \h^2 \sinh^2 r \right)
    \di \tau^2 \right] + \sqrt{k\left(1-2\h^2 \right)}\sinh r\, \di \tau\, \di y
  \label{ads3metdefren}
\end{equation}
and
\begin{equation}
  \omega_{[3]} = \frac{k}{4} \, \cosh r\,  \di r
  \land \di \tau \land \di y.
\end{equation}
For $\h^2_{\vphantom x}$ close to $\h^2_{\text{max}}$, the $y$-direction
factorizes
\begin{equation}
  \di s^2\xrightarrow[\h^2 \to \h^2_{\text{max}}]{}\di y ^2 +  \frac{k}{4}\left[\di
    r^2- \cosh^2 r \, \di \tau^2\right].
\end{equation}
The latter expression captures the phenomenon we were expecting:
\begin{equation}
  \mathrm{AdS}_3\xrightarrow[\h^2 \to \h^2_{\text{max}}]{} \mathbb{R} \times
  \mathrm{AdS}_2.
\end{equation}
It also shows that the two-dimensional anti de Sitter has radius
$\sqrt{k/4}$ and supports entirely the curvature of the limiting
geometry, $R=-8/k$ (see expression (\ref{Rel})).

The above analysis shows that, starting from the $SL(2,\mathbb{R})_k$
\textsc{wzw} model, there is a line of continuous exact deformation
(driven by a (chromo)electric field) that leads to a conformal model
at the boundary of the modulus $\h^2$. This model consists of a free
non-compact boson times a geometric coset $\mathrm{AdS}_2\equiv
SL(2,\mathbb{R})/U(1)$, with a finite electric field:
\begin{equation}
  F=  \sqrt{k \over k_G}\cosh r \, \di r \land \di \tau
\end{equation}
and vanishing \textsc{ns} three-form background. The underlying
geometric structure that makes this phenomenon possible is that
$\mathrm{AdS}_3$ can be considered as a non-trivial $S^1$ fibration
over an $\mathrm{AdS}_2$ base. The radius of the fiber couples to the
electric field, and vanishes at $\h^2_{\text{max}}$.  The important
result is that this enables us to promote the geometric coset
$\mathrm{AdS}_2$ to an exact string vacuum.

We would like finally to comment on the fate of $\mathrm{dS}_2$ and
$H_2$ geometries, which are both $SL(2,\mathbb{R})$-symmetric. De
Sitter and hyperbolic geometries are not expected to appear in
physical regimes of string theory unless Ramond-Ramond fields are
turned on (see Ch.~\ref{cha:hyperbolic-spaces}). The $H_3$
sigma-model, for example, is an exact conformal field theory, with
imaginary antisymmetric tensor background
though~\cite{Gawedzki:1991yu,Teschner:1997ft}.  Similarly, imaginary
\textsc{ns} background is also required for de Sitter vacua to solve
the low-energy equations. It makes sense therefore to investigate
regimes with $\h^2<0$, where the electric or magnetic backgrounds are
indeed imaginary.

\marginlabel{Non-unitary $H_2$ solution}The elliptic (magnetic)
deformation exhibits a critical behaviour in the region of negative
$\h^2$, where the geometry does not contain closed time-like
curves. The critical behaviour appears at the minimum value
$\h^2_{\text{min}} =-1/2$, below which the metric becomes
Euclidean. The vanishing of $\det \mathbf{g}$ at this point of the
deformation line, signals the decoupling of the time direction. The
remaining geometry is nothing but a two-dimensional hyperbolic plane
$H_2$. It is Euclidean with negative curvature $R=-8/k$ (see Eq.
(\ref{curnecogo}) with $L^2 = k$).

All this can be made more precise by introducing a rescaled time
coordinate:
\begin{equation}
  T =\sqrt{ \frac{k}{2} \left( \frac{1}{2}  +\h^2\right)} \, t.
\end{equation}
The metric and volume form now read:
\mathindent=0em
\begin{equation}
  \di s^2= - \di T^2 + \frac{k}{4} \left[ \di \rho^2 + \left( 1 -
      2 \h^2 \sinh^2 \rho \right) \di \phi^2 \right] -
  \sqrt{k\left(1+2\h^2\right)} \sinh \rho \di \phi \di T
  \label{ads3metmagdefren}
\end{equation}
\mathindent=\oldindent
and
\begin{equation}
  \omega_{[3]} =  \frac{k}{4}  \cosh \rho  \di \rho \land \di \phi  \land \di T.
\end{equation}
For $\h^2_{\vphantom x}$ close to $\h^2_{\text{min}}$, the $T$-direction
factorizes
\begin{equation}
  \di s^2\xrightarrow[\h^2_{\vphantom m} \to \h^2_{\text{max}}]{} -\di
  T ^2 +  \frac{k}{4}\left[ \di \rho^2 + \cosh^2 \rho \di \phi^2 \right].
\end{equation}
The latter expression proves the above statement:
\begin{equation}
  \mathrm{AdS}_3\xrightarrow[\h^2_{\vphantom m} \to \h^2_{\text{min}}]{}
  \mathbb{R} \times H_2,
\end{equation}
and the two-dimensional hyperbolic plane has radius $\sqrt{k/4}$.

Our analysis finally shows that the continuous line of exactly
marginal (chromo)magnetic deformation of the $SL(2,\mathbb{R})$
conformal model has a boundary at $\h^2 = -1/2 $ where its target
space is a free time-like coordinate times a hyperbolic plane. The
price to pay for crossing $\h^2 = 0$ is an imaginary magnetic field,
which at $\h^2 = -1/2$ reads:
\begin{equation}
  F= \sqrt{-{k \over k_G}}\cosh \rho \, \di \phi \land \di \rho.
\end{equation}
The \textsc{ns} field strength vanishes at this point, and the
geometric origin of the decoupling at hand is again the Hopf fibration
of the $\mathrm{AdS}_3$ in terms of an $H_2$.

\subsubsection{The $H_2$ spectrum}
\label{sec:h_2-spectrum}

A $H_2 \times \setR_t$ limit geometry can be reached if we allow for
negative values of $\h^2$ which in turn imply the presence of an
imaginary magnetic field. Although this implies that the
corresponding string theory is pathological (in example because of
unitarity problems), we obtain a perfectly respectable \textsc{cft}
for which we can write, using the same technique as above, a
modular-invariant partition function.

\marginlabel{The $H_2$ \textsc{cft}}Let us start from the deformed
partition function (see~\cite{Israel:2003cx}).  The interesting part
for us is:
\begin{multline}
  \label{eq:deform-AdS-partition}
  \int \di^2 t \ Z_{\text{cigar}} \oao{-t_1}{-t_2} \sum_{N,W,n,\bar{n} \in
    \mathbb{Z}} e^{i\pi \left(2Nt_2 + b(n+\frac{a}{2}) - \delta
      (\bar{n}+\frac{\gamma}{2})
    \right)} \\
  \times \ q^{- \left[ \frac{\cos \zeta}{\sqrt{k+2}} \left( \frac{N}{2} +
        \frac{k+4}{2} (W+t_1 ) + n + \frac{a}{2} \right) +
      \frac{\sin \zeta}{\sqrt{2}} \left( \bar{n} + \frac{\gamma}{2} \right)
    \right]^2 + \frac{k+4}{2(k+2)}
    \left( n+\frac{a}{2} +(W+t_1 ) + \frac{N}{k+4} \right)^2} \\
  \times \ \bar{q}^{- \frac{1}{k+4} \left( \frac{N}{2} - \frac{k+4}{2}
      (W+t_1) \right)^2 + \left[ \frac{\cos \zeta}{\sqrt{2}} \left(
        \bar{n} + \frac{\gamma}{2} \right) - \frac{\sin \zeta}{\sqrt{k+2}}
      \left( N + \frac{k+4}{2} (W+t_1) + n + \frac{a}{2} \right)
    \right]^2}
\end{multline}
where
\begin{equation}
  \cos \zeta = \frac{1}{1+2\h^2}  .
\end{equation}
If we consider $\h^2 < 0$ the trigonometric functions became
hyperbolic and there is a critical point $\h^2 = - \nicefrac{1}{2}$
where the boost diverges. Consistency then imposes the following
constraint on the charges:
\begin{equation}
  \frac{1}{\sqrt{k+2}} \left( \frac{N}{2} + \frac{k+4}{2} (W+t_1 ) + n + \frac{a}{2} \right)
  + \frac{1}{\sqrt{2}} \left( \bar{n} + \frac{\gamma}{2}
  \right) = 0
\end{equation}
introducing, for notation convenience
\begin{equation}
  k = 2 p^2 - 2  
\end{equation}
where $p \in \setR$ (there is no reason for quantization), the
constraint can be rewritten as
\begin{equation}
  N + 2\left( p^2 + 1 \right)\left(W + t_1 \right) + 2n + a + p \left( \bar n + 2 \gamma  \right) = 0
\end{equation}
that is equivalent to asking
\begin{equation}
  \begin{cases}
    N + 2n +a = Q \in \setZ \\
    2 \left(p^2 + 1 \right) (W+t_1) + p \left(\bar n + 2 \gamma \right) =
    - Q
  \end{cases}
\end{equation}
whence we can rewrite $t_1 $ as
\begin{equation}
  t_1 = - \frac{Q + p \left(2 \bar n + \gamma  \right)}{2 \left(p^2 + 1 \right)} - W
\end{equation}
and Eq.~\eqref{eq:deform-AdS-partition} becomes:
\begin{multline}
  \int \di t_2 \sum_{N,Q,W,\bar n \in \setZ} Z_{\text{cigar}}
  \oao{\frac{Q + 2 p \left(\bar n + \nicefrac{\gamma}{2} \right)}{2
      \left(p^2 + 1 \right)}+ W}{-t_2} e^{\imath \pi \left( 2 N t_2 +
      b \frac{ Q - N}{2} -
      \delta \left( \bar n + \nicefrac{\gamma}{2} \right) \right) } \times \\
  \times q^{\frac{1}{2 \left(p^2 + 1\right)} \left( p \frac{Q-N}{2} -
      \left( \bar n + \nicefrac{\gamma }{2} \right) \right)^2} \bar
  q^{-\frac{1}{2 \left( p^2 + 1 \right)} \left( \frac{Q + N }{2} + p
      \left( \bar n + \nicefrac{\gamma }{2} \right) \right)^2}
\end{multline}
or, introducing the integers $A, B$ as:
\begin{align}
  A &= Q + N \\
  B &= Q - N
\end{align}
finally can write the $H^2 $ partition function as follows:
\begin{multline}
  \int \di t_2 \sum_{A,B,W,\bar n \in \setZ} Z_{\text{cigar}}
  \oao{\frac{A+B + 4 p \left(\bar n + \nicefrac{\gamma}{2} \right)}{4
      \left(p^2 + 1 \right)}+ W}{-t_2} e^{\imath \pi \left( \left(A-B
      \right) t_2 + b \frac{B}{2} - \delta \left( \bar n +
        \nicefrac{\gamma}{2} \right) \right) } \times\\
  \times q^{\frac{1}{2 \left(p^2 + 1\right)} \left( p \frac{B}{2} -
      \left( \bar n + \nicefrac{\gamma }{2} \right) \right)^2} \bar
  q^{-\frac{1}{2 \left( p^2 + 1 \right)} \left( \frac{A }{2} + p
      \left( \bar n + \nicefrac{\gamma }{2} \right) \right)^2} .
\end{multline}
It is intriguing to find that the partition function for the geometric
coset $H_2 = \mathrm{AdS}_3 / \setR$ is related to the one for the
adjoint coset $\text{cigar}=\mathrm{AdS}_3/ \setR$. One may wonder if
this hints at some operation allowing to pass from the former to the
latter, but we will not speculate further in this direction.

\section{Near horizon geometry for the Bertotti-Robinson black hole}
\label{sec:near-horiz-geom}

  The $\mathrm{AdS}_2 \times S^2$ geometry  appeared first in the
context of Reissner--Nordstr\"om black holes. The latter are
solutions of Maxwell--Einstein theory in four dimensions,
describing charged, spherically symmetric black holes. For a black
hole of mass $M$ and charge $Q$, the solution reads:
\begin{subequations}
  \begin{align}
    \di s^2 &= - \left( 1 - \frac{r_+}{r} \right)\left( 1 - \frac{r_-}{r}
    \right)\di t^2 + \frac{\di r^2}{\left( 1 - \frac{r_+}{r} \right)\left( 1
        - \frac{r_-}{r}
      \right)} + r^2 \di \Omega_{2}^2 \, ,\\
    F &= \frac{Q}{r^2} \ \di t \land \di r \hspace{1em}  \text{with}
    \hspace{1em} r_{\pm} = G_4 \left( M \pm \sqrt{M^2 - Q^2}
    \right);
  \end{align}
\end{subequations}
$r_+$ and $r_-$ are the outer and inner horizons, and $G_4$ is
Newton's constant in four dimensions.

In the extremal case, $r_+ = r_- = r_0$ ($M^2 = Q^2$), and the
metric approaches the $\mathrm{AdS}_2 \times S^2$ geometry in the
near-horizon\footnote{With the near-horizon coordinates
$U=(1-r_0/r)^{-1}$ and $T=t/r_0$, the near-horizon geometry is
$$\di s^2 = r_{0}^2 \left( - \frac{\di T^2 }{U^2} +\frac{\di U^2}{U^2} +  \di
\Omega_{2}^2\right).$$ Both $\mathrm{AdS}_2$ and $S^2$ factors
have the same radius $r_{0}$.} limit $r\to r_0$. This solution can
of course be embedded in various four-dimensional
compactifications of string theory, and will be supersymmetric in
the extremal case (see e.g.~\cite{Youm:1997hw} for a review). In
this context we are dealing with some heterotic compactification.

Notice that the $\mathrm{AdS}_2 \times S^2$ geometry also appears in
type IIB superstring theory, but with \textsc{rr}
backgrounds~\cite{Ferrara:1995ih}. The black hole solution is obtained
by wrapping D3-branes around 3-cycles of a Calabi--Yau three-fold; in
the extremal limit, one obtains the $\mathrm{AdS}_2 \times S^2$
solution, but at the same time the CY moduli freeze to some particular
values. A hybrid Green--Schwarz sigma-model action for this model has
been presented in~\cite{Berkovits:1999zq} (see
also~\cite{Verlinde:2004gt} for $\mathrm{AdS}_2$).  The interest for
$\mathrm{AdS}_2 \times S^2$ space--time is motivated by the fact that
it provides an interesting simplified laboratory for AdS/\textsc{cft}
correspondence~\cite{Maldacena:1998re}. In the present case the dual
theory should correspond to some superconformal quantum
mechanics~\cite{Boonstra:1998yu,Claus:1998ts,Gibbons:1998fa,Cadoni:2000gm}.

\subsection{The spectrum}\label{ads2spec}

As a first step in the computation of the $\mathrm{AdS}_2 \times S^2$
string spectrum, we must determine the spectrum of the
$\mathrm{AdS}_2$ factor, by using the same limiting procedure as in
Sec.~\ref{sec:deformed-su2} for the sphere. The spectrum of the
electrically deformed $\mathrm{AdS}_3$, is displayed in
Eqs.~(\ref{leftspecadsdef}) and
(\ref{rightspecadsdef}). \marginlabel{Boost on the spectrum of
  $\mathrm{AdS}_2$ primaries}The $\mathrm{AdS}_2$ limit is reached for
$\cosh x \to \infty$, which leads to the following constraint on the
charges of the primary fields:
\begin{equation}
\bar{n} + \frac{h}{2} + \sqrt{\frac{2}{k}} \left( \mu
+ n + \frac{a}{2} \right) = 0.
\label{ads2limit}
\end{equation}
In contrast with the $S^2$ case, since $\mu$ is any real number
--~irrespectively of the kind of $SL(2,\mathbb{R})$
representation~-- there is \emph{no extra} quantization
condition for the level to make this limit well-defined. In this
limit, the extra $U(1)$ decompactifies as usual and can be
removed. Plugging the constraint~(\ref{ads2limit}) in the
expressions for the dimensions of the affine primaries, we find
\begin{subequations}
  \label{primAdS2}
  \begin{align}
    L_0 &= -\frac{j(j-1)}{k} - \frac{1}{2} \left( \bar{n} + \frac{h}{2} \right)^2
    - \frac{1}{2} \left( n+ \frac{a}{2} \right)^2  , \label{AdS2L} \\
    \bar{L}_0 &= -\frac{j(j-1)}{k}.\label{AdS2R}
  \end{align}
\end{subequations}

In addition to the original $\mathrm{AdS}_3$ spectrum,
Eqs.~(\ref{leftspecads}) and (\ref{rightspecads}), the right-moving
part contain an extra fermionic lattice describing the states charged
under the electric field. Despite the absence of $N=2$ superconformal
symmetry due to the Lorentzian signature, the theory has a
``fermion-number'' left symmetry, corresponding to the current:
\begin{equation}
J = \imath \psi^1 \psi^3 + \frac{2}{k} \left(J^2 + \imath \psi^1
\psi^3\right).
\end{equation}
The charges of the primaries (\ref{primAdS2}) are
\begin{equation}
\mathcal{Q}_F = n + \frac{a}{2} - \sqrt{\frac{2}{k}} \left( \bar{n} +
\frac{h}{2} \right).
\end{equation}

\subsection{$\mathrm{AdS}_2 \times S^2 \times \mathcal{M}$ and space--time
  supersymmetry} 

Let us now consider the complete heterotic string background which
consists of the $\mathrm{AdS}_2 \times S^2$ space--time times an
$N=2$ internal conformal field theory $\mathcal{M}$, that we will
assume to be of central charge $\hat{c}=6$ and with integral
$R$-charges. Examples of thereof are toroidal or flat-space
compactifications, as well as Gepner models~\cite{Gepner:1988qi}.

\marginlabel{Supersymmetry and level quantization for $\mathrm{AdS}_2
  \times S^2$}The levels $k$ of $SU(2)$ and $\hat{k}$ of
$SL(2,\mathbb{R})$ are such that the string background is critical:
\begin{equation}
  \hat{c} = \frac{2(k-2)}{k} + \frac{2(\hat{k}+2)}{\hat{k}} =
  4 \implies k = \hat{k}.
\end{equation}
This translates into the equality of the radii of the corresponding
$S^2$ and $\mathrm{AdS}_2$ factors, which is in turn necessary for
supersymmetry. Furthermore, the charge quantization condition for the
two-sphere (Sec.~\ref{sec:deformed-su2}) imposes a further restriction
on the level to $k = 2p^2$, $p \in \mathbb{N}$.

In this system the total fermionic charge is
\begin{equation}
\mathcal{Q} = n + \frac{a}{2} - \frac{N-h/2}{p} + n' + \frac{a}{2}
- \frac{\bar{n}' + h/2}{p} + \mathcal{Q}_{\mathcal{M}}.
\end{equation}
Hence, assuming that the internal $N=2$ charge
$\mathcal{Q}_{\mathcal{M}}$ is integral, further constraints on
the electromagnetic charges of the theory are needed in order to
achieve space--time supersymmetry. Namely, we must only keep
states such that
\begin{equation}
  N + \bar{n'}  = 0 \mod p.
\end{equation}
This projection is some kind of generalization of Gepner models.
Usually, such a projection is supplemented in string theory by new
twisted sectors. We then  expect that, by adding on top of this
projection the usual \textsc{gso} projection on odd fermion number, one
will obtain a space--time supersymmetric background. However, the
actual computation would need the knowledge of hyperbolic coset
characters of $SL(2,\mathbb{R})$ (i.e. Lorentzian black-hole
characters), and of their modular properties.  We can already
observe that this ``Gepner-like'' orbifold keeps only states which
are ``dyonic'' with respect to the electromagnetic field background.
Notice that, by switching other fluxes in the internal theory
$\mathcal{M}$ one can describe more general projections.


\section{The three-dimensional black string revisited}
\label{sec:blackstring}

The $\mathrm{AdS}_3$ moduli space contains black hole geometries. This
has been known since the most celebrated of them -- the
two-dimensional $SL (2, \setR )/U(1)$ black hole -- was found by
Witten~\cite{Witten:1991yr,Dijkgraaf:1992ba}.  Generalisations of
these constructions to higher dimensions have been considered
in~\cite{Horne:1991gn,Gershon:1991qp,Horava:1991am,Klimcik:1994wp}.
The three-dimensional black
string~\cite{Horne:1991gn,Horne:1991cn,Horowitz:1993jc} has attracted
much attention, for it provides an alternative to the Schwarzschild
black hole in three-dimensional asymptotically flat
geometries\footnote{Remember that the \emph{no hair} theorem doesn't
  hold in three
  dimensions~\cite{Israel:1967wq,Heusler:1998ua,Gibbons:2002av}.}. In
this section we want to show how this black string can be interpreted
in terms of marginal deformations of $SL ( 2, \setR)$, which will
enable us to give an expression for its string primary states.

\marginlabel{The three-dimensional black string as a current-current
  deformation}In \cite{Horne:1991gn} the black string was obtained as
an $\left( SL (2, \setR) \times \setR \right) /\setR$ gauged
model. More precisely, expressing $g \in SL(2,\setR) \times \setR$ as:
\begin{equation}
  g = \begin{pmatrix}
    a & u & 0 \\
    -v & b & 0 \\
    0 & 0 & { e}^x
  \end{pmatrix},
\end{equation}
the left and right embeddings of the $\setR$ subgroup are identical and given by:
\begin{align}
    \epsilon_{L/R} :& \,  \setR \to SL(2,\setR) \times \setR \\
    \lambda &\mapsto \begin{pmatrix}
      {\mathrm e}^{\frac{1}{\sqrt{\lambda^2 + 2}}} & 0 & 0 \\
      0 & {\mathrm e}^{-\frac{1}{\sqrt{\lambda^2 + 2}}}   & 0 \\
      0 & 0 & {\mathrm e}^{\frac{\lambda}{\sqrt{\lambda^2 + 2}}}
    \end{pmatrix} .
\end{align}
From the discussion in Sec.~\ref{sec:backgr-fields-symm}, we see that
performing this gauging is just one of the possible ways to recover
the $J^2 \bar J^2$ symmetrically deformed $SL(2,\setR)$ geometry. More
specifically, since the gauged symmetry is axial ($g \to h g h$), it
corresponds (in our notation) to the $\kappa_2 < 1$ branch of the
deformed geometry in Eq. \eqref{eq:J2J2-metric}\footnote{The $R
  \gtrless 1$ convention is not univocal in literature.}.  One can
find a coordinate transformation allowing to pass from the usual
black-string solution
\begin{equation}
\label{eq:black-string}
  \begin{cases}
    \di s^2 = \frac{k}{4}\left[-\left(1-\frac{1}{r}\right) \di t^2 +
      \left(1-\frac{\mu^2}{ r}\right) \di x^2 +
      \left(1-\frac{1}{r}\right)^{-1}
      \left(1-\frac{\mu^2}{r}\right)^{-1} \frac{\di r^2}{r^2}\right], \\
    H = \frac{k}{4} \frac{\mu}{r} \di r \land \di x \land \di t, \\
    \mathrm{e}^{2 \Phi} = \frac{\mu}{r}
  \end{cases}
\end{equation}
to our (local) coordinate system, Eq.~\eqref{eq:J2J2-deform}. The
attentive reader might now be puzzled by this equivalence between a
one-parameter model such as the symmetrically deformed model and a
two-parameter one such as the black string in its usual coordinates
(in Eqs.~\eqref{eq:black-string} we redefined the $r$ coordinate as
$r\to r/M$ and then set $\mu = Q/M$ with respect to the conventions
in~\cite{Horne:1991gn}). \marginlabel{Single physical parameter for
  the black string}A point that it is interesting to make here is that
although, out of physical considerations, the black string is usually
described in terms of two parameters (mass and charge), the only
physically distinguishable parameter is their ratio $\mu = Q/M$ that
coincides with our $\kappa_2$ parameter. In the next section we will
introduce a different (double) deformation, this time giving rise to a
black hole geometry depending on two actual parameters (one of which
being related to an additional electric field).


As we remarked above, the axial gauging construction only applies
for $\mu <1 $, while, in order to obtain the other $\kappa_2 > 1$
branch of the $J^2 \bar J^2$ deformation, one should perform a
vector gauging. On the other hand, this operation, that would be
justified by a \textsc{cft} point of view, is not natural when one
takes a more geometrical point of view and writes the black string
metric as in Eq.~\eqref{eq:black-string}. In the latter, one
can study the signature of the metric as a function of $r$ in the
two regions $\mu^2 \gtrless 1 $, and find the physically sensible
regions (see Tab.~\ref{tab:bs-signature}).

\begin{table} 
  \centering
  \begin{tabular}{|c|c|c|c|c|c|c|} \hline 
    $\mu$ & name & $\di t^2$ & $\di x^2$ & $\di r^2$ & range & \textsc{cft} interpretation\\
    \hline \hline \multirow{3}{*}{$\mu^2 >1$}& $\left(
      c^+ \right)$ & $-$ & $+$ & $+$ &$ r
    > \mu^2$ & $J^3 \bar J^3$, $\kappa_3>1$  \\
    \cline{2-2}\cdashline{3-3} \cline{4-7} &$\left( b^+ \right)$ & $-$ & $-$
    & $-$ &$1< r< \mu^2$ &\\ \cline{2-3} \cdashline{4-4} \cline{5-7} &$\left(
      a^+ \right)$ & $+$ & $-$ & $+$ & $0< r<1$ & $J^3 \bar J^3$, $\kappa_3<1$
    \\ \hline \hline \multirow{3}{*}{$\mu^2 < 1$}& $\left( a^- \right)$ & $+$
    & $-$ & $+$ & $0 < r < \mu^2$ & \multirow{3}{*}{$J^2 \bar J^2$, $\kappa_2 <
      1$} \\ \cline{2-2}\cdashline{3-3} \cline{4-6} &$\left( b^- \right)$ &
    $+$ & $+$ & $-$ &$\mu^2 < r < 1 $ & \\ \cline{2-3} \cdashline{4-4}
    \cline{5-6} &$\left( c^- \right)$ & $-$ & $+$ & $+$ &$ r > 1$ & \\
    \hline
  \end{tabular}
  \caption{Signature for the black-string metric as a function of $r$, for
    $\mu^2 \gtrless 1 $.}
  \label{tab:bs-signature}
\end{table}

Our observations are the following:
\begin{itemize}
\item The $\mu^2 < 1$ branch always has the correct $\left( -, +, + \right)
  $ signature for any value of $r$, with the two special values $r = 1 $ and
  $r = \mu^2 $ marking the presence of the horizons that hide the
  singularity in $r=0$.
\item The $\mu^2> 1$ branch is different. In particular we see that there
  are two regions: $\left(a^+\right)$ for $0<r<1$ and $\left( c^+ \right)$
  for $r > \mu^2$ where the signature is that of a physical space.
\end{itemize}
A fact deserves to be emphasized here: one should notice that while for
$\mu^2 < 1$ we obtain three different regions of the same space, for $\mu^2 >
1$ what we show in Tab.~\ref{tab:bs-signature} really are three different
spaces and the proposed ranges for $r$ are just an effect of the chosen
parameterization. The $\left( a^+ \right), \kappa_3 < 1 $ and $\left(
  c^+\right), \kappa_3 > 1 $ branches are different spaces and not different
regions of the same one and one can choose in which one to go when
continuing to $\mu > 1$.

But there is more. The $\mu^2 > 1 $ region is obtained via an
analytic continuation with respect to the other branch, and this
analytic continuation is precisely the one that interchanges the
roles of the $J^2$ and the $J^3$ currents. As a result, we pass
from the $J^2 \bar J^2$ line to the $J^3 \bar J^3$ line. More
precisely the $\left(c^+\right)$ region describes the ``singular''
$\kappa_3 > 1 $ branch of the $J^3 \bar J^3 $ deformation
(\textit{i.e.} the branch that includes the $r=0$ singularity) and
the $\left(a^+\right)$ region describes the regular $\kappa_3<1$
branch that has the \emph{cigar} geometry as $\kappa_3 \to 0$
limit. Also notice that the regions $r<0$ have to be excluded in
order to avoid naked singularities (of the type encountered in the
Schwarzschild black hole with negative mass). The black string
described in~\cite{Horne:1991gn} covers the regions
$\left(a^-\right), \left(b^-\right), \left(c^-\right),
\left(a^+\right)$.

Our last point concerns the expectation of the genuine $\mathrm{AdS}_3$
geometry as a zero-deformation limit of the black-string metric, since the
latter turns out to be a marginal deformation of AdS$_3$ with parameter
$\mu$. The straightforward approach consists in taking the line element in
Eq.~\eqref{eq:black-string} for $\mu = 1$. It is then puzzling that the
resulting extremal black-string geometry \emph{is not} $\mathrm{AdS}_3 $.
This apparent paradox is solved by carefully looking at the coordinate
transformations that relate the black-string coordinates $(r,x,t)$ to either
the Euler coordinates $(\rho, \phi, \tau)$ (\ref{euler}) for the $J^3 \bar
J^3$ line, or the hyperbolic coordinates $(y,x,t)$ for the
$J^2 \bar J^2$ line. These transformations are singular at $\mu = 1$, which
therefore corresponds neither to $\kappa_3 =1$ nor to $\kappa_2 =1$.  Put differently,
$\mu = 1$ is not part of a continuous line of deformed models but marks a
jump from the $J^2 \bar J^2 $ to the $J^3 \bar J^3 $ lines.

The extremal black-string solution is even more peculiar.
Comparing Eqs. (\ref{eq:black-string}) at $\mu = 1$ to Eqs.
(\ref{eq:null-deform}), which describe the symmetrically
null-deformed $SL(2,\mathbb{R})$, we observe that the two
backgrounds at hand are related by a coordinate transformation,
provided $\nu = -1$.

The black string background is therefore entirely described in terms
of $SL(2,\mathbb{R})$ marginal symmetric deformations, and involves
all three of them. The null deformation appears, however, for the
extremal black string only and at a negative value of the parameter
$\nu$. The latter is the density of fundamental strings, when the
deformed $\mathrm{AdS}_3$ is considered within the \textsc{ns5/f1}
system. This might be one more sign pointing towards a
Gregory-Laflamme instability in the black string
\cite{Gregory:1993vy}.

Notice finally that expressions~\eqref{eq:black-string} receive
$1/k$ corrections. Those have been computed
in~\cite{Sfetsos:1992yi}. Once taken into account, they contribute
in making the geometry smoother, as usual in string theory.

\subsection{An interesting mix}
\label{sec:an-interesting-mix}

A particular kind of asymmetric deformation is what we will call in
the following \emph{double deformation}
\cite{Kiritsis:1995iu,Israel:2003cx}. At the Lagrangian level this is
obtained by adding the following marginal perturbation to the
\textsc{wzw} action:
\begin{equation}
  \delta S = \delta \kappa^2 \int \di^2 z \: J \bar J + \h \int \di^2 z \: J \bar
  I;
\end{equation}
$J$ is a holomorphic current in the group, $\bar J$ the corresponding
anti-holomorphic current and $\bar I$ an external (to the group)
anti-holomorphic current (\emph{i.e.} in the right-moving heterotic sector
for example). A possible way to interpret this operator consists in thinking
of the double deformation as the superposition of a symmetric -- or
gravitational -- deformation (the first addend) and of an antisymmetric one
-- the electromagnetic deformation. This mix is consistent because if we
perform the $\kappa $ deformation first, the theory keeps the $U(1) \times U(1) $
symmetry generated by $J$ and $\bar J$ that is needed in order to allow for
the $\h$ deformation. Following this trail, we can read off the background
fields corresponding to the double deformation by using at first one of the
methods outlined in Sec.~\ref{sec:backgr-fields-symm} and then applying the
Kaluza-Klein reduction to the resulting background fields.

The final result consists in a metric, a three-form, a dilaton and a gauge
field. It is in general valid at any order in the deformation parameters $\kappa
$ and $\h$ but only at leading order in $\alpha^\prime$ due to the presence of the
symmetric part.

Double deformations of $\mathrm{AdS}_3$ where $J$ is the time-like
$J^3$ operator have been studied in~\cite{Israel:2003cx}. It was
there shown that the extra gravitational deformation allows to get
rid of the closed time-like curves, which are otherwise present in
the pure $J^3$ asymmetric deformation (Eq.~(\ref{dsnecogo})) --
the latter includes G\"odel space. Here, we will  focus instead on
the case of double deformation generated by space-like operators,
$J^2$ and $\bar J^2$.

\subsection{The hyperbolic double deformation}

\marginlabel{A two-parameter charged black string}In order to follow
the above prescription for reading the background fields in the
double-deformed metric let us start with the fields in
Eqs.~(\ref{eq:J2J2-deform}). We can introduce those fields in the
sigma-model action. Infinitesimal variation of the latter with respect
to the parameter $\kappa^2$ enables us to reach the following
expressions for the chiral currents $J^2_\kappa \left( z \right)$ and
$\bar J^2_\kappa \left( \bar z \right)$ at finite values of
$\kappa^2$:
\begin{align}
  J^2_\kappa \left( z \right) &= \frac{1}{\cos^2 t + \kappa^2 \sin^2
    t}\left(\cos^2 t \: \partial \psi -\sin^2t
    \: \partial \varphi \right), \\
  \bar{J}^2_{\kappa} (\bar z) &= \frac{1}{\cos^2 t + \kappa^2 \sin^2 t}
  \left(\cos^2 t \: \partial \psi + \sin^2 t \: \partial \varphi
  \right).
\end{align}
Note in particular that the corresponding Killing vectors (that
clearly are $\partial_\varphi $ and $\partial_\psi $) are to be
rescaled as $L_2 = \frac {1}{\kappa^2} \partial_\psi
- \partial_\varphi $ and $R_2 = \frac {1}{\kappa^2} \partial_\psi
+ \partial_\varphi $. Once the currents are known, one has to apply
the construction sketched in Sec.~\ref{sec:backgr-fields-asymm-1} and
write the background fields as follows: \mathindent=0em
\begin{small}
  \begin{equation}
    \label{tmetdef}
    \begin{cases}
      \begin{split}
        \frac{1}{k} \di s^2 = - \di t^2 + \cos^2 t \frac{ \left( \kappa^2 - 2
            \h^2 \right) \cos^2 t + \kappa^4 \sin^2 t }{\Delta_\kappa (t)^2} \di \psi^2 -
        4 \h^2 \frac{\cos^2 t \sin^2 t}{\Delta_\kappa (t)^2} \di \psi \di \varphi + \\
        \hspace{6cm}+ \sin^2 t \frac{ \cos^2 t + \left( \kappa^2 - 2 \h^2
          \right) \sin^2 t}{\Delta_\kappa (t)^2} \di \varphi^2
      \end{split}  \\
      \frac{1}{k} B = \frac{\kappa^2 - 2 \h^2 }{\kappa^2} \frac{\cos^2 t}{ \Delta_\kappa
        (t) } \di \varphi \land \di \psi    \\
      F = 2 \h \sqrt{\frac{2 k}{k_g}} \frac{\sin \left( 2 t\right)}{\Delta_\kappa
        (t)^2}
      \left( \kappa^2 \di \psi \land \di t + \di t \land \di \varphi \right) \\
      \mathrm{e}^{- \Phi } = \frac{\sqrt{\kappa^2 - 2 \h^2}}{\Delta_\kappa (t)}
    \end{cases}
  \end{equation}
\end{small}%
\mathindent=\oldindent%
where $\Delta_\kappa ( t ) = \cos^2 t + \kappa^2 \sin^2 t$ as in
App.~\ref{cha:symm-deform-su2}. In particular the dilaton, that can be
obtained by imposing the one-loop beta equation is proportional to the
ratio of the double deformed volume form and the $\mathrm{AdS}_3 $
one.

A first observation about the above background is in order here.  The
electric field is bounded from above since $\h^2 \leq
\frac{\kappa^2}{2}$. As usual in string theory, tachyonic
instabilities occur at large values of electric or magnetic fields,
which is just a way of interpreting the decompactification boundary
value for the deformation parameter. At the critical value of $\h$,
one dimension degenerates and the $B$-field vanishes. We are left with
a two-dimensional space (with non-constant curvature) plus electric
field.

The expression~(\ref{tmetdef}) here above of the metric provides only a
local description of the space-time geometry.  To discuss the global
structure of the whole space it is useful to perform several coordinate
transformations. Firstly let us parametrize by
$\kappa^2=\lambda/(1+\lambda)$ the deformation parameter (with $\kappa <1$
for $\lambda>0$ and $\kappa >1$ for $\lambda<-1$) and introduce a radial
coordinate \emph{\`a la} Horne and Horowitz:
\begin{equation}
  r=\lambda +\cos^2t,
\end{equation}
which obviously varies between $\lambda$ and $\lambda +1$. The
expression of the metric (\ref{tmetdef}) becomes in terms of this
new coordinate:
\mathindent=0em
\begin{small}
  \begin{multline}
    \di s^2 = -\left[\left( 2 \h^2 \left( 1 + \lambda \right)^2 - \lambda \right)
      + \frac{\lambda \left( \lambda - 4 \h^2 \left( 1 + \lambda \right)^2 \right) }{r}
      + \frac{ 2\lambda^2 \h^2 \left( 1 + \lambda
        \right)^2))}{r^2}\right]\di \psi^2 + \\
    - \left( 1 + \lambda \right) \left[ 2 \h^2 \left( 1 + \lambda \right) + 1 -
      \frac{ \left( 1 + \lambda \right) \left( 1 + 4 \h^2 \left( 1 + \lambda
          \right)^2 \right) }{r} + \frac{ 2 \left( 1 + \lambda
        \right)^3 \h^2 }{r^2} \right] \di \varphi^2 + \\
    + 4 \h^2 \left( 1 + \lambda \right)^2 \left[ 1 - \frac{ 1 + 2 \lambda }{r} +
      \frac{\lambda \left( 1 + \lambda \right) }{r^2} \right] \di \psi \di\varphi + \frac
    1{4 \left( r - \lambda \right) \left( r - \lambda - 1 \right) }\di r^2 .
  \end{multline}
\end{small}%
\mathindent=\oldindent%
This expression looks close to the one discussed by Horne and
Horowitz. It also represents a black string. However, it depends
on more physical parameters as the expression of the scalar
curvature shows:
\begin{equation}
  \mathcal{R} = 2\frac{2 r \left( 1 + 2 \lambda \right) - 7  \lambda \left( 1 + \lambda
    \right) - 2  \h^2 \left( 1 + \lambda \right)^2}{r^2} .
\end{equation}
Obviously this metric can be extended behind the initial domain of
definition of the $r$ variable. But before to discuss it, it is interesting
to note that the Killing vector ${\mathbf{k}}=(1+\lambda)\,\partial_\psi + \lambda\,\partial_\phi \propto
R_2 $ is of constant square length
\begin{equation}
  \mathbf{k}.\mathbf{k} = \lambda \left( 1 + \lambda \right) - 2 \h^2
  \left( 1 + \lambda \right)^2 := \omega .
\end{equation}
Note that as $\h^2$ is positive, we have the inequality $\omega <\lambda \left( 1 +
  \lambda \right)$. Moreover, in order to have a Lorentzian signature we must impose
$\omega > 0$. The fact that the Killing vector $\mathbf{k}$ is
space-like and of constant length makes it a candidate to perform
identifications. We shall discuss this point at the end of this
section.

The constancy of the length of the Killing vector $\mathbf{k}$
suggests to make a new coordinate transformation (such that
$\mathbf{k}=\partial_x$) :
\begin{subequations}
  \begin{align}
    \psi &= \left( 1 + \lambda \right)  x + t ,\\
    \varphi &= t + \lambda x ,
  \end{align}
\end{subequations}
which leads to the much simpler expression of the line element:
\mathindent=0em
\begin{equation}
  \label{Met2fois}
  \di s^2 = -\frac{\left( r - \lambda \right) \left( r - \lambda - 1 \right)}{r^2}\di
  t^2 + \omega \left( \di x +\frac{1}{r} \di t \right)^2 +\frac{1}{4 \left( r -
      \lambda \right) \left( r - \lambda - 1 \right)} \di r^2 .
\end{equation}
This metric is singular at $r=0, \lambda, \lambda +1$; $r=0$ being a curvature
singularity. On the other hand, the volume form is
$\nicefrac{\sqrt{\omega}}{\left(2 r \right)} \di t \land \di x \land \di r$, which
indicates that the singularities at $r=\lambda$ and $r=\lambda +1 $ may be merely
coordinate singularities, corresponding to horizons. Indeed, it is the case.
If we expand the metric, around $r=\lambda +1$, for instance, at first order
(\emph{i.e.} for $r=\lambda +1 +\epsilon$) we obtain:
\begin{multline}
  \di s^2 = \frac {\omega}{\left( 1 + \lambda \right)^2}(\di t + \left( 1 + \lambda
  \right) \di x)^2 -\frac \epsilon {\left( 1 + \lambda \right)^2} \di t \left[\di t + 2
    \frac{\omega}{1+\lambda}\left(\di t+ \left( 1 + \lambda \right)\di x\right) \right]
  + \\ +\frac 1 {4\epsilon} \di r^2
\end{multline}
\mathindent=\oldindent
indicating the presence of a horizon. To eliminate the
singularity in the metric, we may introduce Eddington--Finkelstein
like coordinates:
\begin{subequations}
  \begin{align}
    t &= \left( 1 + \lambda \right) \left (u\ \pm \ \frac{1}{2} \ln \epsilon
    \right) - \omega \xi  ,    \\
    x &= \left( 1 + \frac{\omega}{ 1 + \lambda } \right) \xi - \left( u \pm
      \frac{1}{2} \ln \epsilon \right) .
  \end{align}
\end{subequations}
The same analysis can also be done near the horizon located at
$r=\lambda$. Writing $r=\lambda +\epsilon$, the corresponding
regulating coordinate transformation to use is given by:
\begin{subequations}
  \begin{align}
    t &= \lambda \left( u  \pm  \frac{1}{2} \ln \epsilon \right) + \omega \xi ,\\
    x &= \left( 1 - \frac{\omega}{\lambda} \right) \xi - \left( u\ \pm
      \frac{1}{2} \ln \epsilon \right) .
  \end{align}
\end{subequations}
In order to reach the null Eddington--Finkelstein coordinates, we
must use null rays. The geodesic equations read, in terms of a
function $\Sigma^2[E,P,\varepsilon;r]=\left( E r - P \right)^2 -
\left( P^2/\omega \right) - \varepsilon \left( r - \lambda \right)
\left( r - \lambda - 1 \right)$:
\begin{subequations}
  \begin{align}
    \sigma &= \int \frac 1{4\, \Sigma[E,P,\varepsilon;r]}\di r\\
    t &=\int \frac { \left(E r - P \right) r }{2 \left( r - \lambda \right)
      \left( r - \lambda - 1 \right)\, \Sigma[E,P,\varepsilon;r]} \di r    \\
    x &= -\int \frac{\left( E r - P \right) + P / \omega}{2 \left( r -
        \lambda \right) \left( r - \lambda - 1 \right)\,
      \Sigma[E,P,\varepsilon;r]} \di r
  \end{align}
\end{subequations}
where $E$ and $P$ are the constant of motion associated to $\partial_t$ and
$\partial_x$, $\sigma$ is an affine parameter and $\varepsilon$, equal to $1,0,-1$,
characterizes the time-like, null or space-like nature of the geodesic.
Comparing these equations (with $\varepsilon =0$ and $P=0$) with the coordinates
introduced near the horizons, we see that regular coordinates in their
neighbourhoods are given by
\begin{subequations}
  \begin{align}
    t &= T \pm \frac{1}{2} \left( \left( 1 + \lambda \right) \ln\abs{ r -
        \lambda - 1 } - \lambda  \ln \abs{ r - \lambda } \right) ,\\
    x &= X \mp \frac{1}{2} \left( \ln\abs{ r - \lambda - 1 } - \ln \abs{ r - \lambda}
    \right) ,
  \end{align}
\end{subequations}
which leads to the metric
\mathindent=0em
\begin{equation}
  \label{EFmet}
  \di s^2 = \left(-1+\frac{ 1 + 2 \lambda }{r} -\frac{\lambda \left( 1 + \lambda \right) -
      \omega} {r^2}\right) \di T^2 + 2 \frac \omega r \di   X \di T + \omega
  \di X^2 \mp \frac 1 r \di T  \di r
\end{equation}
\mathindent=\oldindent

According to the sign, we obtain incoming or outgoing null
coordinates; to build a Kruskal coordinate system we have still to
exponentiate them.

Obviously, we may choose the $X$ coordinate in the
metric~(\ref{EFmet}) to be periodic without introducing closed causal
curves.  The question of performing more general identifications in
these spaces will be addressed below.

We end this section by computing the conserved charges associated to
the asymptotic symmetries of our field configurations. As is well
known, their expressions provide solutions of the equations of motion
derived from the low energy effective action
\begin{equation}
  S = \int d^d x \sqrt{-g} \, \mathrm{e}^{- 2\Phi } \left[R +
    4(\nabla\Phi)^2 -\frac{1}{12} H^2 -\frac{k_g}{8} F^2 +
    \frac{\delta c}{3} \right] \quad ,
\end{equation}
in which we have choosen the units such that ${\delta c}=12$.

The expression (\ref{Met2fois}) for the metric is particularly
appropriate to describe the asymptotic properties of the solution.
In these coordinates, the various, non gravitational, fields read
as
\begin{subequations}
  \begin{align}
    F &=\pm \frac {\sqrt{2} \h (1+\lambda )}{r^2\sqrt{k_g}}\di t
    \land \di r ,\\
    H &= \mp \frac{\omega}{r^2}\di t\land \di x \land \di r ,\\
    \Phi &= \Phi_\star -\frac 12 \ln r  ,
  \end{align}
\end{subequations}
By setting $\sqrt{\omega}x=\bar x$ and $r=\mathrm{e}^{2\bar \rho}$,
near infinity ($\bar \rho\to \infty$), the metric
asymptotes the standard flat metric: $ds^2 = -dt^2 + d\bar{x}^2 +
d\bar \rho^2 $, while the fields $F$ and $H$ vanish and the
dilaton reads $\Phi=\Phi_\star - \bar \rho$. This allows to
interpret the asymptotic behavior of our solution (\ref{tmetdef})
as a perturbation around the solution given by $F=0$, $H=0$, the
flat metric and a linear dilaton: $\bar\Phi=\Phi_\star + f_\alpha
X^\alpha$, (with here $f_\alpha=(0,0,-1)$). Accordingly, we may
define asymptotic charges associated to each asymptotic
reductibility parameter (see \cite{Glenn:2001}).

\marginlabel{Asymptotic charges for the charged black string}For the
gauge symmetries we obtain as charges, associated to the $H$ field
\begin{equation}
  Q_H=\pm 2\mathrm{e}^{-2\Phi_\star}\sqrt{\omega}
\end{equation}
and to the $F$ field
\begin{equation}
  Q_F=\pm \frac{2\sqrt{2}{\mathrm{e}}^{-2\Phi_\star}\h
    (1+\lambda)}{\sqrt{k_g}} .
\end{equation}
The first one reduces (up to normalization) for $\h =0$ to the
result given in \cite{Horne:1991gn}, while the second one provides
an interpretation of the deformation parameter $\h$.

Moreover, all the Killing vectors of the flat metric defining
isometries preserving the dilaton field allow to define asymptotic
charges. These charges are obtained by integrating on the surface
at infinity the antisymmetric tensor:
\begin{equation}
  \label{kasymp}
  k^{[\mu \nu]}_{\xi}= {\mathrm{e}}^{-2\bar \Phi}\left(\xi_\sigma\
    \partial_\lambda
    {\mathcal H}^{\sigma \lambda\mu \nu} +\frac 12
    \partial_\lambda\xi_\sigma\ {\mathcal H}^{\sigma \lambda\mu
      \nu}+2(\xi^\mu h^\nu_\lambda f^ \lambda - \xi^\nu h^\mu_\lambda
    f^\lambda)\right)
\end{equation}
where
\begin{equation}
  {\mathcal H}^{\sigma \lambda\mu\nu}=\bar h^{\sigma\nu}\eta^{\lambda
    \mu}+\bar h^{\lambda \mu} \eta^{\sigma\nu}-\bar h^{\sigma\mu}\eta^{\lambda \nu}-\bar
  h^{\lambda \nu}\eta^{\sigma\mu}
\end{equation}
is the well known tensor sharing the symmetries of the Riemann
tensor and $ \bar h^{\mu\nu}=h^{\mu\nu}-\frac 12 \eta
^{\mu\nu}\eta^{\alpha\beta}h_{\alpha\beta} $, while the Killing
vector $\xi$ has to verify the invariance condition $\xi_\alpha
f^\alpha = 0$. The expression of the tensor $k^{[\mu \nu]}_{\xi}$
depends only on the perturbation $h_{\mu\nu}$ of the metric tensor
because, on the one hand, the $F$ and $H$ fields appear
quadratically in the lagrangian, and their background values are
zero, while, on the other hand, the perturbation field for the
dilaton vanishes: $\Phi=\bar \Phi$ .

Restricting ourselves to constant Killing vectors, we obtain the
momenta (defined for the indice $\sigma=t$ and $\bar{x}$)
\begin{equation}
  \label{momenta}
  P^\sigma=\int \di \bar x\ { \mathrm {e}}^{-2\bar
    \Phi}\left(\partial_\lambda {\mathcal H}^{\sigma \lambda t \bar
      \rho}-2 \eta ^{\sigma t} h^\nu_{\bar\rho}\right)
\end{equation}

\emph{i.e.} the density of mass ($\mu$) and momentum ($\varpi $)
per unit length:
\begin{align}
  \mu= 2{\mathrm{e}}^{-2\Phi_\star}(1+2\lambda) & \text{and} &
  \varpi=-2{\mathrm{e}}^{-2\Phi_\star}\sqrt{\omega} .
\end{align}

Of course, if we perform identifications such that the string
acquires a finite length, the momenta (\ref{momenta}) become also
finite.

To make an end let us notice that the expressions of $\mu$ and
$\varpi$ that we obtain differ from those given in
\cite{Horne:1991gn} by a normalization factor but also in their
dependance with respect to $\lambda$, even in the limit $\h =0$;
indeed, the asymptotic Minkowskian frames used differ from each
other by a boost.

  \subsection{Discrete identifications}
  \label{sec:btz}

  In the same spirit as the original \textsc{btz} construction
  reminded in the previous section, we would like to investigate to
  what extent discrete identifications could be performed in the
  deformed background.
  Necessary conditions for a solution~\eqref{EFmet} to remain
  ``viable'' black hole can be stated as follows:
  \begin{itemize}
  \item the identifications are to be performed along the orbits of
    some Killing vector $\xi$ of the deformed metric
  \item there must be causally safe asymptotic regions (at spatial
    infinity)
  \item the norm of $\xi$ has to be positive in some region of
    space-time, and chronological pathologies have to be hidden with
    respect to an asymptotic safe region by a horizon.

  \end{itemize}

  The resulting quotient space will exhibit a black hole structure if,
  once the regions where $\norm{\xi}<0$ have been removed, we are left
  with an almost geodesically complete space, the only incomplete
  geodesics being those ending on the locus $\norm{\xi}=0$. It is
  nevertheless worth emphasizing an important difference with the
  \textsc{btz} construction. In our situation, unlike the undeformed
  $\mathrm{AdS}_3$ space, the initial space-time where we are to
  perform identifications do exhibit curvature singularities.

\subsubsection{The BTZ black hole}
\label{sec:btz-black-hole}

In the presence of isometries, discrete identifications provide
alternatives for creating new backgrounds. Those have the
\emph{same} local geometry, but differ with respect to their
global properties. Whether these identifications can be
implemented as orbifolds at the level of the underlying
two-dimensional string model is very much dependent on each
specific case.

For $\mathrm{AdS}_3$, the most celebrated geometry obtained by
discrete identification is certainly the \textsc{btz} black
hole~\cite{Banados:1992wn}. The discrete identifications are made
along the integral lines of the following Killing vectors (see
Eqs. (\ref{eq:Killing-SL2R})):
\begin{subequations}
  \begin{align}
     \text{non-extremal case}&: \ \ \xi =
     \left(r_+ + r_- \right)R_2 -  \left(r_+ - r_- \right)L_2,\label{next}\\
    \text{extremal case}&: \ \ \xi = 2 r_+  R_2 -
     \left( R_1-R_3\right) -
      \left( L_1+L_3\right) \label{ext}.
  \end{align}
\end{subequations}

In the original \textsc{btz} coordinates,
the metric reads:
\begin{equation}
\di s^2 =L^2 \left[ -f^2(r) \, \di t^2 + f^{-2}(r) \, \di r^2 +
r^2 \left( \di \varphi-\frac{r_+  r_-}{r^2} \di t
\right)^2\right] ,
\end{equation}
with
\begin{equation}
f(r) =\frac{1}{r}\sqrt{\left( r^2_{\vphantom +}-r^2_{+} \right)
\left(r^2_{\vphantom -}-r^2_{-}\right)} .
\end{equation}
In this coordinate system,
\begin{equation}
\partial_{\varphi} \equiv \xi \ , \ \
\partial_t \equiv
-\left(r_+ + r_- \right) R_2 - \left(r_+ - r_- \right) L_2\ \ \text{and} \ \ r^2 \equiv \norm{\xi}.
\end{equation}
In AdS$_3$ $\varphi$ is not a compact coordinate. The discrete
identification makes $\varphi$ an angular variable, $\varphi \cong
\varphi+2\pi$, which imposes to remove the region with $r^2<0$.
The \textsc{btz} geometry describes a three-dimensional black
hole, with mass $M$ and angular momentum $J$, in a space--time
that is locally (and asymptotically) anti-de Sitter. The
chronological singularity at $r=0$ is hidden behind an inner
horizon at $r=r_-$, and an outer horizon at $r=r_+$. Between these
two horizons,  $r$ is time-like. The coordinate $t$ becomes
space-like inside the ergosphere, when $r^2_{\vphantom g}<
r^2_{\text{erg}} \equiv r_+^2 + r_-^2$. The relation between $M,J$
and  $r_±$ is as follows:
\begin{equation}
r_±^2  = {ML \over 2}\left[ 1± \sqrt{1-\left({J\over  ML
}\right)^2} \right] .
\end{equation}
Extremal black holes have  $\vert J \vert =ML$ ($r_+ = r_-$). In
the special case $J= ML=0$ one finds the near-horizon geometry of
the five-dimensional  \textsc{ns5/f1} stringy black hole in its
ground state. Global  AdS$_3$  is obtained for $J=0$ and $ML=-1$.

Many subtleties arise, which concern \emph{e.g.} the appearance of
closed time-like curves in the excised region of negative  $r^2$
(where $\partial_\varphi$ would have been time-like) or the
geodesic completion of the manifold; a comprehensive analysis of
these issues can be found in~\cite{Banados:1993gq}. At the
string-theory level, the \textsc{btz} identification is realized
as an \emph{orbifold} projection, which amounts to keeping
invariant states and adding twisted
sectors~\cite{Natsuume:1998ij,Hemming:2001we}.

Besides the \textsc{btz} solution, other locally $\mathrm{AdS}_3$
geometries are obtained, by imposing identification under purely
left (or right) isometries, refereed to as self-dual (or
anti-self-dual) metrics. These were studied
in~\cite{Coussaert:1994tu}. Their classification and isometries
are exactly those of the asymmetric deformations studied in the
present chapter. The Killing vector used for the identification is
(A) time-like (elliptic), (B) space-like (hyperbolic) or (C) null
(parabolic), and the isometry group is $U(1) ×
SL(2,\mathbb{R})$. It was pointed out in~\cite{Coussaert:1994tu}
that the resulting geometry was free of closed time-like curves
only in the case (B).

  \subsubsection{Discrete identifications in asymmetric deformations}
  \label{sec:btzas}

  Our analysis of the residual isometries in purely asymmetric
  deformations (Sec.~\ref{sec:wzw-deformations}) shows that the
  vector $\xi$ (Eq.~(\ref{next})) survives only in the hyperbolic
  deformation, whereas $\xi$ in Eq.~(\ref{ext}) is present in the
  parabolic one. Put differently, non-extremal \textsc{btz} black
  holes allow for electric deformation, while in the extremal ones,
  the deformation can only be induced by an electro-magnetic wave.
  Elliptic deformation is not compatible with \textsc{btz}
  identifications.

  The question that we would like to address is the following: how
  much of the original black hole structure survives the deformation?
  The answer is simple: a new chronological singularity appears in the
  asymptotic region of the black hole. Evaluating the norm of the
  Killing vector shows that a naked singularity appears. Thus the
  deformed black hole is no longer a viable gravitational background.
  Actually, whatever the Killing vector we consider to perform the
  identifications, we are always confronted to such pathologies.

  The fate of the \emph{asymmetric parabolic} deformation of
  $\mathrm{AdS}_3$ is similar: there is no region at infinity free of
  closed time-like curves after performing the identifications.

  \subsubsection{Discrete identifications in symmetric deformations}

  Let us consider the \emph{symmetric hyperbolic} deformation, whose
  metric is given by~\eqref{Met2fois} with $\h = 0$, i.e. $\omega = \lambda
  \left( 1 + \lambda \right)$. This metric has two residual Killing vectors,
  manifestly given by $\partial_t$ and $\partial_x$. We may thus, in general,
  consider identifications along integral lines of
  \begin{equation}
    \label{KillId}
    \xi=a\,\partial_t + \partial_x .
  \end{equation}
  This vector has squared norm:
  \begin{equation}
    \norm{\xi}^2=\left( \lambda \left( 1 + \lambda \right) - a^2 \right) + \frac {a \lambda
      \left( 1 + \lambda \right) + a^2 \left( 1 + 2 \lambda \right) }{r} .
  \end{equation}
  To be space-like at infinity the vector $\xi$ must verify the
  inequality $\lambda \left( 1 + \lambda \right) > a^2$. If $a > 0 $, or $-\sqrt{
    \lambda \left( 1 + \lambda \right) } < a < -2\lambda \left( 1 + \lambda \right) / \left( 1
    + 2 \lambda \right)$, $\xi$ is everywhere space-like. Otherwise, it
  becomes time-like behind the inner horizon ($r = \lambda$), or on this
  horizon if $a=-\lambda$. In this last situation, the quotient space will
  exhibit a structure similar to that of the black string, with a
  time-like singularity (becoming light-like for $a=-\lambda$) and two
  horizons.

  \subsubsection{Discrete identifications in double deformations}
  \label{sec:btz-ident-comb}

\begin{figure}
    \centering
    \includegraphics[width=7cm]{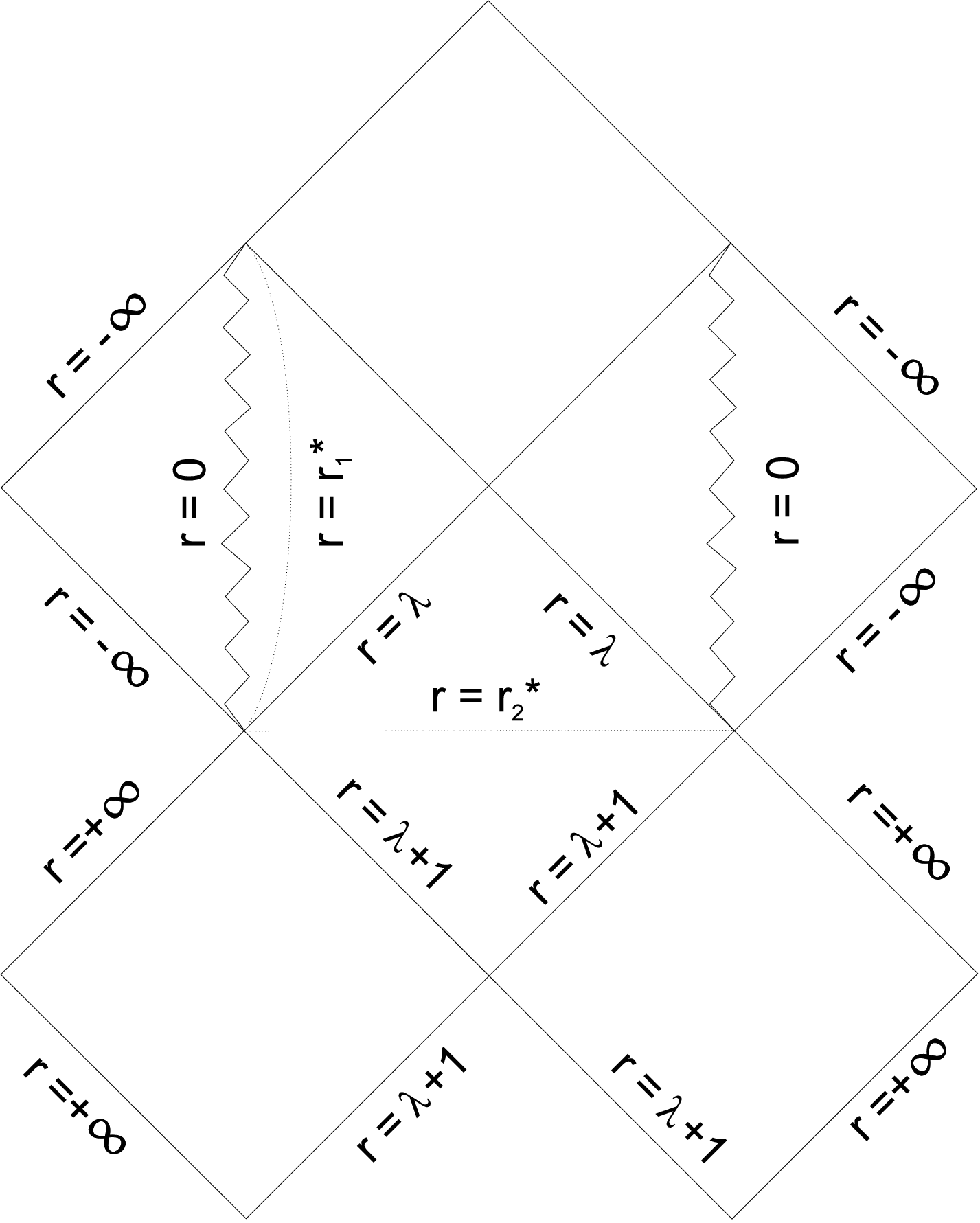}
    \caption{Penrose diagram exhibiting the global structure of the
    double hyperbolic deformation. The time-like curvature singularities
    $r=0$ are represented, as well as the horizons, located at $r=\lambda$ and
    $r=\lambda +1$. When performing identifications along orbits of a Killing
    vector allowing for a causally safe region at infinity, there appears
    chronological singularities, which can be time-like and hidden behind an
    outer and an inner horizon ($r=r^*_1$), or space-like and hidden behind
    a single horizon ($r=r^*_2$), while the regions where $r<r^*$ have to be
    removed.}
    \label{fig:Penrose}
  \end{figure}

  The norm squared of the identification vector (\ref{KillId}) in the
  metric~\eqref{Met2fois} is
  \begin{equation}
    \norm{\xi}^2= \left( \omega - a^2 \right) + 2 \frac{a \omega +
      a^2 \left( 1 + 2 \lambda \right) }{r} - \frac{a^2 \left( \lambda \left( 1+\lambda
        \right) - \omega \right)}{r^2} .
  \end{equation}
  Between $r=0$ and $r=\infty$, this scalar product vanishes once and only
  once (if $a\neq 0$).  To be space-like at infinity we have to restrict
  the time component of $\xi$ to $\abs{a} < \omega $. Near $r=0$ it is
  negative, at the outer horizon ($r=\lambda + 1$) it takes the positive
  value $\omega \left( 1 + \lambda + a \right)^2/\left( 1 + \lambda \right)^2$ and near
  the inner horizon ($r=\lambda$) the non-negative value $\omega \left( \lambda + a
  \right)^2/\lambda^2$. Accordingly, by performing identifications using
  this Killing vector, we will encounter a chronological singularity,
  located at $r=r^*$, with $0<r^*<\lambda+1$. When $r^* < \lambda$, the
  singularity will be of the same type as the one in the symmetric
  case. But when $\lambda < r^* < \lambda + 1$, the chronological singularity will
  be space-like, and the causal structure we get is much like that of
  the Schwarschild black hole, as is shown in Fig.~\ref{fig:Penrose}.

  \subsection{Towards the exact spectra}
  \label{sec:conf-field-theory}

  Let us consider the algebraic point of view. Again as in the
  electric deformation of $SL(2,\setR)$ we can't write the partition
  function but we must content ourselves with the spectrum which will
  generalize what we found in
  Eqs.~\eqref{leftspecadsdef}~and~\eqref{rightspecadsdef}.

  \subsubsection{Deformed Spectrum}
  \label{sec:deformed-spectrum}

  Consider the double deformation described above for a $SL (2, \setR
  )_k$ super-\textsc{wzw} model where $J$ is the hyperbolic
  (space-like) $J_2$ current.

  The evaluation of the spectrum for our deformed model is pretty
  straightforward once one realizes that the deformations act as $O
  \left( 2,2 \right)$ pseudo\hyph orthogonal transformations on the charge
  lattice corresponding to the abelian subgroup of the
  $\mathfrak{sl}(2,\setR)$ heterotic model (as described in
  Sec.~\ref{sec:wzw-deformations}). Left and right weights for the
  relevant lattices are (see Eqs.~\eqref{eq:left-weights}
  and~\eqref{eq:right-weights}):
  \begin{subequations}
    \begin{align}
      L_0 &= \frac{1}{k} \left( \mu +n + \frac{a}{2} \right)^2, \\
      \bar L_0 &= \frac{\bar \mu^2}{k+2} + \frac{1}{k_g} \left( \bar n +
        \frac{\bar a}{2} \right)^2,
    \end{align}
  \end{subequations}
  where the anti-holomorphic part contains the contribution coming
  from a $\mathfrak{u}(1)$ subgroup of the heterotic gauge group.

  At the Lagrangian level, the infinitesimal deformation we want to
  describe is given by the following marginal operator:
  \begin{equation}
    \mathcal{O} = \kappa^2  \frac{\left(J^2 + \imath \psi_1 \psi_3
      \right)}{\sqrt{k}} \frac{\bar J^2}{\sqrt{k+2}} + \h
    \frac{\left(J^2 + \imath \psi_1 \psi_3 \right)}{\sqrt{k}}
    \frac{\bar I}{\sqrt{k_g}}.
  \end{equation}
  This suggests that the actual $O (2,2)$ transformation should be
  obtained as a boost between the holomorphic part and the result of a
  rotation between the two anti-holomorphic components. The deformed
  lattices then read:
  \mathindent=0em
  \begin{subequations}
    \begin{align}
      L_0^{\text{dd}} &= \Set{\frac{1}{\sqrt{k}} \left( \mu + n +
          \frac{a}{2} \right) \cosh x + \left( \frac{\bar \mu
          }{\sqrt{k+2}} \cos \alpha + \frac{1}{\sqrt{k_g}} \left( \bar n +
            \frac{\bar a}{2} \right) \sin \alpha \right) \sinh x }^2,
      \label{eq:rotation-left}\\
      \bar L_0^{\text{dd}} &= \Set{ \left( \frac{\bar \mu }{\sqrt{k+2}}
          \cos \alpha + \frac{1}{\sqrt{k_g}} \left( \bar n + \frac{\bar
              a}{2} \right) \sin \alpha \right) \cosh x +
        \frac{1}{\sqrt{k}} \left( \mu + n + \frac{a}{2} \right) \sinh x
      }^2, \label{eq:rotation-right}
    \end{align}
  \end{subequations}
  \mathindent=\oldindent
where the parameters $x $ and $\alpha$ can be expressed as functions of
  $\kappa$ and $\h $ as follows:
  \begin{equation}
    \begin{cases}
      \kappa^2 = \sinh (2x) \cos \alpha, \\
      \h = \sinh (2x) \sin \alpha.
    \end{cases}
  \end{equation}
  Of course this is a generalization of the expressions in
  Eq.~\eqref{eq:J2spectrum}.

  \subsubsection{Twisting}
  \label{sec:twisting}

  The identification operation we performed in the symmetrically and
  double-deformed metric (as in Sec.~\ref{sec:btz}) is implemented in
  the string theory framework by the orbifold construction. This was
  already obtained in~\cite{Natsuume:1998ij,Hemming:2001we} for the
  ``standard'' \textsc{btz} black hole that was described as a $SL (2,
  \setR )/ \setZ$ orbifold.

  In order to write the spectrum that will contain the twisted
  sectors, the first step consists in writing explicitly the primary
  fields in our theory, distinguishing between the holomorphic and
  anti-holomorphic parts (as it is natural to do since the
  construction is intrinsically heterotic).
  \begin{itemize}
  \item The holomorphic part is written by introducing the charge
    boost of Eq.~\eqref{eq:rotation-left} in
    Eq.~\eqref{eq:superparafermion}:
    \begin{equation}
      \Phi^{\text{dd}}_{j \mu \nu \bar \mu \bar \nu} (z) = U_{j \mu } (z) \exp \left[  \imath \left( \sqrt{\frac{2}{k}} \left( \mu + n + \frac{a}{2} \right) \cosh x +
          \sqrt{2} \bar Q_\alpha \sinh x \right) \vartheta_2 \right],
    \end{equation}
    where $Q_\alpha = \bar \mu \sqrt{\frac{2}{k+2}} \cos \alpha + \bar \nu
    \sqrt{\frac{2}{k_g}} \sin \alpha$ and the dd superscript stands for
    double deformed
  \item To write the anti-holomorphic part we need at first to
    implement the rotation between the $\bar J^3$ and gauge current
    components:
    \begin{multline}
      \bar \Phi_{j \bar \mu \bar \nu} ( \bar z ) = V_{j \mu } (\bar z) e^{\imath
        \bar \mu \sqrt{\nicefrac{2}{k+2}} \bar \theta_2} e^{\imath \bar \nu
        \sqrt{2/k_g}
        \bar X}  = \\
      = V_{j \mu } (\bar z) e^{\imath \sqrt{2} \bar Q_{\alpha} \left( \bar \theta_2
          \cos \alpha + \bar X \sin \alpha \right) } e^{\imath \sqrt{2} \bar Q_{\alpha -
          \nicefrac{\pi}{2}} \left( - \bar \theta_2 \sin \alpha + \bar X \cos \alpha
        \right) },
    \end{multline}
    and then realize the boost in Eq.~\eqref{eq:rotation-right} on the
    involved part:
  \mathindent=0em
    \begin{multline}
      \bar \Phi^{\text{dd}}_{j \mu \bar \mu \nu \bar \nu} ( \bar z ) = V_{j \mu }
      e^{\imath \sqrt{2} \bar Q_{\alpha - \nicefrac{\pi}{2}} \left( - \bar
          \theta_2
          \sin \alpha + \bar X \cos \alpha \right)}  \times \\
      \times \exp \left[ \imath \left( \sqrt{\frac{2}{k}} \left( \mu + n +
            \frac{a}{2} \right) \sinh x + \sqrt{2} \bar Q_\alpha \cosh x
        \right) \left( \bar \theta_2 \cos \alpha + \bar X \sin \alpha \right)
      \right].
    \end{multline}
\mathindent=\oldindent
  \end{itemize}

  Now that we have the primaries, consider the operator $W_w \left( z,
    \bar z \right)$ defined as follows:
  \begin{equation}
    W_w \left( z, \bar z \right) = e^{-\imath \frac{k}{2} w \Delta_-
      \vartheta_2 +\imath \frac{k+2}{2} w\Delta_+ \bar \theta_2 },
  \end{equation}
  where $w \in \setZ $ and $\bar \theta_2 $ the boson corresponding to the $\bar
  J_2 $ current. It is easy to show that the following \textsc{ope}'s
  hold:
  \begin{align}
    \vartheta_2 \left( z \right) W_n \left( 0, \bar z\right) &\sim -\imath w \Delta_- \log
    z W_w
    \left( 0, \bar z\right), \\
    \bar \theta_2 \left( \bar z \right) W_n \left( z, 0 \right) &\sim \imath w \Delta_+
    \log \bar z W_w \left( z, 0\right),
  \end{align}
  showing that $W_w \left( z, \bar z\right)$ acts as twisting operator
  with winding number $w$ ($\vartheta_2 $ and $\bar \theta_2 $ shift by $2 \pi \Delta_- w
  $ and $2 \pi \Delta_+ w $ under $z \to e^{2 \pi \imath } z $). This means that
  the general primary field in the $SL \left( 2, \setR \right)_k / \setZ $
  theory can be written as:
  \begin{equation}
    \Phi^{\text{tw}}_{j \mu \bar \mu \nu \bar \nu  w} \left( z, \bar z \right) =  
    \Phi^{\text{dd}}_{j \mu \bar  \mu \nu \bar \nu  } \left( z, \bar z \right) W_w \left( z, \bar z  
    \right).
  \end{equation}
  where the tw superscript stands for twisted.

  Having the explicit expression for the primary field, it is simple
  to derive the scaling dimensions which are obtained, as before, via
  the \textsc{gko} decomposition of the Virasoro algebra $T \left[
    \mathfrak{sl}\left( 2, \setR \right) \right] = T \left[
    \mathfrak{sl}\left( 2, \setR \right)/\mathfrak{o} \left( 1,1\right)
  \right] + T \left[ \mathfrak{o} \left(1,1 \right) \right] $. Given
  that the $ T \left[ \mathfrak{sl}\left( 2, \setR \right)/\mathfrak{o}
    \left( 1,1\right) \right] $ part remains invariant (and equal to
  $L_0 = - j \left( j+1\right)/ k - \mu^2/\left( k + 2\right)$ as in
  Eq.~\eqref{eq:boson-coset}), the deformed weights read:
  \mathindent=0em
  \begin{subequations}
    \begin{align}
      L^{\text{tw}}_0 &= \Set{ \frac{k}{2\sqrt{2}} w \Delta_- +
        \frac{1}{\sqrt{k}} \left( \mu + n + \frac{a}{2} \right) \cosh x
        +  \bar Q_\alpha \sinh x }^2,\\
      \bar L^{\text{tw}}_0 &= \Set{ - \frac{k+2}{2\sqrt{2}} w \Delta_+ \cos
        \alpha+ \bar Q_\alpha \cosh x + \frac{1}{\sqrt{k}}
        \left( \mu + n + \frac{a}{2} \right) \sinh x }^2 + \nonumber \\
      & \hspace{6.5cm}+ \Set{ \frac{k+2}{2 \sqrt{2}} w \Delta_+ \sin \alpha +
        \bar Q_{\alpha - \nicefrac{\pi}{2}}}^2.
    \end{align}
\mathindent=\oldindent
  \end{subequations}


\section{New compactifications}
\label{sec:new-comp}

Up to this point we have focused on the squashed and coset models under
the underlying hypothesis that they act as parts of larger
ten-dimensional backgrounds. In this section we will study other
examples which are likely to be part of physically sound models. In
particular we will closely study the $SU(3)/U(1)^2$ coset that can be
used as the six-dimensional compact counterpart of an $\mathrm{AdS}_4$
background.

\subsection{The $SU(3)/U(1)^2$ flag space}
\label{sec:su3-u1-u1}

Let us now consider the next example in terms of coset dimensions, $SU
\left( 3 \right) / U \left( 1 \right)^2$. As a possible application
for this construction we may think to associate this manifold to a
four-dimensional $\left( 1,0 \right)$ superconformal field theory
$\mathcal{M}$ so to compactify a critical string theory since $\dim
\left[SU \left( 3 \right)/ U \left( 1 \right)^2 \right]= 8 - 2 =6$.
Our construction gives rise to a whole family of \textsc{cft}'s
depending on two parameters (since $\rank \left[ SU \left( 3 \right)
\right] = 2$) but in this case we are mainly interested to the point
of maximal deformation, where the $U \left( 1 \right)^2 $ torus
decouples and we obtain an exact theory on the $SU \left( 3 \right) /
U \left( 1 \right)^2 $ coset.  Before giving the explicit expressions
for the objects in our construction it is hence useful to recall some
properties of this manifold. The first consideration to be made is the
fact that $SU \left( 3 \right) / U \left( 1 \right)^2 $ is an
asymmetric coset in the mathematical sense defined in
Sec.~\ref{sec:geom-squash-groups} (as we show below). This allows for
the existence of more than one left-invariant Riemann metric. In
particular, in this case, if we just consider structures with constant
Ricci scalar, we find, together with the restriction of the
Cartan-Killing metric on $SU \left( 3 \right)$, the K\"ahler metric of
the flag space $F^3$. The construction we present in this section will
lead to the first one of these two metrics. This is known to admit a
nearly-K\"ahler structure and has already appeared in the superstring
literature as a basis for a cone of $G_2$ holonomy
\cite{Atiyah:2001qf}.

\marginlabel{Gauss decomposition for $SU(3)$}A suitable
parametrisation for the $SU \left( 3 \right)$ group is obtained via
the Gauss decomposition described in App.~\ref{sec:suleft-3right}. In
these terms the general group element is written as: \mathindent=0em
\begin{equation}
  g \left( z_1, z_2, z_3, \psi_1, \psi_2 \right) = \begin{pmatrix}
    \frac{e^{\imath \psi_1/2}}{\sqrt{f_1}} & - \frac{\bar z_1 + z_2 \bar
      z_3}{\sqrt{f1 f2}} e^{\imath \left( \psi_1 - \psi_2 \right)/2} & - \frac{\bar
      z_3 - \bar z_1 \bar z_2}{\sqrt{f_2}} e^{-\imath \psi_2 /2} \\
    \frac{z_1 e^{\imath \psi_1/2}}{\sqrt{f_1}} & - \frac{1+\abs{z_3}^2 - z_1 z_2
      \bar z_3 }{\sqrt{f1 f2}} e^{\imath \left( \psi_1 - \psi_2 \right)/2} & - \frac{
      \bar z_2}{\sqrt{f_2}} e^{-\imath \psi_2 /2} \\
    \frac{z_3 e^{\imath \psi_1/2}}{\sqrt{f_1}} & - \frac{z_2 - \bar z_1 z_3 +
      z_2 \abs{z_1}^2}{\sqrt{f1 f2}} e^{\imath \left( \psi_1 - \psi_2 \right)/2} &
    \frac{1}{\sqrt{f_2}} e^{-\imath \psi_2 /2} \\ 
\end{pmatrix}
\end{equation}
\mathindent=\oldindent
where $z_i $ are three complex parameters, $\psi_j$ are two real
parameters and $f_1 = 1 + \abs{z_1}^2 + \abs{z_3}^2$, $f_2 = 1 +
\abs{z_2}^2 + \abs{z_3 - z_1 z_2}^2$. As for the group, we need also
an explicit parametrisation for the $\mathfrak{su} \left( 3 \right)$
algebra, such as the one provided by the Gell-Mann matrices in
Eq.~\eqref{eq:Gell-Mann-matrices}. It is a well known result that if a
Lie algebra is semi-simple (or, equivalently, if its Killing form is
negative-definite) then all Cartan subalgebras are conjugated by some
inner automorphism\footnote{This is the reason why the study of
  non-semi-simple Lie algebra deformation constitutes a richer
  subject. In example the $SL \left( 2, \setR \right)$ group admits for 3
  different deformations, leading to 3 different families of exact
  \textsc{cft}'s with different physics properties. On the other hand
  the 3 possible deformations in $SU \left( 3 \right)$ are
  equivalent.}. This leaves us the possibility of choosing any couple
of commuting generators, knowing that the final result won't be
influenced by such a choice. In particular, then, we can pick the
subalgebra generated by $\mathfrak{k} = \braket{\lambda_3,
  \lambda_8}$.\footnote{In this explicit parametrisation it is
  straightforward to show that the coset we're considering is not
  symmetric.  It suffices to pick two generators, say $\lambda_2 $ and
  $\lambda_4$, and remark that their commutator $\left[ \lambda_2, \lambda_4 \right] = -
  1/ \sqrt{2} \lambda_6$ doesn't live in the Cartan subalgebra.}

The holomorphic currents of the bosonic $SU \left( 3 \right)_k$
corresponding to the two operators in the Cartan are:
\begin{align}
  \mathcal{J}^3 = - \braket{ \lambda_3 g\left( z_\mu , \psi_a \right)^{-1} \di
    g\left( z_\mu , \psi_a \right)} && \mathcal{J}^8 = - \braket{
    \lambda_8 g\left( z_\mu, \psi_a \right)^{-1} \di g\left( z_\mu , \psi_a \right)}
\end{align}
and in these coordinates they read:
\mathindent=0em
\begin{small}
  \begin{multline}
    \mathcal{J}^3 = - \frac{\imath}{\sqrt{2}} \left\{ \left( \frac{\bar
          z_1}{f_1}+ \frac{ z_2 \left( - \bar z_1 \bar z_2 + \bar z_3
          \right)}{2 f_2} \right)\di z_1 - \frac{ \bar z_2 \left( 1+
          \abs{z_1}^2\right)- z_1 \bar z_3 }{2 f_2} \di z_2 + \left(
        \frac{\bar z_3}{f_1} + \frac{ \bar z_1 \bar z_2 - \bar z_3 }{2
          f_2} \right) \di z_3 \right\} \\+ \text{c.c.} + \frac{\di
      \psi_1}{\sqrt{2}}- \frac{\di \psi_2 }{2 \sqrt{2}}
  \end{multline}
  \begin{equation}
    \mathcal{J}^8 = - \imath \sqrt{\frac{3}{2}} \left\{ \frac{\bar z_1 \bar z_2 -
        \bar z_3 }{2f_2} z_2 \di z_1 + \frac{ \bar z_2 + \abs{z_1}^2
        \bar z_2 - z_1 \bar z_3 }{2f_2} \di z_2 + \frac{ -\bar z_1 \bar
        z_2 + \bar z_3}{2f_2}\di z_3 \right\} + \text{c.c.}+
    \frac{1}{2} \sqrt{\frac{3}{2}} \di \psi_2 .
  \end{equation}
\end{small}%
\mathindent=\oldindent%
Those currents appear in the expression of the exactly marginal
operator that we can add to the $SU \left( 3 \right)$ \textsc{wzw}
model action:
\begin{multline}
    V = \frac{\sqrt{k k_g}}{2 \pi } \h \int \di z^2 \: \h_3 \left( J^3 - \frac{\imath
      }{\sqrt{2} k}\left( 2 : \psi_2 \psi_1 : + : \psi_5 \psi_4 : + :\psi_7 \psi_6 :
      \right) \right) \bar J^3 +\\+ \h_8 \left( J^8 - \frac{\imath }{k}
      \sqrt{\frac{3}{2}} \left( :\psi_5 \psi_4: + :\psi_7 \psi_6:\right)\right) \bar
    J^8
\end{multline}
where $\psi^i $ are the bosonic current superpartners and $\bar J^3,
\bar J^8$ are two currents from the gauge sector both generating a $U
\left( 1 \right)_{k_g}$.

Since $\rank \left[ SU \left( 3 \right) \right] = 2 $ we have a
bidimensional family of deformations parametrised by the two moduli
$\h_3 $ and $\h_8$. The back-reaction on the metric is
given by:
\begin{equation}
  \di s^2 = g_{\alpha \bar \beta } \di z^\alpha \otimes \di \bar z^\beta + \left( 1 - 2 \h_3^2
  \right)  \mathcal{J}^3 \otimes \mathcal{J}^3  + \left( 1 - 2 \h_8^2
  \right)  \mathcal{J}^8 \otimes  \mathcal{J}^8 
\end{equation}
where $g_{\alpha \bar \beta }$ is the restriction of the $SU \left( 3 \right)$
metric on $SU \left( 3 \right)/U\left( 1 \right)^2$. It is worth to
remark that for any value of the deformation parameters $\h_3
$ and $\h_8$ the deformed metric is Einstein with constant
Ricci scalar.

With a procedure that has by now become familiar we introduce the
following reparametrization:
\begin{align}
  \psi_1 = \frac{\hat \psi_1}{\sqrt{1-2\h^2}} && \psi_2 = \frac{\hat \psi_2}{\sqrt{1-2\h^2}}
\end{align}
and take the $\h_3\to1/ \sqrt{2}$, $\h_8\to1/ \sqrt{2}$ limit. The resulting metric
is:
\begin{equation}
  \di s^2 =  g_{\alpha \bar \beta } \di z^\alpha \otimes \di \bar z^\beta + \frac{\di \hat \psi_1 \otimes \di
    \hat \psi_1 - \di \hat \psi_1 \otimes \di \hat \psi_2 + \di \hat \psi_2 \otimes \di \hat \psi_2 }{2}
\end{equation}
that is the metric of the tangent space to the manifold $SU \left( 3
\right)/U\left( 1 \right)^2 \times U\left( 1\right) \times U \left( 1 \right)$.
The coset metric hence obtained has a $\setC$-structure, is Einstein and
has constant Ricci scalar $R=15/k$.  The other background fields at
the boundary of the moduli space read:
\newcommand{\jj}[2]{\ensuremath{\mathcal{J}^{#1} \land \mathcal{J}^{#2}}}
\mathindent=0em
\begin{gather}
  F = \di \mathcal{J}^3 + \di \mathcal{J}^8  \\
  H_{\left[ 3 \right]} =- 3 \sqrt{2} \left\{ \mathcal{J}^1 \land \left( \jj{4}{5} -
      \jj{6}{7}\right) + \sqrt{3} \mathcal{J}^2 \land \left( \jj{4}{5}
    + \jj{6}{7} \right) \right\}
\end{gather}
\mathindent=\oldindent

\marginlabel{Superymmetry properties of $SU(3)/U(1)^2$}If we consider
the supersymmetry properties along the deformation line we can remark
the presence of an interesting phenomenon. The initial $SU \left( 3
\right)$ model has $N=2$ but this symmetry is naively broken to $N=1$
by the deformation. This is true for any value of the deformation
parameter but for the boundary point $\h_3^2 = \h_8^2 =
\nicefrac{1}{2}$ where the $N = 2 $ supersymmetry is restored.
Following~\cite{Gates:1984nk,Kazama:1989qp,Kazama:1989uz} one can see
that a $G/T$ coset admits $N=2$ supersymmetry if it possesses a
complex structure and the corresponding algebra can be decomposed as
$\mathfrak{j} = \mathfrak{j}_+ \oplus \mathfrak{j}_-$ such as
$\comm{\mathfrak{j}_+,\mathfrak{j}_+}=\mathfrak{j}_+$ and
$\comm{\mathfrak{j}_-,\mathfrak{j}_-}=\mathfrak{j}_-$. Explicitly,
this latter condition is equivalent (in complex notation) to $f_{ijk}
= f_{\bar i \bar j \bar k} = f_{a i j} = f_{a \bar i \bar j} = 0$.
These are easily satisfied by the $SU \left( 3 \right)/ U \left( 1
\right)^2 $ coset (and actually by any $G/T $ coset) since the
commutator of two positive (negative) roots can only be proportional
to the positive (negative) root obtained as the sum of the two or
vanish.. Having $N=2$ supersymmetry is equivalent to asking for the
presence of two complex structures. The first one is trivially given
by considering positive roots as holomorphic and negative roots as
anti-holomorphic, the other one by interchanging the role in one out
of the three positive/negative couples (the same flip on two couples
would give again the same structure and on all the three just takes
back to the first structure). The metric is Hermitian with respect to
both structures since it is $SU \left( 3\right)$ invariant. It is
worth to remark that such background is different from the ones
described in~\cite{Kazama:1989uz} because it is not K\"ahler and can't
be decomposed in terms of Hermitian symmetric spaces.

\subsection{Different constructions on $SU (3) /U(1)^2$}
\label{sec:su-3-}

To study the $SU \left( 3 \right)$ case we will use the ``current''
approach of Sec.~\ref{sec:gauging}, since a direct computation in
coordinates would be impractical. As one could expect, the study of
$SU \left( 3 \right)$ deformation is quite richer because of the
presence of an embedded $SU \left( 2 \right)$ group that can be
gauged. Basically this means that we can choose two different
deformation patterns that will lead to the two possible Einstein
structures that can be defined on the $SU \left( 3 \right) / U \left(
  1 \right)^2$ manifold.

\subsubsection{Direct gauging.}

The first possible choice leads to the same model as before by simply
gauging the $U \left( 1 \right)^2 $ Cartan torus.  Consider the
initial $SU \left( 3 \right)_k \times U \left( 1 \right)_{k^\prime } \times U \left(
  1 \right)_{k^{\prime \prime }}$ model. In the $\braket{\mJ_1,\ldots, \mJ_8,
  \mathcal{I}_1, \mathcal{I}_2}$ base ($\set{\mJ_i}$ being the $SU
\left( 3 \right)$ generators and $\set{\mathcal{I}_k}$ the 2 $U\left(
  1 \right)$'s), the initial metric is written as:
\renewcommand{\uni}{\ensuremath{\mathbb{I}}}
\begin{equation}
  g = \left( \begin{tabular}{c|c}
        $k \uni{8} $& $ 0$\\ \hline
        $ 0 $& $
        \begin{matrix}
          k^\prime \\
          & k^{\prime \prime }
        \end{matrix}$
      \end{tabular}\right)
\end{equation}
the natural choice for the Cartan torus is given by the usual
$\braket{\mJ_3, \mJ_8 }$ generators, so we can proceed as before and write
the deformed metric as:
\begin{equation}
  g = 
  \begin{pmatrix}
    k \uni{2} \\
    & \lambda_1 \left( k, k^\prime, \h_3 \right) \\
    & & k \uni{4} \\
    & & & \lambda_1 \left( k, k^{\prime\prime }, \h_8\right) \\ 
    & & & & \lambda_2 \left( k, k^\prime, \h_3 \right) \\
    & & & & & \lambda_2 \left( k, k^{\prime \prime}, \h_8 \right) \\    
  \end{pmatrix}
\end{equation}
where $\h_3 $ and $\h_8 $ are the deformation parameters and
$\lambda_1$ and $\lambda_2 $ are the eigenvalues for the interaction
matrices, given in Eq.~\eqref{eq:Interact-Eigenv}. In particular,
then, in the $\h_3^2 \to 1/2 $, $\h_8^2 \to 1/2$ limit two eigenvalues
vanish, the corresponding directions decouple and we are left with the
following (asymmetrically gauged) model:
\begin{equation}
  g = \left( \begin{tabular}{c|c}
      $k \uni{6} $ \\ \hline
      & $ \begin{matrix}
        k + k^\prime \\
        & k + k^{\prime \prime }
      \end{matrix}$
    \end{tabular}\right)
\end{equation}
in the $\braket{\mJ_1, \mJ_2, \mJ_4, \mJ_5, \mJ_6, \mJ_7,
  \sqrt{k^\prime } \mathcal{I}_1 + \sqrt{k} \mJ_3, \sqrt{k^{\prime
      \prime }} \mathcal{I}_2 + \sqrt{k} \mJ_8 }$ basis that can be
seen as a $U \left( 1 \right)^2$ fibration over an $SU \left( 3 \right)
/ U \left( 1 \right)^2 $ base with metric $\diag \left( 1,1,1,1,1,1
\right)$ (in the current basis). This is precisely the same result we
obtained in the previous section when we read the fibration as a gauge
field living on the base.
\begin{equation}
  \begin{CD}
    U \left( 1 \right)^2 @>>> \mathcal{M} \\
    @.      @VVV\\
    {} @. SU \left( 3 \right)/ U \left( 1 \right)^2
  \end{CD}
\end{equation}
As in the previous example all this construction is valid only if the 
asymmetrically gauged \textsc{wzw} model is anomaly-free. 

\subsubsection{The $F_3$ flag space}

Let us now turn to the other possible choice for the $SU \left( 3 \right)$
gauging, namely the one where we take advantage of the $SU \left( 2 \right)$
embedding. Let us then consider the $SU \left( 3 \right)_{k_3} \times SU \left( 2
\right)_{k_2} \times U \left( 1 \right)_{k^\prime } \times U \left( 1 \right)_{k^{\prime \prime
  }}$ \textsc{wzw} model whose metric is
\begin{equation}
  g =  \left( \begin{tabular}{c|c|cc}
      $k_3 \uni{8}$ & & &\\  \hline
      & $k_2 \uni{3}$ & &\\  \hline
      & & $k^\prime$ & \\
      & &  & $k^{\prime \prime }$      
    \end{tabular}\right)
\end{equation}
in the $\braket{\mJ_1, \ldots, \mJ_8, \mathcal{I}_1, \mathcal{I}_2,
  \mathcal{I}_3, \mathcal{K}_1, \mathcal{K}_2}$ basis, where $\braket{\mJ_i}$ generate
the $SU\left( 3 \right)$, $\braket{\mathcal{I}_i}$ generate the $SU
\left( 2 \right)$ and $\braket{\mathcal{K}_i} $ generate the $U \left( 1
\right)^2 $.

The first step in this case consists in an asymmetric gauging mixing the
$\set{\mJ_1, \mJ_2, \mJ_3 }$ and $\set{\mathcal{I}_1, \mathcal{I}_2, \mathcal{I}_3}$ currents respectively. At
the gauging point, a whole 3-sphere decouples and we obtain the following
metric:
\begin{equation}
  g =  \left( \begin{tabular}{c|c|cc}
      $k_3 \uni{5}$ & & &\\  \hline
      & $\left( k_2 + k_3 \right) \uni{3}$ & &\\  \hline
      & & $k^\prime$ & \\
      & &  & $k^{\prime \prime }$      
    \end{tabular}\right)
\end{equation}
where we have to remember that in order to have an admissible embedding $k_2
= k_3 = k$. Our result is again -- not surprisingly -- a $SU \left( 2
\right)$ fibration over a $SU \left( 3 \right) / SU \left( 2 \right)$ base
(times the two $U \left( 1 \right)$'s).
\begin{equation}
  \begin{CD}
    SU \left( 2 \right) @>>> \mathcal{M} \\
    @.      @VVV\\
    {} @. SU \left( 3 \right)/ SU \left( 2 \right)
  \end{CD}
\end{equation}

Of course one could be tempted to give $\mathcal{M}$ the same
interpretation as before, namely an $SU \left( 3 \right) / SU\left( 2
\right)$ space supported by a chromo-magnetic $SU \left( 2 \right)$
field (or, even better, gauging an additional $U \left( 1 \right)$, of
a $\setC \mathbb{P}^2 $ background with an $SU \left( 2 \right)\times
U\left( 1 \right)$ chromo-magnetic field).  Actually this is not the
case. The main point is the fact that this $SU \left( 3 \right) \times
SU \left( 2 \right)$ model is essentially different from the previous
ones because the $U \left( 1 \right)$ factors were the result of the
bosonization of the right-moving gauge current which in this way
received a (fake) left-moving partner as in
Sec.~\ref{sec:wzw-deformations}.  This is not possible in the
non-abelian case since one can't obtain an $SU \left( 2 \right)$ at
arbitrary level $k$ out of the fermions of the theory\footnote{This
  would be of course be possible if we limited ourselves to small
  values of $k$, but in this case the whole geometric interpretation
  of the background would be questionable. However for Gepner-like
  string compactifications this class of models is relevant.}. In
other words, the $SU\left( 2 \right)$ factor is in this case truly a
constituent of the theory and there is no reason why it should be
decoupled or be given a different interpretation from the $SU \left( 3
\right)$ part.  This is why the structure obtained by the $SU \left( 2
\right)$ asymmetric gauging is to be considered an eight-dimensional
space admitting an $SU \left( 2 \right) \to SU \left( 3 \right)/SU
\left( 2 \right)$ fibration structure, or, equivalently, a deformed
$SU \left( 3 \right)$ where an embedded $SU \left( 2 \right)$ is at a
level double with respect to the other generators.

On the other hand we are still free to gauge away the two $U \left( 1
\right)$ factors just as before. This time we can choose to couple $K_1$
with the $\mJ_8$ factor that was left untouched in the initial $SU \left( 3
\right)$ and $\mathcal{K}_2 $ with the $\mJ_3+  \mathcal{I}_3 $
generator. Again we find a two-parameter family of deformations whose metric
can be written as:
\begin{equation}
  g =  \left( \begin{tabular}{c|c|c|ccc}
      $k \uni{4}$ & & &\\  \hline
      & $\mu_1 $ & &\\ \hline
      & & $2 k \uni{2}$ & \\  \hline
      & & & $\nu_1$ \\
      & & & & $\mu_2$ & \\
      & & & & & $\nu_2$      
    \end{tabular}\right)
\end{equation}
where:
\begin{align}
  \mu &= \lambda \left( k, k^\prime, \h^\prime \right) \\
  \nu &= \lambda \left( 2 k, k^{\prime \prime}, \h^{\prime \prime } \right) .
\end{align}
In particular now we can take the decoupling $\h^\prime = \h^{\prime \prime } \to 1/2 $
limit where we obtain:
\begin{equation}
  g =  \left( \begin{tabular}{c|c|cc}
      $k \uni{4}$ & &  \\  \hline
      & $2 k \uni{2}$ & \\  \hline
      & & $k + k^\prime $ \\
      & & & $2 k + k^{\prime \prime }$
    \end{tabular}\right)
\end{equation}
this structure is once more a $U \left( 1 \right)^2 \to SU \left( 3
\right)/U\left( 1\right)^2 $ fibration but in this case it is
perfectly fine to separate the space components from the gauge field
ones. So we can read out our final background fields as the K\"ahler
metric on $F_3 $ supported by a $U \left( 1 \right)^2 $
(chromo)magnetic field.

To summarize our results we can say that the two Einstein structures that
one can define on $SU \left( 3 \right) / U \left( 1 \right)^2$ are both
exact string theory backgrounds:
\begin{itemize}
\item The first one, obtained as the asymmetric coset $\frac{SU \left(
      3 \right) \times U \left( 1 \right)^2}{U \left( 1 \right)^2}$ is
  supported by an \textsc{ns-ns} field strength and a magnetic field;
\item The second, corresponding to the $\frac{SU \left( 3 \right)
    \times SU \left( 2 \right) \times U \left( 1 \right)^2}{SU \left(
      2 \right) \times U \left( 1\right)^2}$ asymmetric coset is
  K\"ahler and hence supported by the (chromo-)magnetic field alone.
\end{itemize}

\marginlabel{K\"ahler form for $SU(3)/U(1)^2$}This K\"ahler structure
has been deeply studied both from the mathematical and physical points
of view. In particular the K\"ahler form can be written as in
App.~\ref{sec:suleft-3right}:
\begin{equation}
  K \left( \gamma_\mu, \bar \gamma_\mu \right) = \log \left[ 1 + \abs{\gamma_1}^2 + \abs{\gamma_3}^2
  \right] + \log \left[ 1 + \abs{\gamma_2}^2 + \abs{\gamma_3 - \gamma_1 \gamma_2 }^2
  \right] .
\end{equation}
It is immediate to show that this manifold is Einstein and in
particular its Ricci scalar is $R = 12 $. Being K\"ahler, $F_3 $ is
torsionless, that means in turn that there is no \textsc{ns-ns}
form\footnote{To be precise one could define a $B$ field but this
  would have to be closed}. Moreover there is no dilaton by
construction\footnote{The dilaton would basically measure the
  difference between the asymmetric coset volume form and the
  homogeneous space one as it is shown in~\cite{Tseytlin:1994my}}. The
only other field that supports the background comes from the $U \left(
  1 \right)^2$ fibration. Since the manifold is K\"ahler it is useful to
take advantage of the complex structure and write our background
fields in complex formalism. In these terms the metric is written as:
\begin{equation}
  g = \frac{k}{2} \left( \mJ^1 \otimes \mJ^{\bar 1} + \mJ^2 \otimes \mJ^{\bar 2} +
    2 \mJ^3 \otimes \mJ^{\bar 3} \right)  
\end{equation}
where $\mJ^i $ and $\bar \mJ^{\bar i}$ are the Maurer-Cartan corresponding
to positive and negative roots respectively and the field strength is given
by:
\begin{equation}
  F^a = \sqrt{\frac{k}{2 k_g }} \F{a}{\mu \bar \rho } C^{\bar \rho \sigma
  } R_{\sigma \bar \nu } \mJ^\mu \land  \mJ^{\bar \nu}
\end{equation}
where $C $ is the following tensor
\begin{equation}
  C = \sum_\alpha \mJ^\alpha \otimes \mJ^{\bar \alpha }
\end{equation}

\subsection{New linear dilaton backgrounds of Heterotic strings}
\label{sec:heter-strings-sing}

These left-coset superconformal field theories can be used to construct 
various supersymmetric 
exact string backgrounds. The first class are generalizations of 
Gepner models~\cite{Gepner:1988qi} and Kazama-Suzuki 
constructions~\cite{Kazama:1989qp} using the left cosets as building 
blocks for the internal \textsc{scft}. This has already 
been considered in~\cite{Berglund:1996dv} for the $S^2$ coset 
but can be extended using the new theories constructed above. In this 
case there is no geometric interpretation from the sigma model point 
of view since these theories have no semi-classical limit. Indeed 
the levels of the cosets are frozen because their central charge 
must add up to $c=9$ (in the case of four-dimensional compactification). 
However we expect that they correspond to special points in the 
moduli spaces of supersymmetric compactifications, generalizing the 
Gepner points of the CY manifolds.

Another type of models are the left cosets analogues of the NS5-branes
solutions~\cite{Callan:1991dj,Kounnas:1990ud} and of their extensions
to more generic supersymmetric vacua with a dilaton background. It was
shown in~\cite{Giveon:1999zm} that a large class of these linear
dilaton theories are dual to singular CY manifolds in the decoupling
limit. An extensive review of the different possibilities in various
dimensions has been given in~\cite{Eguchi:2003yy} with all the
possible $G/H$ cosets.  The left cosets that we constructed allows to
extend all these solutions to heterotic strings, with a different
geometrical interpretation since our cosets differ from ordinary
gauged \textsc{wzw} model. However the superconformal structure of the
left sector of our models is exactly the same as for the corresponding
gauged \textsc{wzw} --~except that the values of the $N=2$ R-charges
that appear in the spectrum are constrained~-- so we can carry over
all the known constructions to the case of the geometric cosets.

In the generic case these constructions involve non-abelian cosets,
and as we showed the asymmetric deformations and gaugings apply only
to the abelian components.  Thus in general we will get mixed models
which are gauged \textsc{wzw} models w.r.t. the non-abelian part of
$H$ and geometric cosets w.r.t. the abelian components of $H$.  Below
we will focus on purely abelian examples, i.e. corresponding to
geometric cosets. The dual interpretation of these models, in terms of
the decoupling limit of some singular compactification manifolds, is
not known. Note however that by construction there are about
$\sqrt{k}$ times less massless states in our models than in the
standard left-right symmetric solutions. Therefore they may correspond
to some compactifications with fluxes, for which the number of moduli
is reduced. It would be very interesting to investigate this issue
further.

\paragraph{Six-dimensional model.}

Let us take as a first example the critical superstring background:
\begin{equation}
  \setR^{5,1} \times \frac{SL(2,\mathbb{R})_{k+2}\times SO(2)_1}{U\left( 1 \right)_{k}} 
  \times \left[ \sfrac{U\left( 1 \right)_k}{SU(2)_{k-2} \times SO(2)_1} \right]
\end{equation}
the first factor is an ordinary gauged model while the second one is a
left coset \textsc{cft} as discussed in this paper. This is the direct
analogue of the five-brane solution, or more precisely of the double
scaling limit of NS5-branes on a circle~\cite{Giveon:1999px,IKPT}, in
the present case with magnetic flux. This theory has $N=2$ charges
but, in order to achieve spacetime supersymmetry one must project onto
odd-integral $N=2$ charges on the left-moving side, as in the type II
construction~\cite{IKPT}. This can be done in the standard way by
orbifoldizing the left $N=2$ charges of the two cosets.

\paragraph{Four-dimensional model.}

A simple variation of the six-dimensional theory is provided by 
\begin{multline}
\mathbb{R}^{3,1}\ \times\ \frac{SL(2,\mathbb{R})_{k/2+2}\times SO(2)_1}{U\left( 1 \right)_{2k}} \ \times 
\left[ \sfrac{U\left( 1 \right)_k}{SU(2)_{k-2} \times SO(2)_1} \right] \\
\times 
\left[ \sfrac{U\left( 1 \right)_k}{SU(2)_{k-2} \times SO(2)_1} \right] 
\end{multline}
which is the magnetic analogue of the 
(double scaling limit of) intersecting five-branes solution.
Also here an orbi\-foldization of the left $N=2$ charges is needed 
to achieve space-time supersymmetry.

\paragraph{Three-dimensional models: the flagbrane${}^\copyright$.}
 
We can construct the following background of the $G_2$ holonomy type, as in the 
case of symmetric coset~\cite{Eguchi:2001xa}:
\begin{equation}
  \setR^{2,1} \times  \mathbb{R}_Q  \times \left[  \sfrac{U\left( 1 \right)_k \times
      U\left( 1 \right)_{3k}}{SU\left(3\right)_{k-3} \times SO(6)_1} \right]
\end{equation}
and the non-trivial part of the metric is
\begin{equation}
  ds^2 = -\di t^2 + \di x^2 +  \di y^2 + \frac{k}{4r^2} 
  \left[ \di r^2 + 4 r^2 \di s^2 (\nicefrac{SU\left(3\right)}{U\left( 1 \right)^2}) \right].
\label{asyflagbrane}
\end{equation} 
Without the factor of four it would be a direct analogue of the NS5-brane, 
being conformal to a cone over the flag space. 

Another possibility in three dimensions is 
to lift the $SL(2,\mathbb{R})/U(1)$ coset to the group manifold 
$SL(2,\mathbb{R})$. In this case, as for the standard gauged \textsc{wzw} 
construction~\cite{Argurio:2000tg} we will get the following anti-de Sitter 
background:
\begin{equation}
SL(2,\mathbb{R})_{k/4+2}    \times \left[  
\sfrac{U\left( 1 \right)_{3k}}{SU\left(3\right)_{k-3} \times SO(6)_1} \right]
\end{equation}
and the left moving sector of this worldsheet \textsc{cft} defines 
an $N=3$ superconformal algebra in spacetime. 

\paragraph{Two-dimensional model.}

In this case we can construct the background:
\begin{equation}
  \setR^{1,1} \times \frac{SL(2,\mathbb{R})_{k/4+2}\times SO(2)_1}{U\left( 1 \right)_{4k}} \times
  \frac{\sfrac{U\left( 1 \right)_{3k}}{SU\left(3\right)_{k-3} \times SO(6)_1}}{U\left( 1 \right)_k}  
\label{falseflagbrane}
\end{equation}
which corresponds in the classification of~\cite{Eguchi:2003yy} to a
non-compact manifold of $SU(4)$ holonomy once the proper projection is
done on the left $N=2$ charges. This solution can be also be thought
as conformal to a cone over the Einstein space
$SU\left(3\right)/U\left( 1 \right)$.  Using the same methods as for
the NS5-branes in~\cite{IKPT}, we can show that the full solution
corresponding to the model~(\ref{falseflagbrane}) can be obtained
directly as the null super-coset:
\begin{equation}
\frac{SL(2,\mathbb{R})_{k/4} \times \sfrac{U\left( 1 \right)}{SU\left(3\right)_{k}}}{U\left( 1 \right)_L \times U\left( 1 \right)_R}
\end{equation}
where the action is along the elliptic generator in the $SL \left( 2,\setR
\right) $, with a normalization $\braket{t^3}^2  = -4$, and
along the direction $\alpha_1 + 2\alpha_2$ in the coset space $\ssfrac{U\left( 1
  \right)}{SU\left(3\right)}$, with a canonical normalization.  For $r
\to \infty$ the solution asymptotes the cone but when $r \to 0$ the strong
coupling region is smoothly capped by the cigar.




\chapter{Squashed groups in type II}
\label{cha:squashed-groups-type}

\chapterprecis{In this chapter we start deviating from the preceding
  ones because we will no longer deal with \textsc{wzw} models but
  with configurations in which the group manifold geometry is
  sustained by \textsc{rr} fields. In particular, then, we see how the
  squashed geometries can be obtained in type II theories by
  T-dualizing black brane configurations.}

\lettrine{T}{he models} that we have studied so far are intrinsically
heterotic. In fact it is the very presence of a heterotic
electromagnetic field that allows for the solution of the equations of
motion. On the other hand, and this is especially true for purely
asymmetric deformations that only have a constant dilaton, we can
expect them to be mapped via S-duality to type II solutions. In this
chapter we will build such solutions but using a slightly different
path: in particular we will see how using a T-duality it is possible
to modify a fibration geometry in the same way as we did before by
adding a marginal operator, thus recovering the same geometries as
above, but this time in presence of Ramond-Ramond fields.

\section{$SL(2,\setR) \times SU(2)$ as a D-brane solution}
\label{sec:sl2--sectiont}

\oldstuff

Up to this point we have considered \textsc{wzw} models for the sake
of their self-consistency. In other words we have used group manifolds
as part of critical string backgrounds on the ground of the existence
of an underlying \textsc{cft}. On the other hand, at low energies we
should obtain a \textsc{sugra} description, so it is plausible that a
description for the same geometry is available in terms of diverse
ten-dimensional sources.

\marginlabel{Black brane ansatz} The starting point is a
higher-dimensional generalization of the usual four-dimensional
charged black hole. The natural, most symmetric, ansatz for the
geometry in presence of a $p$-dimensional extended object (a black
$D_p$-brane) consists in keeping the Lorentz symmetry in $\left( p + 1
\right)$ dimensions and a spherical symmetry in $\left(9-p
\right)$. In other words breaking the $SO(1,9)$ symmetry to $SO(1,p)
\times SO(9-p)$. Moreover we expect a $C\form{\left(p+1 \right)}$ form
naturally coupled to the $p$-brane. It is possible to show
\cite{Stelle:1998xg} that the solution also contains a dilaton $\Phi
(r)$ and is completely determined in terms of a harmonic function $H_p
(r)$:
\begin{equation}
  \begin{cases}
    \di s^2 = H_p(r)^{-1/2} \left( -\di t^2 + \di x_1^2 + \ldots + \di x_p^2 \right) + H_p(r) \left( \di r^2 + r^2 \di \Omega_{8-p}^2 \right) \\
    C\form{\left( p + 1 \right)} = \left( 1 - H_p (r)^{-1} \right) \di t \wedge \di x^1 \wedge \ldots \wedge \di x^p \\
    e^{\Phi} = H_p(r)^{(3-p)/4}
  \end{cases}
\end{equation}
where $H_P(r)$ is explicitly given by
\begin{equation}
  H_p (r) = 1 + \frac{Q_p}{r^{7-p}}    
\end{equation}

More complicated solutions with intersecting branes can be studied and
in particular the solution for a D1-D5 system reads:
\mathindent=0em
  \begin{equation}
    \begin{cases}
      \di s^2 = H_1(r)^{-1/2} H_5(r)^{-1/2} \left( - \di t^2 + \di x_1^2 \right) +  H_1(r)^{1/2} H_5(r)^{-1/2} \left( \di x_2^2 + \ldots+ \di x_5^2 \right) + \\
      \hspace{16em}+ H_1(r)^{1/2} H_5(r)^{1/2} \left( \di r^ 2 + \di \Omega_3^2 \right) \\
        C\form{2} = \left( 1 - H_1 (r)^{-1} \right) \di t \wedge \di x^1 \\
        C\form{6} = \left( 1 - H_5 (r)^{-1} \right) \di t \wedge \di x^1 \wedge \ldots \wedge \di x^5 \\
        e^{\Phi} = H_1(r)^{1/2} H_5(r)^{-1/2}
    \end{cases}
  \end{equation}
  \mathindent=\oldindent where in this case both $H_1(r)$ and $H_5(r)$
  have the same $1/r^2$ behaviour. It is then simple to see that in
  the near-horizon limit, \emph{i.e.} for $r \to 0$, the geometry we
  find is $\mathrm{AdS}_3 \times S^3 \times \setR^4$, or $SL(2,\setR)
  \times SU(2)$, plus four flat directions.

\section{T duality with \textsc{rr} fields}

\danger In the IIB solutions we consider in this section the r\^ole of
sustaining the geometry previously held by the Kalb--Ramond three-form
is taken by \textsc{rr} field strengths. This has a number of
consequences, first of all the lack of a proper \textsc{cft}
description for such configurations. In particular this means also
that the usual Buscher rules~\cite{Buscher:1987sk} prove insufficient
and we are forced to follow a slightly more involved path to write
T-duals: derive two low-energy effective actions and explicitly write
the transformations relating them (in this we will follow the same
procedure as in~\cite{Duff:1998us,Duff:1998cr}).

\marginlabel{Type II action in nine dimensions} In ten dimensions,
type IIA and IIB are related by a T-duality transformation, stating
that the former theory compactified on a circle of radius $R$ is
equivalent to the latter on a circle of radius $1/R$. This means in
particular that there is only one possible nine-dimensional $N=2$
\textsc{sugra} action. The rules of T-duality are then easily obtained
by explicitly writing the two low-energy actions and identifying the
corresponding terms.

  For sake of clarity let us just consider the bosonic sector of both
  theories. In~\cite{Lu:1995yn,Lu:1996ge} it was found that the
  IIA action in nine dimensions is given by
  \begin{multline}
    e^{-1} L_{IIA} = R - \frac{1}{2} (\partial \phi)^2 - \frac{1}{2}(\partial
    \varphi)^2 - \frac{1}{2} ({\mathcal{F}}\form{1}^{(12)})^2
    e^{\frac{3}{2}\phi +\frac{\sqrt7}{2} \varphi} + \\
    -\frac{1}{48} (F\form{4})^2 e^{\frac{1}{2}\phi +\frac{3}{2\sqrt7} \varphi}
    -\frac{1}{12} (F\form{3}^{(1)})^2 e^{-\phi +\frac{1}{\sqrt7} \varphi} -
    \frac{1}{12} (F\form{3}^{(2)})^2
    e^{\frac{1}{2}\phi - \frac{5}{2\sqrt7} \varphi} + \\
    -\frac{1}{4} (F\form{2}^{(12)})^2 e^{-\phi - \frac{3}{\sqrt7}\varphi}
    -\frac{1}{4} ({\mathcal{F}}\form{2}^{(1)})^2 e^{\frac{3}{2} \phi
      +\frac{1}{2\sqrt7}\varphi} - \frac{1}{4}
    ({\mathcal{F}}\form{2}^{(2)})^2
    e^{\frac{4}{\sqrt7} \varphi} + \\
    -\frac{1}{2e} \tilde F\form{4} \land \tilde F\form{4} \land
    A\form{1}^{(12)} - \frac{1}{e} \tilde F\form{3}^{(1)} \land \tilde
    F\form{3}^{(2)} \land A\form{3}\ ,
  \end{multline}
  where $\phi$ is the original dilaton, $\varphi $ is a scalar measuring the
  compact circle, defined by the reduction (in string frame)
  \begin{equation}
    \di s^2 = e^{\phi/2} \di s^2_{10} = e^{\phi/2} \left( e^{-\varphi/ \left(2 \sqrt{7} \right)} \di s^2_9 + e^{\sqrt{7} \varphi /2} \left( \di z + \mathcal{A}\form{1} \right)^2 \right)    
  \end{equation}
  and $F\form{n}$ are $n$-form field strengths defined as
  \begin{subequations}
    \begin{align}
      F\form{4} &=\tilde F\form{4} - \tilde F\form{3}^{(1)}\land
      \mathcal{A}\form{1}^{(1)} - \tilde F\form{3}^{(2)}\land
      \mathcal{A}\form{1}^{(2)} - \frac{1}{2} \tilde F\form{2}^{(12)}
      \land
      \mathcal{A}\form{1}^{(1)} \land  \mathcal{A}\form{1}^{(2)} \\
      F^{(1)}\form{3} &= \tilde F^{(1)}\form{3} - \tilde F\form{2}^{(12)} \land \mathcal{A}\form{1}^{(2)} \\
      F\form{3}^{(2)} &= \tilde F\form{3}^{(2)} + F\form{2}^{(12)} \land
      \mathcal{A}\form{1}^{(1)} - \mathcal{A}\form{0}^{(12)} \left( \tilde F^{(1)} \form{3} - F\form{2}^{(12)}\land \mathcal{A}\form{1}^{(2)} \right) \\
      F\form{2}^{(12)} &= \tilde F^{(12)} \form{2}\\
      \mathcal{F}\form{2}^{(1)} &= \mathcal{F}\form{2}^{(1)}
      + \mathcal{A}\form{0}^{(12)} \mathcal{F}\form{1}^{(2)} \\
      \mathcal{F}\form{2}^{(2)} &= \tilde{\mathcal{F}}\form{2}^{(2)} \\
      \mathcal{F}\form{1}^{(12)} &= \tilde{\mathcal{F}}\form{1}^{(12)}
      \ .
    \end{align}
  \end{subequations}
  In the same way, starting from the IIB action one obtains
  the following nine-dimensional IIB Lagrangian:
  \begin{multline}
    e^{-1} L_{IIB} = R - \frac{1}{2} (\partial \phi)^2 -\frac{1}{2}
    (\partial
    \varphi)^2 - \frac{1}{2} e^{2\phi} (\partial  \chi)^2 + \\
    -\frac{1}{48} e^{-\frac{2}{\sqrt7} \varphi} F\form{4}^2 -\frac{1}{12}
    e^{-\phi+\frac{1}{\sqrt7}\varphi} (F\form{3}^{({\textsc{ns}})})^2
    -\frac{1}{2} e^{\phi
      +\frac{1}{\sqrt7} \varphi} (F\form{3}^{({\textsc{r}})})^2 + \\
    -\frac{1}{4} e^{\frac{4}{\sqrt7} \varphi} ({\mathcal{F}}\form{2})^2 -
    \frac{1}{4} e^{\phi - \frac{3}{\sqrt7}\varphi}
    (F\form{2}^{({\textsc{r}})})^2 -
    \frac{1}{4} e^{-\phi - \frac{3}{\sqrt7} \varphi} (F\form{2}^{({\textsc{ns}})})^2 + \\
    -\frac{1}{2e} \tilde F\form{4} \land \tilde F\form{4} \land
    {\mathcal{A}}\form{1} - \frac{1}{e}\, \tilde F_3^{({\textsc{ns}})} \land \tilde
      F\form{3}^{({\textsc{r}})} \land A\form{3}\ .
  \end{multline}  
  Knowing that both describe the same theory we easily obtain the
  conversion table in Tab.~\ref{tab:T-duality} which acts as a
  dictionary between IIA and IIB in ten dimensions
  plus the following relation between the scalar fields
  \begin{equation}
    \begin{pmatrix}
      \phi \\
      \varphi 
    \end{pmatrix}_{IIA} =
    \begin{pmatrix}
      3/4  & - \sqrt{7} /4 \\
      -\sqrt{7} / 4 & - 3 / 4
    \end{pmatrix} \begin{pmatrix}
      \phi \\
      \varphi 
    \end{pmatrix}_{IIB} 
  \end{equation}
  This completes the T-duality relations generalizing the usual
  ones~\cite{Buscher:1987sk} valid in the \textsc{ns-ns} sector.

  \begin{table}
    \centering
    \begin{tabular}{|c|c|c|c|c|c|}\hline
      &\multicolumn{2}{|c|}{IIA} &
      &\multicolumn{2}{c|}{IIB} \\ \cline{2-6}
      & $D=10$ & $D=9$ &T-duality & $D=9$ & $D=10$ \\ \hline\hline
      & $A_3$ & $A_3$ & $\longleftrightarrow$ &
      $A_3$ & $B_4$ \\ \cline{3-6}
      R-R & &  $A_2^{(2)}$& $\longleftrightarrow$
      & $A_2^{\textsc{r}}$ & $A_2^{\textsc{r}}$
      \\ \cline{2-5}
      fields& ${\mathcal{A}}_1^{(1)}$ & ${\mathcal{A}}_1^{(1)}$ &
      $\longleftrightarrow$ &
      $A_1^{\textsc{r}}$ & \\ \cline{3-6}
      & & ${\mathcal{A}}_0^{(12)}$ & $\longleftrightarrow$
      & $\chi$ &$\chi$
      \\ \hline\hline
      NS-NS & $G_{\mu\nu}$ & ${\mathcal{A}}_1^{(2)}$
      & $\longleftrightarrow$ &
      $A_1^{\textsc{ns}}$ & $A_2^{\textsc{ns}}$ \\ \cline{2-5}
      fields& $A_2^{(1)}$ & $A_2^{(1)}$ &
      $\longleftrightarrow$ & $A_2^{\textsc{ns}}$ &
      \\ \cline{3-6}
      & & $A_1^{(12)}$ & $\longleftrightarrow$ &
      ${\mathcal{A}}_1$ & $G_{\mu\nu}$
      \\ \hline
\end{tabular}
    \caption{T-duality dictionary with \textsc{rr} fields}
    \label{tab:T-duality}
  \end{table}

\section{The squashed sphere}
\label{sec:squashed-sphere}

Start with a $D_1 - D_5$ system in type IIB described in
Sec.~\ref{sec:sl2--sectiont}. The near-horizon geometry is
\begin{equation}
  \di s_{10}^2 = \mathrm{AdS}_3 \times S^3 \times \setR^4  ,
\end{equation}
with a three-form flux
\begin{equation}
  F_3 = \sqrt{2} m \left( \vol_{\mathrm{AdS}} + \vol_S \right) ,
\end{equation}
where $\vol$ is the volume form ans the constant $m $ is fixed by
demanding:
\begin{align}
  \left. Ric \right|_{\mathrm{AdS}} &= -m^2 \left. g\right|_{\mathrm{AdS}} \\
  \left. Ric \right|_{S} &= m^2 \left. g\right|_{S}   .
\end{align}
Now, introduce the coordinates $\left( \vartheta, \varphi, \psi,
  x\right)$ on $S^3 \times S^1$ and write explicitly:
\begin{align}
  \di s_{10}^2 &= \mathrm{AdS}_3 \times \setR^3 + \frac{1}{2m^2} \left[ \di \vartheta^2
    + \di \varphi^2 + \di \psi^2 + 2 \cos \vartheta \di \varphi \di \psi \right] + \di x^2 \\
  F_3 &= m \sqrt{2} \vol_{\mathrm{AdS}} + \frac{\sin \vartheta }{2 m^2} \di \vartheta
  \land \di \varphi \land \di \psi ,
\end{align}
where $x$ is a periodic variable with period
\begin{equation}
  x \sim x + \frac{4 \pi}{\lambda} n . 
\end{equation}
If we change the variable $\psi $ to $\psi = \alpha + \lambda x $ we
still have a $4 \pi$-periodic direction $\alpha $ and can rewrite the
metric as:
\mathindent=0em
\begin{align}
  \begin{split}
    \di s_{10}^2 &= \mathrm{AdS}_3 \times \setR^3 + \frac{1}{2m^2} \left[ \di \vartheta^2
      + \di \varphi^2 + \di \alpha^2 + 2 \cos \vartheta \di \varphi \di \alpha \right] + \left( 1
      + \frac{\lambda^2}{2m^2} \right) \di x^2 + \\
    &  + \frac{\lambda }{m^2 } \left( \di \alpha
      + \cos \vartheta \di \varphi \right) \di x \end{split} \\
  F_3 &= m \sqrt{2} \vol_{\mathrm{AdS}} + \frac{\sin \vartheta }{2 m^2} \di \vartheta
  \land \di \varphi \land \di \alpha + \lambda \frac{\sin \vartheta }{2 m^2} \di \vartheta \land \di \varphi \land \di x . 
\end{align}
\mathindent=\oldindent
Redefining
\begin{align}
  z = \sqrt{1+ \frac{\lambda^2}{2m^2}} x && h = \frac{\lambda }{2m}
  \frac{1}{\sqrt{1+\nicefrac{\lambda^2}{2m^2} }} ,
\end{align}
the fields read
\begin{align}
  \begin{split}
    \di s_{10}^2 &= \mathrm{AdS}_3 \times \setR^3 + \frac{1}{2m^2} \left[ \di \vartheta^2
      + \sin^2 \vartheta \di \varphi^2 + \left( 1- 2h^2 \right) \left( \di \alpha + \cos \vartheta
        \di \varphi \right)^2 \right] + \\
    & +\left[ \di z + \frac{h}{m} \left( \di
      \alpha + \cos \vartheta \di \varphi \right) \right]^2     
  \end{split} \\
  F_3 &= m \sqrt{2} \vol_{\mathrm{AdS}} + \frac{\sin \vartheta }{2 m^2} \di \vartheta
  \land \di \varphi \land \di \alpha + \frac{h}{m} \sin \vartheta \di \vartheta \land \di \varphi \land \di z ,
\end{align}
and we can perform a Kaluza-Klein reduction on $z$ and go to nine
dimensions. The metric reads:
\begin{equation}
  \di s_{9}^2 = \mathrm{AdS}_3 \times \setR^3 + \frac{1}{2m^2} \left[ \di \vartheta^2
    + \sin^2 \vartheta \di \varphi^2 + \left( 1- 2h^2 \right) \left( \di \alpha + \cos \vartheta
      \di \varphi \right)^2 \right] ,
\end{equation}
and the gauge fields are obtained from:
\begin{equation}
  F_3 = F^{(3)}_3 + F_2^{(3)} \land \left( \di z + A \right) ,   
\end{equation}
where $F_m^{(n)}$ is the $m$-form obtained from the reduction of a
$n$-form and $A$ is the one-form
\begin{equation}
  A = \frac{h}{m} \left( \di  \alpha + \cos \vartheta \di \varphi \right) .
\end{equation}
Explicitly, adding the extra Kaluza-Klein two-form:
\begin{align}
  F^{(3)}_3 &= m \sqrt{2} \vol_{\mathrm{AdS}} +\left( 1 - 2h^2\right)
  \frac{\sin \vartheta }{2m^2} \di \vartheta \land \di \varphi \land \di \alpha \\
  F_2^{(3)} &= \frac{h}{m} \sin \vartheta \di \vartheta \land \di \varphi \\
  F_2^{(g)} &= \di A = \frac{h}{m} \sin \vartheta \di \vartheta \land \di \varphi  .
\end{align}
For the moment this is just a rewriting. Now, let us perform a
T-duality to go to type IIA. The fields keep their expressions but the
interpretation changes according to Tab.~\ref{tab:T-duality}:
$F_3^{(3)} $ now comes from the reduction of a four-form in ten
dimensions, $F_2^{(2)} $ from a two-form and $F_2^{(g)}$ is now
obtained as the result of the reduction of the Kalb-Ramond field:
\begin{align}
  F_3^{(4)} = F_3^{(3)} && F_2^{(2)} = F_2^{(3)} && F_2^{(B)} = F_3^{(g)} .
\end{align}
We can oxidise back to ten dimensions and get a IIA background:
\begin{align}
  \di s_{10}^2 &= \mathrm{AdS}_3 \times \setR^3 + \frac{1}{2m^2} \left[ \di \vartheta^2
    + \sin^2 \vartheta \di \varphi^2 + \left( 1- 2h^2 \right) \left( \di \alpha + \cos \vartheta
      \di \varphi \right)^2 \right] + \di \xi^2 \\
  F_4 &= F_3^{(4)} \land \di \xi = \left[ m \sqrt{2} \vol_{\mathrm{AdS}} +
    \left( 1- 2h^2 \right) \frac{\sin \vartheta }{2 m^2} \di \vartheta \land \di \varphi \land \di \alpha
  \right] \land \di \xi \\
  F_2 &= F_2^{(2)} = \frac{h}{m} \sin \vartheta \di \vartheta \land \di \varphi \\
  H_3 &= F_2^{B} \land \di \xi = \frac{h}{m} \sin \vartheta \di \vartheta \land \di \varphi \land \di \xi . 
\end{align}

It is worthwhile to emphasize that by construction $\alpha $ is $4
\pi$-periodic and then the geometry is the one of a respectable
squashed three-sphere. A very similar construction was considered
in~\cite{Duff:1998cr}. In that case, though, the authors start with
the same $\mathrm{AdS}_3 \times S^3$ geometry with both \textsc{rr}
and \textsc{ns-ns} fields and then by reducing on one of the sphere
isometries, find the Lens space $S^3/\setZ_p$ or a squashed version,
where $p$ and the squashing depend on the values of the charges. This
is clearly an orbifold of the solutions above.

In principle these constructions can be extended to other group
manifold geometries (\emph{e.g.} the obvious choice leading to a
squashed $\mathrm{AdS}_3$) but in any case one should start from a
configuration with \textsc{rr} fields (typically S-dual to the
\textsc{wzw} models we described in great detail previously), since
the absence of \textsc{ns-ns} antisymmetric fields is the key
ingredient for the trivialization of the fiber bundle. More general
geometries can be obtained by starting with a mixed
\textsc{rr}-\textsc{ns-ns} configuration.


\chapter{Out of the conformal point: Renormalization Group Flows}
\label{cha:renorm-group-flows}

\chapterprecis{This chapter is devoted to the study of the relaxation
  of squashed \textsc{wzw} models further deformed by the insertion of
  non-marginal operators. The calculation is carried from both the
  target space and world-sheet points of view, once more highlighting
  the interplay between the two complementary descriptions. In the
  last part such techniques are used to outline the connection between
  the time evolution and the \textsc{rg}-flow which is seen as a large
  friction limit description; we are hence naturally led to a
  \textsc{frw}-type cosmological model.}

\lettrine{S}{tring theory}, at least in its world-sheet formulation,
is most easily studied on-shell. Thanks to the power of conformal
field theory, this permits a profound analysis of specific
backgrounds.  At the same time, though, it makes it difficult to
describe more general effects that require a less local knowledge of
the theory and its moduli space. In particular it is not obvious how
to describe transitions between two different solutions or even the
relaxation of an unstable background towards an on-shell solution.

In this chapter we deviate from conformality by adding non-marginal
deformations on the top of exact solutions, such as
\textsc{wzw}-models or squashed-group models. The resulting
\textsc{rg}-flow then drives those systems back to or away from the
conformal point, depending on the character of the deformation. As it
is usually the case, these calculations can be faced from two
complementary points of view: either in terms of the target space
description or in terms of world-sheet two-dimensional theory. We will
consider both approaches and show how they do really complete each
other, in the sense that they can be considered as two different
series expansions of the same quantity. As such, each side contains
more information than the other at any given order in perturbation.
This will allow us in particular to make a prediction on the outcome
of a technically involved one-loop calculation in the \textsc{wzw} and
squashed group \textsc{cft}s on the basis of a two-loop result on
target space renormalization.

In the last part of this chapter we use the technology developed so
far to show how an \textsc{rg}-flow analysis can allow for further
insights on the issue of time-dependent solutions. More precisely we
will see how for a given class of systems whose geometry is the warped
product of a constant curvature space and a time direction the
\textsc{rg}-flow equations are a sort of large-friction approximation
with the central charge deficit playing the r\^ole of an effective
friction coefficient.

\section{The target space point of view}
\label{sec:target-space-renormalization}

\subsection{Renormalization in a dimensional regularization scheme}
\label{sec:renorm-dimens-regul}

\danger Consider the $\sigma$-model with Lagrangian density
\begin{equation}
  \mathcal{L} = \frac{1}{2 \lambda} \left( g_{ij} + B_{ij} \right) 
  \Xi^{ij} ,
\end{equation}
where $g_{ij} $ is a metric, $B_{ij}$ a two-form and $\Xi^{ij} = \partial_\mu X^i
\partial^\mu X^j + \epsilon_{\mu \nu} \partial^\mu X^i \partial^\nu X^j$. We will say that the model is
renormalizable if the corresponding counterterms at any given order in
the loop expansion can be reabsorbed into a renormalization of the
coupling constant and other parameters that appear in the expressions
for $g_{ij}$ and $B_{ij}$.\footnote{When this is not the case the
  model might nevertheless be renormalizable in a more general sense,
  in the infinite-dimensional space of metrics and torsions}

The standard technique for dealing with this kind of Lagrangian
consists in incorporating the Kalb--Ramond field (or, equivalently,
the \textsc{wz} term) into the geometry by reading it as a
torsion. This means that instead of the usual Levi-Civita connection
one uses the connection $\Gamma^-$ defined as
\begin{equation}
    {\Gamma^-}^{i}_{\phantom{i}jk}  = {i
  \atopwithdelims\{\} j k } - \frac{1}{2} H^i_{\phantom{i}jk } .  
\end{equation}
where ${i \atopwithdelims\{\} j k }$ is the Christoffel symbol and with
respect to this connection one defines the Riemann and Ricci tensors
$R^-$ and $Ric^-$.

\marginlabel{Two-loop target space \textsc{rg}-flow}Now, using the
background field method in a dimensional regularization scheme (see
\cite{Osborn:1989bu,Alvarez-Gaume:1981hn,Friedan:1980jm,Friedan:1980jf,Braaten:1985is,Hull:1987pc}
and for various applications
\cite{Balog:1996im,Balog:1998br,Sfetsos:1998kr,Sfetsos:1999zm}) we can
calculate the one- and two\nb-loop counterterms that turn out to be:
\begin{align}
  \mu^\varepsilon \mathcal{L}^{(1)} &= \frac{1}{\pi \varepsilon}
  T^{(1)} = \frac{\alpha^\prime}{2 \varepsilon \lambda } {Ric^-}_{ij}
  \Xi^{ij} ,\\
  \mu^\varepsilon \mathcal{L}^{(2)} &= \frac{\lambda}{8 \pi^2
    \varepsilon} T^{(2)} = \frac{\alpha^{\prime 2}}{16 \varepsilon
    \lambda } Y^{lmk}_{\phantom{lmk}j}{R^-}_{iklm} \Xi^{ij} ,
\end{align}
where $Y$ is given by
 \begin{equation}
  Y_{i j k l } = -2 {R^-}_{i j k l } + 3 {R^-}_{[k i j ] l } +
  \frac{1}{2} \left( {H^2}_{k i } g_{j l } - {H^2}_{k j } g_{i l } 
  \right) ,
\end{equation}
and
\begin{equation}
  {H^2}_{ij} = {H_{[3]}}_{ilm} {H_{[3]}}_j^{\phantom{j}lm} .
\end{equation}

In general the metric and the Kalb--Ramond field depend on a set of
bare parameters $a^{(0)}_k$. In this case we can convert the
counterterms given above into coupling and parameter renormalizations
if we write perturbatively the bare quantities as:
\begin{equation}
\label{eq:bare-stuff}
  \begin{cases}
    \lambda^{(0)} = \mu^\varepsilon \lambda \left( 1 + \frac{J_1
        (a)}{\pi \varepsilon} \lambda + \ldots \right) =
    \mu^\varepsilon \lambda \left( 1 + \frac{y_\lambda}{\epsilon} +
      \ldots \right) ,\\
    a^{(0)}_k = a_k + \frac{a_k^{(1)}(a)}{\pi \varepsilon} + \ldots =
    a_k \left( 1 + \frac{y_{a_k}}{\epsilon} + \ldots \right) ,\\
    X^{(0) \mu } = X^\mu + \frac{X^{(1) \mu }( X, a)}{\pi \varepsilon} + \ldots , 
  \end{cases}
\end{equation}
where we allowed for a slight generalization with respect to the
definition of renormalization given above in terms of a coordinate
reparametrization of the target space\footnote{This redefinition is in
  general non-linear; in the special case when the $X^{(i)}$'s depend
  linearly on $X$ the last equation of the system~\ref{eq:bare-stuff}
  reduces to a multiplicative wavefunction renormalization. The only
  condition is that $X^{(i)}$ shouldn't depend on the derivatives of
  $X$. In more geometric terms we are just using the diffeomorphism
  invariance of the renormalized theory.}. Then the one- and
two\nb-loop $\beta$\nb-functions are given by:
\begin{equation}
  \begin{cases}
    \beta_\lambda = \displaystyle{\Deriv{\lambda}{t} = \lambda^2
    \deriv{y_\lambda}{\lambda}} = \frac{\lambda^2 }{\pi} \left(
      J^{(1)} (a) + \frac{\lambda}{4 \pi} J^{(2)} (a) \right) ,\\
    \beta_{a_k} = \displaystyle{\Deriv{a_k}{t}= \lambda a_k
    {\deriv{y_{a_k}}{\lambda}}} = \frac{\lambda }{\pi} \left( a^{(1)}_k
      (a) + \frac{\lambda}{4 \pi} a^{(2)}_k (a) \right) .
  \end{cases}
\end{equation}

The unknown functions $J^{(i)}, a^{(i)}, X^{(i)\mu}$ are determined by
the equation
\begin{equation}
\label{eq:renorm-counter}
  T^{(i)} = - J^{(i)} \mathcal{L} + \deriv{\mathcal{L}}{a_k} a_k^{(i)} 
  + \deriv{\mathcal{L}}{X^{\mu}} X^{(i)\mu}.
\end{equation}
This corresponds to demanding the generalized quantum effective action
$\Gamma[X]$ to be finite order by order.

\subsection{Two-loop $\beta$ equations for a \textsc{wzw} model.}
\label{sec:two-loop-equations}

As we have already announced in Ch.~\ref{cha:wess-zumino-witten}, the
normalization for the \textsc{wz} term in a \textsc{wzw} action can be
fixed by an \textsc{rg}-flow calculation. This is precisely what we
will do in the following at two-loop order using the dimensional
regularization scheme outlined above. In this way we will find a new
apparent non-trivial solution that doesn't show up at first order (and
which will prove to be an artifact as we'll see in the following, by
using a \textsc{cft} description in Sec.~\ref{sec:cft-approach-1}).
Moreover we will see how the roles of the \textsc{ir} and \textsc{uv}
limits are interchanged between the compact and the non-compact case,
\emph{ie} how the same kind of deformation is relevant or irrelevant
depending on the compactness of the starting model.

Consider the following action
\begin{equation}
  S_{\lambda,H} = \frac{1}{2 \lambda } \int_\Sigma \di^2 z \:
  \left( g_{ab} + \h B_{ab}\right) J^a_\mu J^b_\nu \partial
  X^\mu \bar \partial X^\nu 
\end{equation}
where $J^a$ are the Maurer--Cartan one-forms for a group $G$ whose
algebra has structure constants $f_{abc}$, $g_{ab} = -1/(2 g^\ast)
\F{t}{as} \F{s}{bt}$ ($g^\ast $ is the dual Coxeter number) and
$B_{ab}$ is an antisymmetric matrix satisfying $ \di \left(B_{ab} J^a
  \land J^b\right) = 1/3! f_{abc} J^a \land J^b \land J^c$ as in
Sec.~\ref{sec:target-space-wzw}. Since the deformation (parameterized
by having $\h \neq 1$) doesn't affect the geometric part (but for the
overall normalization) we can still express the curvature in terms of
the Lie algebra structure constants.  In particular it is easy to
recover that the Riemann tensor is written as:
\begin{equation}
  R^a_{\phantom{a}bcd} = \frac{1}{4}\F{a}{be} \F{e}{cd}
\end{equation}
and the Ricci tensor is obtained by contracting:
\begin{equation}
  Ric_{ab} = \frac{g^\ast }{2} g_{ab}.
\end{equation}
as in Eqs.~\eqref{eq:group-curvatures}.

In order to write the beta equations as described above we need to
incorporate the \textsc{wz} term (or, more precisely, the Kalb--Ramond
field) into the geometry. The most natural approach is to consider
$H^a_{\phantom{a}bc}$ as a torsion and include it in the
connection~\cite{Braaten:1985is}.  We hence define:
\begin{equation}
  \label{eq:minus-connection}
  {\Gamma^-}^{a}_{\phantom{a}bc}  = {a
  \atopwithdelims\{\} b c } - \frac{1}{2} H^a_{\phantom{a}bc } .
\end{equation}
The covariant derivative of a one-form is then defined as:
\begin{equation}
  {\nabla^-}_a V_b = \partial_a V_b - {\Gamma^-}^c_{\phantom{c}ab} V_c =
  \nabla_a V_b + \frac{1}{2} H^c_{\phantom{c}ab } V_c
\end{equation}
where $\nabla_a$ is the covariant derivative with respect to the
Levi--Civita connection. Similarly we define the curvature:
\begin{equation}
  \comm{{\nabla^-}_a, {\nabla^-}_b} V_c = {R^-}_{c \phantom{d} a b}^{\phantom{c}d} V_d +
  H^d_{\phantom{d}ab} {\nabla^-}_d V_c 
\end{equation}
and it is straightforward to show that:
\begin{equation}
  {R^-}_{abcd} = R_{abcd} + \frac{1}{2} \nabla_c H_{abd} - \frac{1}{2} \nabla_d
  H_{abc} + \frac{1}{4} H_{fac} H^f_{\phantom{f}db} - \frac{1}{4} H_{fad}
  H^{f}_{\phantom{f}cb}. 
\end{equation}

Let us now specialize this general relation to our particular
deformation. Since $H_{abc} = \h f_{abc} $ it is immediate that
$\nabla_a H_{bcd} = 0$ and that the Jacobi identity holds.  We
then derive:
\begin{equation}
  {R^-}_{abcd} = \left( 1 - \h^2 \right)R_{abcd} ,
\end{equation}
whence in particular:
\begin{equation}
  {Ric^-}_{ab}  = \left( 1 - \h^2\right) Ric_{ab} = \left( 1 - \h^2 \right)
  \frac{g^\ast }{2} g_{ab}.
\end{equation}
The one-loop counterterm becomes:
\begin{equation}
  T^{(1)}_{ab} = \frac{g^\ast }{4} \left( 1 - \h^2 \right) g_{ab} .
\end{equation}

The evaluation of the two-loop counterterm is lengthy but straightforward
once $R_{abcd}$ is written in terms of the structure constants. The result
is:
\begin{equation}
   T^{(2)}_{ab} = \frac{{g^\ast}^2 }{8} \left( 1-\h^2\right)
   \left( 1-3 \h^2\right) g_{ab} .
\end{equation}

Substituting these expressions in Eq.~(\ref{eq:renorm-counter}) (and
using the fact that $g_{ab}$ and $\epsilon_{ab}$ are orthogonal) one
sees that they become identities for the following choice of
parameters:
\begin{align}
  \begin{cases}
    J^{(1)} = - \frac{g^\ast }{4} \left( 1 - \h^2\right) \\
    a_\h^{(1)} = - \frac{g^\ast }{4} \h \left( 1 - \h^2\right)
  \end{cases} &&
  \begin{cases}
    J^{(2)} = - \frac{{g^\ast}^2 }{8}  \left( 1 - \h^2 \right)
    \left( 1 - 3 \h^2 \right) \\
    a_\h^{(2)} = - \frac{{g^\ast}^2 }{8} \h \left( 1 - \h^2 \right)
    \left( 1 - 3 \h^2 \right)
  \end{cases}
\end{align}
corresponding to the following beta equations:
\begin{align}
  \beta_\h &= - \frac{1}{4 \pi } \lambda^\ast \h \left( 1 - \h^2 \right)
  \left( 1 +
    \frac{1 }{8 \pi } \lambda^\ast \left( 1-3 \h^2 \right)\right) \\
  \beta_{\lambda^\ast} &= - \frac{1 }{4 \pi } {\lambda^\ast}^2 \left( 1 - \h^2
  \right) \left( 1 + \frac{1 }{8 \pi } \lambda^\ast \left( 1-3 \h^2
    \right)\right)
\end{align}
where $\lambda^\ast = g^\ast \lambda$ is the effective coupling
constant (this is precisely the fixed parameter in a 't~Hooft limit
since for a $SU(N) $ group $g^\ast = N$). The difference between a
compact and a non-compact group lies in the sign of the dual Coxeter
number that is respectively positive/negative. In both cases we remark
that $\h / \lambda^\ast$ remains constant, which is a nice check of
our construction, since in the notation of
Ch.~\ref{cha:wess-zumino-witten} this is just the level of the model
that, in the compact case, is quantized and hence is not expected to
receive any perturbative correction. On the other hand non
perturbative effects do eventually lead to the $k \to k+g^\ast $ shift
which is the reason for the two-loop behaviour of the flow. 

\begin{figure}
  \includegraphics[width=.8\linewidth]{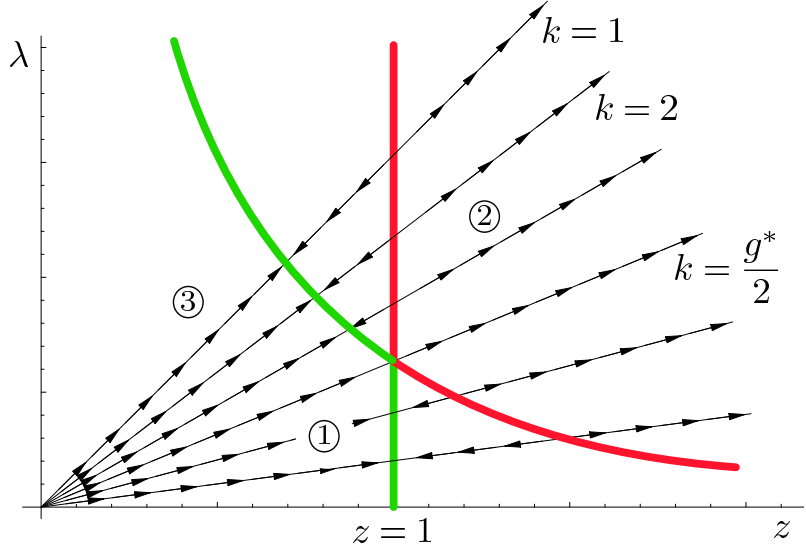}
  \caption{Two-loop \textsc{rg}-flow diagram for compact groups.}
  \label{fig:flow-lines-2l-compact}
\end{figure}

Let us analyze the flow in detail:
\begin{itemize}
\item The flow diagram for the compact case is drawn in
  Fig.\ref{fig:flow-lines-2l-compact} where we see the presence of
  three phases:
  \begin{itemize}
  \item region {1} is the basin of attraction for the \textsc{wzw} model
    ($z = 1$);
  \item the points in region {2} describe systems that flow towards
    asymptotic freedom;
  \item region {3} seems to be the basin of attraction for a different theory,
    always with a group manifold geometry but with a
    differently-normalized \textsc{wz} term.
  \end{itemize}
  only a discrete set of trajectories is allowed and, in particular,
  region {3} -- separated from region {1} by the line
  $\lambda^\ast = 4 \pi \h$ -- is only accessible for levels $k <
  g^\ast /2 $.
\item The flow diagram for the non\nb-compact case is drawn in Fig...;
  again we see three different phases:
  \begin{itemize}
  \item region {1} describe theories flowing to asymptotic freedom;
  \item region {2} looks like the basin of attraction for the non\nb-trivial
    solution with the group manifold metric and a new normalization
    topological term;
  \item region {3} describe theories flowing to a strong coupling
    regime.
  \end{itemize}
  In particular it is interesting to remark that the roles of the
  \textsc{uv} and \textsc{ir} are somehow inverted. The
  \textsc{wzw} model appears as a \textsc{uv} fixed point and thus an
  unstable solution from the point of view of dynamical systems.
\end{itemize}

\subsection{Renormalization group-flow in squashed compact groups}

The models that we have presented in Ch.~\ref{cha:deformations} are
conformal; for this reason we expect to find them as fixed points in
an \textsc{rg} flow. To verify this claim let us introduce a
two-parameter family of $\sigma$~models generalizing the exact
backgrounds; a possible choice consists in adding an extra magnetic
field on the top of the one responsible for the squashing, but now
coming from a higher-dimensional right sector. Explicitly
\begin{equation}
  \label{eq:general-deformed}
  \begin{cases}
    \di s^2 = \displaystyle{\sum_{\mu \in G/T}} J^\mu J^\mu + \left( 1 - \h^2 \right)
    \displaystyle{\sum_{a \in T}} J^a J^a ,\\
    H_{[3]} = \frac{\hb}{2\h} f_{\mu \nu\rho} J^\mu \land J^\nu \land
    J^\rho & \mu \in G/T, \\
    F^a = \frac{\h + \hb}{2} \sqrt{\frac{k}{k_g}} \F{a}{\mu \nu} J^\mu \land
    J^\nu & \text{$\mu \in G/T$, $a \in T$}, \\
    \bar F^a = \frac{\h - \hb}{2} \sqrt{\frac{k}{k_g}} \F{a}{\mu \nu} J^\mu \land J^\nu & \text{$\mu \in G/T$, $a \in T$}
  \end{cases}
\end{equation}
and in particular for $SU(2)$:
\begin{equation}
\label{eq:2par-su2}
  \begin{cases}
    \di s^2 = \di \theta^2 + \di \psi^2 + \di \phi^2
+ \cos \theta \di \psi \di \phi - \h^2 \left( \di \psi + \cos \theta \di \phi \right)^2\ ,\\
    B = \frac{\hb}{\h} \cos \theta \di \psi \land \di \phi\ ,\\
    A = \left( \h + \hb \right) \left( \di \psi + \cos \theta \di \phi \right)\ ,\\
    \bar A = \left( \h - \hb \right) \left( \di \psi + \cos \theta \di \phi \right)\ ,
  \end{cases}
\end{equation}
where $\hb $ is a new parameter, describing the deviation from the
conformal point. It is clear that the above background reduces to the
one we are used to in the $\hb \to \h $ limit.  In particular we see
that the metric is unchanged, the Kalb--Ramond field has a different
normalization and a new field $\bar A$ appears.  This configuration
can be described in a different way: the geometry of a squashed sphere
supports two covariantly constant magnetic fields with charge $Q = \h
+ \hb$ and $\bar Q =\h - \hb$; the \textsc{rg} flow will describe the
evolution of these two charges from a generic $\left( Q, \bar Q
\right)$ to $\left( 2 \h, 0 \right)$, while the sum $Q + \bar Q = 2
\h$ remains constant.  In this sense the phenomenon can be interpreted
as a charge transmutation of $\bar Q $ into $Q$.  The conservation of
the total charge is in fact a consequence of having chosen a
perturbation that keeps the metric and only changes the antisymmetric
part of the background.

We can also see the background in Eq.\eqref{eq:general-deformed} from
a higher dimensional perspective where only the metric and the
Kalb-Ramond field are switched on. Pictorially:
\begin{align}
  \label{eq:4d-2par-su2}
  g = \left(
    \begin{tabular}{ccc|c}
      & & & \\
      & $g_{\textsc{wzw}}$ & & $\h J_a$  \\
      & & &  \\ \hline
      & $\h J_a$ & & 1
    \end{tabular} \right) &&
  B = \left(
    \begin{tabular}{ccc|c}
      & & & \\
      & $\frac{\hb}{\h} B_{\textsc{wzw}}$ & & $\hb J_a$  \\
      & & &  \\ \hline
      & $-\hb J_a$ & & 0
    \end{tabular} \right)
\end{align}
where $g_{\textsc{wzw}}$ and $B_{\textsc{wzw}}$ are the usual metric
and Kalb--Ramond fields for the \textsc{wzw} model on the group $G$.
More explicitly in the $SU(2)$ case:
\begin{align}
  g = \begin{pmatrix}
    1 & 0 & 0 & 0 \\
    0 & 1 & \cos \theta & \h \\
    0 & \cos \theta & 1 & \h \cos \theta \\
    0 & \h & \h \cos \theta & 1
  \end{pmatrix} &&
  B = \begin{pmatrix}
    0 & 0 & 0 & 0 \\
    0 & 0 & \frac{\hb}{\h} \cos \theta & \hb \\
    0 & -\frac{\hb}{\h} \cos \theta & 0 & \hb \cos \theta \\
    0 & -\hb & -\hb \cos \theta & 0
  \end{pmatrix}
\end{align}
where the fourth entry represents the bosonized internal current.  In
particular this clarifies the stated right-sector origin for the new
gauge field $\bar A$. This higher dimensional formalism is the one we
will use in the following \textsc{rg} analysis.

\bigskip 

The beta-equations at two-loop order in the expansion in powers of the
overall coupling constant $\lambda$ and the field redefinitions for
the internal coordinates $X^i$ turn out to be:
\begin{equation}
\label{eq:beta-compact-2loops}
  \begin{cases}
    \beta_{\lambda^\ast} =
 \Deriv{\lambda^\ast}{t} = - \frac{\lambda^{\ast 2}}{4 \pi} \left( 1-
      \frac{\hb^2}{\h^2} \right) \left( 1+ \frac{\lambda^\ast}{8 \pi}
\left( 1 - 3 \frac{\hb^2}{\h^2} \right) \right), \\
    \beta_{\h} = \Deriv{\h}{t} = \frac{\lambda^\ast \h}{8 \pi} \left( 1- \h^2 \right)
    \left( 1- \frac{\hb^2}{\h^2} \right)
\left( 1+ \frac{\lambda^\ast}{8 \pi} \left( 1 - 3 \frac{\hb^2}{\h^2} \right) \right) , \\
    \beta_{\hb} = \Deriv{\hb}{t} = - \frac{\lambda^\ast \hb}{8 \pi} \left( 1 + \h^2
    \right) \left( 1- \frac{\hb^2}{\h^2} \right)
\left( 1+ \frac{\lambda^\ast}{8 \pi} \left( 1 - 3 \frac{\hb^2}{\h^2} \right) \right) , \\
    X^i = X^i - \frac{\lambda^\ast}{16}
\left( 1 - \h^2 \right) \left( 1- 4 \frac{\hb^2}{\h^2} + 3 \frac{\hb^4}{\h^4} \right) ,
  \end{cases}
\end{equation}
where $\lambda^\ast = \lambda g^\ast$, $g^\ast$ being the dual Coxeter
number, is the effective coupling constant ($\lambda^\ast = N \lambda
$ for $G = SU(N)$). The contributions at one- and two-loop order are
clearly separated. In the following we will concentrate on the
one-loop part and we will comment on the two-loop result later.
Let us then consider the system:
\begin{equation}
\label{eq:compact-beta}
  \begin{cases}
    \beta_{\lambda^\ast} = \Deriv{\lambda^\ast}{t} = - \frac{\lambda^{\ast 2}}{4 \pi} \left( 1-
      \frac{\hb^2}{\h^2} \right) , \\
    \beta_{\h} = \Deriv{\h}{t} = \frac{\lambda^\ast \h}{8 \pi} \left( 1- \h^2 \right)
    \left( 1- \frac{\hb^2}{\h^2} \right), \\
    \beta_{\hb} = \Deriv{\hb}{t} = - \frac{\lambda^\ast \hb}{8 \pi} \left( 1 + \h^2
    \right) \left( 1- \frac{\hb^2}{\h^2} \right)\ .
  \end{cases}
\end{equation}
This can be integrated by introducing the parameter $z = \hb / \h$ which makes one
of the equations redundant. The other two become:
\begin{equation}
\label{eq:compact-red-system}
  \begin{cases}
    \dot \lambda^\ast = - \frac{\lambda^{\ast 2}}{4 \pi} ( 1 - z^2 ) ,\\
    \dot z = - \frac{\lambda^\ast z}{4 \pi} ( 1 - z^2 )\ .
  \end{cases}
\end{equation}
By inspection one easily sees that $\dot \lambda / \lambda = \dot z /
z $, implying $ \lambda (t) = C z (t)$, where $C$ is a constant.
This was to be expected since $C$ is proportional to the normalization
of the topological \textsc{wz} term. Since we are dealing with a
compact group it turns out that $C$ is, as in \cite{Witten:1983ar},
quantized with:
\begin{equation}
  C_k = \frac{2 \pi}{k},\ \ k \in \setN \ .
\end{equation}
Now it's immediate to separate the system and find that $z (t)$ is
defined as the solution to the implicit equation:
\begin{equation}
  - \frac{t}{2 k} = \frac{1}{z_0} - \frac{1}{z(t)} + \log \left[ \frac{
      \left( z(t) + 1 \right) \left( z_0 - 1 \right) }{\left( z(t) - 1
      \right) \left( z_0 + 1 \right)} \right]
\end{equation}
with the initial condition $z(0) = z_0$. A similar expression was found in
\cite{Braaten:1985is,Witten:1983ar}. The reason for this is, as pointed out
previously~\cite{Kiritsis:1995iu}, that the conformal model ($\hb = \h$) in its
higher-dimensional representation (the one in Eq.~(\ref{eq:4d-2par-su2}))
coincides with a $G \times H$ \textsc{wzw} model after a suitable local
field redefinition.

\begin{figure}
  \subfigure[$(z, \lambda) $ plane]{\includegraphics[width=7cm]{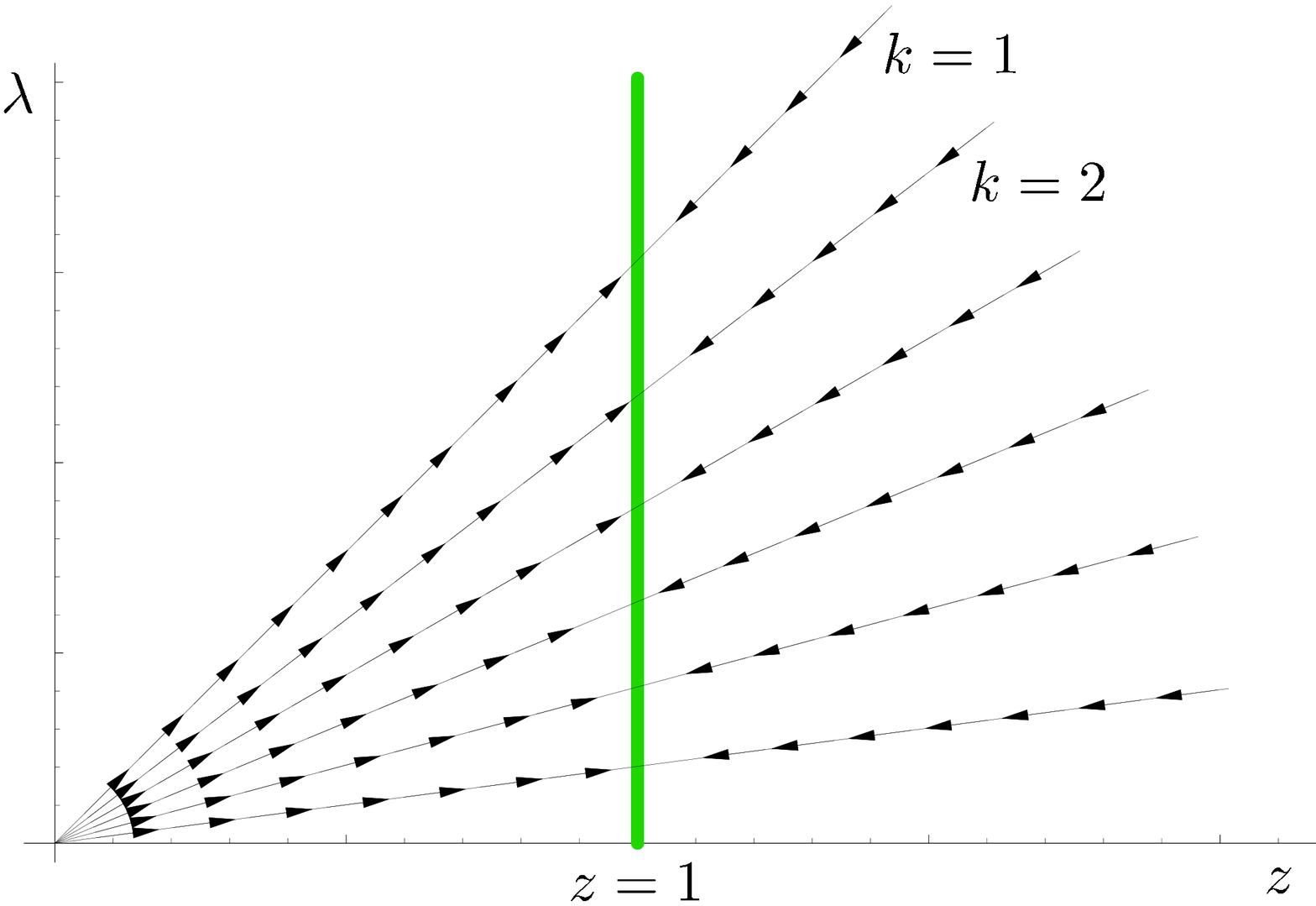}}
  \subfigure[$(\h, \hb)$ plane]{\includegraphics[width=7cm]{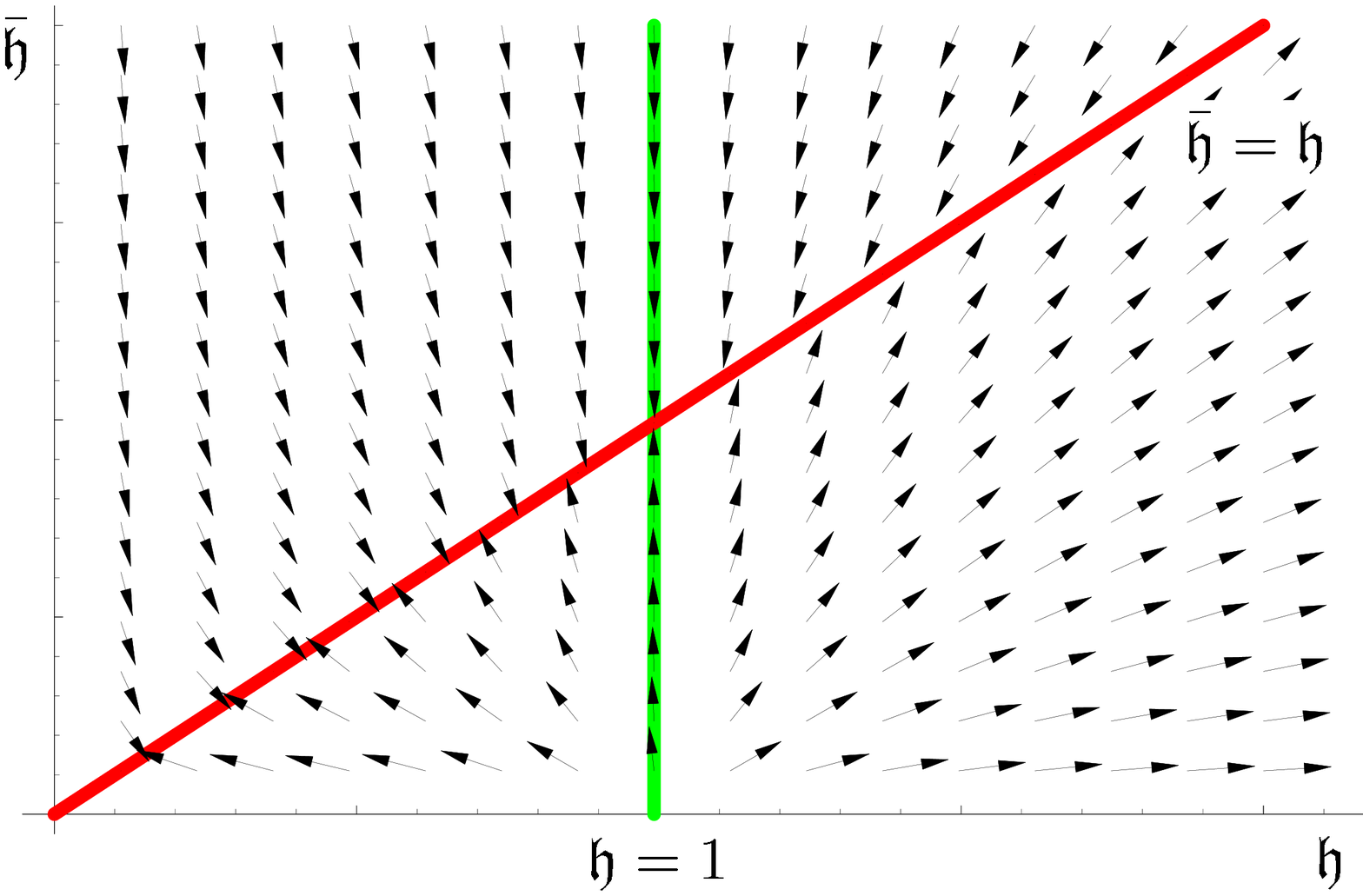}}
  \caption{Flow lines for the deformed (non-conformal) squashed
    \textsc{wzw} model in (a) the $(z, \lambda )$ and (b) the $(\h, \hb)$
    planes. The arrows point in the negative $t$ direction, \emph{i.e.}
    towards the infrared; in (a) we see how the squashed \textsc{wzw} model
    $z=1$ appears as an \textsc{ir} fixed point, in (b) how perturbing the
    conformal $\hb = \h $ model by increasing $\hb$ leads to a a new fixed
    point corresponding to a value of $\h$ closer to $1$.}
  \label{fig:flow-lines-su2}
\end{figure}

As it is usually the case in the study of non-linear dynamics, a
better understanding of the solution is obtained by drawing the
\textsc{rg} flow. In a $\left( z, \lambda^\ast \right)$ plane, the
trajectories are straight lines through the origin and only a discrete
set of them are allowed.  Moreover the line $z = 1 $ is an \textsc{ir}
fixed-point locus. This situation is sketched in
Fig.~\ref{fig:flow-lines-su2}(a). Just as expected the $z = \hb / \h =
1 $ point, corresponding to the initial exact model described in
Ch.~\ref{cha:deformations}, is an \textsc{ir} fixed point for the
\textsc{rg} flow.

Further insights can  be gained if
we substitute  the condition $\lambda^\ast  = C_k  \hb /  \h$ into the
system (\ref{eq:compact-beta}) thus getting:
\begin{equation}
  \label{eq:compact-HHb}
  \begin{cases}
    \Deriv{\h}{t} =  \frac{\hb}{4 k} \left( 1- \h^2 \right)
    \left( 1 - \frac{\hb^2}{\h^2} \right) , \\
    \Deriv{\hb}{t} = - \frac{\hb^2}{4 k \h} \left( 1 + \h^2 \right)
    \left( 1 - \frac{\hb^2}{\h^2} \right) .
  \end{cases}
\end{equation}
The flow diagram for this system in the $\left( \h, \hb \right)$
plane, Fig.~\ref{fig:flow-lines-su2}(b), shows how the system relaxes
to equilibrium after a perturbation. In particular we can see how
increasing $\hb$ leads to a a new fixed point corresponding to a value
of $\h$ closer to $1$.

We would like to pause for a moment and put the above results in
perspective. Consider for simplicity the $SU(2)$ case: the
target-space of the sigma-model under consideration is a squashed
three-sphere with two different magnetic fields. Along the flow, a
transmutation of the two magnetic charges occurs: the system is driven
to a point where one of the magnetic charges vanishes. This fixed
point is an ordinary squashed-\textsc{wzw} (of the type studied in
Ch.~\ref{cha:deformations}), that supports a single magnetic charge.

As we pointed out in Ch.~\ref{cha:deformations}, in the
squashed-\textsc{wzw}, the magnetic field is bounded by a critical
value, $\h = 1$. As long as $\h \leq 1$, the geometry is a genuine
squashed three-sphere. For $\h > 1$, the signature becomes Lorentzian
and the geometry exhibits closed time-like curves.  Although of
limited physical interest, such a background can be used as a
laboratory for investigating the fate of chronological pathologies
along the lines described above.  In the case under consideration and
under the perturbation we are considering the model shows a symmetry
between the $\h >1 $ and $\h<1$ regions. The presence of closed
time-like curves doesn't seem to effect the stability (note that
regions with different signatures are disconnected, \emph{i.e.}  the
signature of the metric is preserved under the \textsc{rg} flow).  It
is clear however that these results are preliminary. To get a more
reliable picture for the fate of closed time-like curves, one should
repeat the above analysis in a wider parameter space, where other
\textsc{rg} motions might appear and deliver a more refined stability
landscape.

\bigskip

A final remark concerns the fact that we find the same \textsc{rg}
flow behaviour as for a compact (non-squashed) group. We have already
made extensive use of the fact that formally the squashed $SU(2)$
behaves like a $SU(2) \times U(1)$ \textsc{wzw} model, in particular
in Sec.~\ref{sec:no-renorm-theor} where this was at the root of the
no-renormalization theorem. In some sense, then, the present
calculation is just a perturbative confirmation of that statement.

\subsection{Renormalization group-flow in squashed anti de Sitter}
\label{sec:non-compact}

As we've already discussed in Sec.~\ref{sec:deformed-sl2}, sigma
models based on non-compact group offer richer (\emph{i.e.}  more
complex) phase diagrams than the compact ones. In our particular
models this is because the possible choices for a Cartan torus are not
pairwise conjugated by inner automorphisms and this is why different
choices correspond to inequivalent backgrounds, exhibiting different
physical properties. If we concentrate our attention on the $SL (2,
\setR )$ \textsc{wzw} model (that is the only non-compact case with
just one time direction), we see that the three possible choices for
the Cartan generator (elliptic, parabolic, hyperbolic) respectively
lead to the exact backgrounds we introduced in
Sec.~\ref{sec:deformed-sl2} and we report here for convenience:
\begin{align}
  &\begin{cases}
    \di s^2 = \di \rho^2 - \di t^2 + \di \phi^2 - 2 \sinh \rho \di t
    \di \phi - \h^2 \left( \di t + \sinh \rho \di \phi \right)^2 ,\\
    B = \sinh \rho \di t \land \di \phi ,\\
    A = 2 \h \left( \di t + \sinh \rho \di \phi \right) .
  \end{cases} \\
  &\begin{cases}
    \di s^2 = \frac{\di u^2}{u^2} + \frac{\di x^+ \di x^-}{u^2} - \h^2
    \frac{\di x^+ \di x^+}{u^4} , \\
    B = \frac{\di x^+ \land \di x^-}{u^2} ,\\
    A = 2 \h \frac{\di x^+}{u^2}.
  \end{cases}\\
  &\begin{cases}
    \di s^2 = \di r^2 + \di x^2 - \di \tau^2 + 2 \sinh r \di x \di
    \tau - \h^2 \left( \di x + \sinh r \di \tau \right)^2 \\
    B = \sinh r \di x \land \di \tau ,\\
    A = 2 \h \left( \di x + \sinh r \di \tau \right).
  \end{cases}
\end{align}

Since these solutions are exact \textsc{cft} backgrounds, we expect
them to appear as fixed points for an \textsc{rg} flow, like the
compact configuration described in the previous section. As we will
see in the following this is actually the case, but with a difference
regarding the role of the \textsc{uv} and \textsc{ir} which is proper
to non\nb-compact groups (as explained in
Sec.~\ref{sec:two-loop-equations}). Using the same technique as above,
the first step consists in generalizing the three backgrounds by
introducing the following three families of low energy configurations:
\begin{align}
  &\begin{cases} \di s^2 = \di \rho^2 - \di t^2 + \di \phi^2 - 2 \sinh
    \rho \di t \di \phi - \h^2 \left( \di t + \sinh \rho \di \phi
    \right)^2  \\
    B = \frac{\hb}{\h} \sinh \rho \di t \land \di \phi \\
    A = \left( \h + \hb \right) \left( \di t + \sinh \rho \di \phi \right) \\
    \bar A = \left( \h - \hb \right) \left( \di t + \sinh \rho \di
      \phi \right)
  \end{cases} \\
  &\begin{cases} \di s^2 = \frac{\di u^2}{u^2} + \frac{\di x^+ \di
      x^-}{u^2} - \h^2 \frac{\di x^+ \di x^+}{u^4}  \\
    B = \frac{\hb}{\h} \frac{\di x^+ \land \di x^-}{u^2} \\
    A = \left( \h + \hb \right) \frac{\di x^+}{u^2} \\
    \bar A = \left( \h - \hb \right) \frac{\di x^+}{u^2}
  \end{cases} \\
  &\begin{cases} \di s^2 = \di r^2 + \di x^2 - \di \tau^2 + 2 \sinh r
    \di x \di \tau - \h^2 \left( \di x + \sinh r \di \tau \right)^2 \\
    B = \frac{\hb}{\h} \sinh r \di x \land \di \tau \\
    A = \left( \h + \hb \right) \left( \di x + \sinh r \di \tau \right) \\
    \bar A = \left( \h - \hb \right)\left( \di x + \sinh r \di \tau
    \right)
  \end{cases}
\end{align}
The guiding principle remains the same, \emph{i.e.} keep the same
geometry, rescale the \textsc{kr} field and introduce a new
electromagnetic field, coming (in a four-dimensional perspective) from
the right\nb-moving sector. Again we will observe the same
charge-transmutation effect as before, this time in terms of charge
density (or charge at infinity).

\begin{figure}
  \centering
  \includegraphics[width=.8\linewidth]{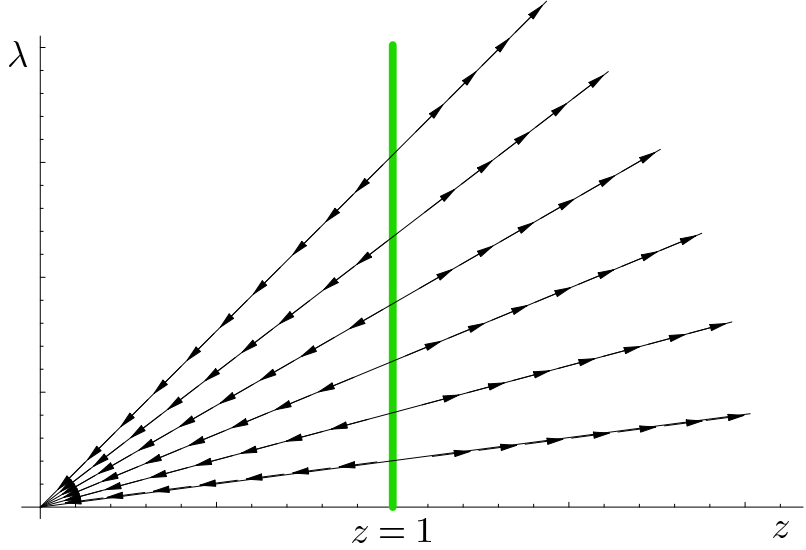}
  \caption{Flow diagram for the system in Eq.~(\ref{eq:ad3-one-loop}).
    The arrows point from the \textsc{uv} to the \textsc{ir} and $z=1$
    appears as a \textsc{uv} stable solution locus.}
  \label{fig:flow-lines-sl2}  
\end{figure}

The backgrounds above can be equivalently described in four dimensions
by a metric and a \textsc{kr} field as follows:
\begin{small}
  \begin{align}
    \hspace{-1cm}g &=
    \begin{pmatrix}
      1 & 0 & 0 & 0 \\
      0 & -1 & -\sinh \rho & \h \\
      0 & -\sinh \rho & 1 & \h \sinh \rho \\
      0 & \h & \h \sinh \rho & 1
    \end{pmatrix} &
    B &= \begin{pmatrix}
      0 & 0 & 0 & 0 \\
      0 & 0 & \frac{\hb}{\h} \sinh \rho & \hb \\
      0 & -\frac{\hb}{\h} \sinh \rho & 0 & \hb \sinh \rho \\
      0 & -\hb & -\hb \sinh \rho & 0
    \end{pmatrix} \\
    \hspace{-1cm}g & = \begin{pmatrix}
      \frac{1}{u^2} & 0 & 0 & 0 \\
      0 & 0 & \frac{1}{2u^2} & \frac{\h}{u^2} \\
      0 & -\frac{1}{2u^2} & 0 & 0 \\
      0 & \frac{\h}{u^2} & 0 & 1
    \end{pmatrix} &
    B &= \begin{pmatrix}
      0 & 0 & 0 & 0 \\
      0 & 0 & \frac{\hb}{\h} \frac{1}{2u^2} & \frac{\hb}{u^2} \\
      0 & \frac{\hb}{\h} \frac{1}{2u^2} & 0 & 0 \\
      0 & -\frac{\hb}{u^2} & 0 & 0
    \end{pmatrix} \\
    \hspace{-1cm}g &= \begin{pmatrix}
      1 & 0 & 0 & 0 \\
      0 & 1 & \sinh r & \h \\
      0 & \sinh r & -1 & \h \sinh r \\
      0 & \h & \h \sinh r & 1
    \end{pmatrix} &
    B &= \begin{pmatrix}
      0 & 0 & 0 & 0 \\
      0 & 0 & \frac{\hb}{\h} \sinh r & \hb \\
      0 & -\frac{\hb}{\h} \sinh r & 0 & \hb \sinh r \\
      0 & -\hb & -\hb \sinh r & 0
    \end{pmatrix}
  \end{align}  
\end{small}
We must now evaluate the $R^-$ tensor (\emph{i.e.}  the Ricci tensor
with respect to the connection $\Gamma^- = \Gamma + 1/2 H$) and read
the counterterms in a dimensional regularization scheme as described
in Eq.~\eqref{eq:bare-stuff}:
\begin{small}
  \begin{align}
    \hspace{-1cm}\begin{cases}
      J^{(1)} = \frac{1}{4} \left( 1 - \frac{\hb^2}{\h^2} \right) ,\\
      a_{\h}^{(1)} = - \frac{1+\h^2}{8} \h \left( 1 - \frac{\hb^2}{\h^2} \right) ,\\
      a_{\hb}^{(1)} = \frac{1 - \h^2}{8} \hb \left( 1 - \frac{\hb^2}{\h^2} \right), \\
      X^{(1)}_X = \frac{ 1 + \h^2 }{8} \left( 1 - \frac{\hb^2}{\h^2} \right) X .
    \end{cases} 
    \begin{cases}
      J^{(1)} = \frac{1}{4} \left( 1 - \frac{\hb^2}{\h^2} \right) ,\\
      a_{\h}^{(1)} = - \frac{1}{8} \h \left( 1 - \frac{\hb^2}{\h^2} \right) ,\\
      a_{\hb}^{(1)} = \frac{1}{8} \hb \left( 1 - \frac{\hb^2}{\h^2} \right) ,\\
      X^{(1)}_X = \frac{ 1 }{8} \left( 1 - \frac{\hb^2}{\h^2} \right) X .
    \end{cases} &&
    \begin{cases}
      J^{(1)} = \frac{1}{4} \left( 1 - \frac{\hb^2}{\h^2} \right) ,\\
      a_{\h}^{(1)} = - \frac{1 - \h^2}{8} \h \left( 1 - \frac{\hb^2}{\h^2} \right) ,\\
      a_{\hb}^{(1)} = \frac{1 + \h^2}{8} \hb \left( 1 - \frac{\hb^2}{\h^2} \right) ,\\
      X^{(1)}_X = \frac{ 1 - \h^2 }{8} \left( 1 - \frac{\hb^2}{\h^2} \right) X .
    \end{cases}
  \end{align}
\end{small}
The analogies among the three configurations are clear, but become
striking when we introduce the parameter $z = \hb / \h $ and all 
three $\beta$\nb-functions systems all reduce to the following:
\begin{equation}
  \label{eq:ad3-one-loop}
  \begin{cases}
    \dot \lambda = \frac{\lambda^2}{4 \pi} ( 1 - z^2 ) ,\\
    \dot z = \frac{\lambda z}{4 \pi} ( 1 - z^2 ) .
  \end{cases}
\end{equation}
This is (up to a sign) the same system we found in the compact case
and it is hence immediate to write the solution
\begin{gather}
  \lambda (t) = C z(t) \\
  \frac{C t}{4 \pi } = \frac{1}{z_0} - \frac{1}{z (t)} + \log \left[
    \frac{ \left( z(t) + 1 \right) \left( z_0 - 1 \right) }{\left( z(t) - 1
      \right) \left( z_0 + 1 \right)} \right] .
\end{gather}
Although, as expected, $z = 1$ is a fixed point (corresponding to the
conformal points) some differences are important. First of all the
background is non\nb-compact, so $C$ is not quantized and, although
the flow trajectories are still straight lines through the origin, the
angular parameter is now arbitrary. The other difference is that $z =
1$ is a fixed point, but it doesn't correspond to a \textsc{ir} stable
configuration but to a \textsc{uv} stable one. This is precisely the
same behaviour that one encounters for non\nb-compact \textsc{wzw}
models when varying the normalization of the \textsc{wz} term (as in
Sec.~\ref{sec:two-loop-equations}). Again the flow diagram is the same
as for the original $SL(2,\setR)$ group and is summarized in
Fig.~\ref{fig:flow-lines-sl2}.

\section{The CFT approach}
\label{sec:cft-approach-1}

In order to make contact with genuine \textsc{cft} techniques, we must
identify the relevant operators which are responsible for the $(\h,
\hb)$ deformation of the $G \times H$ original \textsc{wzw} model ($H
= U(1)^{\rank G}$). At lowest approximation, all we need is their
conformal dimensions in the unperturbed theory.

Following~\cite{Zamolodchikov:1986gt}, let $S_0$ be the unperturbed
(conformal) action and $ \mathcal{O}_i$ the relevant operators of
conformal dimension $\Delta_i = 1 - \epsilon_i$. Consider the
perturbed model,
\begin{equation}
  S = S_0 + g^i \mathcal{O}_i.
\end{equation}
The tree-level beta-functions read:
\begin{equation}
  \label{eq:cft-beta}
  \beta^i (g) = - \epsilon_i g^i ,
\end{equation}
where $g^i$ is supposed to be small, for the perturbative
expansion of $\beta^i$ to hold\footnote{One should be very careful in the
  choice of signs in these formulae. In~\cite{Zamolodchikov:1986gt} the time
  variable, in fact, describes the evolution of the system towards the infrared
  and as such it is opposite with respect to the $t = \log \mu $ convention
  that we used in the previous section (as in \cite{Witten:1983ar}).}.

The $G \times H$ primary operator we need can be written as follows:
\begin{equation}
  \mathcal{O} = \sum_{\ssc{a,b}} \braket{ t^{\ssc{a}} g t^{\ssc{b}} g^{-1}} \braket{ t^{\ssc{a}} \partial g g^{-1}  } \braket{ t^{\ssc{b}} g^{-1} \bar \partial g} =  \sum_{\ssc{a,b}} \Phi^{\ssc{ab}}  J^{\ssc{a}} \bar J^{\ssc{b}} ,
\end{equation}
where $\Phi^{\ssc{ab}}$ is a primary field transforming in the adjoint
representation of the left and right groups $G$. As such, the total
conformal dimensions (as we've seen in Sec.~\ref{sec:cft-approach})
are
\begin{equation}
  \Delta = \bar \Delta = 1 + \frac{g^\ast}{g^\ast + k} ,
\end{equation}
where $g^\ast$ is the dual Coxeter number and as such the operator is
irrelevant (in the infrared).

Specializing this general construction to our case we find that the
action for the fields in Eq.~\eqref{eq:KK-action} is:
\mathindent=0em
\begin{equation}
  \label{eq:cft-action}
  S = \frac{k}{4 \pi} \left\{ S_0 + \left( \frac{\h}{\hb} - 1\right)  \sum_{\ssc{a,b}} \Phi^{\ssc{ab}} J^{\ssc{a}} \bar J^{\ssc{b}} + \frac{\h}{\hb} \left( \h + \hb \right) \sum_i J^{a_i} \bar J^i +  \frac{\h}{\hb} \left( \h - \hb \right) \sum_{i,\ssc{a}} J^i \Phi^{a_i \ssc{a}} \bar J^{\ssc{a}}  \right\}  .
\end{equation}
\mathindent=\oldindent
where $\ssc{a}$ runs over all currents, $i$ over the internal currents
(in $H$) and $J^{a_i}$ is the \textsc{wzw} current of the Cartan
subalgebra of $G$ coupled to the internal $\bar J^i$.  The extra terms
can be interpreted as relevant combinations of operators in the $G \times
H$ model. The beta-functions are thus computed following
Eq.~(\ref{eq:cft-beta}):
\begin{equation} \label{eq:hhb-beta}
  \begin{cases}
    \left. \Deriv{}{t} \log \left[ \left( \frac{\h}{\hb} -1 \right) \right]
    \right|_{\hb = \h} = \frac{ g^\ast}{g^\ast + k } = \frac{g^\ast}{k } - 
    \frac{g^{\ast 2}}{k^2} + \mathcal{O} \left(\frac{1}{k^3}\right), \\
    \left. \Deriv{}{t} \log \left[ \frac{\h}{\hb} \left( \h + \hb
        \right) \right] \right|_{\hb = \h} =  \mathcal{O} \left(\frac{1}{k^3}\right) , \\
    \left. \Deriv{}{t} \log \left[ \frac{\h}{\hb} \left( \h - \hb \right)
      \right] \right|_{\hb = \h}=  \frac{ g^\ast}{g^\ast + k } =
    \frac{g^\ast}{k } -  \frac{g^{\ast 2}}{k^2} + \mathcal{O}
    \left(\frac{1}{k^3}\right).
  \end{cases}
\end{equation}

Equations~(\ref{eq:hhb-beta}) agree with the results of the
field-theoretical approach (up to the overall normalization), at least
in the regime where~(\ref{eq:hhb-beta}) are valid, namely for small
$\h$ and $\hb$ perturbations. But there's more: as pointed out before
the conformal model ($\hb = \h$) is exact because it coincides with a
$G \times H$ \textsc{wzw} model after a suitable field redefinition
for any value of $\h$.  As a consequence the equations remain valid
for any finite $\h$. This is reassuring both for the validity of the
geometrical approach\footnote{There is no doubt on the method
  itself. It could simply fail to describe the desired phenomenon due
  to an inappropriate ansatz for the off-criticality excursion in
  parameter space.} and for the conclusions on the stability picture
of the models under consideration.

The extra information that we obtain from this calculation is about
the interpretation for the two-loop $\beta$-function we described
in the previous section. In fact it is now clear that with the
target-space approach we just describe the Taylor expansion of the
tree-level \textsc{cft} result:
\begin{equation}
  \frac{g^\ast}{g^\ast + k } = \frac{g^\ast}{k} - \frac{g^{\ast 2}}{k^2} + \mathcal{O} \left(\frac{1}{k^3} \right).
\end{equation}
This is not surprising since the would-be non-trivial fixed point of
the two-loop expansion lay out of the validity range for our
approximation. If we really want to go beyond the large $k$ limit, we
need to push the analysis from this, \textsc{cft}, side. 

\marginlabel{Target space \emph{vs} \textsc{cft} renormalization}From
the target space view point, the renormalization approach remains
valid in the large $k$ limit for any value of $\h / \hb$. This enables
us to use Eq.~\eqref{eq:beta-compact-2loops} and push (for $k \to
\infty$) Eq.~\eqref{eq:hhb-beta} at least to the next leading order in
$\left( 1 - \h / \hb \right)$ so to get \mathindent=0em
\begin{equation} 
  \label{eq:hhb-beta1}
  \left. \Deriv{}{t} \left( \frac{\h}{\hb} -1 \right)
  \right|_{\hb = \h} = \left(\frac{g^\ast}{k } -
    \frac{g^{\ast 2}}{k^2}\right)\left(\frac{\h }{\hb } - 1 \right)+
  \frac{1}{2} \left(- \frac{g^\ast }{k} + 7 \frac{{g^\ast}^2}{k^2} \right)\left(\frac{\h}{\hb} - 1\right)^2
  + \ldots 
\end{equation}
\mathindent=\oldindent
that obviously agrees to first order in the coupling $\left(\h / \hb-1
\right)$ with the expression above.

The extra information that we obtain from this calculation is about
the interpretation for the two-loop beta-function we described in the
previous section. The one-loop corrections to~(\ref{eq:cft-beta}) are
of the form $ C_{ijk} \, g^i \, g^j$, where $ C_{ijk}$ are related to
the three-point function of the unperturbed theory
\cite{Zamolodchikov:1986gt}.  This coefficient is a measure of the
dimension of the operator $\mathcal{O}_i$ in the theory perturbed by
the set of all operators. Eq.(\ref{eq:hhb-beta1}), based on the
target-space approach, precisely predicts the coefficient of the term
$\left(\h / \hb - 1 \right)^2$ to second order in the
$1/k$-expansion. It seems that such a computation is feasible from the
\textsc{cft} viewpoint at least as a series expansion for large
$k$. This would allow for a genuine two-loop comparison of the two
methods, and is left for future investigation.

\section{RG flow and friction}
\label{sec:rg-flow-friction}

It has already been noted in literature \cite{Gutperle:2002ki} that a
deep link exists between the equations of motion and the
\textsc{rg}-flow. In an oversimplified toy model one can consider the
equations of motion for a system with friction:
\begin{equation}
\label{eq:point-dynamics}
  \frac{\di^2 r}{\di t^2} = - V^\prime (r) - k \frac{\di r}{\di t} .
\end{equation}
Large friction corresponds to the $k\to \infty $ limit where the
dynamics described by this second order equation is well approximated
by a first order one:
\begin{equation}
\label{eq:large-friction}
  \frac{\di r}{\di t^\prime } = - V^\prime (r), \hspace{2cm} t^\prime = \frac{t}{k}.
\end{equation}
At least in some cases the same link exists between the second order
equations of motion and the first order \textsc{rg} flow equation: the
latter provide a good approximation for the dynamics of the system in
some region of the moduli space. In this section we will provide a
class of systems (with constant-curvature metrics and no dilaton)
where this can be explicitly verified and the ``friction'' identified
with the expectation value for the dilaton which appears out of
equilibrium.  More precisely we will consider the \textsc{rg} flow for
the coupling constant of the metric with respect to the Kalb-Ramond
field and then show that the equations that one obtains in this way
are an approximation of those for a system in which the constant is a
field depending on an extra time direction.

\subsection{The \textsc{rg}-flow approach}
\label{sec:textscrg-flow-appr}

As announced above we would like to study the \textsc{rg}-flow for the
coupling of the metric in a system without dilaton, that is for the
sigma model
\begin{equation}
\label{eq:low-dimensional-sigma}
  S = \frac{1}{2 \lambda } \int \di^2 z \:  \left( c g_{\mu \nu} + B_{\mu \nu}  \right) \partial X^\mu \bar \partial X^\nu 
\end{equation}
knowing that for $c = 1$ the model is conformal. Using the geometric
\textsc{rg}-flow approach developed in
Sec.~\ref{sec:target-space-renormalization} we find that Riemann
tensor with respect to the connection of
Eq.~\eqref{eq:minus-connection}:
\begin{equation}
  {R^-}\ud{\mu}{\nu \rho \sigma} = \left( 1 - \frac{1}{c^2} \right) R\ud{\mu}{ \nu \rho \sigma} .
\end{equation}
It follows that the one-loop counterterm is given by
\begin{equation}
  T_{\mu \nu} = \frac{R}{d} \left( 1 - \frac{1 }{c^2} \right) g_{\mu \nu}  
\end{equation}
where for simplicity we supposed the manifold to be Einstein, which is
consistent with the fact that the conformal model with fields $g$ and
$B$ doesn't include a dilaton. Hence we immediately find the
parameters
\begin{align}
  J(c) = 0 && a (c) = \frac{R}{d} \left(1 - \frac{1}{c^2} \right)
\end{align}
and the corresponding beta equations
\begin{equation}
  \begin{cases}
    \beta_{\lambda} = 0 ,\\
    \beta_{c} = \frac{\lambda }{\pi } a_c = \frac{\lambda R }{d \pi } \left(1 -
      \frac{1}{c^2} \right) .
  \end{cases}
\end{equation}

In order to compare this result with what we will find in the
following we can write
\begin{equation}
  c (\mu) = e^{2\sigma (\mu)}  
\end{equation}
where $\mu $ is the energy scale. Then the energy evolution of $\sigma
(\mu)$ (going towards the infrared) gives:
\begin{equation}
  \label{eq:energy-evol-sigma}
  \frac{\di \sigma }{\di \mu } = -\frac{\lambda R }{2 d \pi } e^{-2\sigma (\mu)}\left(1 - e^{-4\sigma (\mu)}  \right) = - V^\prime (\sigma(\mu)) 
\end{equation}
which admits the implicit solution
\begin{equation}
  \log \mu = - \frac{1}{4} \left( 2 e^{2\sigma(\mu)} + \log (\tanh \sigma(\mu)) \right) .
\end{equation}

This is for us the equivalent of Eq.~\eqref{eq:large-friction}. Now we
move to the $\left( d+1 \right)$-dimensional spacetime to find the
corresponding Eq.~\eqref{eq:point-dynamics}.

\subsection{Spacetime interpretation}
\label{sec:equations-motion}

\paragraph{Equations of motion.}

As we said above we want to describe the same system by introducing an
extra time dimension and reading the coupling as a time-dependent
field. In other words we would like to write the equations of motion
for the following sigma model:
\begin{equation}
  S = \int \di^2 z \: \left[ - \partial t \bar \partial t + \left( c(t) g_{\mu \nu} + B_{\mu \nu}  \right) \partial X^\mu \bar \partial X^\nu \right] 
\end{equation}
where, $g$ and $B$ are background fields solving the low-energy string
equations of motion. In order to write the equations of motion let us
rewrite the $d+1$ dimensional metric in terms of a Weyl rescaling as:
\begin{equation}
\label{eq:g-bar}
  \bar g_{\textsc{mn}} = e^{2 \sigma(t)}
  \begin{pmatrix}
    - e^{-2\sigma(t)} & 0 \\
    0 & g_{\mu \nu}
  \end{pmatrix} = e^{2 \sigma (t)} g_{\ssc{mn}}
\end{equation}
where $c(t) = e^{2 \sigma (t)}$. This means in particular that the
Ricci tensor (this time with respect to the standard Levi-Civita
connection) can be written as
\begin{equation}
  \overline{Ric}_{\ssc{mn}} = Ric_{\ssc{mn}} - g_{\ssc{mn}} K\du{\ssc{l}}{\ssc{l}} - \left(d - 1 \right) K_{\ssc{mn}}  
\end{equation}
where $K_{\ssc{mn}}$ is defined as
\begin{align}
  K\du{\ssc{m}}{\ssc{n}} &= - \partial_{\ssc{m}} \sigma
  g^{\ssc{nl}} \partial_{\ssc{l}} \sigma + g^{\ssc{nl}}
  \left( \partial_{\ssc{m}} \partial_{\ssc{l}} \sigma -
    \Gamma\ud{\ssc{p}}{\ssc{ml}} \partial_{\ssc{p}} \sigma \right) +
  \frac{1}{2} g^{\ssc{lp}} \partial_{\ssc{l}}
  \sigma \partial_{\ssc{p}} \sigma
  \delta_{\ssc{m}}^{\phantom{\ssc{m}}\ssc{n}} \\
  K_{\ssc{mn}} &= g_{\ssc{nl}} B\du{\ssc{m}}{\ssc{l}}
\end{align}
After some algebra one finds that
\begin{align}
  \Gamma\ud{t}{tt} = - \dot \sigma(t) && \Gamma\ud{t}{t\mu} = 0 && \Gamma\ud{t}{\mu \nu} = 0 
\end{align}
\begin{subequations}
  \begin{align}
    K\du{t}{t} = - e^{2 \sigma(t)} \left( \ddot \sigma(t) + \frac{\dot
        \sigma^2(t)}{2} \right) &&
    K\du{\mu}{\nu} = -\frac{1}{2} e^{2\sigma(t)} \dot \sigma^2(t) \delta\du{\mu}{\nu} \\
    K_{tt} = \left(\ddot \sigma(t) + \frac{\dot \sigma^2(t)}{2} \right) && K_{\mu
      \nu} = - \frac{e^{2 \sigma(t) }}{2} \dot \sigma^2(t) g_{\mu \nu}
  \end{align}
\end{subequations}
where $\dot \sigma(t) $ is the notation for
\begin{equation}
  \dot \sigma(t) = \frac{\di \sigma (t)}{\di t}  
\end{equation}
In particular this implies that
\begin{equation}
  K\du{\ssc{l}}{\ssc{l}} = - e^{2\sigma(t)} \left( \frac{d+1}{2} \dot \sigma^2(t) + \ddot \sigma(t)  \right)  .
\end{equation}
It then easily follows that
\begin{subequations}
\label{eq:Ric-bar}
  \begin{align}
    \overline{Ric}_{tt} &= -d \left( \ddot \sigma (t)+ \dot \sigma^2(t) \right) \\
    \overline{Ric}_{t\mu } &= 0 \\
    \overline{Ric}_{\mu \nu } &= Ric_{\mu \nu } + \bar g_{\mu \nu} \left( d \dot
      \sigma^2(t) + \ddot \sigma(t) \right) .
  \end{align}
\end{subequations}
The other terms in the equations of motion read
\begin{align}
  \bar H_{\mu \nu}^2 &= H_{\mu \alpha \beta } H_{\nu \gamma \delta} \bar g^{\alpha \gamma} \bar g^{\beta \delta} = e^{-4 \sigma(t)} H_{\mu \nu}^2 \\
  \bar \nabla_{\ssc{m}} \bar \nabla_{\ssc{n}} \Phi &= \partial_{\ssc{m}} \partial_{\ssc{n}} \Phi -
  \bar \Gamma\ud{\lambda}{\ssc{mn}} \partial_\lambda \Phi
\end{align}
now, $\bar \Gamma\ud{t}{\mu \nu } = - \dot \sigma(t) \bar g_{\mu \nu}$ so
\begin{subequations}
  \begin{align}
    \bar \nabla_t \bar \nabla_t \Phi  &= \ddot \Phi(t) \\
    \bar \nabla_\mu  \bar \nabla_t \Phi  &= 0 \\
    \bar \nabla_\mu \bar \nabla_\nu \Phi &= \dot \sigma(t) \dot
    \Phi(t) \bar g_{\mu \nu} .
  \end{align}
\end{subequations}
These are all the ingredients we need to write the equations of motion:
\begin{equation}
  \overline{Ric}_{\ssc{mn}} - \frac{1}{4} \bar H_{\ssc{mn}}^2 + 2  \bar \nabla_{\ssc{m}} \bar \nabla_{\ssc{n}} \Phi = 0  .
\end{equation}
Splitting the time component we obtain
\begin{subequations}
  \begin{align}
    Ric_{tt} &+ 2 \partial_t \partial_t \Phi(t) = -d \left( \ddot \sigma(t) + \dot \sigma^2(t) \right) + 2 \ddot \Phi(t) = 0 \\
    Ric_{\mu \nu} &\left( 1 - e^{-4\sigma(t)} \right) + \bar g_{\mu
      \nu} \left( d \dot \sigma^2(t) + \ddot \sigma(t) - 2 \dot
      \sigma(t) \dot \Phi(t) \right) = 0
  \end{align}
\end{subequations}
where we have used the equations of motion for the system in $\sigma = 0$:
\begin{equation}
  Ric_{\mu \nu} = \frac{1}{4} H_{\mu\nu}^2 
\end{equation}
The system admits a solution if and only if $g_{\mu \nu}$ is Einstein
(since the original system didn't have any dilaton). Taking the trace
with $\bar g^{\ssc{mn}}$ we obtain the system:
\begin{equation}
  \begin{cases}
    d \left( \ddot \sigma(t)  + \dot \sigma^2(t) \right) - 2 \ddot \Phi(t) = 0 \\
    R e^{-2\sigma(t)} \left( 1 - e^{-4\sigma(t)} \right) + d \left( d \dot
      \sigma^2(t) + \ddot \sigma(t) - 2 \dot \sigma(t) \dot \Phi(t) \right) = 0
  \end{cases}
\end{equation}
Introducing
\begin{equation}
  \label{eq:defin-Psi}
  Q (t) = -\dot \Phi (t) + \frac{d}{2} \dot \sigma (t)
\end{equation}
the equations become:
\begin{equation}
\label{eq:friction-syst}
  \begin{cases}
    \dot Q(t) = - d \dot \sigma^2(t) \\
    \ddot \sigma(t) = - \frac{R}{d} e^{-2\sigma(t)} \left( 1 - e^{-4 \sigma(t)} \right) - 2
    \dot \sigma(t) Q(t)
  \end{cases}
\end{equation}
This second equation has precisely the structure of the motion in a potential
\begin{equation}
  V (\sigma) - V(0) = \frac{R}{6d} e^{-6 \sigma} \left( 1 - 3 e^{4\sigma} \right) \sim - \frac{R}{3d} +  \frac{2 R }{d} \sigma^2 .
\end{equation}
and with a time-dependent friction coefficient $Q(t)$. In the limit of
$Q \to \infty $ we clearly recover Eq.~\eqref{eq:energy-evol-sigma}
with the same potential $V(\sigma)$ when identifying the \emph{energy
  scale} $\mu$ for the off-shell system with the \emph{time direction}
here following
\begin{equation}
  \log \mu = \frac{\pi \bar Q}{\lambda} t .  
\end{equation}

\paragraph{Linearization.}

\begin{figure}
  \begin{center}
    \subfigure[Small $Q(0)$]{
      \includegraphics[width=.4\linewidth]{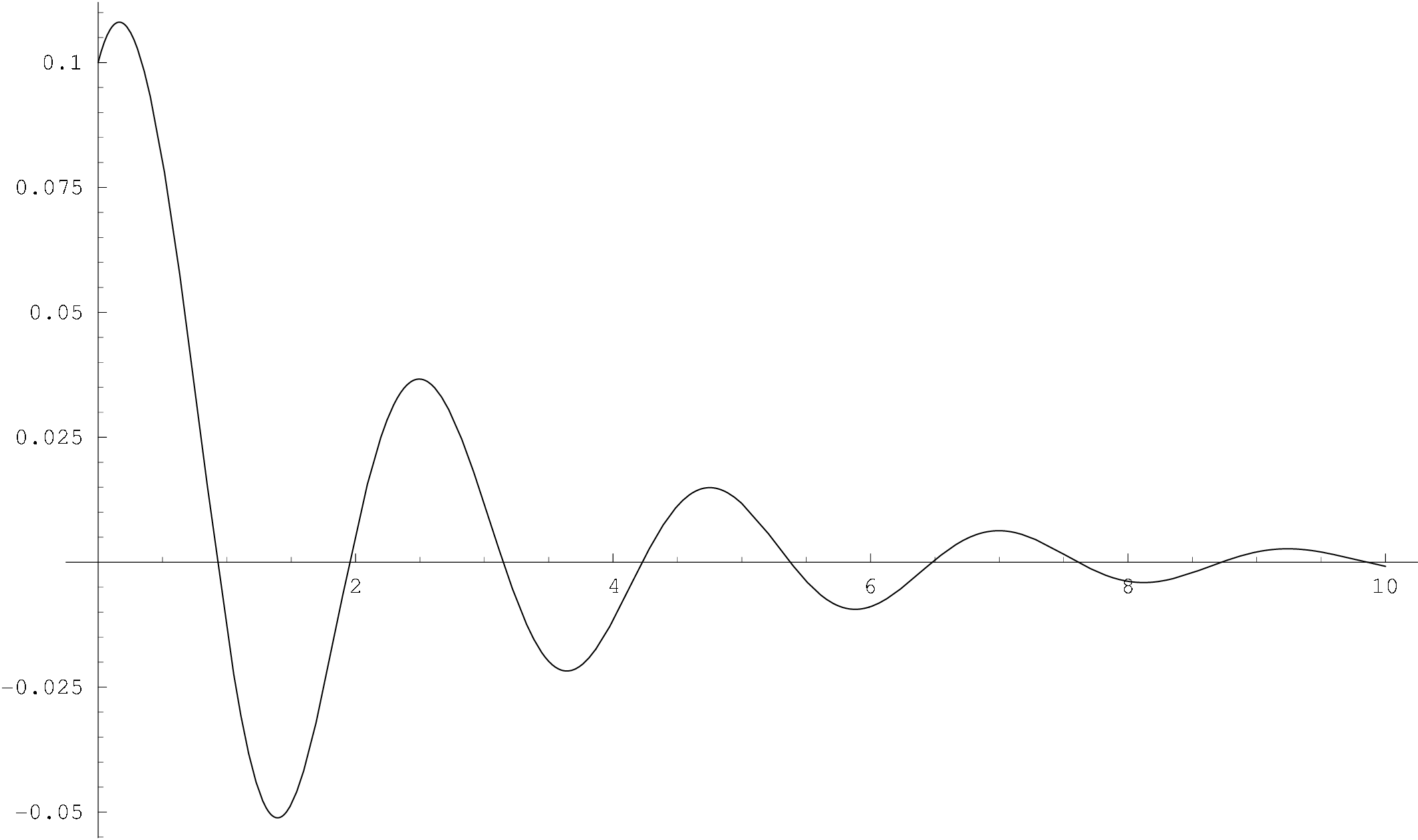}}
    \subfigure[Large $Q(0)$]{
      \includegraphics[width=.4\linewidth]{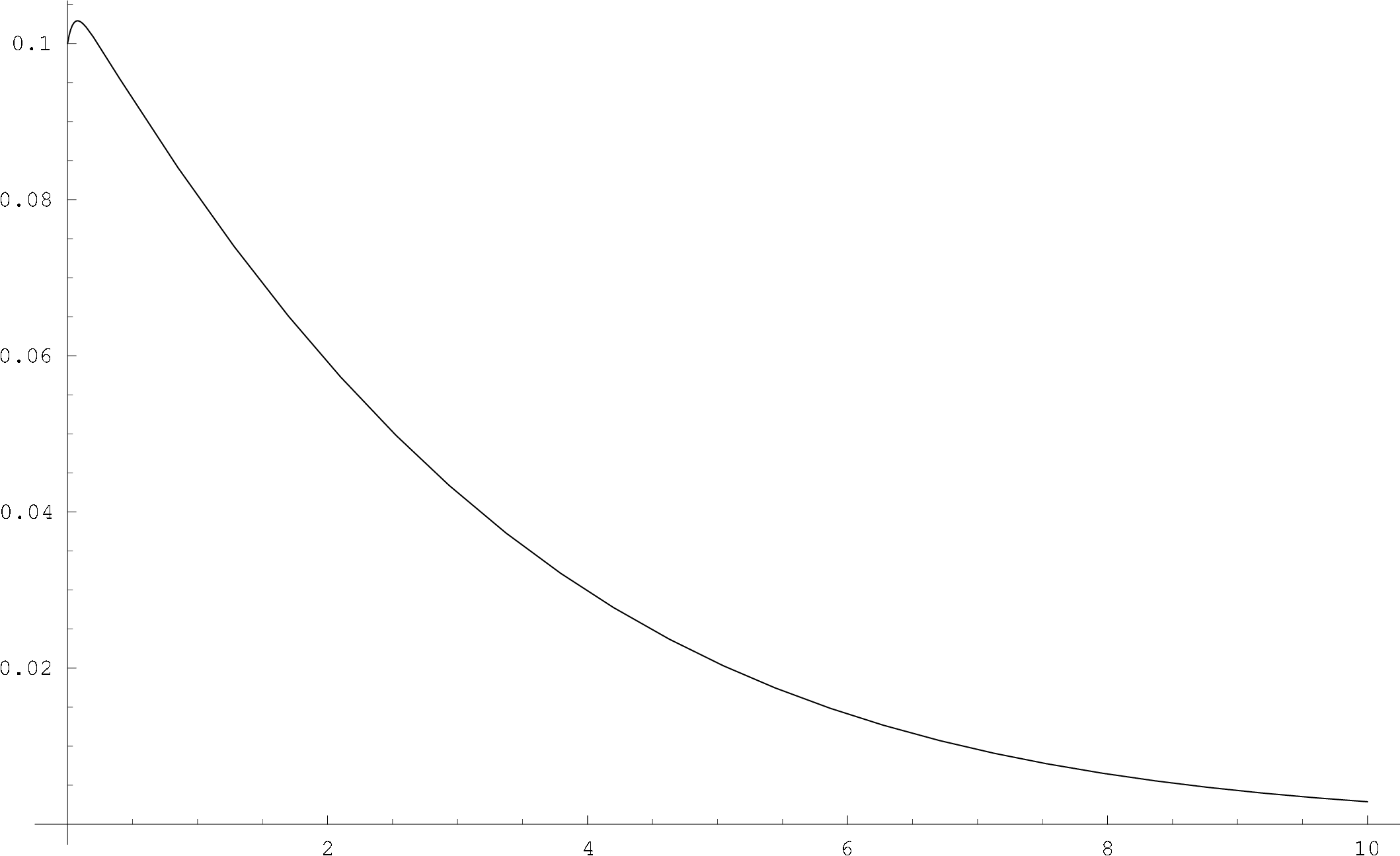}}
  \end{center}
  \caption{Typical behaviour for $\sigma(t)$ in the
    system~\eqref{eq:friction-syst} for (a)~small and (b)~large
    (positive) initial values of $Q(t)$.}
  \label{fig:num-solution-Psi}
\end{figure}

The system~\eqref{eq:friction-syst} can be solved numerically and
typical results for large $Q (0)$ and small $Q(0)$ are presented in
Fig.~\ref{fig:num-solution-Psi}.

A further step can be made by linearization. Introduce
\begin{equation}
  \Sigma (t)= \dot \sigma (t)   
\end{equation}
the system becomes a first order one:
\begin{equation}
  \begin{cases}
    \dot  Q(t) = - d \Sigma^2(t) \\
    \dot \sigma(t) = \Sigma(t) \\
    \dot \Sigma(t) = - V^\prime (\sigma(t)) - 2 \Sigma(t) Q(t)
  \end{cases}
\end{equation}
which has a fixed point for $\left( Q, \sigma, \Sigma \right) = \left(
  \bar Q, 0, 0 \right)$ where $\bar Q$ is a constant. Around this
point the equations read:
\begin{equation}
\label{eq:linear-system}
  \begin{cases}
    \dot Q(t) = 0 \\
    \dot \sigma(t) = \Sigma(t) \\
    \dot \Sigma(t) = - V^{\prime \prime} (0) \sigma(t) - 2 \bar Q \Sigma(t) = - \frac{4 R}{d}
    \sigma(t) - 2 \bar Q \Sigma(t)
  \end{cases}
\end{equation}
so, $Q $ decouples (and remains constant) and the equation of motion
around the fixed point is
\begin{equation}
  \frac{\di^2 \sigma (t)}{\di t^2} = - \frac{4 R}{d}
  \sigma(t) - 2 \bar Q \frac{\di \sigma (t)}{\di t} ,
\end{equation}
which can be integrated giving
\begin{equation}
  \sigma (t) = C_1 \exp \left[ -\left(  \bar Q + \sqrt{\bar Q^2 -
        \frac{4R}{d}}  \right) t \right] + C_2 \exp 
  \left[ -\left( \bar Q - \sqrt{\bar  Q^2 - \frac{4R}{d}}  \right)
    t \right]
\end{equation}
with $C_1 $ and $C_2$ integration constants.

For positive $\bar Q$ the solution converges to $\sigma = 0$ with or
without oscillations if $\bar Q^2 \lessgtr 4R/d$.  In terms of $\sigma(t) $
and $\Phi(t)$ this limit solution is
\begin{align}
  \sigma(t) \xrightarrow[t\to \infty ]{} 0 && \Phi (t) \sim - \bar Q t ,
\end{align}
which is not surprisingly the initial conformal model in
Eq.~\eqref{eq:low-dimensional-sigma} plus a linear dilaton.

\paragraph{The meaning of $\bar Q$.}

$\bar Q $ is linked to the dilaton: larger values correspond to
negative and larger absolute values for $\Phi$, \emph{i.e.} moving
further inside the perturbative regime.  On the other hand, negative
values of $\bar Q $ give diverging solutions, but in this case the
dilaton grows (see Eq.~\ref{eq:defin-Psi}) and the very underlying
perturbative approach collapses. It is worth to remark that if we make
an hypothesis of uniqueness for the system~\eqref{eq:friction-syst},
$Q$ can't change sign because $Q(t) = 0$, $\sigma(t) = 1$ is a solution
(the starting conformal model with constant dilaton).

A better understanding of the actual meaning of this parameter can be
obtained if we consider the limiting situation of linear dilaton. In
this case, in fact, it is immediate to derive the central-charge of
the overall system:
\begin{equation}
  c = \left( d + 1 \right) - 3 \bar Q^2 - c_d + c_I = 0 ,
\end{equation}
where $c_d$ is the central-charge of the conformal system in
Eq.~\eqref{eq:low-dimensional-sigma} (\emph{e.g.} $6 / \left(k+2
\right)$ for the $SU(2)$ \textsc{wzw} model) and $c_I$ is the internal
central-charge. If follows that for a critical model
\begin{equation}
  \bar Q^2 = \frac{1}{3} \left( d + 1 - c_d + c_I \right)  
\end{equation}
and $\bar Q$ is essentially a measure of the deficit.

A final remark regards the consistency of the approximation for the
dynamics one obtains from the \textsc{rg}-flow
equation~\eqref{eq:energy-evol-sigma}, corresponding to a $Q \to \infty$
limit. The linearized system~\eqref{eq:linear-system} provides a
justification for such limit: in fact the time scale for $Q (t)$ is
comparably larger than $\sigma(t)$'s -- to the point that the former
decouples around the fixed point. For this reason it can be taken as a
constant (fixed by the initial conditions) if we just concentrate on
the evolution of the warping factor $\sigma (t)$.

\section{Cosmological interpretation}
\label{sec:friedm-roberts-walk}

The type of backgrounds we are studying are time-dependent and as such
can have a cosmological interest. For this reason, since there is a
non-trivial dilaton, one should better move to the Einstein frame (as
opposed to the string frame we've been using thus far). This means
that the metric is written as:
\begin{equation}
  \tilde g_{\ssc{mn}} = e^{- \Phi (t) /2 } \bar g_{\ssc{mn}}
\end{equation}
and after a coordinate change
\begin{equation}
  \tau(t) = \int  e^{-\Phi(t)/4} \di t
\end{equation}
can be put back to the same warped product form as in
Eq.~\eqref{eq:g-bar}:
\begin{multline}
  \label{eq:Einst-metric}
  \widetilde{\di s^2} = \tilde g_{\ssc{mn}} \di x^{\ssc{m}} \di
  x^{\ssc{n}} = -\di \tau^2 + \left. e^{2 \sigma(t) - \Phi (t)/2}
  \right|_{t = t(\tau)} \left( g_{\mu\nu} \di x^\mu \di x^\nu \right)
  = \\ = -\di \tau^2 + w(\tau) \left( g_{\mu\nu} \di x^\mu \di x^\nu
  \right).
\end{multline}

Cosmologically interesting solutions are obtained when $d=3$. In this
case the $H$ field is proportional to the volume form on $g$. This
implies that $H_{\mu \nu}^2 \propto g_{\mu \nu}$ and then the
equations reduce to
\begin{equation}
  Ric_{\mu \nu} = \Lambda^2 g_{\mu \nu}
\end{equation}
\emph{ie} $g_{\mu \nu}$ is to be the metric of an Einstein
three-manifold (the most simple case being a three-sphere). What we
get then is a typical example of Friedmann-Robertson-Walker
(\textsc{frw}) spacetime such as those already studied in
\cite{Tseytlin:1991ss,Tseytlin:1992ye,Goldwirth:1993ha,Copeland:1994vi}.
As such it describes the time evolution of an isotropic spacetime (or
more in general of a spacetime with the symmetries of the conformal
theory in Eq.~\eqref{eq:low-dimensional-sigma}). Some intuition about
the time evolution can be developed if we take the linearized system
in Eq.~\eqref{eq:linear-system} and consider the large $t$ limit. In
fact, as remarked above the solution asymptotically approaches a
linear dilaton background (which was already studied from this point
of view in \cite{Antoniadis:1988vi}):
\begin{align}
  \sigma (t) \xrightarrow[t\to \infty ]{} 0 && Q (t) = \bar Q && \Phi(t) \sim - \bar Q t
\end{align}
hence one verifies that the metric in the Einstein frame is
asymptotically
\begin{equation}
  \widetilde{\di s^2} \sim - \di \tau^2 + \bar Q^2 \tau^2 \left( g_{\mu\nu} \di x^\mu \di x^\nu \right) 
\end{equation}
which corresponds to an expanding universe with curvature
\begin{equation}
  \tilde R \sim \frac{R + \bar Q^2 d \left( d - 1 \right)}{\bar Q^2 \tau^2} .
\end{equation}

A similar result, with a polynomial expansion is found if we consider
an exponential decrease for $\sigma(t)$, or better for $c(t)$ (in the
linear limit $c(t) - 1$ obeys the same equations as
$\sigma(t)$). After a redefinition of the variables we can let
\begin{equation}
  c(t) = e^{-t} + 1  .
\end{equation}
It is easy to check that in general\footnote{On a side note, since $c(t)>0$ by construction the relation $\tau = \tau (t)$ is always invertible.}
\begin{equation}
  \tau (t) = \int c(t)^{-d/16} e^{1/4\int Q(t^\prime) \di t^\prime} \di t
\end{equation}
and in this linearized approximation the latter becomes
\begin{equation}
  \tau (t)= \int \left( e^{-t} + 1 \right)^{-d/16} e^{\bar Q t / 4} \di t .
\end{equation}
This integral can be solved analytically: \mathindent=0em
\begin{equation}
  \tau (u) = \frac{16}{d + 4 \bar Q} \left( 1 + \frac{1}{u} \right)^{-d/16} u^{\bar Q/4} \left( 1 + u \right)^{d/16} {}_2F_1 \left( \frac{d}{16} , \frac{d+4 \bar Q}{16}; \frac{d + 4 \bar Q}{16} + 1, - u\right) ,  
\end{equation}
\mathindent=\oldindent
where $u = e^t$ and ${}_2F_1$ is an hypergeometric
function\footnote{The hypergeometric function ${}_2F_1$ is defined as
  follows:
  \begin{equation}
    {}_2F_1 (a,b;c,u) = \sum_{k=0}^\infty \frac{ \left( a \right)_k \left( b \right)_k}{\left(c \right)_k} \frac{z^k}{k!}    
  \end{equation}
  where $\left(a \right)_k$ is the Pochhammer symbol
  \begin{equation}
    \left( a \right)_k = \frac{\Gamma (a+k)}{\Gamma (a)}
  \end{equation}%
}. It is better however to consider the asymptotic behaviours. For $u
\to \infty$ one finds that $\tau (u)$ and the warping factor $w(u)$ go
like:
\begin{align}
  \tau (u) \sim \frac{4}{\bar Q} u^{\bar Q /4} , && w(u) \sim u^{\bar Q /2} ,
\end{align}
and consistently with the results above for the linear dilaton case (which is
precisely the large-$u$ limit):
\begin{equation}
  w (\tau) \sim \tau^2 ;
\end{equation}
similarly for small $u$:
\begin{align}
  \tau (u) \sim \frac{16}{d+4 \bar Q} u^{\left( d+4 \bar Q\right)/16} , && w(u) \sim u^{d/4+2+\bar Q/2}
\end{align}
and then
\begin{equation}
  w(\tau) \sim \tau^{4 + 8 \left( 4 - \bar Q \right)/\left( d + 4 \bar Q \right)} .
\end{equation}
Note that this behaviour precisely measures the effect of a finite
value for $\bar Q$ and in fact for $\bar Q \to \infty$ we recover
again $w (\tau) \sim \tau^2$.  Summarizing, just as advertised, we get
again a polynomially expanding universe (a so-called big-bang
solution).

\begin{figure}
  \begin{center}
    \subfigure[$\bar Q \sim 0$]{
      \includegraphics[width=.45\linewidth]{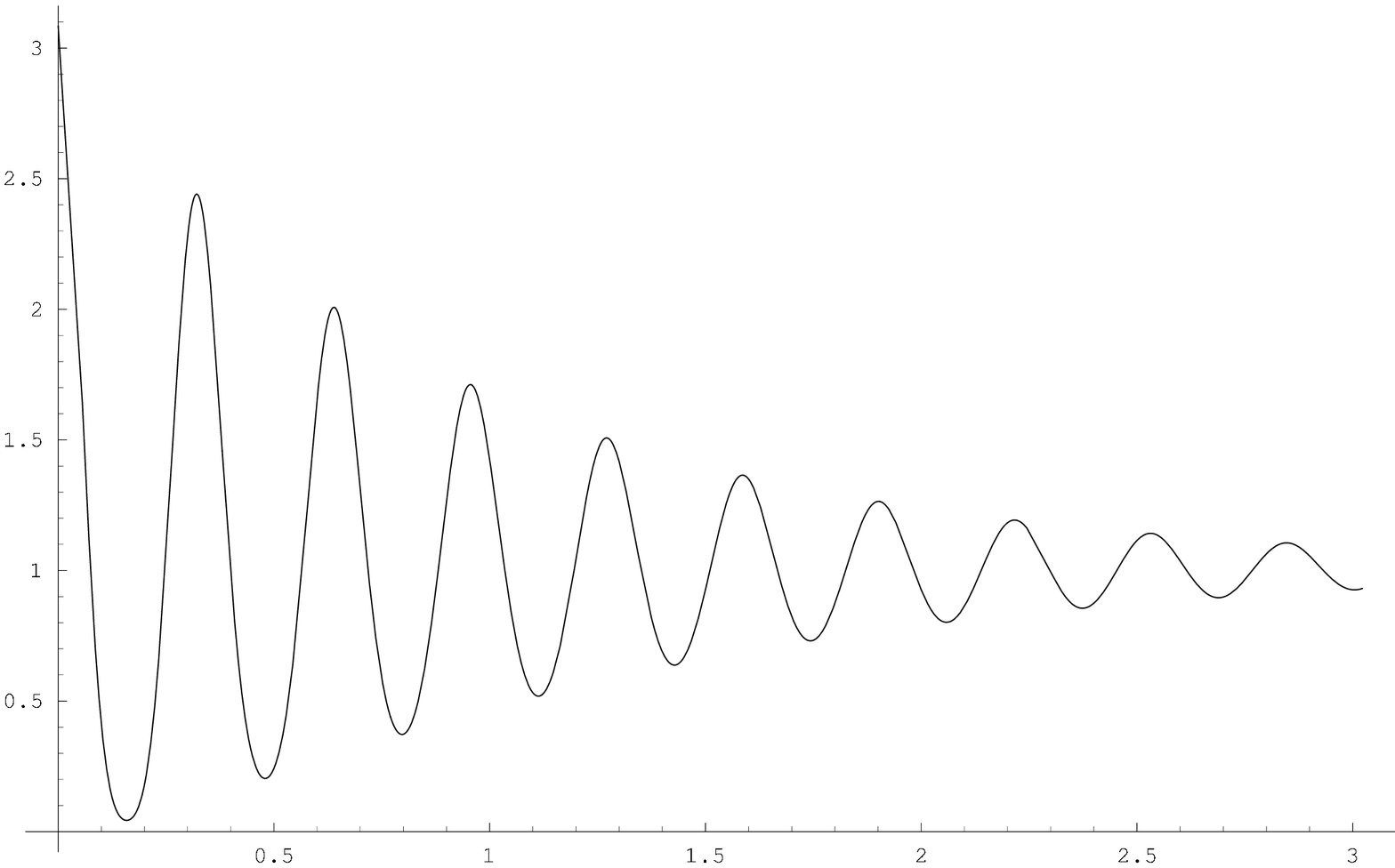}}
    \subfigure[Small $\bar Q$]{
      \includegraphics[width=.45\linewidth]{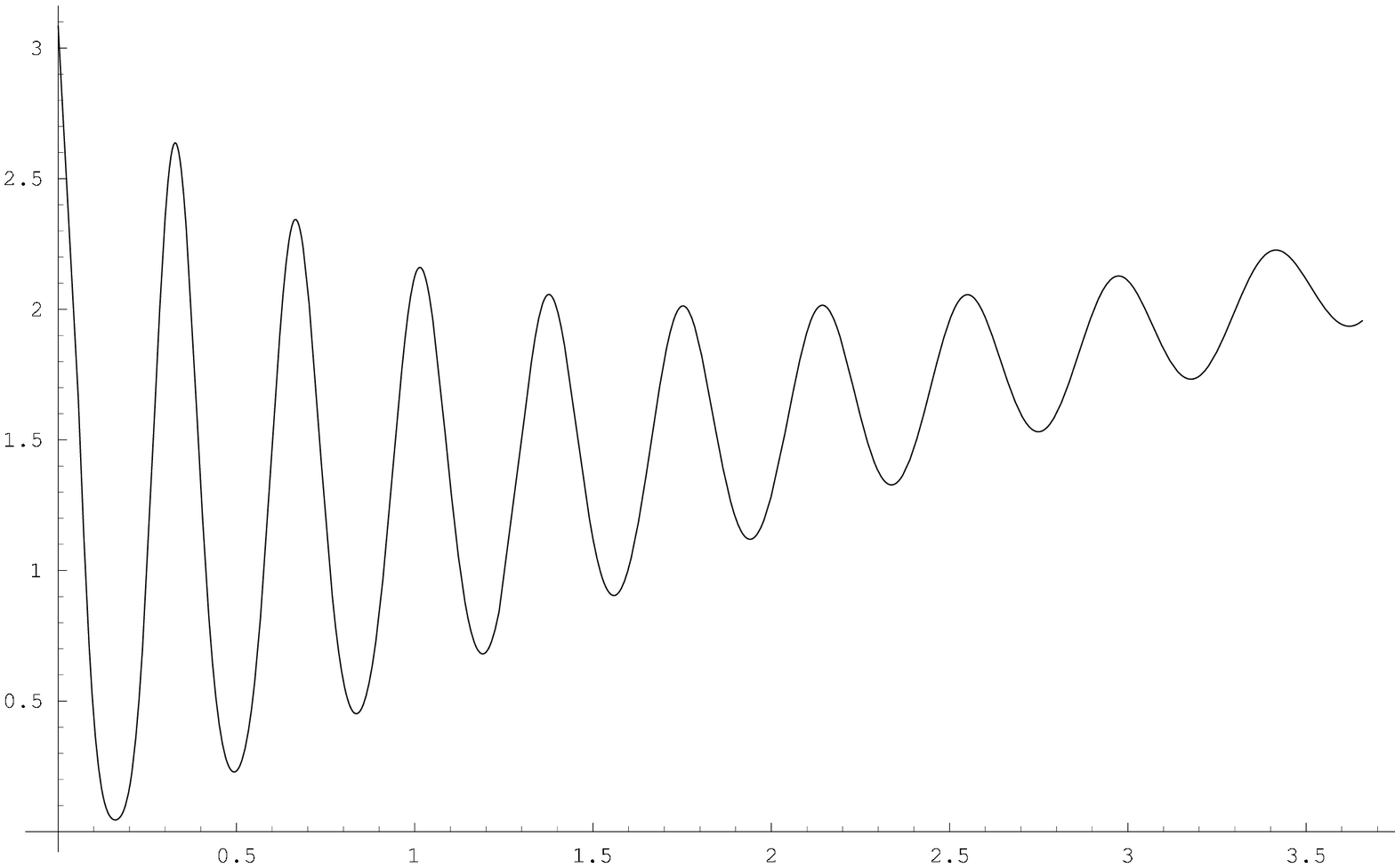}}
    \subfigure[Medium $\bar Q$]{
      \includegraphics[width=.45\linewidth]{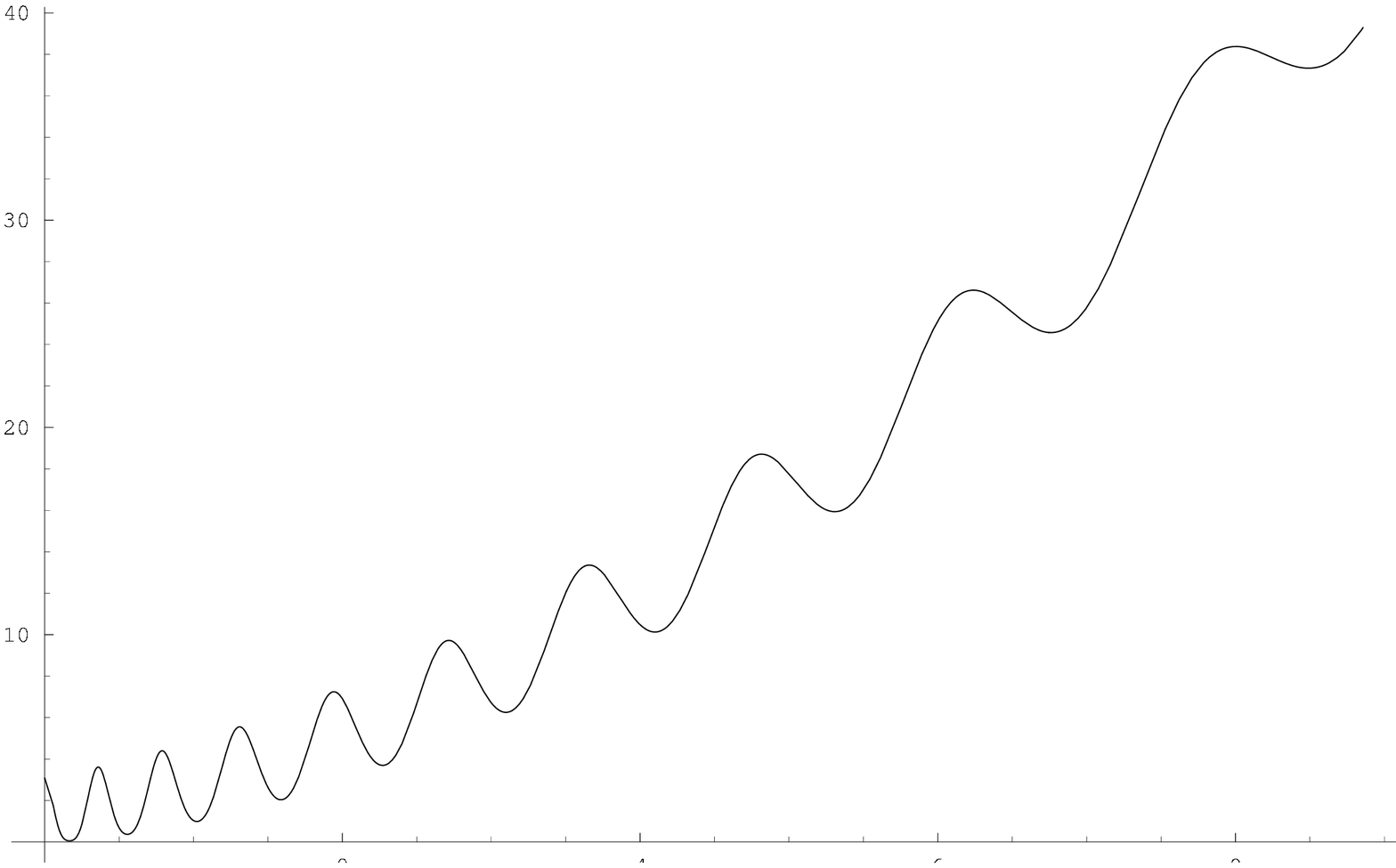}}
  \end{center}
  \caption{Typical behaviour for the warping factor in the small-$\bar
    Q$ regime. We consider (a) $\bar Q$ very small but not
    vanishing, (b) small $\bar Q$ and (c) larger $\bar Q$ (but
    still compatible with oscillations). }
  \label{fig:num-solution-warp}
\end{figure}

The analysis for the small-$\bar Q$ regime is clearly more difficult
to be carried out analytically. Apart from numerical solutions (see
Fig.~\ref{fig:num-solution-warp}), in general we can study $w(\tau)$
as a parametric curve in the $\left(w,\tau\right)$ plane defined by
$\left(w(t),\tau(t) \right)$. Then $\tau(t)$ appears to be a
monotonically increasing function since $c(t)>0$ which implies that
$w(\tau)$ has an extremum for each extremum in $w(t)$. This means that
we expect the superposition of a polynomial expansion and a damped
oscillation. The limiting situation is obtained when $\bar Q$ is small
(but not vanishing), and for large $t$, $\tau (t) \sim t$ so that
$w(\tau)$ slowly converges, oscillating, to a constant value.

\chapter{Hyperbolic Spaces}
\label{cha:hyperbolic-spaces}

\chapterprecis{In this chapter we investigate type II and M-theory
  geometries written as direct products of constant-curvature spaces.
  We find in particular a class of backgrounds with hyperbolic
  components and we study their stability with respect to small
  fluctuations.}

\lettrine{T}{his chapter} does in some sense deviate from the main
theme we developed in this thesis. In fact we will deal with type II
theories in presence of Ramond-Ramond fields, that -- to this moment
-- still elude a precise \textsc{cft} treatment. For this reason our
analysis will be mostly confined to supergravity considerations. On
the other hand we still continue to follow one of the main guiding
threads, \emph{i.e.} look for backgrounds with simple geometric
interpretation, which in this case means (maximally) symmetric spaces,
with special emphasis on hyperbolic, negative curvature, Poincar\'e
spaces. We will show in fact that these spaces can be used as building
blocks for M-theory and type II backgrounds, in genuinely perturbative
configurations or in presence of D~branes, both in the non-compact and
in the compact part after discrete identifications. In particular we
find a series of M-theory solutions that can be obtained starting with
the usual $\mathrm{AdS}_7 \times S^4$ by splitting the anti-de Sitter
in a $\mathrm{AdS} \times H$ product and verify their stability with
respect to small fluctuations. Finally we also show how obtaining
negative-curvature Euclidean signature spaces is not in general an
easy task and in particular show that the presence of orientifold
planes, giving a negative contribution to the stress-energy tensor is
not enough to allow for $H_3$ spaces in type IIB theories.

  \section{M-theory solutions}
  \label{sec:m-theory-solutions}
  
  Let us start with the action of the bosonic sector of
  eleven-dimensional supergravity:
  \begin{equation}
    S = \frac{1}{2 \kappa_{11}^2} \int \di^{11} x \: \sqrt{-g} \left( R - \frac{1}{2} \abs{F\form{4}}^2 \right) + \frac{1}{12 \kappa_{11}^2} \int A\form{3} \land F\form{4} \land F\form{4} ,
  \end{equation}
  and the corresponding equations of motion
  \begin{equation}
    R_{\mu \nu } - \frac{1}{2} R g_{\mu \nu } - \frac{1}{2} \abs{F\form{4}^2}_{\mu \nu } + \frac{1}{4} \abs{F\form{4}^2} g_{\mu \nu } = 0 ,
  \end{equation}
  where with the notation $\abs{F\form{n}}^2$ we mean
  \begin{equation}
    \abs{F\form{p}}^2 = \frac{1}{p!} F_{\mu_1 \mu_2 \ldots \mu_p} F^{\mu_1 \mu_2 \ldots \mu_p} .
  \end{equation}
  
  The ansatz we want to study is the following. We consider direct
  products of symmetric spaces of the form $M_{11} = M_0 \times M_1
  \times M_2 \times \ldots $ where $M_0$ has signature $-,+,\ldots,+$
  and all the other spaces are Riemann.  Since we assume that they are
  all symmetric spaces, we can split the Ricci tensor in blocks and
  each block will be proportional to the metric of the corresponding
  submanifold. To fix the notation we can introduce the parameters
  $\Lambda_i$ as
  \begin{equation}
    \left. R_{\mu \nu} \right|_{i} = \Lambda_i \left. g_{\mu \nu} \right|_i ,
  \end{equation}
  so that the Ricci scalars are given by
  \begin{equation}
    R_i = \Lambda_i \dim M_i  = \Lambda_i d_i .
  \end{equation}
  In particular we can raise an index and rewrite
  \begin{equation}
    \left. R\ud{\mu}{\nu} \right|_{i} = \Lambda_i \left. \delta\ud{\mu}{\nu} \right|_i .
  \end{equation}
  
  The Poincar\'e invariance constraint fixes the allowed gauge fields
  to be proportional to the volume form of each submanifold. It is
  always possible to perform an electric/magnetic duality so that
  there's no field on the Minkowskian submanifold.  This means that we
  can consider gauge fields having the form:
  \begin{equation}
    F\form{d_i} = F_i = Q_i \tilde \omega_i ,
  \end{equation}
  where $F_i$ is a $d_i$-form, $Q_i \in \setN$ and $\tilde \omega_i =
  \widetilde{\vol_{M_i}}$ is the volume form on $M_i$, normalized to one.
  It is useful to rewrite the expression above as
  \begin{equation}
    F_i = \sqrt{2} k_i Q_i  \abs{\Lambda_i}^{d_i/2} \omega_i
  \end{equation}
  where $\omega_i$ is the volume form on $M_i$ and $k_i$ is a constant whose value is
  \begin{itemize}
  \item $k_i = \Gamma (i/2) / \left( 2 \sqrt{2} \pi^{i/2} \right) $ for a sphere $S^i$
  \item $k_i = 1/\left(4 \pi \left(g-1 \right) \right)$ for a genus $g$
    Riemann surface $H^2/ \Gamma$
  \item some value that completely identifies the lattice in a $H^3/ \Gamma
    $ compactification (rigidity theorem for three-manifolds
    [Thurston]).
  \end{itemize}
  In coordinates
  \begin{equation}
    \left. F_i \right|_{\mu_1 \ldots \mu_{d_i}} = k_i Q_i  \Lambda_i^{d_i/2}\sqrt{ 2\det g_{i}} \epsilon_{\mu_1 \ldots \mu_{d_i}}
  \end{equation}
  which implies (as one can verify in a non-coordinate basis):
  \begin{equation}
    \left. F_i \right|_{\mu \mu_2 \ldots \mu_{d_i}} \left. F_i \right|^{\nu \mu_2 \ldots \mu_{d_i}} = 2 \left( n - 1 \right)! k_i^2 Q_i^2 \Lambda^{d_i} \delta\ud{\mu}{\nu}
  \end{equation}
  Furthermore
  \begin{align}
    \abs{F_i^2}\ud{\mu}{\nu} &=  2 k_i^2 Q_i^2 \abs{\Lambda_i}^{d_i} \delta\ud{\mu}{\nu} \\
    \abs{F_i^2} &=  2 k_i^2 Q_i^2 \abs{\Lambda_i}^{d_i} .
  \end{align}

  We are now in position to write the equations of motion that will
  simply translate into an algebraic system for the $\left( \Lambda_0,
    \Lambda_i \right)$:
  \begin{equation}
    \begin{cases}
      \Lambda_0 - \frac{1}{2} R = - \frac{1}{2} \displaystyle{ \sum_j k_j^2 Q_j^2 \abs{\Lambda_j}^{d_j}} , \\
      \Lambda_i - \frac{1}{2} R = - \frac{1}{2} \displaystyle{ \sum_j
        \left(-\right)^{\delta_{ij}} k_j^2 Q_j^2 \abs{\Lambda_j}^{d_j}} ,
    \end{cases}
  \end{equation}
  where
  \begin{equation}
    R = d_0 \Lambda_0 + \sum_i d_i \Lambda_i  .
  \end{equation}
  Let us now turn to study some particular examples.

  \subsubsection{$M^{1,6-d} \times M^d \times M^4 $}
  \label{sec:m1-6-dt}

  Let us consider as an example the case $M_{11} = M_0 \times M_1 \times M_2 $
  where $\left(d_0,d_1,d_2 \right) = \left(7-d, d, 4 \right)$. If we
  turn $F_2$ in the equations of motion read:
  \begin{equation}
    \begin{cases}
      \Lambda_0 - \frac{1}{2} R= -\frac{1}{2} k_2^2 Q_2^2 \Lambda_2^4 ,\\
      \Lambda_1 - \frac{1}{2} R = -\frac{1}{2} k_2^2 Q_2^2 \Lambda_2^4 ,\\
      \Lambda_2 - \frac{1}{2} R = \frac{1}{2} k_2^2 Q_2^2 \Lambda_2^4 ,
    \end{cases}
  \end{equation}
  with $R = \left(7-d \right) \Lambda_0 + d \Lambda_1 + 4
  \Lambda_2$. The solution is:
  \begin{align}
    \Lambda_0 = \Lambda_1 = - \frac{1}{2}\left(\frac{3}{2}
    \right)^{1/3} \frac{1}{\left(k_2 Q_2 \right)^{2/3}} , && \Lambda_2
    = \left(\frac{3}{2} \right)^{1/3} \frac{1}{\left(k_2 Q_2
      \right)^{2/3}} ,
  \end{align}
  and given the curvatures, this describes an $\mathrm{AdS}_{7-d} \times
  H_{d} \times S^4$ space.

  A few remarks are in order. First of all, the result doesn't depend
  on $d$ and in particular it would be the same for $d=0$ (which is
  some sort of limit case). In other words, at the level of the
  equations of motion, we can't distinguish between an
  $\mathrm{AdS}_7$ space and any product of the form
  $\mathrm{AdS}_{7-d} \times H_d$ once the respective curvatures are
  such as
  \begin{equation}
    \frac{R_{\mathrm{AdS}}}{7-d} = \frac{R_H}{d} .
  \end{equation}
  This calculation is generalizable to any product of the form $M_0
  \times M_1 \times \ldots \times M_n$ with dimensions $\left(d_0,
    d_1, \ldots, d_{n-1}, 4 \right)$. In fact the equations of motion
  read
  \begin{equation}
    \begin{cases}
      \Lambda_0 - \frac{1}{2} R= -\frac{1}{2} k_n^2 Q_n^2 \Lambda_n^4 ,\\
      \Lambda_1 - \frac{1}{2} R = -\frac{1}{2} k_n^2 Q_n^2 \Lambda_n^4 ,\\
      \ldots\\
      \Lambda_{n-1} - \frac{1}{2} R = -\frac{1}{2} k_n^2 Q_n^2 \Lambda_n^4 ,\\
      \Lambda_n - \frac{1}{2} R = \frac{1}{2} k_2^2 Q_2^2 \Lambda_2^4 ,
    \end{cases}
  \end{equation}
  with $R= d_0 \Lambda_0 + d_1 \Lambda_1 + \ldots + d_n \Lambda_n$.
  From the system above we conclude that $\Lambda_0 = \Lambda_1 =
  \ldots = \Lambda_{n-1}$ and then we are back to the situation above:
  \mathindent=0em
  \begin{align}
    \Lambda_0 = \Lambda_1 = \ldots =\Lambda_{n-1} = -
    \left(\frac{3}{2} \right)^{1/3} \frac{1}{2\left(k_n Q_n
      \right)^{2/3}} , && \Lambda_n = \left(\frac{3}{2} \right)^{1/3}
    \frac{1}{\left(k_n Q_n \right)^{2/3}} .
  \end{align}
  \mathindent=\oldindent
 We have thus found a series of possible
  M-theory backgrounds where the anti-de~Sitter component is split
  into two or more subspaces of the form
  \begin{equation}
    \mathrm{AdS}_n \to \mathrm{AdS}_{n-p} \times H_p  ,
  \end{equation}
  with Ricci scalars obeying   \begin{equation}
    \frac{R^{(n)}}{n} = \frac{R^{(n-p)}}{n-p} = \frac{R^{(p)}}{p} .
  \end{equation}
  In particular we get the direct products $\mathrm{AdS}_{2} \times
  H_2 \times H_3 \times S^4$, $\mathrm{AdS}_{2} \times H_5 \times
  S^4$, $\mathrm{AdS}_{3} \times H_2 \times H_2 \times S^4$,
  $\mathrm{AdS}_{3} \times H_4 \times S^4$, $\mathrm{AdS}_{4} \times
  H_3 \times S^4$, $\mathrm{AdS}_{5} \times H_2 \times S^4$.
  
  \subsubsection{$M^{1,3} \times M^d \times M^{7-d}$}

  The dual situation is obtained for $\left(d_0, d_1, d_2 \right) =
  \left(4,d,7-d \right)$. In this case we can turn on the $7$-form field
  \begin{equation}
    F_{[7]} = \sqrt{2} k_0 Q_0 \Lambda_1^{d/2} \Lambda_2^{\left(7-d \right)/2}      
  \end{equation}
  and the equations of motion read:
  \begin{equation}
    \begin{cases}
      \Lambda_0 - \frac{1}{2} R= -\frac{1}{2} k_0^2 Q_0^2
      \abs{\Lambda_1^{d} \Lambda_2^{7-d }} , \\
      \Lambda_1 - \frac{1}{2} R = \frac{1}{2} k_0^2 Q_0^2
      \abs{\Lambda_1^{d} \Lambda_2^{7-d }} ,\\
      \Lambda_2 - \frac{1}{2} R = \frac{1}{2} k_0^2 Q_0^2
      \abs{\Lambda_1^{d} \Lambda_2^{7-d }} ,
    \end{cases}
  \end{equation}
  with $R= 4 \Lambda_0 + d \Lambda_1 + \left(7-d \right) \Lambda_2
  $. Again the solution is easily found
  \begin{align}
    \Lambda_0 = - 2\frac{3^{1/6}}{\left( k_0 Q_0 \right)^{2/3}} &&
    \Lambda_1 = \Lambda_2 = \frac{3^{1/6}}{\left( k_0 Q_0
      \right)^{2/3}} .
  \end{align}
  Just as before this does not depend on $d$ and shows that at this
  level an $\mathrm{AdS}_4 \times S^d \times S^{7-d}$ space is not
  distinguishable from a $\mathrm{AdS}_4 \times S^7 $ one.
  Again we can consider more general configurations with dimensions $\left(4, d_1, \ldots, d_n \right)$ to find that the solution remains the same
  \begin{align}
    \Lambda_0 = - 2\frac{3^{1/6}}{\left( k_0 Q_0 \right)^{2/3}} , && \Lambda_1 =
    \Lambda_2 = \ldots = \Lambda_n = \frac{3^{1/6}}{\left( k_0 Q_0 \right)^{2/3}} .
  \end{align}

  This describes a second series of solutions in which a sphere is
  split into the product of two smaller spheres according to
  \begin{equation}
    S^n \to S^{n-p} \times S^p  
  \end{equation}
  with Ricci scalars obeying as above
  \begin{equation}
    \frac{R^{(n)}}{n} = \frac{R^{(n-p)}}{n-p} = \frac{R^{(p)}}{p} ,
  \end{equation}
  and then we recover $\mathrm{AdS}_4 \times S^{2} \times S^2 \times S^3$,
  $\mathrm{AdS}_4 \times S^{2} \times S^5$, $\mathrm{AdS}_4 \times S^{3} \times S^2 \times
  S^2$, $\mathrm{AdS}_4 \times S^{3} \times S^4 $, $\mathrm{AdS}_{4} \times S^{5} \times
  S^2$.

  Finally we can combine the two types of splitting and put
  $\mathrm{AdS}_2 \times H^2$ instead of $\mathrm{AdS}_4$ and $S^2 \times S^2 $
  instead of $S^4$ in the former series.

  \bigskip

  The key ingredient to these constructions is the fact that with a
  careful choice of radii the product spaces still remain Einstein
  and this is all one needs to satisfy the equations of motion. This
  means of course that in general one can use any Einstein manifold
  with the proper curvature. In particular, then, instead of the
  five-sphere $S^5$ one can put a generalization of the $S^3 \times
  S^2$ product, such as any of the representatives of the
  two-parameter class of spaces $T^{p,q}$ obtained as $S^1$ fibrations
  over $S^2 \times S^2$ or equivalently as the coset $\left(SU (2)
    \times SU(2) \right)/U(1)$, the parameter being the embedding
  indexes of $U(1)$ in the $SU(2)$'s\footnote{This clearly is the
    basis of the conifold.}. Similarly, instead of $\mathrm{AdS}^5$
  one can put a space written as a time-fibration over $H_2 \times H_2
  $ or as the coset $\left(SL(2,\setR) \times SL(2,\setR) \right)/
  \setR$. The metric of such $L_{Q_1,Q_2}$ space can be written as
  \begin{multline}
    \di s^2 = Q_1 \left( \frac{\di x^2}{1+x^2} + \left(1+x^2 \right) \di
      u^2 \right) + Q_2 \left( \frac{\di y^2}{1+y^2} + \left(1+y^2
      \right) \di v^2 \right) +\\
    + \frac{2 Q_1 \left( Q_1 - 2 Q_2 \right)}{Q_2} \left( \di t + x \di
      u + \frac{Q_2}{Q_1} \sqrt{\frac{Q_2 - 2 Q_1}{Q_1 - 2 Q_2}} y \di v
    \right)^2 .
  \end{multline}
  
  Such cosets were studied in~\cite{PandoZayas:2000he} where they were
  found to be exact string backgrounds by using a construction very
  close to the asymmetric cosets of Ch.~\ref{cha:deformations}.

  \bigskip

  Of course since in general these geometries don't preserve any
  supersymmetry we should address the problem of their stability,
  which we do in the following section.

\section{Stability}
\label{sec:stability}

The set of solutions we found above are not in general protected by
supersymmetry.  This implies in particular that we should care about
their stability. In our analysis we will deal with the breathing modes
of the compact $H_n$ and $S^n$ internal manifolds which, in an
effective action, are to be described by scalar fields. The stability
(with respect to small fluctuations) will then translate intro the
positivity of the squared mass for such fields, condition that can be
relaxed into the respect of a Breitenlohner-Freedman bound when the
spacetime is of the anti-de~Sitter type.

First of all then we give an explicit derivation for the bound, so to
completely fix the notation and then construct the general expression
for the spacetime effective action for the breathing modes -- hence
finding again the same solutions as above as stationary points for a
potential whose Hessian matrix encodes the stability for the
background.

\subsection{Breitenlohner-Freedman bound}
\label{sec:bf-bound}

\paragraph{Anti-de Sitter}

Consider an action of the kind
\begin{equation}
  S = \int \di^d x \sqrt{-g^d}\left( R - \frac{1}{2} \partial_\mu \phi \partial^\mu \phi - V(\phi) \right)  
\end{equation}
the equation of motion for $\phi$ reads
\begin{equation}
  - \nabla_\mu \partial^\mu \phi = - \frac{\partial V}{\partial \phi}  
\end{equation}
The relations that we obtain are all tensorial so we can just choose a
suitable coordinate system, knowing that the result will remain
invariant. In $\mathrm{AdS}_d$ a good choice would be
\begin{equation}
  \di s^2 = \di r^2 + e^{2 H r} \left( -\di t^2 + \di x_1^2 + \dots + \di x_{d-2}^2 \right)  
\end{equation}
and we can consider a potential $V$ with a minimum in $\phi = 0$:
\begin{equation}
  V (\phi) = V_0 + \frac{m^2}{2} \phi^2 .
\end{equation}
The equation of motion for an $r$-dependent field $\phi$ reads
\begin{equation}
  \phi^{\prime \prime} + \left( d - 1 \right) H \phi^{\prime} - m^2 \phi = 0 .
\end{equation}
Solving it one can see that the presence of the friction term
effectively changes the mass to
\begin{equation}
  M^2 = \left( \frac{d-1}{2} H\right)^2 + m^2 \geq 0
\end{equation}
or, given that $R = - d \left( d - 1 \right) H^2$
\begin{equation}
  M^2 = - \frac{d-1}{4 d} R + m^2 \geq 0
\end{equation}
or, again, in terms of minimum of the potential:
\begin{equation}
  \label{eq:FB-bound}
  M^2 = - \frac{d-1}{4 \left(d - 2 \right)} \braket{V} + m^2 \geq 0 .
\end{equation}

Positivity of the effective mass squared, and thus stability,
translate therefore into a less stringent constraint for $m^2$. This
is the \textsc{bf} bound.

\paragraph{$L_{Q_1,Q_2}$ spaces.}

The presence of the bound is due to the curvature of the manifold. It
can be restated by saying that in an appropriate coordinate system the
Klein--Gordon equation can be put in the form
\begin{equation}
  - \Box \phi + m^2 \phi = \left( - \tilde \Box + \Delta \right) \phi + m^2 \phi = - \tilde \Box \phi  + M^2 \phi = 0 ,
\end{equation}
where $\Box$ is the d'Alembertian for the curved space, $\tilde \Box$
is the d'Alembertian for flat Minkowski space and $\Delta $ is some
constant depending on the curvature and other details of the geometry
($\Delta = - \left( d-1 \right)/ \left( 4 d \right) R$ in the case of
$\mathrm{AdS}_d$). It is natural to expect a similar behaviour for
other negative-curvature spaces, but the precise value of $\Delta $
will depend on the details. In particular it is interesting to
consider the $L_{Q_,1, Q_2}$ spaces introduced above.  Again, as
before, we can choose a coordinate system and then use the fact that
the equations we get are tensor relations and hence invariant. Take
into example the following metric:
\begin{multline}
  \di s^2 = Q_1 \left( \frac{\di x^2}{1+x^2} + \left(1+x^2 \right) \di
    u^2 \right) + Q_2 \left( \frac{\di y^2}{1+y^2} + \left(1+y^2
    \right) \di v^2 \right) +\\
  + \frac{2 Q_1 \left( Q_1 - 2 Q_2 \right)}{Q_2} \left( \di t + x \di
    u + \frac{Q_2}{Q_1} \sqrt{\frac{Q_2 - 2 Q_1}{Q_1 - 2 Q_2}} y \di v
  \right)^2 ,
\end{multline}
which describes an $L_{Q_1, Q_2} $ space with Ricci scalar
\begin{equation}
\label{eq:Ricci-LQQ}
  R = - \frac{5}{3} \frac{Q_1 + Q_2}{Q_1 Q_2} .
\end{equation}
The d'Alembertian on $\phi (x,y)$ gives:
\begin{equation}
  \Box \phi (x,y) = \frac{1}{Q_1} \left[ \left(1+x^2 \right) \phi_{xx} + 2 x \phi_x \right] + \frac{1}{Q_2} \left[ \left(1+y^2 \right) \phi_{yy} + 2 y \phi_y \right] .
\end{equation}
This is the same expression we would have got by considering the
d'Alembertian in an $\mathrm{AdS}_3 \times \mathrm{AdS}_3$ space with
coordinates \mathindent=0em
\begin{equation}
  \di s^2 =  \frac{1}{Q_1} \left[ \frac{\di x^2}{1+x^2} - \di t^2 - 2 x \di t \di u + \di u^2 \right] +  \frac{1}{Q_2} \left[ \frac{\di y^2}{1+y^2} - \di \tau^2 - 2 y \di \tau  \di v + \di v^2 \right] .
\end{equation}
\mathindent=\oldindent
The two subspaces have curvature
\begin{equation}
  R_i = -\frac{3}{2Q_i}  
\end{equation}
and since the shifts compose linearly this gives an overall shift
\begin{equation}
  M^2 = m^2 + \frac{1}{4Q_1} + \frac{1}{4Q_2} = m^2 + \frac{1}{4} \frac{Q_1 + Q_2}{Q_1 Q_2} ,
\end{equation}
which is therefore the \textsc{bf} bound for $L_{Q_1,Q_2}$ space. This
can be compared with an $\mathrm{AdS}_5$ space with the same curvature
as in Eq.~\eqref{eq:Ricci-LQQ}, where the shift would be given by
\begin{equation}
  M^2 = m^2 + \frac{1}{3}  \frac{Q_1 + Q_2}{Q_1 Q_2} . 
\end{equation}
In this case the shift is larger: in some sense then, as one might
have expected, an anti de-Sitter space is more stable with respect to
small fluctuations than a $L_{Q_1,Q_2}$ space with the same scalar
curvature.

\subsection{Effective low-dimensional description}
\label{sec:effect-low-dimens}

We want to write the $d$-dimensional effective action for a
$\mathrm{AdS}_{d} \times M_1 \times M_2$ background, where $M_i$ are
constant curvature spaces, starting from the following sector of the
M-theory action
\begin{equation}
  \label{eq:Initial action}
  S = \frac{1}{2\kappa_{11}^2}\int \di^{11} x \: \sqrt{-g^{(11)}} \left( R^{11} - V \right)
\end{equation}
where $V$ takes into account the presence of the fluxes. In order to
study the stability of the product background at hand let us consider
the following ansatz:
\begin{equation}
  \di s_{11}^2 = \di s^2 (M^{(d)} ) + e^{2 \varphi_1 (x)} \di s^2 (M^{(1)}) + e^{2 \varphi_2 (x)} \di s^2 (M^{(2)})  
\end{equation}
where the fields $\varphi_i$ depend only on the coordinates of
$M^{(d)}$. The strategy is the following:
\begin{enumerate}
\item write the curvature $R^{(11)}$ in terms of the curvature for $M^{(d)}$
\item separate the determinant $g^{(11)}$ in its compact and non-compact part
\item Weyl-rescale the four-dimensional metric to get a canonical
  Einstein-Hilbert action
\item rescale the scalar fields $\varphi_i$ to get a canonical kinetic term
\item study the potential and verify the stability taking into account
  the \textsc{bf} bound.
\end{enumerate}

\paragraph{Step a}
\label{sec:step-1}
 
For a warped product
\begin{equation}
  \di s^2 = \di s^2 (M^{(d)} (x)) + e^{2 \varphi(x)} \di s^2 (M^{(1)})    
\end{equation}
where $M^{(1)}$ has dimension $d_1$, the Ricci scalar is
\begin{equation}
  R = R^{(1)} + e^{-2\varphi(x)} R^{(2)} - 2 d_1 \nabla_\mu \partial^\mu \varphi(x) + d_1 \left(d_1 - 1 \right) \partial_\mu \varphi(x) \partial^\mu \varphi (x)    
\end{equation}
where the covariant derivative and the raising is done with the warped
product metric. Note that the result only depends on $d_1$ and not on
$d$. This easily generalizes to a warped product with three factors
like the ones we study
\begin{equation}
  \di s^2 = \di s^2 (M^{(d)} (x)) + e^{2 \varphi_1(x)} \di s^2 (M^{(1)}) +  e^{2 \varphi_2(x)} \di s^2 (M^{(2)})    
\end{equation}
and one obtains
\begin{multline}
  R^{(11)} = R^{(d)} + e^{-2\varphi_1(x)} R^{(1)} + e^{-2\varphi_2(x)} R^{(2)}- 2 d_1 \nabla_\mu
  \partial^\mu \varphi_1(x) + \\+ d_1 \left(d_1 - 1 \right) \partial_\mu \varphi_1(x) \partial^\mu \varphi_1(x) + 
  - 2 d_2 \nabla_\mu \partial^\mu \varphi_2(x) + \\ + d_2 \left( d_2 - 1 \right) \partial_\mu \varphi_2(x) \partial^\mu
  \varphi_2(x) + 2 d_1 d_2 \partial_\mu \varphi_1 (x) \partial^\mu \varphi_2(x)
\end{multline}
where $d_i = \dim M^{(i)}$.

Since the covariant derivative is with respect to the whole $g^{(11)}$
metric, two of the terms above are total derivatives and the action is
then equivalent to:
\begin{multline}
  S = \frac{1}{2\kappa_{11}^2} \int \di^{11} x \: \sqrt{-g^{(11)}} \left( R^{(11)} - V \right) \sim \\
  \sim  \frac{1}{2\kappa_{11}^2} \int \di^{11} x \: \sqrt{-g^{(11)}} \left( R^{(d)} +
    e^{-2\varphi_1(x)} R^{(1)} + e^{-2\varphi_2(x)} R^{(2)} +  \right. \\  + d_1
    \left(d_1 - 1
    \right) \partial_\mu \varphi_1(x) \partial^\mu \varphi_1(x) +
     d_2 \left( d_2 - 1 \right) \partial_\mu
    \varphi_2(x) \partial^\mu \varphi_2(x) + \\ \left. \phantom{R^{(d)}} 2 d_1 d_2 \partial_\mu
    \varphi_1 (x) \partial^\mu \varphi_2(x) - V \right)
\end{multline}

\paragraph{Step b}
\label{sec:step-2}

The eleven-dimensional determinant can be written as
\begin{equation}
  \det g^{(11)} = \det g^{(1)} \det g^{(2)} \det g^{(d)} e^{d_1 \varphi_1 + d_2 \varphi_2}
\end{equation}
in particular we can integrate over the internal coordinates and get:
\begin{multline}
  S = \frac{V_1 V_2}{2\kappa_{11}^2} \int \di^{d} x \: \sqrt{-g^{(d)}}
  e^{d_1 \varphi_1 + d_2 \varphi_2} \left( R^{(d)} +
    e^{-2\varphi_1(x)} R^{(1)} +
    e^{-2\varphi_2(x)} R^{(2)} + \right. \\
   + d_1 \left(d_1 - 1 \right) \partial_\mu
    \varphi_1(x) \partial^\mu \varphi_1(x) + d_2 \left( d_2 - 1
    \right) \partial_\mu \varphi_2(x) \partial^\mu \varphi_2(x) + \\
    \left. \phantom{R^{(d)}} + 2 d_1 d_2 \partial_\mu \varphi_1
      (x) \partial^\mu \varphi_2(x) - V\right)
\end{multline}
We will introduce
\begin{equation}
  \Psi = d_1 \varphi_1 + d_2 \varphi_2
\end{equation}
for future purposes.

\paragraph{Step c}
\label{sec:step-3}

The action above is not in the usual Hilbert-Einstein form because of
the $e^{d_1 \varphi_1 + d_2 \varphi_2}$ factor. For this reason we
perform a Weyl rescaling
\begin{equation}
  \bar g_{\mu \nu} = e^{2 \sigma(x)} g_{\mu \nu}
\end{equation}
under which the curvature becomes:
\begin{equation}
  \bar R = e^{-2\sigma(x)} R - 2 \left(d -1 \right) \bar \nabla_\mu \partial^\mu \sigma(x) +
  \left( d - 1 \right) \left( d - 2 \right) \partial_\mu \sigma (x) \partial^\mu \sigma(x) 
\end{equation}
where $\bar \nabla$ is the covariant derivative with respect to $\bar
g$ and $\mu$ is raised with $\bar g$.

In $d$ dimensions a term $\sqrt{\det g} e^{\Psi} R $ is brought to the
standard form by the Weyl rescaling
\begin{equation}
  \label{eq:Weyl-d-dim}
  \bar g_{\mu \nu} = \exp \left[ \frac{2}{d-2} \Psi \right]  g_{\mu \nu}
\end{equation}
and
\begin{equation}
  \sqrt{g} e^\Psi R = \sqrt{\bar g} \left( \bar R + 2 \frac{d-1}{d-2} \bar \nabla_\mu \partial^\mu \Psi - \frac{ d - 1 }{ d - 2} \partial_\mu \Psi \partial^\mu \Psi  \right)  .
\end{equation}
Discarding the total derivatives, the action now reads
\mathindent=0em
\begin{multline}
  S = \frac{V_1 V_2}{2 \kappa_{11}^2} \int \di^{d} x \: \sqrt{-\bar
    g^{(d)}} \left[ \bar R^{(d)} - \frac{ d - 1 }{ d - 2} \partial_\mu
    \Psi \partial^\mu \Psi + \right.  \\+ d_1 \left(d_1 - 1
  \right) \partial_\mu \varphi_1(x) \partial^\mu \varphi_1(x) + d_2
  \left( d_2 - 1 \right) \partial_\mu \varphi_2(x) \partial^\mu
  \varphi_2(x) +  \\ \left. +2 d_1 d_2 \partial_\mu \varphi_1 (x) \partial^\mu
  \varphi_2(x) + e^{-2\Psi_1 /\left( d-2 \right)} \left(
      e^{-2\varphi_1(x)} R^{(1)} + e^{-2\varphi_2(x)} R^{(2)} - V
      (\varphi_1, \varphi_2) \right) \right]
\end{multline}
\mathindent=\oldindent
where indices are raised with $\bar g^{(d)}$.

\paragraph{Step d}
\label{sec:step-4}

Now let us collect all the $\varphi $ terms in the action:
\begin{multline}
  S = \frac{V_1 V_2}{2\kappa_{11}^2} \int \di^d x \: \sqrt{-\bar g^{(d)}} \left[ \bar R^{(d)}
    - d_1 \left(\frac{d_1}{d-2} + 1 \right) \partial_\mu
    \varphi_1 \partial^\mu \varphi_1 + \right. \\
  \left. - d_2 \left(\frac{d_2}{d-2} + 1
    \right) \partial_\mu \varphi_2 \partial^\mu \varphi_2 - \frac{2 d_1 d_2}{d-2} \partial_\mu \varphi_1 \partial^\mu
    \varphi_2 - \bar V (\varphi_1, \varphi_2) \right]
\end{multline}
where
\begin{equation}
\label{eq:bar-potential}
  \bar V (\varphi_1, \varphi_2) =  e^{-2 \left( d_1 \varphi_1 + d_2 \varphi_2 \right) /\left( d-2
    \right)} \left( - e^{-2\varphi_1(x)} R^{(1)} - e^{-2\varphi_2(x)}
    R^{(2)} + V (\varphi_1, \varphi_2) \right) .
\end{equation}
To bring the kinetic terms to the standard form we introduce:
\begin{equation}
  \begin{cases}
    \Phi_1 = \sqrt{\frac{2 \left( D-2 \right)}{\left(d_1 + d_2 \right)
        \left( d - 2 \right)}} \left( d_1 \varphi_1 + d_2 \varphi_2
    \right) \\
    \Phi_2 = \sqrt{\frac{2 d_1 d_2}{d_1+d_2}} \left( \varphi_1 -
      \varphi_2 \right)
  \end{cases}
\end{equation}
or, the other way round:
\begin{equation}
\label{eq:phi-Phi}
  \begin{cases}
    \varphi_1 = \frac{1}{\sqrt{2 \left( d_1 + d_2 \right)}} \left(
      \sqrt{\frac{d-2}{D-2}} \Phi_1 + \sqrt{\frac{d_2}{d_1}}
      \Phi_2 \right) \\
    \varphi_1 = \frac{1}{\sqrt{2 \left( d_1 + d_2 \right)}} \left(
      \sqrt{\frac{d-2}{D-2}} \Phi_1 - \sqrt{\frac{d_1}{d_2}} \Phi_2
    \right)
 \end{cases}
\end{equation}
where $D= d+d_1+d_2 = 11$, and we finally obtain
\begin{equation}
  S = \frac{V_1 V_2}{2\kappa_{11}^2} \int \di^d x \: \sqrt{- g^{(d)}} \left[ 
    R^{(d)} - \frac{1}{2} \partial_\mu \Phi_1 \partial^\mu \Phi_1 -
    \frac{1}{2} \partial_\mu \Phi_2 \partial^\mu \Phi_2 -
    \bar V (\Phi_1, \Phi_2) \right]
\end{equation}

\paragraph{Step e}
\label{sec:step-5}

The type of backgrounds we obtain after compactification are
$\mathrm{AdS}$. Therefore, negative-$m^2$ modes are not tachyonic if
they don't cross the Breitenlohner-Freedman bound.

Using the results of the previous section this means:
\begin{equation}
  \label{eq:FB-bound}
  M^2 = - \frac{d-1}{4 \left(d - 2 \right)} \braket{V} + m^2 \geq 0
\end{equation}
that must be respected by each eigenvalue $m^2$ of the Hessian matrix
\begin{equation}
  m_{ij}^2 =  \left. \frac{\partial^2 V}{\partial \Phi_i \partial \Phi_j} \right|_{V = V_0}
\end{equation}

These general considerations apply to any choice of background gauge
fields. Then one should write the potential for each case at hand in
terms of $\Phi_1$ and $\Phi_2$ (and possibly the dilaton in type II) and
check the mass matrix against the \textsc{bf} bound.

\subsection{$\mathrm{AdS}_4 \times H^3 \times S^4$}
\label{sec:mathrmads_4-times-h3}

As a first example let us consider $\mathrm{AdS}_4 \times H^3 \times
S^4$. The potential $V$ in Eq.~\eqref{eq:Initial action} is due to the
presence of a four-form field on the $S^4$ part.
\begin{equation}
  V ( \varphi_1, \varphi_2 ) = \frac{Q^2}{2} e^{- 8 \varphi_2}
\end{equation}
the  dimensions are $d=4$, $d_1 = 3$ and $d_2 = 4$ and the curvatures
read $R^{(1)} = - 3/2$ and $R^{(2)} = 2$, so the expression in
Eq.~\eqref{eq:bar-potential} becomes:
\begin{equation}
  \bar V ( \varphi_1, \varphi_2 ) = e^{-3 \varphi_1 -4 \varphi_2} \left( \frac{3}{2} e^{-2 \varphi_1} - 2 e^{\varphi_2} + \frac{Q^2}{2} e^{-8 \varphi_2} \right) .
\end{equation}
To get the canonical scalar fields $\Phi_i$ and $\Phi_2$ we can use
Eq.~\eqref{eq:phi-Phi}:
\begin{equation}
  \begin{cases}
    \varphi_1 = \frac{1}{3 \sqrt{7}} \left( \Phi_1 + \sqrt{6} \Phi_2 \right) \\
    \varphi_2 = \frac{1}{12 \sqrt{7}} \left( 4 \Phi_1 -3 \sqrt{6} \Phi_2 \right)
  \end{cases}
\end{equation}
and this leads to the effective potential
\begin{equation}
  V(\Phi_1,\Phi_2) = \frac{1}{2} e^{-5 \Phi_1/ \sqrt{7} - 2 \Phi_2 \sqrt{2/21}} \left( 3 e^{2\Phi_1/ \sqrt{7}} - 4 e^{2 \Phi_1 / \sqrt{7} + \Phi_2 \sqrt{7/6}} + Q^2 e^{8 \Phi_2 \sqrt{2/21}} \right)   
\end{equation}
which has a minimum for
\begin{align}
  \Phi_1 = \sqrt{7} \log \frac{2^{8/7} Q}{\sqrt{3}} && \Phi_2 = \sqrt{\frac{6}{7}} \log 2  
\end{align}
and in correspondence of this point
\begin{equation}
  \braket{V} = - \frac{3 \sqrt{3}}{16 Q^3}  .
\end{equation}
The Hessian matrix  on the minimum is:
\begin{equation}
  \frac{\partial^2 V}{\partial \Phi_i \partial \Phi_j} = \frac{3}{28 Q^3}
  \begin{pmatrix}
    \frac{15 \sqrt{3}}{4} & - \frac{9}{\sqrt{2}} \\
    - \frac{9}{\sqrt{2}} & \frac{17 \sqrt{3}}{2}
  \end{pmatrix} .
\end{equation}
Both eigenvalues are positive
\begin{align}
  \left. m^2 \right|_1 = \frac{3 \sqrt{3}}{16 Q^3} && \left. m^2 \right|_2 = \frac{9 \sqrt{3}}{8 Q^3} 
\end{align}
and the solution is stable.


\subsection{$\mathrm{AdS}_{7-d_1} \times H^{d_1} \times S^4$}
\label{sec:mathrm-d_1-times}

All the backgrounds of the form $\mathrm{AdS}_{7-d_1} \times H^{d_1}
\times S^4$ can be treated similarly.
The potential in Eq.~\eqref{eq:bar-potential} becomes:
\begin{equation}
  \bar V ( \varphi_1, \varphi_2 ) = e^{-2 \left( d_1 \varphi_1 -4 \varphi_2 \right)/\left(5-d_1 \right)} \left( - \frac{d_1}{2} e^{-2 \varphi_1} - 2 e^{\varphi_2} + \frac{Q^2}{2} e^{ - 8 \varphi_2} \right) 
\end{equation}
and one finds that in each case there is a minimum (stable) solution.
The actual values are in Tab.~\ref{tab:AdS-7-d-H-S}. In any case, not
surprisingly one can see that for any $d_1$ the warping factors $\varphi_1 $
and $\varphi _2 $ always take the values
\begin{align}
  \varphi_1 = \frac{1}{6} \left( \log \frac{16 Q^2}{3} \right) && \varphi_2 =
  \frac{1}{6} \left( \log \frac{3 Q^2}{2} \right)
\end{align}
which agrees with $\Lambda_2 $ and $\Lambda_3 $ being
\begin{align}
  \Lambda_2 = - \left( \frac{3}{2} \right)^{1/3}\frac{1}{2 Q^{2/3}} &&
  \Lambda_3 = \left( \frac{3}{2} \right)^{1/3}\frac{1}{ Q^{2/3}}
\end{align}
as we have already seen by directly solving the equations of motion in
eleven dimensions.

\begin{table}
  \centering
  \begin{tabular}{|c|c|c|c|c|c|} \hline
    $d_1$ & $\braket{\Phi_1}$ & $\braket{\Phi_2}$ & $\braket{V}$ & $\frac{\left. m^2 \right|_1}{\abs{\braket{V}}}$ & $\frac{\left. m^2 \right|_2}{\abs{\braket{V}}}$ \\ \hline \hline
    $2$ & $ \log \frac{4 Q^2}{3}$ & $\sqrt{\frac{2}{3}} \log 2$ & $- \frac{9}{8 \times 2^{2/3} Q^2}$ & $2/3$ & $4$ \\ \hline
    $3$ & $ \frac{\sqrt{7}}{2} \log \frac{4^{8/7} Q^2}{3}$ & $\sqrt{\frac{6}{7}} \log 2$ & $- \frac{3 \sqrt{3}}{16 Q^{3}}$ & $1$ & $6$ \\ \hline
    $4$ & $2 \log \frac{4 \sqrt{2} Q^2}{3}$ & $\log 2$ & $- \frac{27}{512 Q^{6}}$ & $2$ & $12$ \\ \hline
  \end{tabular}
  \caption{Minima and masses for $\mathrm{AdS} \times H$ backgrounds}
  \label{tab:AdS-7-d-H-S}
\end{table}


\subsection{$\mathrm{AdS}_4 \times S^3 \times S^4$}
\label{sec:mathrmads_4-times-h3}

Consider now $\mathrm{AdS}_4 \times S^3 \times S^4$. The potential $V$
in Eq.~\eqref{eq:Initial action} is due to the presence of a
seven-form field on the $S^3 \times S^4$ factor.
\begin{equation}
  V ( \varphi_1, \varphi_2 ) = \frac{Q^2}{2} e^{- 6 \varphi_1 - 8 \varphi_2} .
\end{equation}
The dimensions are $d=4$, $d_1 = 3$ and $d_2 = 4$ and the curvatures
read $R^{(1)} = 3/2$ and $R^{(2)} = 2$, so the expression in
Eq.~\eqref{eq:bar-potential} becomes:
\begin{equation}
  \bar V ( \varphi_1, \varphi_2 ) = e^{-3 \varphi_1 -4 \varphi_2} \left( - \frac{3}{2} e^{-2 \varphi_1} - 2 e^{\varphi_2} + \frac{Q^2}{2} e^{-6 \varphi_1 - 8 \varphi_2} \right) .
\end{equation}
To get the canonical scalar fields $\Phi_i$ and $\Phi_2$ we can use
Eq.~\eqref{eq:phi-Phi}:
\begin{equation}
  \begin{cases}
    \varphi_1 = \frac{1}{3 \sqrt{7}} \left( \Phi_1 + \sqrt{6} \Phi_2 \right) \\
    \varphi_2 = \frac{1}{12 \sqrt{7}} \left( 4 \Phi_1 -3 \sqrt{6} \Phi_2 \right)
  \end{cases}
\end{equation}
and this leads to the effective potential
\begin{equation}
  V (\Phi_1, \Phi_2 ) = \frac{1}{2} e^{-\sqrt{7} \Phi_1 } \left( -3 e^{4 \Phi_1 / \sqrt{7} - 2 \Phi_2 \sqrt{2/21}} - 4 e^{4\Phi_1/ \sqrt{7} + \Phi_2 \sqrt{3/14}} + Q^2 \right) .
\end{equation}
In this case there is an extremum for
\begin{align}
  \Phi_1 = \frac{\sqrt{7}}{4} \log \frac{Q^2}{3} && \Phi_2 = 0
\end{align}
and in correspondence of this point:
\begin{equation}
  \braket{V} = - 2 \frac{3^{3/4}}{Q^{3/2}} .
\end{equation}
To verify the stability we write the Hessian matrix:
\begin{equation}
  \frac{\partial^2 V}{\partial \Phi_i \partial \Phi_j} = 2 \frac{3^{3/4}}{Q^{3/2}}
  \begin{pmatrix}
    3 & 0 \\
    0 & - \frac{1}{2}
  \end{pmatrix}
\end{equation}
which possesses a negative eigenvalue
\begin{align}
  \left. m^2 \right|_1 = 3 \abs{\braket{V}} && \left. m^2 \right|_2 = - \frac{1}{2} \abs{\braket{V}} .
\end{align}
This must be confronted with the BF bound we found in
Eq.~\eqref{eq:FB-bound}:
\begin{equation}
  M^2 = \frac{3}{8} \abs{\braket{V}} + m^2 = \left( \frac{3}{8} - \frac{1}{2} \right) \abs{\braket{V}} = - \frac{1}{8} \abs{\braket{V}} < 0 .
\end{equation}
The mode, corresponding to the ratio of the two radii $\Phi_2$, is
actually unstable. This is not so surprising since in this case the
flux is proportional to the total volume of the compact directions and
the system is intrinsically unstable with respect to a perturbation
that woukd keep this volume constant while changing the ratio between
the two radii.


\subsection{$\mathrm{AdS}_4 \times S^{d_1} \times S^{7-d_1}$}
\label{sec:mathrm-times-sd_1}

The field choice we made above proves to be very useful when dealing
with general $\mathrm{AdS}_4 \times S^{d_1} \times S^{7-d_1}$
backgrounds. In fact the potential reads:
\begin{multline}
  V (\Phi_1, \Phi_2 ) = \frac{1}{2} e^{-\sqrt{7} \Phi_1 } \left( -d_1
    e^{4 \Phi_1 / \sqrt{7} - \Phi_2 \sqrt{\left( -2 d_1 + 14 \right)/
        \left(7d_1 \right)}} + \right. \\ \left. - \left( 7-d_1
    \right) e^{4\Phi_1/ \sqrt{7} + \Phi_2 \sqrt{2 d_1/ \left( 7 \left(
            d_1 - 7 \right) \right)}} + Q^2 \right)
\end{multline}
where $d_1$ appears only as a coefficient for $\Phi_2$. Therefore the
solution is independent of the dimension $d_1$ and the stationary
point is again
\begin{align}
  \Phi_1 = \frac{\sqrt{7}}{4} \log \frac{Q^2}{3} && \Phi_2 = 0
\end{align}
with
\begin{equation}
  \braket{V} = - 2 \frac{3^{3/4}}{Q^{3/2}} .
\end{equation}

This solution is again unstable, since the square masses are again
\begin{align}
  \left. m^2 \right|_1 = 3 \abs{\braket{V}} && \left. m^2 \right|_2 = - \frac{1}{2} \abs{\braket{V}}
\end{align}
and the BF bound is always for $-3/8 \abs{\braket{V}}$ since it only
depends on the non-compact dimensions that are $d=4$.

\section{Type IIB backgrounds}
\label{sec:type-iib-backgrounds}

Some of the solutions we found thus far can be naturally reduced to
type II. This is the case when they contain factors of
$\mathrm{AdS}_3$ or odd-dimensional spheres for they can be
respectively written as space-like fibrations over $\mathrm{AdS}_2 $
or complex projective planes and as such, when compactified on the
fiber don't give rise to dilaton fields.

On the other hand one might also directly look for type II solutions
with the same type of geometry factorized in constant curvature
spaces. In this section we will in particular concentrate on type IIB
solutions with structure $V_4 \times M_3 \times \tilde M_3$. Before
starting one can try to make some educated guesses about the expected
kind of solutions. It would then appear rather natural to expect
perturbative $\mathrm{AdS}_4 \times S^3 \times S^3$ solutions that
might prove to be unstable (in the same spirit as
in~\cite{DeWolfe:2001nz}) and one might further imagine that adding
non-perturbative objects with negative tension -- orientifold planes
-- the no-go theorem for de Sitter \cite{Maldacena:2000mw} can be
contoured, thus allowing for internal hyperbolic manifolds and for de
Sitter solutions (as it was suggested in~\cite{Saltman:2004jh}). In
fact we will prove that both guesses are ultimately wrong by carefully
studying the effective potential in four dimensions obtained by taking
into account dilaton, \textsc{rr} zero-form and the two breathing
modes for the internal manifolds. More precisely we will show that:
\begin{itemize}
\item no truly perturbative solution exists, \emph{i.e.} the presence
  of D-branes is necessary;
\item in presence of orientifolds the only possible compactification
  happens on a $T^6$;
\item no solution is allowed with hyperbolic or de~Sitter components;
\item the only allowed solution is $\mathrm{AdS}_4 \times S^3 \times S^3$ and
  this solution is perturbatively stable, thanks to the \textsc{bf}
  bound, although it can't be completely trusted since it belongs to
  an intermediate-coupling regime.
\end{itemize}

Let us consider a type IIB background of the form $V_4 \times M_3
\times \tilde M_3$.  The action in Einstein frame reads:
\begin{multline}
  S = S_0 + S_{\text{loc}} = \frac{1}{2 \kappa_{10}^2} \int \di^{10} x \sqrt{-g^{(10)}} \left[ R^{(10)} - \frac{\partial_\mu \tau \partial^\mu \bar \tau}{2 \left( \im \tau \right)^2 } - \frac{1}{2 } M_{ij} F_3^i F_3^j + \right. \\ \left. - \frac{1}{2} C_4 \land F_3^1 \land F_3^2 \right] + S_{\text{loc}}  
\end{multline}
where we didn't include the 1-form and the 5-form which are not
compatible with the symmetries of the metric ansatz. As usual $\tau = C_0 +
\imath e^{-\Phi}$ is the dilaton-axion field, $M_{ij}$ is
\begin{equation}
  M_{ij} = \frac{1}{\im \tau}
  \begin{pmatrix}
    \abs{\tau}^2 & - \re \tau \\
    -\re \tau & 1
  \end{pmatrix}
\end{equation}
and $S_{\text{loc}}$ is the contribution due to $D_3$-branes and $O_3$-planes:
\begin{equation}
  S_{\text{loc}} = N T_3 \int \di^4 x \sqrt{-g^{(4)}} .
\end{equation}

As before we look for a solution of the kind
\begin{equation}
  \di s_{(10)}^2 = \di s^2_{(4)} + e^{2 \varphi(x)} \di s_3^2 + e^{2 \tilde \varphi(x)} \di s^2_{(3)}  
\end{equation}
where for the moment being the two internal manifolds can have
positive or negative curvature. With respect to the M-theory
situation, in this case we will have to pay attention to the extra
$\tau$ complex scalar field and to the extra-term in the action
$S_{\text{loc}}$. 

Let's start with the latter. This already has the form of a
four-dimensional integral, so to evaluate the corresponding
contribution to the potential we only need to take into account the
Weyl rescaling in Eq.~\eqref{eq:Weyl-d-dim} and the presence of the
internal volumes and Newton constant:
\begin{equation}
  S_{\text{loc}} = N T_3 \int \di^4 x \: \sqrt{-g^{(4)}} = 
  \frac{V_3 \tilde V_3}{2 \kappa_{10}^2}\frac{2 N T_3 \kappa_{10}^2}{V_3 \tilde V_3} 
  \int \di^4 x \: e^{-2\Psi_1} \sqrt{\bar g^{(4)}}  
\end{equation}
and then
\begin{equation}
  V_{\text{loc}} = \frac{2 N T_3 \kappa_{10}^2}{V_3 \tilde V_3} e^{-6 \varphi - 6 \tilde \varphi} .
\end{equation}

For reasons that will appear more clear in the following let us
rescale the $\tau $ field as:
\begin{equation}
  z = \frac{\tau}{k}  
\end{equation}
$k $ being a real constant. This does not affect the kinetic
term:
\begin{equation}
  \frac{\partial_\mu \tau \partial^\mu  \bar \tau}{2 \left( \im \tau  \right)^2} =   \frac{\partial_\mu z \partial^\mu  \bar z}{2 \left( \im z  \right)^2}
\end{equation}
but makes it canonical when studying the small perturbations around
the equilibrium solutions.

The most general three-field configuration compatible with the
symmetries is
\begin{equation}
  F_3^i = 4 \pi^2 \alpha^{\prime } \left( n_i \omega_3 + \tilde n_i \tilde \omega_3 \right)  
\end{equation}
where $n_i$ and $\tilde n_i$ are integers and $\omega_3$ and $\tilde
\omega_3$ are the normalized volume forms. The subscript $1$ stands
for \textsc{rr} field, $2$ for \textsc{ns}. The corresponding potential reads:
\begin{equation}
  V_F =  \frac{\left(4 \pi^2 \right)^2 \alpha^{\prime 2} }{2 k \im z V_3^2} \abs{n_1 k z - n_2}^2 e^{-6 \varphi }  + \frac{\left(4 \pi^2 \right)^2 \alpha^{\prime 2} }{2 k \im z \tilde V_3^2} \abs{\tilde n_1 k z - \tilde n_2}^2 e^{-6 \tilde \varphi }
\end{equation}
in order to clean the notation let us write the volumes as
\begin{align}
  V_3 = 2 x \pi^2 && \tilde V_3 = 2 \tilde x \pi^2 .  
\end{align}
Adding the contribution from the curvatures of the internal manifolds:
\begin{align}
  R = \frac{3}{2} \epsilon && \tilde R = \frac{3}{2} \tilde \epsilon   
\end{align}
where $\epsilon = 1$ for spheres and $\epsilon = -1$ for hyperbolic spaces, we can write the potential as
\begin{multline}
  V (\varphi, \tilde \varphi, z) = e^{-3 \varphi -3 \tilde \varphi}
  \left( - \frac{3\epsilon}{2} e^{-2 \varphi } - \frac{3 \tilde
      \epsilon }{2} e^{-2 \tilde \varphi} + \frac{2 \alpha^{\prime 2} }{x^2 k \im z}
    \abs{n_1 k z - n_2}^2 e^{-6 \varphi } + \right. \\ \left. + \frac{2 \alpha^{\prime 2}
    }{x^2 k \im z} \abs{\tilde n_1 k z - \tilde n_2}^2 e^{-6 \tilde
      \varphi }\right)  + \frac{2 N T_3 \kappa_{10}^2}{x \tilde x
    \left(2 \pi^2 \right)^2} e^{-6 \varphi - 6 \tilde \varphi} .
\end{multline}
As a first step we normalize the scalars $\varphi $ and $\tilde \varphi $:
\begin{align}
  \varphi = \frac{\Phi_1 + 2 \Phi_2}{4 \sqrt{3}} && \tilde \varphi = \frac{\Phi_1 -2 \Phi_2}{4 \sqrt{3}}  
\end{align}
so to rewrite the potential as
\begin{multline}
  V (\Phi_1, \Phi_2, z ) = e^{- \Phi_1 \sqrt{3}/2} \left( -
    \frac{3}{2} \epsilon e^{-\left( \Phi_1 + 2 \Phi_2 \right)/\left(2
        \sqrt{3} \right)} - \frac{3}{2} \epsilon e^{-\left( \Phi_1 - 2
        \Phi_2 \right)/\left(2 \sqrt{3} \right)} + \right. \\
  + \frac{2 \alpha^{\prime 2} }{x^2 k \im z} \abs{n_1 k z - n_2}^2
  e^{- \left( \Phi_1 + 2 \Phi_2 \right) \sqrt{3}/2} + \\ \left.
    \frac{2 \alpha^{\prime 2} }{x^2 k \im z} \abs{\tilde n_1 k z -
      \tilde n_2}^2 e^{- \left( \Phi_1 - 2 \Phi_2 \right) \sqrt{3}/2}
  \right) + \frac{2 N T_3 \kappa_{10}^2}{x \tilde x \left(2 \pi^2
    \right)^2} e^{-\Phi_1 \sqrt{3}} .
\end{multline}
We have to take into account the anomaly cancellation condition
\begin{equation}
  N = n_1 \tilde n_2 - n_2 \tilde n_1  
\end{equation}
and express all the constants in terms of $\alpha^{\prime}$:
\begin{align}
  T_3 = \abs{\mu_3} = \frac{1}{\left(2 \pi \right)^2 \alpha^{\prime 2}} && 2 \kappa_{10}^2 = \left(2 \pi \right)^7 \alpha^{\prime 4}  
\end{align}
so that $2 \kappa_{10}^2 T_3 = \left(2 \pi \right)^4 \alpha^{\prime 2}$.

Up to this moment the sign of $N$ is arbitrary, so one might think
that different kinds of solution are possible. Actually this is not
the case. If we differentiate with respect to $\im z $ and $\re z$ we
obtain the values corresponding to the stationary point
\begin{align}
  \im z = \frac{e^{\Phi_2 \sqrt{3}} \abs{N} x \tilde x}{k \left( e^{2
        \Phi_2 \sqrt{3}} \tilde n_1^2 x^2 + n_1^2 \tilde x^2 \right)}
  && \re z = \frac{ e^{2 \Phi_2 \sqrt{3}} \tilde n_1 \tilde n_2 x^2 +
    n_1 n_2 \tilde x^2 }{k \left( e^{2 \Phi_2 \sqrt{3}} \tilde n_1^2
      x^2 + n_1^2 \tilde x^2 \right)} .
\end{align}
Putting these back into the potential one gets:
\mathindent=0em
\begin{equation}
  V (\Phi_1, \Phi_2 ) = \frac{e^{\left( 3 \Phi_1 + \Phi_2 \right)/ \sqrt{3}}}{2 x \tilde x} \left( 8 e^{\Phi_2 / \sqrt{3}} \left( N + \abs{N} \right) - 3 x \tilde x e^{\Phi_1/ \sqrt{3}} \left( \epsilon + \tilde \epsilon e^{\left( \Phi_1 + 2 \Phi_2 \right)/ \sqrt{3}} \right) \right)
\end{equation}
\mathindent=\oldindent and for $N \gtrless 0 $ this gives
respectively:
\begin{align}
  V^+ ( \Phi_1, \Phi_2 ) &= \frac{e^{\left( 3 \Phi_1 + \Phi_2 \right)/ \sqrt{3}}}{2 x \tilde x} \left( 16 e^{\Phi_2 / \sqrt{3}} N - 3 x \tilde x e^{\Phi_1/ \sqrt{3}} \left( \epsilon + \tilde \epsilon e^{\left( \Phi_1 + 2 \Phi_2 \right)/ \sqrt{3}} \right) \right) \\
  V^- ( \Phi_1, \Phi_2 ) &= - \frac{3}{2} e^{-\left(2 \Phi_1 + \Phi_2
    \right)/ \sqrt{3}} \left(\epsilon + \tilde \epsilon e^{2 \Phi_2 /
      \sqrt{3}} \right) .
\end{align}
The second potential only admits a solution if $\epsilon = \tilde
\epsilon =0$, but this corresponds to a flat internal space
(Polchinski).

Choosing the first $N>0$ case one finds a stationary point for
\begin{align}
  \braket{\Phi_1} = \sqrt{3} \log \frac{4 N \alpha^{\prime 2} }{x
    \tilde x \tilde \epsilon} \sqrt{ \frac{\tilde \epsilon}{\epsilon}}
  && \braket{\Phi_2} = \frac{\sqrt{3}}{2} \log \frac{\epsilon}{\tilde
    \epsilon}
\end{align}
which imposes $\epsilon = \tilde \epsilon = 1$, \emph{ie} the only
solution has an internal $S^3 \times S^3 $ space, as advertized above.

To summarize, the stationary point corresponds to:
\begin{align}
  \braket{\im z} = \frac{N x \tilde x}{k \left( \tilde n_1^2 x^2 +
      n_1^2 \tilde x^2 \right)} && \braket{\re z} = \frac{\tilde n_1
    \tilde n_2 x^2 + n_1 n_2 \tilde x^2}{k \left( \tilde n_1^2 x^2 +
      n_1^2 \tilde x^2 \right)} \\ \braket{\Phi_1} = \sqrt{3} \log \frac{4 N
    \alpha^{\prime 2}}{x \tilde x} && \braket{ \Phi_2} = 0
\end{align}
now we can choose
\begin{equation}
  k = \frac{N x \tilde x}{\tilde n_1^2 x^2 + n_1^2 \tilde x^2}  
\end{equation}
and putting $x =\tilde x =1$ (as we must in the case of spheres) we
have
\begin{align}
  \braket{\im z} = 1 && \braket{\re z} = \frac{\tilde n_1 \tilde n_2 +
    n_1 n_2}{n_1 \tilde n_2 - n_2 \tilde n_1} && \braket{\Phi_1} =
  \sqrt{3} \log \left( 4 N \alpha^{\prime 2} \right) && \braket{\Phi_2
  }= 0
\end{align}
and the corresponding potential is
\begin{equation}
  \braket{V} = - \frac{1}{16 N^2 \alpha^{\prime 4}} .
\end{equation}

To check the stability we write the Hessian matrix and compute its
eigenvalues. They turn out to be
\begin{align}
  \left. m^2 \right|_1 = \abs{\braket{V}} && \left. m^2 \right|_2 = 2 \abs{\braket{V}} \\
  \left. m^2 \right|_3 = \frac{\sqrt{13}+3}{2} \abs{\braket{V}} && \left. m^2 \right|_4 = -
  \frac{\sqrt{13}-3}{2} \abs{\braket{V}}
\end{align}
and the only negative one $\left. m^2 \right|_4$ doesn't cause any instability since
it doesn't cross the \textsc{bf} bound:
\begin{equation}
  \left. m^2 \right|_4 + \frac{3}{8} \abs{\braket{V}} = \frac{15 - 4 \sqrt{13}}{8} \abs{\braket{V}} > 0  
\end{equation}

This result about the stability of the product of two three-spheres
can at first sight be puzzling since, after analyzing the results of
Sec.~\ref{sec:mathrm-times-sd_1}, we've grown to expect such
configurations to be unstable under the mode in which one of the two
spheres shrinks and the other grows, keeping constant the overall
volume. Here it is not the case and this can be easily understood in
terms of gauge fields. In the M-theory configurations, in fact, only
one field was turned on and it spanned over the whole internal
manifold, so that it effected only the total volume. Here, on the
other hand, \textsc{rr} and \textsc{ns-ns} fields are turned on
separately on the two submanifolds, thus contributing to the
stabilization of each of the radii.


\chapter{Conclusions and further perspectives}

\setlength{\epigraphwidth}{18em}
\begin{epigraphs}
  \qitem{Beauty is truth, truth beauty, -- that is all\\
    Ye know on earth, and all ye need to know.}{Ode on a Grecian Urn\\
    \textsc{John Keats}}
\end{epigraphs}

The search for exact string solutions is a fascinating subject by
itself. It is based, as most of the wonders of string theory, on the
interplay between the two-dimensional conformal field theory
description on the world-sheet and the ten-dimensional low-energy
interpretation in terms of spacetime fields. 

In this thesis we have dealt with a new class of string backgrounds
living in the moduli space of \textsc{wzw} models. They enjoy at the
same time nice supergravity properties, all geometrical quantities
being naturally expressed in terms of algebraic invariants, and a
clear \textsc{cft} characterization, inherited from the beautiful
theory of group manifolds.

Apart from their intrinsic elegance, those new backgrounds also find
interesting physical applications as compactification manifolds,
laboratories for the analysis of string propagation in classically
pathological backgrounds with closed time-like curves, in black hole
configurations with non-trivial topology. Laboratories in which we can
keep higher order effects under control and write down a
modular-invariant partition function, or at least the spectra of
primary operators.

Starting from this solid \textsc{cft} ground we are then allowed to
peep into the off-shell physics using \textsc{rg} techniques, both
from a two-dimensional and field-theoretical perspective. We can thus
observe the relaxation of out-of-equilibrium vacua described by charge
transmutation, \emph{i.e.}  two gauge fields eventually collapsing
into a single one, while the total charge is conserved. A new change
of viewpoint then allows us to recast the problem in terms of a
cosmological time-dependent solution. The \textsc{rg} dynamics becomes
an approximate description -- valid in a certain region of the moduli
space depending on the central charge deficit -- for the behaviour of
a Big-bang-like isotropic \textsc{frw} universe.

We have emphasized many times the importance of having a \textsc{cft}
description and exact models, but this is not possible in general. For
example we completely lack such a kind of interpretation for type II
string or M-theory, but -- and maybe for this very reason -- it is
important to look for new insights in these frameworks. In this spirit
we have studied compactifications involving maximally symmetric spaces
which in general do not preserve supersymmetry, looking in particular
for hyperbolic solutions, in the not-too-concealed hope of reaching
de~Sitter-like spacetimes.

\bigskip

So, what remains to do? By its very nature this is a work in
progress. The path to further developments is full of technical and
conceptual obstacles but one can easily name some of the natural
possible directions. First of all it would be interesting to have a
non-Abelian counterpart for our asymmetric gauging. This would be
allowed by the heterotic string framework and it does indeed work at
the supergravity level; on the other hand it is not clear how it could
be implemented in a \textsc{cft} framework -- the evidence at hand
pointing towards a discrete structure for the deformations. Then, one
would like to reach a better understanding of non-rational
\textsc{cft}s and, in particular, of the $SL(2,\setR)$ \textsc{wzw}
model; this would allow us to write the partition function for
$\mathrm{AdS}_2$ spacetime, Bertotti-Robinson black hole and (charged)
black string. This, in turn, might prove useful for obtaining a
non-trivial microscopic description for the thermodynamics of such
singular objects. Another more phenomenological direction would be to
study the low energy field theory consequence of a compactification on
geometric cosets, again using the complete knowledge which we have of
the spectrum and partition function in this case. As soon as we move
away from the familiar \textsc{cft} framework things become more
difficult and much more interesting. Even if a theory for the
two-dimensional \textsc{rg} flow is established, there are very few
cases in which one can really work out non-trivial examples such as
the flow one expects to link \textsc{wzw} models at different
levels. The spacetime description might then prove useful for getting
new hints at the world-sheet physics.


\printindex

\appendix

\begin{small}
  
\chapter{Table of conventions}
\label{cha:table-conventions}

\begin{tabular}{lll}
  Christoffel symbol & ${\kappa \atopwithdelims\{\} \mu \nu  }$ & $\frac{1}{2}g^{\kappa \lambda} \left( \partial_\mu g_{\lambda \nu} + \partial_\nu g_{\lambda \mu} - \partial_\lambda g_{\mu \nu} \right)$ \\ 
  covariant derivative & $\nabla$ & $\nabla_\nu t^{\lambda_1\ldots\lambda_p}_{\mu_1\ldots\mu_q} = \partial_\nu  t^{\lambda_1\ldots\lambda_p}_{\mu_1\ldots\mu_q} + \Gamma\ud{\lambda_1}{\nu \kappa } t^{\kappa \lambda_2\ldots\lambda_p}_{\mu_1\ldots\mu_q} + \ldots - \Gamma\ud{\kappa}{\nu \mu_1}  t^{\lambda_1\ldots\lambda_p}_{\kappa \mu_2\ldots\mu_q} -\ldots$ \\
  dual Coxeter number & $g^\ast $ & $g^\ast =\frac{1}{2 \dim G} f_{\alpha \beta \gamma} f^{\alpha \beta \gamma} $ \\
  exterior derivative & $\di$ & $\di \omega = \frac{1}{n!} \deriv{\omega_{\mu_1 \mu_2 \ldots \mu_n }}{x^\lambda }  \di x^\lambda \land \di x^1 \land \di x^2 \land \ldots \land \di x^n$ \\
  Maurer-Cartan one-form & $\mJ^a$ & $\mJ^a = \braket{t^a g^{-1} \di g}$ \\
  $n$-form & $\omega $ & $\omega = \frac{1}{n!} \omega_{\mu_1 \mu_2 \ldots \mu_n } \di x^1 \land \di x^2 \land \ldots \land \di x^n $ \\
  structure constants & $f\ud{\alpha}{\beta \gamma} $ & $\comm{t^\alpha , t^\beta} = f\ud{\alpha \beta}{\gamma} t^\gamma$ \\
  trace & $\braket{}$ & $\braket{M_{ij}} = \sum_i M_{ii}$
\end{tabular}


\chapter{Explict parametrizations for some Lie groups}
\label{cha:geometry-arond-group}

\chapterprecishere{In this appendix we collect the explicit
  parametrizations used for the $SU(2)$, $SL(2,R)$, $SU(3)$ and
  $USp(4) $ groups.}

\section{The three-sphere}
\label{sph}

The commutation relations for the generators of $SU(2)$ are
\begin{align}
 \label{eq:comm-su2}
  \comm{ J^1 , J^2} = \imath J^3 && \comm{ J^2 , J^3} = \imath J^1 &&
  \comm{ J^3 , J^1 } = \imath J^2.
\end{align}
A two-dimensional realization is obtained by using the standard
Pauli matrices\footnote{The normalization of the generators with
respect to the Killing product in $\mathfrak{su} \left( 2\right)$:
$\kappa \left( X, Y\right) = \tr \left( X Y\right)$ is such that
$\kappa \left( J^a, J^b \right) = 1/2$ and correspondingly the
root has length squared $\psi = 2$.}$\sigma^a$: $J^a =  \sigma^a /
2$.

The Euler-angle parameterization for $SU(2)$ is defined as:
\begin{equation}
g = { e}^{\imath {\gamma \over 2} \sigma^3} {e}^{\imath
{\beta \over 2} \sigma^1} {e}^{\imath {\alpha \over 2}
\sigma^3}.
\end{equation}
The $SU(2)$ group manifold is a unit-radius
three-sphere. A three-sphere can be embedded in flat Euclidean
four-dimensional space with coordinates $(x^1,x^2,x^3,x^4)$, as
$(x^1)^2 + (x^2)^2 + (x^3)^2 + (x^4)^2 = L^2$. The corresponding
$SU(2)$ element $g$ is the following:
\begin{equation}
  g  =  L^{-1}\
  \begin{pmatrix}
    x^4 + \imath x^2 &  x^3 + \imath x^1 \\ -
      x^3 + \imath x^1 & x^4 - \imath x^2
  \end{pmatrix}. \label{4embS3}
\end{equation}

In general, the invariant metric of a group manifold can be
expressed in terms of the left-invariant Cartan--Maurer one-forms.
In the $SU(2)$ case under consideration (unit-radius $S^3$),
\begin{align}
  \mathcal{J}^1 =\frac{1}{2}\tr \left( \sigma^1 g^{-1} \di g \right),&&
  \mathcal{J}^2 = \frac{1}{2}\tr \left( \sigma^2 g^{-1} \di g
  \right),&& \mathcal{J}^3 = \frac{1}{2}\tr \left( \sigma^3 g^{-1} \di g
  \right)
\end{align}
and
\begin{equation}
  \label{eq:ads-metrix}
  \di s^2 = \sum_{i=1}^3 \mathcal{J}^i \otimes \mathcal{J}^i
\end{equation}
The volume form reads:
\begin{equation}
  \label{eq:ads-vf}
  \omega_{[3]} = \mathcal{J}^1 \land \mathcal{J}^2 \land
  \mathcal{J}^3.
\end{equation}

In the Euler-angle parameterization, Eq.~(\ref{eq:ads-metrix})
reads (for a radius-L three-sphere):
\begin{equation}
  \di s^2 =
  \frac{L^2}{4} \left( \di \alpha^2 + \di \gamma^2 + 2 \cos \beta
    \di \alpha \di \gamma + \di \beta^2\right),
\end{equation}
whereas (\ref{eq:ads-vf}) leads to
\begin{equation}
  \label{eq:su2-vf}
  \omega_{[3]} =\frac{L^3}{8} \sin \beta \di \alpha \land \di \beta \land \di
  \gamma.
\end{equation}
The Levi--Civita connection has scalar curvature $R = 6/L^2$.

The isometry group of the $SU(2)$ group manifold is generated by
left or right actions on $g$: $g\to hg$ or $g\to gh$ $\forall h
\in SU(2)$. From the four-dimensional point of view, it is
generated by the rotations $\zeta_{ab}= i\left( x_a\partial_b -
x_b
  \partial_a\right)$ with $x_a=\delta_{ab}x^b$. We list here explicitly the
six generators, as well as the group action they correspond to:
\begin{subequations}
  \label{eq:Killing-SU2}
  \begin{align}
    L_1 &= \frac{1}{2}\left(-\zeta_{32} + \zeta_{41}\right), & g &\to
    \mathrm{e}^{-\imath\frac{\lambda}{2}\sigma^1}g,
    \label{SL1} \\
    L_2 &= \frac{1}{2}\left(-\zeta_{43}-\zeta_{12} \right), &  g &\to
    \mathrm{e}^{\imath\frac{\lambda}{2}\sigma^2}g,
    \label{SL2} \\
    L_3 &= \frac{1}{2}\left(-\zeta_{31} - \zeta_{42}\right), & g&\to
    \mathrm{e}^{\imath\frac{\lambda}{2}\sigma^3}g,
    \label{SL3} \\
    R_1 &= \frac{1}{2}\left( \zeta_{41} + \zeta_{32}\right), & g&\to
    g\mathrm{e}^{\imath\frac{\lambda}{2}\sigma^1},
    \label{SR1} \\
    R_2 &= \frac{1}{2}\left(-\zeta_{43} + \zeta_{12}\right), &  g&\to
    g\mathrm{e}^{\imath\frac{\lambda}{2}\sigma^2},
    \label{SR2} \\
    R_3 &= \frac{1}{2}\left(\zeta_{31} - \zeta_{42}\right), & g&\to
    g\mathrm{e}^{\imath\frac{\lambda}{2}\sigma^3}.
    \label{SR3}
  \end{align}
\end{subequations}
Both sets satisfy the algebra (\ref{eq:comm-su2}). The norms
squared of the Killing vectors are all equal to $L^2/4$.

The currents of the $SU \left( 2 \right)_k $ \textsc{wzw} model
are easily obtained as:
\begin{align}
\label{eq:su2-currents}
  J^i = - k \tr \left(\imath \sigma^i \partial g g^{-1} \right) &&
  \bar J^i = -k \tr \left(\imath \sigma^i  g^{-1} \bar \partial g
  \right),
\end{align}
where $L=\sqrt{k}$, at the classical level. Explicit expressions
are given in Tab. \ref{tab:su2-currents}.
\newcommand{\LSU}[0]{
  \begin{minipage}{.4\textwidth}
    \vspace{-.7em}
    \begin{small}
      \begin{gather*}
        \frac{\sin \gamma }{\sin \beta} \partial_\alpha + \cos \gamma \partial_\beta -
        \frac{\sin \gamma }{\tan \beta} \partial_\gamma \\
        \frac{\cos \gamma }{\sin \beta} \partial_\alpha - \sin \gamma \partial_\beta -
        \frac{\cos \gamma }{\tan \beta} \partial_\gamma \\
        \partial_\gamma
      \end{gather*}
    \end{small}
    \vspace{-1.5em}
  \end{minipage}
}
\newcommand{\RSU}[0]{
  \begin{minipage}{.4\textwidth}
    \vspace{-.7em}
    \begin{small}
      \begin{gather*}
        - \frac{\sin \alpha }{\tan \beta } \partial_\alpha + \cos \alpha \partial_\beta
        + \frac{\sin \alpha }{\sin \beta} \partial_\gamma \\
        \frac{\cos \alpha}{\tan \beta} \partial_\alpha + \sin \alpha \partial_\beta -
        \frac{\cos \alpha }{\sin \beta} \partial_\gamma \\
        \partial_\alpha
      \end{gather*}
    \end{small}
    \vspace{-1.5em}
  \end{minipage}
}
\newcommand{\JSU}[0]{
  \begin{minipage}{.4\textwidth}
    \vspace{-.7em}
    \begin{small}
      \begin{gather*}
         k \left( \sin \beta \sin \gamma \partial \alpha + \cos
          \gamma \partial b\right) \\[.5em]
         k \left( \cos \gamma \sin \beta \partial \alpha - \sin \gamma \partial \beta \right) \\[.5em]
         k \left(\partial \gamma + \cos \beta \partial \alpha \right)
      \end{gather*}
    \end{small}
    \vspace{-1.5em}
  \end{minipage}
}
\newcommand{\JbSU}[0]{
  \begin{minipage}{.4\textwidth}
    \vspace{-.7em}
    \begin{small}
      \begin{gather*}
         k \left( \cos \alpha \bar \partial \beta + \sin \alpha
          \sin \beta \bar \partial \gamma\right) \\[.5em]
         k \left( \sin \alpha \bar \partial \beta - \cos \alpha \sin \beta
          \bar \partial \gamma\right)\\[.5em]
         k \left( \bar \partial \alpha + \cos \beta \bar \partial \gamma\right)
      \end{gather*}
    \end{small}
    \vspace{-1.5em}
  \end{minipage}
}
\mathindent=0em

\begin{table}
  \centering
  \begin{tabular}{|c|c|c|}
    \hline sector & Killing vector& Current \\ \hline \hline
    \begin{rotate}{90}\hspace{-2em}left moving\end{rotate} & \LSU &
    \JSU \\ \hline
    \begin{rotate}{90}\hspace{-2.5em}right moving\end{rotate} & \RSU &
    \JbSU \\\hline
  \end{tabular}
  \caption{Killing vectors $\{ \imath L_1,\imath L_2,\imath L_3 \} $ and $\{ \imath R_1,\imath R_2, \imath
    R_3 \} $, and holomorphic and anti-holomorphic currents for $SU(2)$ in Euler angles.}
  \label{tab:su2-currents}
\end{table}
\mathindent=\oldindent


\section{$\mathrm{AdS}_3$}
\label{antids}

The commutation relations for the generators of the $SL(2,\mathbb{R})$
algebra are
\begin{align}
  \comm{ J^1 , J^2} = - \imath J^3 && \comm{ J^2 , J^3} = \imath J^1 &&
  \comm{ J^3 , J^1 } = \imath J^2.
  \label{comm}
\end{align}
The sign in the first relation is the only difference with respect
to the $SU(2)$ in Eq.~\eqref{eq:comm-su2}.

The three-dimensional anti-de-Sitter space is the universal
covering of the $SL(2,\mathbb{R})$ group manifold. The latter can
be embedded in a Lorentzian flat space with signature $(-,+,+,-)$
and coordinates $(x^0,x^1,x^2,x^3)$:
\begin{equation}
  g  =  L^{-1}\
  \begin{pmatrix}
    x^0 + x^2 & x^1 + x^3 \\ x^1 -
      x^3 & x^0 - x^2
  \end{pmatrix}, \label{4emb}
\end{equation}
where $L$ is the radius of $\mathrm{AdS}_3$. On can again introduce
Euler-like angles
\begin{equation} g = \mathrm{e}^{\imath (\tau+\phi) \sigma_2/2}
\mathrm{e}^{\rho \sigma_1} \mathrm{e}^{\imath(\tau-\phi)
\sigma_2/2} , \label{euler}
\end{equation}
which provide good global coordinates for $\mathrm{AdS}_3$ when $\tau\in
\left(-\infty,+\infty \right)$, $\rho\in \left[ 0,\infty\right)$, and $\phi\in\left[0,2\pi
\right)$.

An invariant metric (see Eq.~\eqref{eq:ads-metrix}) can be
introduced on $\mathrm{AdS}_3$. In Euler angles, the latter reads:
\begin{equation}
\di s^2= L^2\left[- \cosh ^2 \rho  \, \di \tau ^2 +\di \rho^2 +
\sinh^2 \rho \, \di \phi^2\right]. \label{dseul}
\end{equation}
The Ricci scalar of the corresponding Levi--Civita connection is
$R=-6/L^2$.

The isometry group of the $SL(2,\mathbb{R})$ group manifold is
generated by left or right actions on $g$: $g\to hg$ or $g\to gh$
$\forall h \in SL(2,\mathbb{R})$. From the four-dimensional point
of view, it is generated by the Lorentz boosts or rotations
$\zeta_{ab}= i\left( x_a\partial_b - x_b
  \partial_a\right)$ with $x_a=\eta_{ab}x^b$. We list here explicitly the
six generators, as well as the group action they correspond to:
\begin{subequations}
  \label{eq:Killing-SL2R}
  \begin{align}
    L_1 &= \frac{1}{2}\left(\zeta_{32} - \zeta_{01}\right), & g &\to
    \mathrm{e}^{-\frac{\lambda}{2}\sigma^1}g,
    \label{L1} \\
    L_2 &= \frac{1}{2}\left(-\zeta_{31}-\zeta_{02} \right), &  g &\to
    \mathrm{e}^{-\frac{\lambda}{2}\sigma^3}g,
    \label{L2} \\
    L_3 &= \frac{1}{2}\left(\zeta_{03} - \zeta_{12}\right), & g&\to
    \mathrm{e}^{\imath\frac{\lambda}{2}\sigma^2}g,
    \label{L3} \\
    R_1 &= \frac{1}{2}\left( \zeta_{01} + \zeta_{32}\right), & g&\to
    g\mathrm{e}^{\frac{\lambda}{2}\sigma^1},
    \label{R1} \\
    R_2 &= \frac{1}{2}\left(\zeta_{31} - \zeta_{02}\right), &  g&\to
    g\mathrm{e}^{-\frac{\lambda}{2}\sigma^3},
    \label{R2} \\
    R_3 &= \frac{1}{2}\left(\zeta_{03} + \zeta_{12}\right), & g&\to
    g\mathrm{e}^{\imath\frac{\lambda}{2}\sigma^2}.
    \label{R3}
  \end{align}
\end{subequations}

Both sets satisfy the algebra (\ref{comm}). The norms of the Killing vectors
are the following:
\begin{equation}
  \norm{\imath L_1}^2 = \norm{ \imath R_1}^2 = \norm{\imath L_2}^2 =
  \norm{\imath R_2}^2 =- \norm{ \imath L_3}^2=-\norm{\imath R_3}^2 = \frac{L^2}{4}.
\end{equation}
Moreover $L_i \cdot L_j = 0$ for $i\neq j$ and similarly for the
right set. Left vectors are not orthogonal to right ones.

The isometries of the $SL(2,\mathbb{R})$ group manifold turn into symmetries
of the $SL(2,\mathbb{R})_k$ \textsc{wzw} model, where they are realized in
terms of conserved currents\footnote{When writing actions a choice of gauge
  for the \textsc{ns} potential is implicitly made, which breaks part of the
  symmetry: boundary terms appear in the transformations.  These must be
  properly taken into account in order to reach the conserved currents.
  Although the expressions for the latter are not unique, they can be put in
  an improved-Noether form, in which they have only holomorphic (for
  $L_i$'s) or anti-holomorphic (for $R_j$'s) components.}:
\begin{subequations}
  \label{eq:currents-SL2R}
\begin{align}
  J^1 \left( z \right) \pm J^3 \left( z \right)&= -k \tr \left(
  \left(\sigma^1 \mp \imath \sigma^2\right)
  \partial g
    g^{-1}\right), &
  J^2 \left( z \right) &= - k \tr \left( \sigma^3 \partial g
    g^{-1}\right),
  \\
  \bar J^1 \left( \bar z \right) \pm \bar J^3 \left( \bar z \right)&=
  k \tr \left( \left(\sigma^1 \pm \imath \sigma^2\right) g^{-1}
    \bar \partial g \right), &
  \bar J^2 \left( \bar z \right) &= - k \tr \left( \sigma^3
    g^{-1}\bar \partial g \right).
\end{align}
\end{subequations}

At the quantum level, these currents, when properly normalized,
satisfy the following affine $SL(2,\mathbb{R})_k$
\textsc{opa}\footnote{In some conventions the level is $x=-k$. This
  allows to unify commutation relations for the affine
  $SL(2,\mathbb{R})_x$ and $SU(2)_x$ algebras. Unitarity demands
  $x<-2$ for the former and $0<x$ with integer $x$ for the latter.}:
\begin{subequations}
  \label{LOPA}
  \begin{align}
    J^3(z) J^3(0) & \sim - \frac{k}{2z^2},  \\
    J^3(z) J^{\pm}(0)& \sim \pm \frac{J^{\pm}}{z},   \\
    J^+(z) J^-(0) & \sim \frac{2J^3}{z}-\frac{k}{z^2},
  \end{align}
\end{subequations}
and similarly for the right movers. The central charge of the enveloping
Virasoro algebra is $c= 3+ 6/(k-2)$.

We will introduce three different coordinate systems where the
structure of $\mathrm{AdS}_3 $ as a Hopf fibration is more transparent.
They are explicitly described in the following.
\begin{itemize}
\item The $\left( \rho, t, \phi \right)$ coordinate system used to describe
  the magnetic deformation is defined as follows:
  \newcommand{\CR}[0]{\cosh \frac{\rho}{2}}
  \newcommand{\SR}[0]{\sinh \frac{\rho}{2}}
  \newcommand{\CPHI}[0]{\cosh \frac{\phi}{2}}
  \newcommand{\SPHI}[0]{\sinh \frac{\phi}{2}}
  \newcommand{\CT}[0]{\cos \frac{t}{2}}
  \newcommand{\ST}[0]{\sin \frac{t}{2}}
  \begin{equation}
    \begin{cases}
      \frac{x_0}{L} &= \CR \CPHI \CT - \SR \SPHI \ST \\
      \frac{x_1}{L} &= -\SR \SPHI \CT - \CR \SPHI \ST \\
      \frac{x_2}{L} &= -\CR \SPHI \CT + \SR \CPHI \ST \\
      \frac{x_3}{L} &= -\SR \SPHI \CT - \CR \CPHI \ST .
    \end{cases}
  \end{equation}
  The metric (\ref{eq:ads-metrix}) reads:
   \begin{equation}
     \label{eq:ads-rhotphi-metric}
     \di s^2 = \frac{L^2}{4} \left( \di \rho^2 + \di \phi^2 - \di t^2 -
       2 \sinh \rho \di t \di \phi\right)
   \end{equation}
   and the corresponding volume form is:
   \begin{equation}
     \label{eq:ads-rhotphi-vf}
     \omega_{[3]} = \frac{L^3}{8}\cosh \rho .
     \di \rho \land \di \phi  \land \di t
  \end{equation}
  Killing vectors and currents are given in Tab.
  \ref{tab:currents-timelike}. It is worth to remark that this coordinate
  system is such that the $t$-coordinate lines coincide with the integral
  curves of the Killing vector $\imath L_3$, whereas the $\phi$-lines are
  the curves of $\imath R_2$.
\item The $\left( r, x, \tau \right)$ coordinate system used to
  describe the electric deformation is defined as follows:
  \renewcommand{\CR}[0]{\cosh \frac{r}{2}} \renewcommand{\SR}[0]{\sinh
    \frac{r}{2}} \newcommand{\CX}[0]{\cosh \frac{x}{2}}
  \newcommand{\SX}[0]{\sinh \frac{x}{2}} \renewcommand{\CT}[0]{\cos
    \frac{\tau}{2}} \renewcommand{\ST}[0]{\sin \frac{\tau}{2}}
  \begin{equation}
    \label{eq:ads-rxt-coo}
    \begin{cases}
      \frac{x_0}{L} &= \CR \CX \CT + \SR \SX \ST \\
      \frac{x_1}{L} &= - \SR \CX \CT + \CR \SX \ST \\
      \frac{x_2}{L} &= - \CR \SX \CT - \SR \CX \ST \\
      \frac{x_3}{L} &= \SR \SX \CT - \CR \CX \ST .
    \end{cases}
  \end{equation}
  For $\set{r,x,\tau} \in \mathbb{R}^3$, this patch covers exactly once the
  whole $\mathrm{AdS}_3$, and is regular
  everywhere~\cite{Coussaert:1994tu}.  The metric is then given by
  \begin{equation}
     \label{eq:ads-rxt-met}
    \di s^2 = \frac{L^2}{4} \left( \di r^2 + \di x^2 - \di \tau^2 +
      2 \sinh r \di x \di \tau \right)
  \end{equation}
  and correspondingly the volume form is
  \begin{equation}
    \label{eq:ads-rxt-vf}
    \omega_{[3]} = \frac{L^3}{8} \cosh r \di r \land \di x \land \di \tau .
  \end{equation}
  Killing vectors and currents are given in Tab.
  \ref{tab:currents-spacelike}. In this case the $x$-coordinate lines
  coincide with the integral curves of the Killing vector $\imath R_2$,
  whereas the $\tau$-lines are the curves of $\imath R_3$.
\item The Poincar\'e coordinate system used to obtain the
  electromagnetic-wave background is defined by
  \begin{equation}
    \begin{cases}
    x^0 + x^2 &=\frac{L}{u}\\
    x^0 - x^2 &= Lu + \frac{L x^+ x^-}{u}\\
    x^1 \pm x^3 &= \frac{L x^\pm}{u} .
    \end{cases}
  \end{equation}
  For $\set{u,x^+,x^-} \in \mathbb{R}^3$, the Poincar\'e
  coordinates cover once the $SL(2\mathbb{R})$ group manifold. Its
  universal covering, $\mathrm{AdS}_3$, requires an infinite
  number of such patches. Moreover, these coordinates exhibit
  a Rindler horizon at $\vert u \vert \to \infty$; the conformal
  boundary is at $\vert u \vert \to 0$.
  Now the metric reads:
  \begin{equation}
\label{eq:ads-poinc-met}
    \di s^2 = \frac{L^2}{u^2} \left( \di u^2 + \di x^+ \di x^- \right)
  \end{equation}
  and the volume form:
  \begin{equation}
    \label{eq:ads-poinc-volume}
    \omega_{[3]} = \frac{L^3}{2u^3} \di u \land \di x^- \land \di x^+ .
  \end{equation}
  In these coordinates it is simple to write certain a linear combination of
  the Killing vector so to obtain explicitly a light-like isometry
  generator. For this reason in Tab. \ref{tab:currents-poincare} we report
  the $\set{L_1+L_3, L_1-L_3, L_2, R_1+R_3, R_1-R_3, R_2 } $ isometry
  generators and the corresponding $\set{J_1+J_3, J_1-J_3, J_2, \bar J_1 +
    \bar J_3, \bar J_1 - \bar J_3, \bar J_2} $ currents.
\end{itemize}

Finally, another useful although not global, set of coordinates is
defined by
\begin{equation}
  g = {e}^{{\frac{\psi - \varphi}{2}} \sigma^3} {e}^{\imath t \sigma^1}
  {e}^{\frac{\psi + \varphi }{2} \sigma^3},\label{sphan}
\end{equation}
($\psi $ and $\varphi$ \emph{are not} compact coordinates).  The metric reads:
\begin{equation}
  \di s^2 = L^2\left[ \cos ^2 t  \di \psi^2 -\di t^2 + \sin^2 t\,
    \di \varphi^2\right], \label{dssphan}
\end{equation}
with volume form
\begin{equation}
  \label{vfsphan}
  \omega_{[3]} = \frac{L^3}{2}\sin 2t \di t \land \di \psi  \land \di \varphi.
\end{equation}
Now $L_2 = \frac{1}{2}\left( \partial_\psi - \partial_\varphi\right)$ and $R_2 =
\frac{1}{2}\left( \partial_\psi + \partial_\varphi \right)$.


\newcommand{\Lrhotphi}[0]{
  \begin{minipage}{.45\textwidth}
    \vspace{-.7em}
    \begin{small}
      \begin{gather*}
        \cos t \partial_\rho + \frac{\sin t}{\cosh \rho} \partial_\phi - \sin t \tanh
        \rho \partial_t\\
        -\sin t \partial_\rho + \frac{\cos t}{\cosh \rho} \partial_\phi - \cos t \tanh
        \rho \partial_t \\
        - \partial_t
      \end{gather*}
    \end{small}
    \vspace{-1.5em}
  \end{minipage}
}
\newcommand{\Rrhotphi}[0]{
  \begin{minipage}{.45\textwidth}
    \vspace{-.7em}
    \begin{small}
      \begin{gather*}
        \cosh \phi \partial_\rho - \sinh \phi \tanh \rho \partial_\phi - \frac{\sinh
          \phi }{\cosh \rho} \partial_t \\
        \partial_\phi \\
        \sinh \phi \partial_\rho - \cosh \phi \tanh \rho \partial_\phi - \frac{\cosh
          \phi }{\cosh \rho} \partial_t
      \end{gather*}
    \end{small}
    \vspace{-1.5em}
  \end{minipage}
}
\newcommand{\Jrhotphi}[0]{
  \begin{minipage}{.35\textwidth}
    \vspace{-.7em}
    \begin{small}
      \begin{gather*}
        k\left( \cos t \partial \rho + \cosh \rho \sin t \partial \phi
        \right)\\
        k \left( \cos t \cosh \rho \partial \phi - \sin t \partial \rho
        \right)\\
        k \left( \partial t + \sinh \rho \partial \phi\right)
      \end{gather*}
    \end{small}
    \vspace{-1.2em}
  \end{minipage}
}
\newcommand{\Jbrhotphi}[0]{
  \begin{minipage}{.37\textwidth}
    \vspace{-.7em}
    \begin{small}
      \begin{gather*}
        - k \left( \cosh \phi \bar \partial
          \rho + \cosh \rho \sinh \phi \bar \partial t \right)\\
        k \left( \bar \partial \phi - \sinh \rho \bar \partial
          t\right)\\
        k \left( \cosh \rho \cosh \phi \bar \partial t + \sinh \phi \bar \partial
          \rho \right)
      \end{gather*}
    \end{small}
    \vspace{-1.2em}
  \end{minipage}
}

\mathindent=0em

\begin{table}
  \centering
  \begin{tabular}{|c|c|c|}
    \hline sector & Killing vector& Current \\ \hline \hline
    \begin{rotate}{90}\hspace{-2em}left moving\end{rotate} & \Lrhotphi &
    \Jrhotphi\\ \hline
    \begin{rotate}{90}\hspace{-2.5em}right moving\end{rotate} & \Rrhotphi &
    \Jbrhotphi \\\hline
  \end{tabular}
  \caption{Killing
    vectors $\{ \imath L_1,\imath L_2,\imath L_3\} $ and $\{ \imath R_1,\imath R_2,\imath R_3\} $,
    and holomorphic and anti-holomorphic currents  for the $
    \left( \rho, t, \phi \right)$ coordinate system (elliptic base).}
  \label{tab:currents-timelike}
\end{table}
\mathindent=\oldindent


\newcommand{\Lrxtau}[0]{
  \begin{minipage}{.52\textwidth}
    \vspace{-.7em}
    \begin{small}
      \begin{gather*}
        \cosh x \partial_r - \sinh x \tanh r \partial_x + \frac{\sinh x}{\cosh r} \partial_\tau \\
        \partial_x \\
        - \sinh x \partial_r + \cosh x \tanh r \partial_x - \frac{\cosh x}{\cosh r}
        \partial_\tau
      \end{gather*}
    \end{small}
    \vspace{-1.3em}
  \end{minipage}
}
\newcommand{\Rrxtau}[0]{
  \begin{minipage}{.57\textwidth}
    \vspace{-.7em}
    \begin{small}
      \begin{gather*}
        - \cos \tau \partial_r + \frac{\sin \tau}{\cosh \tau } \partial_x - \sin \tau
        \tanh \tau \partial_\tau\\
        \frac{\left(\cos \tau + \sin \tau \tanh r\right) \partial_x +\left( \cos
            \tau \sinh r - \frac{\sin \tau}{\cosh r}\right) \partial_\tau}{\cosh
          r}\\
        - \partial_\tau
      \end{gather*}
    \end{small}
    \vspace{-1.3em}
  \end{minipage}
}
\newcommand{\Jrxtau}[0]{
  \begin{minipage}{.35\textwidth}
    \vspace{-.7em}
    \begin{small}
      \begin{gather*}
        k \left( \cosh x \partial r - \cosh r \sinh x \partial \tau \right)\\
        k \left( \partial x + \sinh r \partial \tau \right)  \\
        k \left( \cosh r \cosh x \partial \tau - \sinh x \partial r \right)
      \end{gather*}
    \end{small}
    \vspace{-1.3em}
  \end{minipage}
}
\newcommand{\Jbrxtau}[0]{
  \begin{minipage}{.35\textwidth}
    \vspace{-.7em}
    \begin{small}
      \begin{gather*}
        k \left( -\cos \tau \bar \partial r + \cosh r
          \sin \tau \bar \partial r\right)    \\
        k \left( \cos \tau \cosh r \bar \partial x + \sin \tau
          \bar \partial r\right)    \\
        k \left( \bar \partial \tau - \sinh r \bar \partial x\right)
      \end{gather*}
    \end{small}
    \vspace{-1.3em}
  \end{minipage}
}

\mathindent=0em

\begin{table}
  \centering
  \begin{tabular}{|c|c|c|}
    \hline sector & Killing vector& Current \\ \hline \hline
    \begin{rotate}{90}\hspace{-2em}left moving\end{rotate} & \Lrxtau &
    \Jrxtau\\ \hline
    \begin{rotate}{90}\hspace{-2.5em}right moving\end{rotate} & \Rrxtau &
    \Jbrxtau \\\hline
  \end{tabular}
  \caption{Killing
    vectors $\{ \imath L_1,\imath L_2,\imath L_3\} $ and $\{ \imath R_1,\imath R_2,\imath
    R_3\} $, and holomorphic and anti-holomorphic currents for
    the $\left( r, x, \tau \right)$ coordinate system (hyperbolic base).}
  \label{tab:currents-spacelike}
\end{table}
\mathindent=\oldindent


\newcommand{\LPoin}[0]{
  \begin{minipage}{.35\textwidth}
    \vspace{-.7em}
    \begin{small}
      \begin{gather*}
        - \partial_- \\[1em]
        u x^- \partial_u -  u^2 \partial_+ + \left( x^-\right)^2 \partial_-\\[1em]
        \frac{u}{2} \partial_u + x^- \partial_-
      \end{gather*}
    \end{small}
    \vspace{-1.3em}
  \end{minipage}
}
\newcommand{\RPoin}[0]{
  \begin{minipage}{.35\textwidth}
    \vspace{-.7em}
    \begin{small}
      \begin{gather*}
        \partial_+ \\[1em]
        - u x^+ \partial_u - \left( x^+\right)^2 \partial_+ + u^2 \partial_- \\[1em]
         \frac{u}{2} \partial_u +  x^+ \partial_+
      \end{gather*}
    \end{small}
    \vspace{-1.3em}
  \end{minipage}
}
\newcommand{\JPoin}[0]{
  \begin{minipage}{.4\textwidth}
    \vspace{-.7em}
    \begin{small}
      \begin{gather*}
        -2k\frac{\partial x^+}{u^2} \\
         2k\left( 2 x^- \frac{\partial u}{u}-\partial x^- +
          (x^-)^2\frac{\partial x^+}{u^2} \right)\\
         2k\left( \frac{\partial u}{u} + x^- \frac{\partial
            x^+}{u^2}\right)
      \end{gather*}
    \end{small}
    \vspace{-1em}
  \end{minipage}
}
\newcommand{\JbPoin}[0]{
  \begin{minipage}{.4\textwidth}
    \vspace{-.7em}
    \begin{small}
      \begin{gather*}
         2k\frac{\bar\partial x^-}{u^2}\\
         2k\left(- 2 x^+ \frac{\bar\partial u}{u}+ \bar\partial
          x^+ - (x^+)^2\frac{\bar\partial x^-}{u^2} \right)\\
         2k\left( \frac{\bar\partial u}{u} + x^+
          \frac{\bar\partial x^-}{u^2}\right)
      \end{gather*}
    \end{small}
    \vspace{-1em}
  \end{minipage}
}

\mathindent=0em

\begin{table}
  \centering
  \begin{tabular}{|c|c|c|}
    \hline sector & Killing vector& Current \\ \hline \hline
    \begin{rotate}{90}\hspace{-2em}left moving\end{rotate} & \LPoin &
    \JPoin\\ \hline
    \begin{rotate}{90}\hspace{-2.5em}right moving\end{rotate} & \RPoin &
    \JbPoin \\\hline
  \end{tabular}
  \caption{Killing vectors, and holomorphic and anti-holomorphic
    currents  in Poincar\'e coordinates (parabolic
    base). The $\{ \imath L_1+\imath L_3, \imath L_1-\imath L_3, \imath L_2, \imath R_1+\imath R_3, \imath
    R_1-\imath R_3, \imath R_2 \}  $ isometry generators and the corresponding
    $\{ J_1+J_3,  J_1-J_3, J_2, \bar J_1 + \bar J_3, \bar J_1 - 
    \bar J_3, \bar J_2\}  $
    currents are represented so to explicitly obtain light-like isometry
    generators.}
  \label{tab:currents-poincare}
\end{table}

\mathindent=\oldindent

\section{$SU\left( 3 \right)$ }
\label{sec:suleft-3right}

To obtain the the Cartan-Weyl basis $\set{H_a, E^{\alpha_j}}$ for the
$\mathfrak{su} \left( 3\right)$ algebra we need to choose the positive
roots as follows:
\begin{align}
  \alpha_1 = \comm{ \sqrt{2}, 0 } && \alpha_2 = \comm{ - \nicefrac{1}{\sqrt{2}},
    \sqrt{\nicefrac{3}{2}} } && \alpha_3 = \comm{ \nicefrac{1}{\sqrt{2}},
    \sqrt{\nicefrac{3}{2}}}
\end{align}
The usual choice for the defining representation is:
\begin{small}
  \begin{equation}
    \begin{array}{ccc}
      H_1 = 
      \frac{1}{\sqrt{2}}\begin{pmatrix}
        1 & 0 & 0 \\
        0 & -1 & 0 \\
        0 & 0 & 0
      \end{pmatrix}& H_2 = 
      \frac{1}{\sqrt{6}}\begin{pmatrix}
        1 & 0 & 0 \\
        0 & 1 & 0 \\
        0 & 0 & -2
      \end{pmatrix}&
      E_1^+ = 
      \begin{pmatrix}
        0 & 1 & 0 \\
        0 & 0 & 0 \\
        0 & 0 & 0
      \end{pmatrix}\\
      E_2^+ = 
      \begin{pmatrix}
        0 & 0 & 0 \\
        0 & 0 & 1 \\
        0 & 0 & 0
      \end{pmatrix} & E_3^+ = 
      \begin{pmatrix}
        0 & 0 & 1 \\
        0 & 0 & 0 \\
        0 & 0 & 0
      \end{pmatrix} 
    \end{array}        
  \end{equation}
\end{small}
and $E^-_j = \left( E_j^+ \right)^t$. 

A good parametrisation for the $SU \left( 3\right)$ group can be obtained
via the Gauss decomposition: every matrix $g \in SU \left( 3 \right)$ is
written as the product:
\begin{equation}
  g = b_- d b_+  
\end{equation}
where $b_-$ is a lower triangular matrix with unit diagonal elements, $b_+$
is a upper triangular matrix with unit diagonal elements and $d$ is a
diagonal matrix with unit determinant. The element $g$ is written as:
\begin{multline}
  g \left( z_1, z_2, z_3, \psi_1, \psi_2 \right) = \exp \left[ z_1 E_1^- +
    z_2 E_3^- + \left( z_3 - \frac{z_1 z_2}{2}\right)E_2^- \right] \times \\
  \times e^{ -F_1 H_1 - F_2 H_2 } \exp \left[ \bar w_1 E_1^+ +
   \bar w_2 E_3^+ + \left( \bar w_3 - \frac{ \bar w_1 \bar
        w_2}{2}\right) E_2^+ \right]  e^{ \imath \psi_1 H_1 + \imath \psi_2
    H_2}
\end{multline}
where $z_\mu $ are 3 complex parameters, $\psi_i$ are two real and $F_1 $ and
$F_2 $ are positive real functions of the $z_\mu$'s:
\begin{equation}
  \begin{cases}
  F_1 = \log f_1 = \log \left( 1 + \abs{z_1}^2 + \abs{z_3}^2 \right)\\
  F_2 = \log f_2 = \log \left( 1 + \abs{z_2}^2 + \abs{z_3 - z_1 z_2 }^2 \right)
  \end{cases}
\end{equation}
By imposing $g \left( z_\mu , \psi_a  \right)$ to be unitary we find that the $w_\mu
$'s are complex functions of the $z_\mu$'s:
\begin{equation}
  \begin{cases}
    w_1 = - \frac{z_1 + \bar z_2 z_3}{\sqrt{f2}} \\
    w_2 = \frac{\bar z_1 z_3 - z_2 \left( 1+ \abs{z_1}^2
      \right)}{\sqrt{f_1}}\\
    w_3 = - \left( z_3 - z_1 z_2 \right)\sqrt{\frac{f_1}{f_2}}
  \end{cases}
\end{equation}
and the defining element $g \left( z_\mu, \psi_a \right)$ can then be written
explicitly as:
\begin{multline}
  g \left( z_1, z_2, z_3, \psi_1, \psi_2 \right) = 
  \begin{pmatrix}
    1 & 0 & 0 \\
    z_1 & 1 & 0 \\
    z_3 & z_2 & 1
  \end{pmatrix}
  \begin{pmatrix}
    \frac{1}{\sqrt{f_1}} & 0 & 0 \\
    0 & \sqrt{\nicefrac{f_1}{f_2}} & 0 \\
    0 & 0 & \sqrt{f_2}
  \end{pmatrix}\times  \\
  \times \begin{pmatrix}
    1 & \bar w_1 & \bar w_3 \\
    0 & 1 & \bar w_2 \\
    0 & 0 & 1
  \end{pmatrix}
  \begin{pmatrix}
    e^{\imath \psi_1/2} & 0 & 0 \\
    0 & e^{-\imath \nicefrac{\left( \psi_1 - \psi_2 \right)}{2}}& 0\\
    0 & 0 & e^{\imath \nicefrac{\psi_2}{2}}
  \end{pmatrix}
\end{multline}
Now, to build a metric for the tangent space to $SU \left( 3 \right)$ we can
define the 1-form $\Omega\left( \mathbf{z}, \mathbf{\psi} \right) = g^{-1} \left(
  \mathbf{z }, \mathbf{\psi }\right) \di g \left( \mathbf{z }, \mathbf{\psi
  }\right)$ and write the Killing-Cartan metric tensor as $g_{\textsc{kc}} =
\tr \left( \Omega^\dag{} \Omega \right) = - \tr \left( \Omega \Omega \right)$ where we have used
explicitly the property of anti-Hermiticity of $\Omega $ (that lives in the
$\mathfrak{su} \left( 3 \right)$ algebra). The explicit calculation is lengthy
but straightforward.  The main advantage of this parametrization from our
point of view is that it allows for a ``natural'' embedding of the $SU
\left( 3 \right) / U\left( 1\right)^2 $ coset (see \emph{e.g.}
\cite{Gnutzmann:1998JP} or \cite{Kondo:1999tj}): in fact in these
coordinates the K\"ahler potential is
\begin{multline}
\label{eq:Kahler-SU3}
  K \left( z_\mu, \bar z_\mu  \right) = \log \left( f_1\left( z_\mu \right)
    f_2\left( z_\mu \right) \right) = \\ = \log \left[ \left( 1 +
      \abs{z_1}^2 + 
      \abs{z_3}^2 \right) \left( 1 + \abs{z_2}^2 + \abs{z_3 - z_1 z_2
      }^2 \right) \right]  
\end{multline}
and the coset K\"ahler metric is hence simply obtained as:
\begin{equation}
  g_{\alpha \bar \beta} \ \di z^\alpha \otimes \di \bar z^\beta  = \frac{\partial^2}{\partial z_\alpha \partial \bar z_\beta} K \left( z_\mu,
    \bar z_\mu  \right) \di z^\alpha \otimes \di \bar z^\beta   
\end{equation}

Another commonly used $\mathfrak{su} \left( 3 \right)$ basis is given by the
Gell-Mann matrices:
\begin{small}
  \begin{equation}
    \label{eq:Gell-Mann-matrices}
    \begin{array}{ccc}
      \gamma_1 = 
      \frac{1}{\sqrt{2}}\begin{pmatrix}
        0 & \imath & 0 \\
        \imath & 0 & 0 \\
        0 & 0 & 0
      \end{pmatrix} &  \gamma_2 = 
      \frac{1}{\sqrt{2}}\begin{pmatrix}
        0 & 1 & 0 \\
        -1 & 0 & 0 \\
        0 & 0 & 0
      \end{pmatrix} &  \gamma_3 = 
      \frac{1}{\sqrt{2}}\begin{pmatrix}
        \imath  & 0 & 0 \\
        0 & -\imath  & 0 \\
        0 & 0 & 0
      \end{pmatrix} \\
      \gamma_4 = 
      \frac{1}{\sqrt{2}}\begin{pmatrix}
        0 & 0 & \imath  \\
        0 & 0 & 0 \\
        \imath  & 0 & 0
      \end{pmatrix} & 
      \gamma_5 = 
      \frac{1}{\sqrt{2}}\begin{pmatrix}
        0 & 0 & 1 \\
        0 & 0 & 0 \\
        -1 & 0 & 0
      \end{pmatrix} &  \gamma_6 = 
      \frac{1}{\sqrt{2}}\begin{pmatrix}
        0 & 0 & 0 \\
        0 & 0 & \imath  \\
        0 & \imath  & 0
      \end{pmatrix} \\
      \gamma_7 = 
      \frac{1}{\sqrt{2}}\begin{pmatrix}
        0 & 0 & 0 \\
        0 & 0 & 1 \\
        0 & -1 & 0
      \end{pmatrix} & \gamma_8 = 
      \frac{1}{\sqrt{6}}\begin{pmatrix}
        \imath  & 0 & 0 \\
        0 & \imath  & 0 \\
        0 & 0 & -2 \imath 
      \end{pmatrix}
    \end{array}
  \end{equation}
\end{small}
which presents the advantage of being orthonormal $\kappa \left( \lambda_i, \lambda_j
\right) = \delta_{ij}$. In this case the Cartan subalgebra is generated by
$\mathfrak{k} = \braket{\lambda_3, \lambda_8}$.

\section{$USp \left( 4 \right)$}
\label{sec:usp-left-4}

The symplectic group $Sp \left( 4, \setC  \right)$ is the set of $4 \times 4 $ complex
matrices that preserve the symplectic form $J$:
\begin{equation}
  J = \left( \begin{tabular}{c|c}
      $0$& $\mathbb{I}_{2\times2} $\\ \hline
      $-\mathbb{I}_{2\times2} $& $0$
    \end{tabular}\right)  
\end{equation}
that is
\begin{equation}
  Sp\left( 4, \setC \right) = \set{g | g^t J g = J}  
\end{equation}
The unitary symplectic group $USp \left( 4, \setC  \right)$ is the compact
group obtained as the intersection of $Sp \left( 4, \setC  \right)$ with $U
\left( 4 \right)$:
\begin{equation}
  USp\left( 4, \setC \right) = Sp\left( 4, \setC \right) \cap U \left( 4\right)
\end{equation}
It follows easily that the Lie algebra $\mathfrak{usp}\left( 4\right)$ is the
set of complex matrices $X$ such that:
\begin{equation}
  \mathfrak{usp}\left( 4\right) = \set{X | X^t J + J X^t = 0}
\end{equation}

To obtain the the Cartan-Weyl basis $\set{H_a, E^{\alpha_j}}$ we need
to choose the positive roots
\begin{align}
  \alpha_1 = \comm{\nicefrac{\sqrt{2}}{2},-\nicefrac{\sqrt{2}}{2}} && \alpha_2 =
  \comm{0,\sqrt{2}} && \alpha_3 = \comm{\nicefrac{\sqrt{2}}{2},\nicefrac{\sqrt{2}}{2}}
  && \alpha_4 = \comm{\sqrt{2},0}
\end{align}

\begin{figure}[htbp]
  \begin{center}
    \subfigure[$SU(2)$]{
      \includegraphics[width=.4\linewidth]{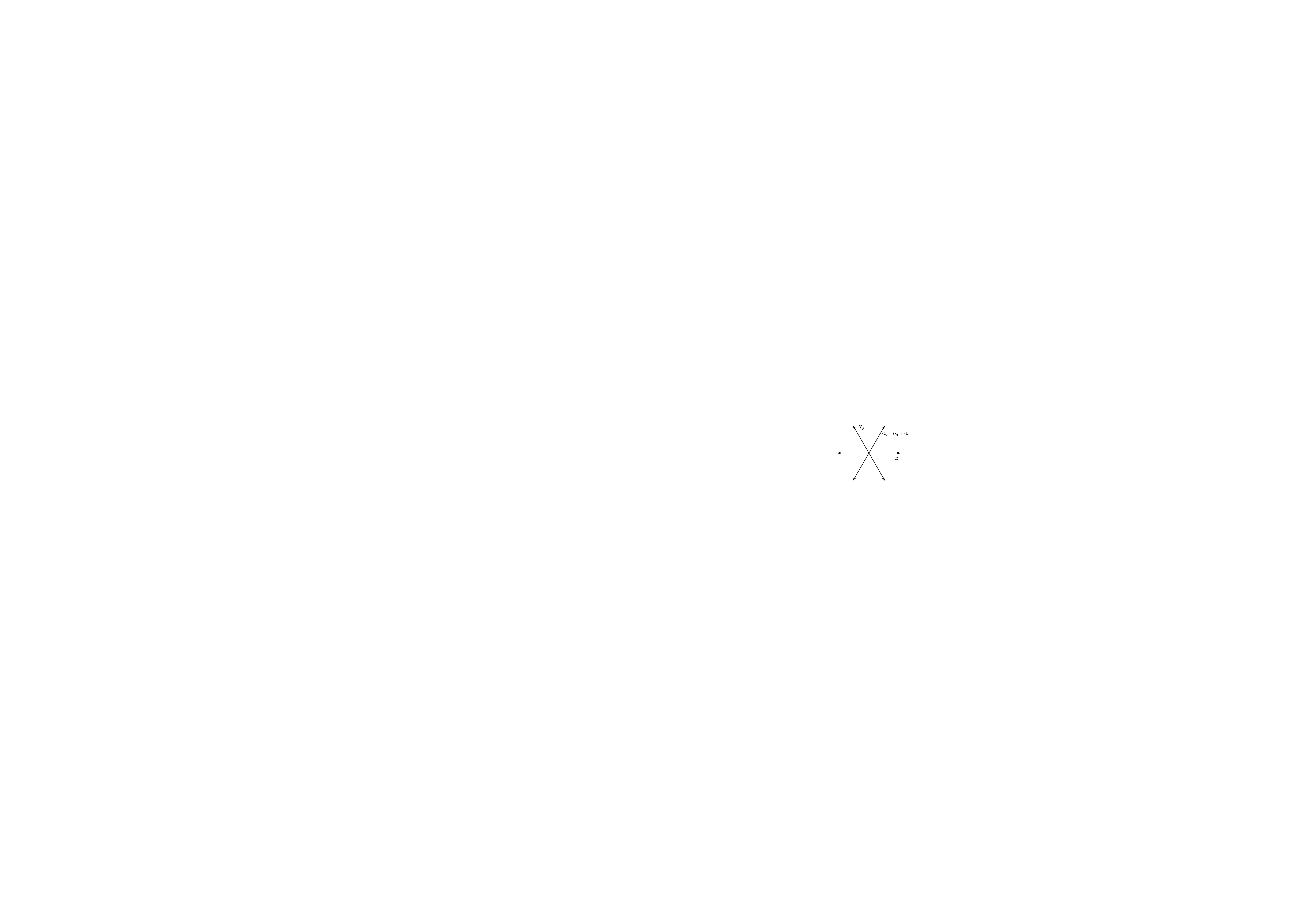}}
    \subfigure[$USp(4)$]{
      \includegraphics[width=.4\linewidth]{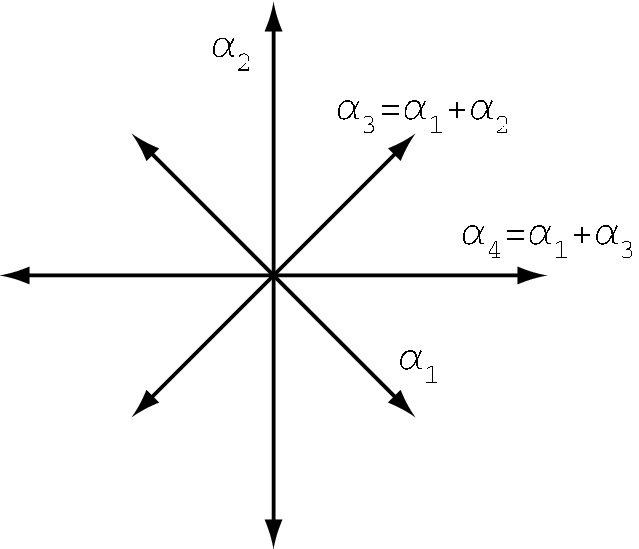}}
  \end{center}
  \caption{Root system for $\mathfrak{su}\left( 3\right)$ and
    $\mathfrak{sp}(4)$.}
  \label{fig:Root-sp4}
\end{figure}

and the $N_{\mu, \nu }$ coefficients:
\begin{align}
  N_{1,2} = 1 && N_{1,3} = 2  
\end{align}
The defining realization is given by the following choice:
\mathindent=0em
\begin{small}
  \begin{equation}
    \begin{array}{ccc}
      H_1  = \frac{1}{\sqrt{2}} 
      \begin{pmatrix}
        1 & 0 & 0 & 0 \\
        0 & 0 & 0& 0 \\
        0 & 0 & -1 & 0 \\
        0 & 0 & 0 & 0
      \end{pmatrix} & H_2 = \frac{1}{\sqrt{2}}
      \begin{pmatrix}
        0 & 0 & 0 & 0 \\
        0 & 1  & 0& 0 \\
        0 & 0 & 0 & 0 \\
        0 & 0 & 0 & -1 
      \end{pmatrix} & E_1^+ =
      \begin{pmatrix}
        0 & 1 & 0 & 0 \\
        0  & 0 & 0& 0 \\
        0 & 0 & 0 & 0  \\
        0 & 0 & -1  & 0
      \end{pmatrix} \\
      E_2^+  =
      \begin{pmatrix}
        0 & 0 & 0 & 0 \\
        0 & 0 & 0 & 1 \\
        0 & 0 & 0 & 0 \\
        0 & 0 & 0 & 0
      \end{pmatrix} & E_3^+  =
      \begin{pmatrix}
        0 & 0 & 0 & 1 \\
        0 & 0 & 1 & 0  \\
        0 & 0 & 0 & 0 \\
        0 & 0 & 0 & 0
      \end{pmatrix} & E_4^+  =
      \begin{pmatrix}
        0 & 0 & 1 & 0 \\
        0 & 0 & 0 & 0  \\
        0 & 0 & 0 & 0 \\
        0 & 0  & 0 & 0
      \end{pmatrix} 
    \end{array}
  \end{equation}
\end{small}
and $E^{-\alpha_\mu } = \left( E^{\alpha_\mu } \right)^t$. 

Just like in the case of $SU \left( 3 \right)$, the general element in
$USp\left( 4\right)$ is written as:
\begin{small}
  \begin{multline}
    g \left( \gamma_\mu, \psi_a \right) = \exp \left[ \gamma_1 E_1^- + \frac{
        \gamma_2}{\sqrt{2}} E_2^- + \frac{ 2\gamma_3 -\gamma_1 \gamma_2}{2} E_3^- + \frac{
        \gamma_1^2 \gamma_2 - \gamma_1 \gamma_3 + \gamma_4 }{\sqrt{2}} E_4^- \right] e^{-F_1
      H_1 -F_2 H_2} \\
    \exp \left[ \bar \beta_1 E_1^+ + \frac{\bar \beta_2}{\sqrt{2}} E_2^+ +
      \frac{2\bar \beta_3 -\bar \beta_1 \bar \beta_2 }{2} E_3^+ + \frac{\bar \beta_1^2
        \bar \beta_2 - \bar \beta_1 \bar \beta_3 + \bar \beta_4 }{\sqrt{2}} E_4^+
    \right] e^{\imath \psi_1 H_1 + \imath \psi_2 H_2}= \\ =
  \begin{pmatrix}
    1 & 0 & 0 & 0 \\
    \gamma_1 & 1 & 0 & 0 \\
    \gamma_4 & -\gamma_1 \gamma_2 + \gamma_3  & 1 & -\gamma_1 \\
     \gamma_3 & \gamma_2 & 0 & 1
  \end{pmatrix} 
  \begin{pmatrix}
    f_1 & 0 & 0 & 0\\
    0 & f_2 & 0 & 0\\
    0 & 0 & 1/f_1 & 0\\
    0 & 0 & 0 & 1/f_2
  \end{pmatrix} 
  \begin{pmatrix}
    1 & \bar \beta_1 &\bar  \beta_4 & \bar \beta_3 \\
    0 & 1 & - \bar \beta_1 \bar \beta_2 + \bar \beta_3 & \bar \beta_2 \\
    0 & 0 & 1 & 0 \\
    0 & 0 & - \bar \beta_1 & 1
  \end{pmatrix} \times \\ \times 
  \begin{pmatrix}
    e^{\imath \psi_1} & 0 & 0 & 0 \\
    0 & e^{\imath \psi_2} & 0 & 0 \\
    0 & 0 & e^{-\imath \psi_1} & 0 \\
    0 & 0 & 0 & e^{-\imath \psi_2} 
  \end{pmatrix}
\end{multline}
\end{small}

A orthonormal basis for the $\mathfrak{usp}\left( 4 \right)$, similar to the
Gell-Mann matrices system is given by the following set of matrices:

\begin{small}
  \begin{equation}
    \label{eq:sp4-ortho-gener}
    \begin{array}{ccc}
      T_1  = \frac{1}{\sqrt{2}} 
      \begin{pmatrix}
        \imath & 0 & 0 & 0 \\
        0 & 0 & 0& 0 \\
        0 & 0 & -\imath & 0 \\
        0 & 0 & 0 & 0
      \end{pmatrix} & T_2  = \frac{1}{\sqrt{2}} 
      \begin{pmatrix}
        0 & 0 & 0 & 0 \\
        0 & \imath  & 0& 0 \\
        0 & 0 & 0 & 0 \\
        0 & 0 & 0 & -\imath 
      \end{pmatrix} & T_3  = \frac{1}{2} 
      \begin{pmatrix}
        0 & \imath & 0 & 0 \\
        \imath  & 0 & 0& 0 \\
        0 & 0 & 0 & - \imath  \\
        0 & 0 & -\imath  & 0
      \end{pmatrix} \\
      T_4  = \frac{1}{2}
      \begin{pmatrix}
        0 & 1 & 0 & 0 \\
        -1 & 0 & 0 &0 \\
        0 & 0 & 0 & 1 \\
        0 & 0 & -1 & 0
      \end{pmatrix} & T_5 = \frac{1}{\sqrt{2}} 
      \begin{pmatrix}
        0 & 0 & 0 & 0 \\
        0 & 0 & 0 & \imath  \\
        0 & 0 & 0 & 0 \\
        0 & \imath & 0 & 0
      \end{pmatrix} & T_6 = \frac{1}{\sqrt{2}} 
      \begin{pmatrix}
        0 & 0 & 0 & 0 \\
        0 & 0 & 0 & 1  \\
        0 & 0 & 0 & 0 \\
        0 & -1  & 0 & 0
      \end{pmatrix} \\
      T_7 = \frac{1}{2}
      \begin{pmatrix}
        0 & 0 & 0 & \imath  \\
        0 & 0 & \imath & 0 \\
        0 & \imath  & 0 & 0 \\
        \imath  & 0 & 0 & 0
      \end{pmatrix} & T_8  = \frac{1}{2} 
      \begin{pmatrix}
        0 & 0 & 0 & 1 \\
        0 & 0 & 1 & 0 \\
        0 & -1 & 0 & 0 \\
        -1 & 0 & 0 & 0
      \end{pmatrix} & T_9  = \frac{1}{\sqrt{2}} 
      \begin{pmatrix}
        0 & 0 & \imath  & 0 \\
        0 & 0 & 0 &0 \\
        \imath  & 0 & 0 & 0 \\
        0 & 0 & 0 & 0
      \end{pmatrix} \\  T_{10}  = \frac{1}{\sqrt{2}} 
      \begin{pmatrix}
        0 & 0 & 1 & 0 \\
        0 & 0 & 0 &0 \\
        -1 & 0 & 0 & 0 \\
        0 & 0 & 0 & 0
      \end{pmatrix}
    \end{array}
  \end{equation}
\end{small}

\mathindent=\oldindent


\chapter{Symmetric deformations of $SL(2, \setR)$}
\label{cha:symm-deform-su2}

The group manifold of $SL(2,\mathbb{R})$ is anti de Sitter in three
dimensions. Metric and antisymmetric tensor read (in Euler
coordinates, see App.~\ref{cha:geometry-arond-group}):
\begin{subequations}
  \begin{align}
    \di s^2&= L^2\left[\di \rho^2 +
    \sinh^2 \rho \, \di \phi^2 - \cosh ^2 \rho  \, \di \tau ^2 \right],     \\
    H_{[3]} &= L^2\sinh 2\rho
      \di \rho \land \di \phi  \land \di \tau,
  \end{align}
\end{subequations}
with $L$ related to the level of $SL(2,\mathbb{R})_k$ as usual:
$L=\sqrt{k+2}$. In the case at hand, three different
lines of symmetric deformations arise due to the presence of
time-like ($J^3$, $\bar J^3$), space-like ($J^1$, $\bar J^1$,
$J^2$, $\bar J^2$), or null generators
\cite{Forste:2003km,Forste:1994wp,Israel:2003ry}. The residual
isometry is $U(1) \times U(1)$ that can be time-like $(L_3, R_3 )$,
space-like $(L_2, R_2 )$ or null $(L_1 + L_3, R_1 + R_3 )$
depending on the deformation under consideration.


The \emph{elliptic deformation} is driven by the $J^3\bar J^3$
bilinear. At first order in $\alpha^{\prime}$ the background
fields are given by\footnote{The extra index ``3'' in the
  deformation parameter $\kappa$ reminds that the deformation refers here to $J^3
  \bar J^3$.}:
\begin{subequations}
  \begin{align}
    \di s^2&= k \left[\di \rho^2 + \frac{\sinh^2 \rho \, \di \phi^2 -\kappa_3^2 \cosh
        ^2 \rho \, \di \tau ^2
      }{\Theta_{\kappa_3} (\rho )} \right],\label{eq:J3J3-metric}\\
    H_{[3]} & = k \frac{\kappa_3^2\sinh 2\rho}{\Theta_{\kappa_3} (\rho )^2 }
    \di \rho \land \di \phi  \land \di \tau,\\
    { \mathrm e}^{\Phi} &= \frac{\Theta_{\kappa_3} (\rho )}{\kappa_3}.
  \end{align}
\end{subequations}
where $\Theta_{\kappa_3} (\rho ) = \cosh ^2 \rho -\kappa_3 \sinh^2 \rho$.  At extreme
deformation ($\kappa_3^2 \to 0$), a time-like direction decouples and we are left
with the axial\footnote{The deformation parameter has two T-dual branches.
  The extreme values of deformation correspond to the axial or vector
  gaugings. The vector gauging leads to the \emph{trumpet}.  For the
  $SU(2)_k / U(1)$, both gaugings correspond to the \emph{bell}.}
${SL(2,\mathbb{R})_k / U(1)_{\text{time}}}$. The target space of the latter
is the \emph{cigar} geometry (also called Euclidean two-dimensional black
hole):
\begin{eqnarray}
{\mathrm e}^{\Phi}&\sim & \cosh^2 \rho,\\
\di s^2&=&k \left[ \di \rho^2+\tanh^2\rho \, \di \phi ^2 \right],
\end{eqnarray}
($0\leq \rho < \infty$ and $0\leq \phi \leq 2\pi$).


Similarly, with $J^2\bar J^2$ one generates the \emph{hyperbolic
deformation}. This allows to reach the Lorentzian two-dimensional
black hole times a free space-like line. Using the coordinates
defined in Eq.~(\ref{sphan}), we find:
\begin{subequations}
\label{eq:J2J2-deform}
  \begin{align}
    \di s^2&= k \left[- \di t^2 + \frac{\sin^2 t \, \di \varphi^2
    + \kappa_2^2 \cos ^2 t  \, \di \psi^2}{\Delta_{\kappa_2} (t)}\right], \label{eq:J2J2-metric}\\
    H_{[3]} & = k \frac{\kappa_2^2\sin 2t}{\Delta_{\kappa_2} (t)^2 }
      \di t \land \di \psi  \land \di \phi,\\
  { \mathrm e}^{\Phi} &= \frac{\Delta_{\kappa_2} (t)}{\kappa_2},
  \end{align}
\end{subequations}
where $\Delta_{\kappa_2} (t) = \cos ^2 t + \kappa_2^2 \sin^2 t$.  This coordinate patch
does not cover the full $\mathrm{AdS}_3$. We will expand on this line in
Sec.~\ref{sec:blackstring}.


Finally, the bilinear $\left(J^1 + J^3\right)\left(\bar J^1 + \bar
  J^3\right)$ generates the \emph{parabolic deformation}. Using
Poincar\'e coordinates in Eq.~\eqref{eq:ads-poinc-met}\footnote{Note
  that $x^\pm = X \pm T$.} we obtain:
\begin{subequations}
\label{eq:null-deform}
  \begin{align}
    \di s^2&= k \left[\frac{\di u^2}{u^2}+
    \frac{\di X ^2- \di T^2}{u^2 + 1/\nu}
    \right], \label{eq:null-metric}\\
    H_{[3]} & = k \frac{2u}{\left(u^2 + 1/ \nu \right)^2}
      \di u \land \di T  \land \di X,\\
  { \mathrm e}^{\Phi} &= \frac{u^2 + 1/\nu }{u^2}.
  \end{align}
\end{subequations}
The deformation parameter is $1/\nu$. At infinite value of the
parameter $\nu$, we recover pure AdS$_3$; for $\nu \to 0$, a whole
light-cone decouples and we are left with a single direction and a
dilaton field, linear in this direction.

The physical interpretation of the parabolic deformation is far reaching,
when AdS$_3$ is considered in the framework of the \textsc{ns5/f1}
near-horizon background, AdS$_3\times S^3 \times T^4$. In this physical
set-up, the parameter $\nu$ is the density of \textsc{f1}'s (number of
fundamental strings over the volume of the four-torus
$T^4$)~\cite{Israel:2003ry,Kiritsis:2003cx}\footnote{Our present convention
  for the normalization of the dilaton results from
  Eq.~(\ref{eq:OddDilaton}). It differs by a factor $-2$ with respect to the
  one used in those papers.}. At infinite density, the background is indeed
AdS$_3\times S^3 \times T^4$. At null density, the geometry becomes
$\mathbb{R}^{1,2}\times S^3 \times T^4$ plus a linear dilaton and a
three-form on the $S^3$.


\chapter{Spectrum of the $ SL(2,\setR)$ super-\textsc{wzw} model}
\label{sec:initial-spectrum}

In this appendix we give a reminder of the superconformal \textsc{wzw} model
on $SL \left( 2, \setR \right)_k$ (for a recent discussion see
\cite{Giveon:2003wn}). The affine extension of the $\mathfrak{sl} \left( 2,
  \setR \right)$ algebra at level $k$ is obtained by considering two sets of
holomorphic and anti-holomorphic currents of dimension one, defined as
\begin{align}
  J^{\textsc{m}} \left( z \right) = k \braket{T^{\textsc{m}}, \mathrm{Ad}_g
    g^{-1} \partial g}, && \bar J^{\textsc{m}} \left( \bar z \right) = k
  \braket{T^{\textsc{m}}, g^{-1} \bar \partial g},
\end{align}
where $\braket{\cdot, \cdot } $ is the scalar product (Killing
form) in $\mathfrak{sl} \left( 2, \setR \right)$,
$\set{T^{\textsc{m}}} $ is a set of generators of the algebra that
for concreteness we can choose as follows:
\begin{align}
  T^1 = \sigma^1, && T^2 = \sigma^3, && T^3 = \sigma^2.
\end{align}
Each set satisfies the \textsc{ope}
\begin{equation}
  J^{\textsc{m}} \left( z \right) J^{\textsc{n}} \left( w \right) \sim \frac{k
    \delta^{\textsc{mn}}}{2 \left( z - w\right)^2} +
  \frac{f^{\textsc{mn}}_{\phantom{\textsc{mn}}\textsc{p}} J^{\textsc{p}}
    \left( w \right)}{z-w},
\end{equation}
where $f^{\textsc{mn}}_{\phantom{\textsc{mn}}\textsc{p}}$ are the
structure constants of the $\mathfrak{sl} \left( 2, \setR \right)$
algebra.  The chiral algebra contains the Virasoro operator
(stress tensor) obtained by the usual Sugawara construction:
\begin{equation}
  T \left( z \right) = \sum_{\textsc{m}} \frac{: J^{\textsc{m}} J^{\textsc{m}}
    :  }{k-2}.
\end{equation}

A heterotic model is built if we consider a left-moving
$\mathcal{N} = 1 $ extension, obtained by adding 3
free fermions which transform in the adjoint representation. More
explicitly:
\begin{align}
  T \left( z \right) &= \sum_{\textsc{m}} \frac{: J^{\textsc{m}} J^{\textsc{m}}
    :  }{k-2} + : \psi_{\textsc{m}} \partial \psi_{\textsc{m}}:, \\
  G \left( z \right) &= \frac{2}{k} \left( \sum_{\textsc{m}} J^{\textsc{m}}
    \psi_{\textsc{m}} - \frac{\imath }{3k } \sum_{\textsc{mnp}} f^{\textsc{mnp}}:
    \psi_{\textsc{m}} \psi_{\textsc{n}} \psi_{\textsc{p}} :
    \right).
\end{align}
On the right side, instead of superpartners, we add a right-moving current
with total central charge $c=16$.

Let us focus on the left-moving part. The supercurrents are given by
$\psi_{\textsc{m}} + \theta \sqrt{2/k} \mathcal{J}_{\textsc{m}}$ where:
\begin{equation}
  \mathcal{J}_{\textsc{m}} = J^{\textsc{m}} - \frac{\imath }{2}
\sum_{\textsc{np}}\epsilon^{\textsc{mnp}}
  \psi_{\textsc{n}} \psi_{\textsc{p}};
\end{equation}
it should be noted that the bosonic $J^{\textsc{m}} $ currents generate an
affine $\mathfrak{sl}\left( 2, \setR \right)$ algebra at level $k+2$, while
the level for the total $\mathcal{J}_{\textsc{m}}$ currents is $k$.

Let us now single out the operator that we used for both the deformation
(Eqs.~\eqref{tmetdef}) and the identifications
(Sec.~\ref{sec:btz-ident-comb}):
\begin{equation}
  \mathcal{J}_2 = J^2 + \imath \psi_1 \psi_3.
\end{equation}
Let us now bosonize these currents as follows:
\begin{align}
  \mathcal{J}_2 &= - \sqrt{ \frac{k}{2}} \partial \vartheta_2,\\
  J^2 &= - \sqrt{\frac{k+2}{2}} \partial \theta_2, \\
   \psi_1 \psi_3 &= \partial H,
\end{align}
and introduce a fourth free boson $X$ so to separate the $\vartheta_2 $
components
both in $\theta_2 $ and $H$:
\begin{align}
  \imath H &= \sqrt{ \frac{2}{k}} \vartheta_2 +  \imath \sqrt{\frac{k+2}{k}}
  X,
\\
  \theta_2 &= \sqrt{\frac{2}{k}} \left( \sqrt{\frac{k+2}{2}} \vartheta_2 +
\imath X\right).
\end{align}

A primary field $\Phi_{j \mu \tilde \mu }$ of the bosonic $SL
\left( 2, \setR \right)_{k+2}$ with eigenvalue $\mu $ with respect
to $J^2 $ and $\bar \mu $ with respect to $\bar J^2$ obeys by
definition
\begin{subequations}
  \begin{align}
    J^2 \left( z \right) \Phi_{j \mu \bar \mu } \left( w, \bar w  \right) &\sim
\frac{\mu
      \Phi_{j \mu \bar \mu } \left( w, \bar w  \right)}{z - w}, \\
    \bar J^2 \left( \bar z \right) \Phi_{j \mu \bar \mu } \left(  w, \bar w
\right)
    &\sim \frac{ \bar \mu
      \Phi_{j \mu \bar \mu } \left(  w, \bar w  \right)}{\bar z - \bar
      w}.
  \end{align}
\end{subequations}
Since $\Phi_{j \mu \bar \mu }$ is purely bosonic, the same
relation holds for the supercurrent:
\begin{equation}
  \mathcal{J}_2 \left( z \right) \Phi_{j \mu \bar \mu } \left( w , \bar w
\right) \sim \frac{\mu
    \Phi_{j \mu \bar \mu } \left( w, \bar w  \right)}{z - w}.
\end{equation}
Consider now the holomorphic part of $\Phi_{j \mu \bar \mu }
\left( z, \bar z \right)$. If $\Phi_{j \mu } $ is viewed as a
primary in the \textsc{swzw} model, we can use the parafermion
decomposition as follows:
\begin{equation}
  \label{eq:superparafermion}
  \Phi_{j\mu } \left( z \right) = U_{j \mu } \left( z \right) {e}^{\imath \mu
\sqrt{2/k} \vartheta_2},
\end{equation}
where $U_{j \mu } \left( z \right)$ is a primary of the superconformal
$\nicefrac{SL \left( 2, \setR \right)_k}{U \left( 1 \right)}$. On the other
hand, we can just consider the bosonic \textsc{wzw} and write:
\begin{equation}
  \Phi_{j\mu } \left( z \right) = V_{j \mu } \left( z \right) {e}^{\imath \mu
\sqrt{2/
      \left( k+2\right)} \theta_2} = V_{j \mu} \left( z \right){e}^{\imath
\frac{2m}{k+2}
    \sqrt{\frac{k+2}{k}} X + \imath \mu \sqrt{2/k} \vartheta_2},
\end{equation}
where now $V_{j \mu } \left( z \right)$ is a primary of the bosonic
$\nicefrac{SL \left( 2, \setR\right)_{k+2}}{U \left( 1 \right)}$. The
scaling dimension for this latter operator (\emph{i.e.} its eigenvalue with
respect to $L_0$) is then given by:
\begin{equation}
\label{eq:boson-coset}
  \Delta \left( V_{j \mu} \right) = - \frac{j \left( j+1\right)}{k} -
\frac{\mu^2}{k+2}.
\end{equation}
An operator in the full supersymmetric $SL \left( 2, \setR \right)_k$ theory
is then obtained by adding the $\psi^1 \psi^3$ fermionic superpartner
contribution:
\begin{equation}
  \Phi_{j \mu \nu } \left( z \right) = \Phi_{j \mu } \left( z\right) {e}^{\imath
\nu H} =
  V_{j \mu } \left( z \right) {e}^{\imath \left( \frac{2 \mu }{k+2} + \nu \right)
    \sqrt{\frac{k+2}{k}} X } {e}^{\imath
    \sqrt{2/k} \left( \mu
      +  \nu \right) \vartheta_2}
\end{equation}
that is an eigenvector of $\mathcal{J}_2 $ with eigenvalue $\mu +\nu $ where
$\mu \in \setR$ and $\nu $ can be decomposed as $\nu = n + a/2$ with $n \in
\setN$ and $a \in \setZ_2 $ depending on whether we consider the \textsc{ns}
or \textsc{r} sector. The resulting spectrum can be read directly as:
\begin{multline}
\label{eq:left-weights}
  \Delta \left( \Phi_{j \mu n} \left( z \right)\right) = - \frac{j \left(
      j+1\right)}{k} - \frac{\mu^2}{k+2} - \frac{k+2}{2k} \left( \frac{2\mu
    }{k+2} + n+\frac{a}{2} \right)^2 + \frac{1}{k} \left( \mu + n
    + \frac{a}{2}\right)^2 = \\
 = - \frac{j \left( j+1\right)}{k} - \frac{1}{2} \left( n +
    \frac{a}{2}\right)^2.
\end{multline}
Of course the last expression was to be expected since it is the sum of the
$\mathfrak{sl} \left( 2, \setR \right)_{k+2}$ Casimir and the contribution
of a light-cone fermion. Nevertheless the preceding construcion is useful
since it allowed us to isolate the $\mathcal{J}_2 $ contribution to the
spectrum $\left( \mu + \nu \right)^2/k$.

The right-moving part of the spectrum is somewhat simpler since there are no
superpartners. This means that we can repeat our construction above and the
eigenvalue of the $\bar L_0 $ operator is simply obtained by adding to the
dimension in Eq.~\eqref{eq:boson-coset} the contribution of the $\bar J^2 $
operator and of some $U \left( 1 \right)$ coming from the gauge sector:
\begin{equation}
\label{eq:right-weights}
  \bar \Delta \left( \bar \Phi_{j \bar \mu \bar n } \left( \bar z\right) \right)
=
  - \frac{j \left(
      j+1\right)}{k} - \frac{\bar \mu^2}{ k + 2 } + \left\{\frac{\bar \mu^2}{ k
+ 2 }
  + \frac{1}{k_g} \left( \bar n + \frac{\bar a}{2}\right)^2
  \right\},
\end{equation}
where again $\bar n \in \setN$ and $\bar a \in \setZ_2$ depending on the
sector.


\end{small}

\begin{small}
  \bibliography{Biblia}
\end{small}

\end{document}